\documentclass[twoside]{article}
\usepackage[german,british]{babel}
\usepackage{amsmath,amssymb,bbold,feynmp,IEEEtrantools,bm,slashed,fancyhdr}
\usepackage[hidelinks,breaklinks=true]{hyperref}
\usepackage[dvips]{graphicx}

\usepackage[sorting=none,doi=false]{biblatex}
\renewbibmacro{in:}{,}

\usepackage{cleveref}

\title{Phenomenology of a left-right-symmetric model inspired by the trinification model}
\author{}
\date{}

\begin{document}
\newcommand{\tr}[1]{\mathrm{Tr}\left\{#1\right\}} 
\newcommand{\vev}[1]{\langle{#1}\rangle{}} 
\newcommand{\ord}[1]{\mathcal{O}\left(#1\right)} 
\newcommand{\lag}{\mathcal{L}} 
\newcommand{\reals}{\mathbb{R}} 
\newcommand{\unity}{\mathbb{1}} 
\renewcommand{\d}{\mathrm{d}} 
\newcommand{\D}{\mathcal{D}} 
\newcommand{\plushc}{+ \text{ h.c. }} 
\newcommand{\gamfu}[1]{\Gamma\left(#1\right)} 
\newcommand{\dint}[1]{\int\frac{\d^d#1}{(2\pi)^d}} 
\newcommand{\dd}[2]{\frac{\mathrm{d}#1}{\mathrm{d}#2}} 
\newcommand{\ddp}[2]{\frac{\partial{}#1}{\partial{}#2}} 
\newcommand{\rep}[1]{\mathbf{#1}} 
\newcommand{\brep}[1]{\mathbf{\overline{#1}}} 
\newcommand{\SM}{{}SU(3)_C\times SU(2)_L\times U(1)_Y{}} 
\newcommand{\LRgroup}{{}SU(3)_C\times SU(2)_L\times SU(2)_R\times U(1)_{B-L}{}} 
\newcommand{\trini}{{}SU(3)_C\times SU(3)_L\times SU(3)_R{}} 
\newcommand{\atCL}[1]{@#1\%\text{ C.L.}} 

\pagestyle{empty}
\begin{center}
Dissertation \\[2ex]
submitted to the \\[2ex]
Combined Faculties of the Natural Sciences and Mathematics \\[2ex]
of the Ruperto-Carola-University of Heidelberg. Germany \\[2ex]
for the degree of \\[2ex]
Doctor of Natural Sciences

\vskip10cm

Put forward by \\[2ex]
\emph{Jamil Hetzel \\[2ex]
born in: Rheden, the Netherlands \\[2ex]
Oral examination: February 4th, 2015}
\end{center}

\cleardoublepage
\maketitle
\thispagestyle{empty}

\vskip10cm

\begin{align*}
\text{Referees:}\hspace{2cm} & \text{Prof.~dr.~Tilman Plehn} \\
& \text{Prof.~dr.~Arthur Hebecker}
\end{align*}

\cleardoublepage

\begin{otherlanguage}{german}
\subsubsection*{\centering Ph\"anomenologie eines links-rechts-symmetrisches Modells basierend auf dem Trinifikationsmodell}
Das Trinifikationsmodell ist eine interessante Erweiterung des Standardmodells, die auf der Eichgruppe $\trini$ basiert.
Das Modell beschreibt die Parit\"atsverletzung durch die spontane Brechung der Eichsymmetrie, und die gemessenen Fermionenmassen und -mischungswinkel k\"onnen mit wenigen Parametern reproduziert werden.
Wir untersuchen die Ph\"anomenologie des Trinifikationsmodells bei niedrigen Energien, um seine Voraussagungen mit Experimenten vergleichen zu k\"onnen.
Zu diesem Zweck konstruieren wir eine effektive Feldtheorie die es erlaubt, mit einer geringeren Anzahl von Teilchen und freien Parametern auszukommen.
Die Modellparameter werden mittels den bereits vorliegenden Pr\"azisionsmessungen und experimentellen Grenzen eingeschr\"ankt.
Der Skalarsektor des Modells erm\"oglicht verschiedene ph\"anomenologische Szenarien, zum Beispiel ein leichtes fermiophobisches Skalarteilchen zus\"atzlich zu einem standardmodellartigen Higgs, oder die Existenz eines entarteten (Zwillings-)Higgsbosons bei 126 GeV.
Wir zeigen wie die Messung der Higgskopplungen es erlaubt, zwischen solchen Szenarien und dem Standardmodell zu unterscheiden.
Es stellt sich heraus, dass das Trinifikationsmodell mehrere neue Skalarteilchen mit Massen im $\ord{100\text{ GeV}}$-Bereich vorhersagt.
Au\ss{}erdem werden in gro\ss{}en Teilen des Parameterraums messbare Abweichungen der Higgskopplungskonstanten von den Standardmodellwerten erwartet.
Das Trinifikationsmodell erwartet daher in den n\"achsten Jahren entscheidende Tests am Large Hadron Collider.
\end{otherlanguage}

\subsubsection*{\centering Phenomenology of a left-right-symmetric model inspired by the trinification model}
The trinification model is an interesting extension of the Standard Model based on the gauge group $\trini$.
It naturally explains parity violation as a result of spontaneous symmetry breaking, and the observed fermion masses and mixings can be reproduced using only a few parameters.
We study the low-energy phenomenology of the trinification model in order to compare its predictions to experiment.
To this end, we construct a low-energy effective field theory, thereby reducing the number of particles and free parameters that need to be studied.
We constrain the model parameters using limits from new-particle searches as well as precision measurements.
The scalar sector of the model allows for various phenomenological scenarios, such as the presence of a light fermiophobic scalar in addition to a Standard-Model-like Higgs, or a degenerate (twin) Higgs state at 126 GeV.
We show how a measurement of the Higgs couplings can be used to distinguish such scenarios from the Standard Model.
We find that the trinification model predicts that several new scalar particles have masses in the $\ord{100\text{ GeV}}$ range.
Moreover, large regions of the parameter space lead to measurable deviations from Standard-Model predictions of the Higgs couplings.
Hence the trinification model awaits crucial tests at the Large Hadron Collider in the coming years.

\cleardoublepage
\par\vspace*{.35\textheight}{\centering \LARGE to Sabine\par}

\clearpage
\section*{\centering Acknowledgements}
I am indebted to my supervisor Tilman Plehn for providing me the opportunity to perform my doctoral study in Heidelberg.
This work would not have been possible without him providing a wonderful research group, financial support, help with both time management and physics, and guidance of the project towards its completion.

I am also grateful to Berthold Stech for a wonderful collaboration.
This thesis would not have existed without all his suggestions and insights that have steered our project to its completion, his endless patience while discussing physics, and his tireless proofreading during the final stage of this work.

Furthermore, I thank Torben Schell, Jamie Tattersall and Johann Brehmer for proofreading parts of this thesis and suggesting numerous improvements.
I also thank my second referee Arthur Hebecker as well as my examiners Ulrich Uwer and Bernd J\"ahne for reading my thesis and being part of my examination committee.
I thank everyone from the phenomenology group at the ITP for providing a great environment that has resulted in three fruitful years filled with plenty of cakes, nuts, and scenic walks on Philosophenweg.

Last but certainly not least, I thank Sabine Keller with all my heart for her unlimited personal support.
I would not have gotten this far without her neverending encouragement, love, and care.

\cleardoublepage
\pagestyle{fancy}
\pagenumbering{arabic}
\fancyhead[LE,RO]{\thepage}
\fancyhead[LO]{\rightmark}
\fancyhead[RE]{\leftmark}
\fancyfoot{}
\pagenumbering{Roman}
\tableofcontents

\cleardoublepage
\pagenumbering{arabic}
\begin{fmffile}{SU2xSU2xU1}

\section{Introduction}
The discovery of the Higgs boson \cite{Aad:2012tfa,Chatrchyan:2012ufa} marks the establishment of the Standard Model (SM) of particle physics as the model that correctly describes physics at energies available at particle colliders to date.
All particles of the SM have been found experimentally, and the experimental data gathered at particle colliders match the predictions of the SM to good precision \cite{Baak:2013ppa}.

Yet, the Standard Model is not regarded to be a complete theory of nature.
Firstly, it does not describe correctly all observations made outside particle colliders.
The Standard Model lacks a description of gravity: as a quantum field theory, it is incompatible with the theory of General Relativity.
Also, the SM does not include any particles that could be viable dark matter candidates, and it is incompatible with the observation of non-zero neutrino masses (for a review, see e.g.~\cite{Gonzalez-Garcia:2013jma}).
Secondly, the Standard Model is unsatisfactory from a theoretician's perspective: the fermion masses and mixings are free parameters that display hierarchical patterns, parity violation has to be introduced by hand, and the Higgs mass is much smaller than the Planck scale despite quadratically divergent loop corrections.

Therefore our quest towards a better theory of nature requires us to extend the Standard Model.
Several extensions exist, such as supersymmetry, superstring models, axion models, and extra dimensions.
Every model attempts to provide a solution to one or more of the aforementioned problems.
Grand Unified Theories (GUTs) are interesting extensions of the Standard Model.
In such theories, the gauge group $\SM$ of the Standard Model is embedded in a larger simple gauge group, such as $SU(5)$ \cite{Georgi:1974sy}, $SO(10)$ \cite{Fritzsch:1974nn}, or $E_6$ \cite{Gursey:1975ki,Achiman:1978vg,Shafi:1978gg}.
At high energy scales, the gauge couplings are unified into a single gauge coupling, and the matter content of the Standard Model is grouped together in one or more matter fields in representations of the GUT group.
GUTs are attractive new-physics models because they provide an origin for the observed structure of the Standard Model: the quantum numbers of the fermions are related to one another by the GUT symmetries, and the fermion masses (which are free parameters in the SM) can be explained using only a few parameters.
Since the GUT group is always larger than the SM gauge group, these models introduce additional gauge bosons.
Also, the representations of these groups have more components than the SM fields can fill up, so new fermions and scalars are introduced as well.
Since we want to test these models experimentally, we are interested in the properties of these new particles.

The exceptional group $E_6$ is an attractive example of a GUT group in which the SM gauge group can be embedded \cite{Gursey:1975ki,Achiman:1978vg,Shafi:1978gg}.
It is anomaly-free and left-right-symmetric (LR-symmetric), and as such it provides an explanation for parity violation in the Standard Model by spontaneous symmetry breaking.
It appears in the compactification of string theories, which leads to either four-dimensional $E_6$ gauge symmetry or one of $E_6$'s maximal subgroups \cite{Candelas:1985en,Witten:1985xc}.
The maximal subgroups of $E_6$ are $SO(10)\times U(1)$, $SU(6)\times SU(2)$  and $SU(3)\times SU(3)\times SU(3)$.
The latter is the so-called `trinification group' $G_{333} \equiv \trini$, and is the one we are interested in.
All SM fermions are grouped into a matter field in the fundamental $\rep{27}$ representation of $E_6$.
When the gauge symmetry is broken from $E_6$ to $G_{333}$, the fermions decompose into representations of $G_{333}$ in a way that displays the cyclical symmetry of $E_6$:
\begin{equation}
\begin{array}{ll}
\text{quarks:} & (\brep{3},\rep{3},\rep{1}), \\
\text{leptons:} & (\rep{1},\brep{3}, \rep{3}), \\
\text{antiquarks:}\quad & (\rep{3},\rep{1},\brep{3}).
\end{array} \label{eq:cyclicalfermionreps}
\end{equation}
The Higgs field is in the $\rep{27}$ representation of $E_6$ as well.
Its $(\rep{1},\brep{3},\rep{3})$-component obtains a vacuum expectation value (vev) and is responsible for the breakdown of the gauge symmetry to the electroweak symmetry of the Standard Model.
The coloured components of the Higgs field cannot obtain vevs, as these would break the $SU(3)_C$ symmetry that should hold at low energy scales.
They are assumed to acquire masses of the order of the GUT scale, and are therefore left out of consideration.

In order to compare the trinification model with experiment, a study of the low-energy phenomenology is necessary.
The problem with such a study is the fact that the model contains many scalars: a single Higgs field contains 18 real component fields, and the trinification model needs at least two of them.
The masses of the physical states and their mixing angles are determined by the eigenvalues and eigenvectors of the tree-level mass matrix.
It is very challenging to calculate these because the mass matrix is at least a $36\times36$ matrix.
However, several of the scalar fields will obtain very large masses when the trinification symmetry is broken, and therefore they can be integrated out from the theory.
The result is an effective field theory with the LR-symmetric gauge group $\LRgroup$, and fewer scalar fields than in the trinification model.
This model has the same low-energy properties as the trinification model, but is easier to study.
We will refer to this model as the low-energy trinification (LET) model.

Left-right symmetric models based on the gauge group $\LRgroup$ \cite{Pati:1974yy,Mohapatra:1974hk,Senjanovic:1975rk} have been studied extensively in the literature.
Moreover, these models have many features in common with the two-Higgs-doublet model (2HDM) \cite{Lee:1973iz,}, which has been studied extensively as well in various contexts.
However, the LET model has properties that distinguish it from more general LR-symmetric models and the 2HDM, due to the trinification origin at high energy scales.
A LR-symmetric model in the context of the trinification model has not been studied before to the best of our knowledge.
Therefore the LET model merits a study.

In this thesis, we explore the low-energy phenomenology of the LET model.
We start with an overview of the trinification model in \cref{s:trinificationmodel}, and subsequently derive the properties of the LET model from this in \cref{s:LETmodel}.
A comparison of the LET model to similar models of beyond-the-Standard-Model physics is given in \cref{s:comparison}.
The subsequent sections are used to constrain the parameter space of the LET model, starting with a simplified version of the model.
In \cref{s:gaugebosonconstraints} we consider constraints from searches for heavy vector bosons, whereas the remaining chapters are focused on the scalar sector.
We show how the LET model may be distinguished from the Standard Model by studying the modifications of the Standard-Model-Higgs couplings in \cref{s:lightHiggs}.
The prospects for detection of the new scalar states of the LET model are discussed in \cref{s:newHiggses}.
Finally, we examine how the results from the previous chapters would change in the complete LET model in \cref{s:completeLET}.
Our conclusions are presented in \cref{s:conclusions}.

\clearpage
\section{The trinification model}\label{s:trinificationmodel}
The exceptional group $E_6$ is an attractive gauge-group candidate for a GUT \cite{Gursey:1975ki,Achiman:1978vg,Shafi:1978gg}.
It is the only exceptional group that has non-real representations, which are necessary to account for the inequivalent colour representations of quarks and antiquarks \cite{Okubo:1977sd}.
The trinification model \cite{Gursey:1975ki,Achiman:1978vg,Shafi:1978gg,Stech:2003sb} is based on the gauge group $G_{333} \equiv \trini$ (the `trinification group'), which is a maximal subgroup of $E_6$.
In $E_6$-based GUTs, all known fermions are in the fundamental $\rep{27}$ representation of $E_6$.
The trinification model makes the cyclical symmetry of $E_6$ manifest by assigning the fermions to the representations of $G_{333}$ as in \cref{eq:cyclicalfermionreps}.
The generators of the cyclic symmetry of $E_6$ transform quarks into leptons, leptons into antiquarks, and antiquarks into quarks.
This symmetry also ensures that the three gauge couplings are equal above the scale where the $E_6$-symmetry is restored.

Although the trinification model merits a study in itself, it is also interesting because it appears in various contexts.
The model can arise from the compactification of $E_8\times E_8$ heterotic superstring theories \cite{Candelas:1985en,Witten:1985xc}: the compactification process leads to either four-dimensional $E_6$ gauge symmetry or one of $E_6$'s maximal subgroups.\footnote{The maximal subgroups of $E_6$ are $SO(10)\times U(1)$, $SU(6)\times SU(2)$, and $SU(3)\times SU(3)\times SU(3)$.}
The model also appears in $N=8$ supergravity \cite{Cremmer:1979uq} and brane-world scenarios \cite{Demaria:2005gka}.
The trinification model we study is non-supersymmetric, but supersymmetric models based on the trinification group exist as well, which naturally give rise to the Minimal Supersymmetric Standard Model (MSSM) at low energies \cite{Lazarides:1993uw,Lazarides:1994px}.
In non-supersymmetric versions, trinification can be combined with large extra dimensions, a large number of copies of the SM states, or AdS/CFT complementarity in order to solve the hierarchy problem \cite{PhysRevD.83.093008}.

Besides having a possible origin from a more fundamental theory, trinification models have several attractive features.
The gauge group is anomaly-free and has a cyclic symmetry that implies left-right symmetry.
As such, these models allow us to explain parity violation in the Standard Model by the spontaneous breakdown of the gauge symmetry.
Proton decay can only be mediated by the decay of heavy coloured Higgs bosons, and is therefore suppressed \cite{Babu:1985gi}.
The gauge symmetry conserves baryon number, so that proton decay cannot be mediated by gauge boson exchange.
This fact allows for gauge-coupling unification without supersymmetry at relatively low unification scales $M_U \sim 10^{14}$ GeV, where effects of Planck-scale physics can be neglected \cite{Sayre:2006ma}.
Trinification models can also account for the baryon-antibaryon asymmetry in the universe through heavy-Higgs decays at the one-loop level \cite{He:1986cs}.

In this section, we will give an overview of the trinification model.
Here and in the rest of this work, `trinification model' will refer to the setup described in refs.~\cite{Achiman:1978vg,Stech:2003sb,Stech:2008wd,Stech:2010gf,Stech:2012zr,Stech:2014tla,}; the rest of this section is based on those references.
The setup described there is interesting for several reasons: fermion masses and mixings of the Standard Model can be reproduced using only a few parameters, with a satisfactory fit for the solar neutrino mass difference and the neutrino mixing pattern.
Also, a Standard-Model-like Higgs with a mass of 126 GeV appears in a large region of parameter space of the model.
Furthermore, it gives predictions for the matrix element of neutrinoless double-beta decay and the neutrino masses, which allow the model to be tested with low-energy experiments.
It also allows for various interesting phenomenological scenarios, such as the presence of a light fermiophobic Higgs in addition to the Standard-Model-like Higgs, or even a degenerate Higgs state at 126 GeV.
We will subsequently discuss the gauge boson sector, the scalar sector, and the fermion sector of this setup.
Then we will discuss how the $G_{333}$-symmetry is broken down to QCD and electromagnetism.

\subsection{Gauge sector}
The Standard Model has gauge group $\SM$.
As is well-known, the factor $SU(3)_C$ comes with eight vector bosons $G^a$, $a=1,\ldots,8$.
The trinification group $G_{333}$ has three factors of $SU(3)$, so the trinification model comes with $3\times8=24$ gauge bosons:
\begin{align}
G^1,\ldots,G^8 & \text{ for } SU(3)_C \text{ with gauge coupling } g_C, \notag\\
W_L^1,\ldots,W_L^8 & \text{ for } SU(3)_L \text{ with gauge coupling } g_L, \notag\\
W_R^,\ldots,W_R^8 & \text{ for } SU(3)_R \text{ with gauge coupling } g_R. \label{eq:G333generators}
\end{align}
We denote the corresponding gauge-group generators by $T_C^a$, $T_L^a$, and $T_R^a$ respectively.
They are traceless and Hermitian, and they satisfy the normalisation condition $\tr{T^aT^b} = \delta^{ab}/2$.
In the trinification model, the scalars and fermions are always in singlet or (anti)triplet representations of the gauge group.
The generators of the triplet representation $\rep{3}$ are given by $T_C^a = T_L^a = T_R^a = \frac{\lambda^a}{2}$, where the $\lambda^a$ are the Gell-Mann matrices:
\begin{IEEEeqnarray}{rClrClrCl}
\lambda^1 &=& \begin{pmatrix} 0 & 1 & 0 \\ 1 & 0 & 0 \\ 0 & 0 & 0 \end{pmatrix},\quad &
\lambda^2 &=& \begin{pmatrix} 0 & -i & 0 \\ i & 0 & 0 \\ 0 & 0 & 0 \end{pmatrix},\quad &
\lambda^3 &=& \begin{pmatrix} 1 & 0 & 0 \\ 0 & -1 & 0 \\ 0 & 0 & 0 \end{pmatrix}, \nonumber\\
\lambda^4 &=& \begin{pmatrix} 0 & 0 & 1 \\ 0 & 0 & 0 \\ 1 & 0 & 0 \end{pmatrix},\quad &
\lambda^5 &=& \begin{pmatrix} 0 & 0 & -i \\ 0 & 0 & 0 \\ i & 0 & 0 \end{pmatrix},\quad &
&& \nonumber\\
\lambda^6 &=& \begin{pmatrix} 0 & 0 & 0 \\ 0 & 0 & 1 \\ 0 & 1 & 0 \end{pmatrix},\quad &
\lambda^7 &=& \begin{pmatrix} 0 & 0 & 0 \\ 0 & 0 & -i \\ 0 & i & 0 \end{pmatrix},\quad &
\lambda^8 &=& \frac{1}{\sqrt3}\begin{pmatrix} 1 & 0 & 0 \\ 0 & 1 & 0 \\ 0 & 0 & -2 \end{pmatrix}. \label{eq:GellMann}
\end{IEEEeqnarray}
The generators of the antitriplet representation $\brep{3}$ are given by $\overline{T}_{C,L,R}^a \equiv -(T_{C,L,R}^a)^* = -(T_{C,L,R}^a)^T$.

\subsection{Scalar sector}\label{s:triniscalarsector}
In $E_6$-based models, the Higgs field is a complex scalar in the fundamental representation $\rep{27}$ of $E_6$.
When the gauge symmetry is broken from $E_6$ to $G_{333}$, the Higgs decomposes as
\begin{equation}
\rep{27} \rightarrow (\rep{1},\brep{3},\rep{3}) \oplus (\rep{3},\rep{1},\brep{3}) \oplus (\brep{3},\rep{3},\rep{1}). \label{eq:27}
\end{equation}
That is, the Higgs field decomposes into three bitriplets, two of which are coloured.
The coloured Higgs fields cannot obtain vevs, since the $SU(3)_C$-symmetry should remain unbroken at low energies.
Hence only the colour-singlet Higgs field plays a role in spontaneous symmetry breaking.
In the trinification model, the coloured Higgs fields are assumed to obtain large masses, so that they can be integrated out when the $E_6$ symmetry is broken.

Thus consider a complex scalar field $H_1 \sim (\rep{1},\brep{3},\rep{3})$.
We employ a matrix notation in which $SU(3)_L$ indices run vertically and $SU(3)_R$ indices run horizontally.
In this notation, $H_1$ can be represented as a $3\times3$ matrix.
We can use the $SU(3)_L\times SU(3)_R$ gauge symmetry to bring the vev of $H_1$ into diagonal form:
\begin{equation}
\vev{H_1} = \frac{1}{\sqrt2}\begin{pmatrix} v_1 & 0 & 0 \\ 0 & b_1 & 0 \\ 0 & 0 & M_1 \end{pmatrix}. \label{eq:H1vev}
\end{equation}
Here $v_1$, $b_1$ and $M_1$ are real parameters.
This vev defines the electromagnetic charge operator $Q_\text{em}$, which is a linear combination of the $G_{333}$ generators from \cref{eq:G333generators} that annihilates the vev in \cref{eq:H1vev}.
It is given by
\begin{equation}
Q_\text{em} \equiv T_L^3 + T_R^3 + \frac{1}{\sqrt3}(T_L^8 + T_R^8). \label{eq:Qem}
\end{equation}
The charges of all fields in the trinification model can be read off from the action of $Q_\text{em}$.
It is straightforward to check that the (1,2) and (1,3) components of $H_1$ have charge $-1$, the (2,1) and (3,1) components have charge $+1$, and the other five components are neutral.

If the model is to describe our world at low energies, we need a vacuum that breaks the $G_{333}$ symmetry to $SU(3)_C\times U(1)_\text{em}$.
However, since we can always find a gauge in which the vev in \cref{eq:H1vev} is diagonal, it is not possible to have it break the left-right-symmetry of $G_{333}$.
We need to introduce a second complex scalar field $H_2 \sim (\rep{1},\brep{3},\rep{3})$ that obtains a vev as well, since we cannot simultaneously diagonalise both vevs in general.
The most general vev for $H_2$ that respects electromagnetic gauge invariance is
\begin{equation}
\vev{H_2} = \frac{1}{\sqrt2}\begin{pmatrix} v_2 & 0 & 0 \\ 0 & b_2 & b_3 \\ 0 & M & M_2 \end{pmatrix}. \label{eq:H2vev}
\end{equation}
Here we take $v_2$, $b_2$, $b_3$, $M$, $M_2$ to be real in order to avoid $CP$-violating vacua.
The off-diagonal parameters $M$ and $b_3$ are taken to be unequal, and as such they break the left-right symmetry.
We assume the presence of large hierarchies among the vev parameters.
In this setup, $M_1,M_2 \sim 10^{13} \text{ GeV}$ are of the order of the scale where the Standard-Model gauge couplings $g_1$ and $g_2$ unify.
The off-diagonal vev $M$ is taken to be an intermediate scale of order $10^{10}$ GeV.
The gauge couplings $g_L$, $g_R$ are equal above this scale, whereas below $M$ the left-right symmetry is broken.
The other vev parameters contribute to the $W$-boson mass and are therefore constrained by the relation $v_1^2+v_2^2+b_1^2+b_2^2+b_3^2 = v^2 = (246\text{ GeV})^2$.
This implies that they are much smaller than $M_1$, $M_2$, and $M$.

The scalar-vector interactions are determined by the gauge-invariant kinetic terms of the Lagrangian.
These are given by
\begin{equation}
\lag_s = \tr{(D^\mu H_1)^\dagger(D_\mu H_1)} + \tr{(D^\mu H_2)^\dagger(D_\mu H_2)}, \label{eq:3x3gbmass}
\end{equation}
where the covariant derivatives are given by
\begin{equation}
D_\mu H_i = \partial_\mu H_i - ig_LW_{L\mu}^a\overline{T}_L^aH_i - ig_RW_{R\mu}^aH_iT_R^{aT},\quad i=1,2. \label{eq:DmuHi}
\end{equation}
Note that the $SU(3)_R$-generators are transposed and appear on the right of $H_i$.
The reason for this is our matrix notation, which allows us to omit many indices from our expressions, but obscures the difference between $SU(3)_L$ and $SU(3)_R$ indices.
For the Higgs fields, rows and columns correspond to left- and right-handed indices respectively, whereas for the $SU(3)_L$ and $SU(3)_R$ generators, rows and columns correspond to either both left-handed or both right-handed indices.
Writing left-handed (right-handed) indices as upper (lower) indices, it is easy to see that this notation leads to the placement of the generators as in \cref{eq:DmuHi}:
\begin{align}
(D_\mu H_i)^j_l \equiv& (\partial_\mu H_i)^j_l - ig_LW_{L\mu}^a(\overline{T}_L^a)^{jk}(H_i)^k_l - ig_RW_{R\mu}^a(T_R^a)_{lm}(H_i)^j_m \notag\\
=& (\partial_\mu H_i)^j_l - ig_LW_{L\mu}^a(\overline{T}_L^a)^{jk}(H_i)^k_l - ig_RW_{R\mu}^a(H_i)^j_m(T_R^{aT})_{ml} \notag\\
\rightarrow& (\partial_\mu H_i)^j_l - ig_LW_{L\mu}^a(\overline{T}_L^aH_i)^j_l - ig_RW_{R\mu}^a(H_iT_R^{aT})^j_l.
\end{align}
Here the arrow denotes that we switch to matrix notation.

The scalar masses, mixings and interactions are determined by the scalar potential.
This is a function of $2\times2\times9=36$ real component fields.
Of the 16 gauge bosons for $SU(3)_L\times SU(3)_R$, only one remains massless below the electroweak scale, namely the photon.
Hence 15 of the real scalars become Goldstone bosons that give mass to the massive gauge bosons, and the 21 remaining scalars should become massive.
In order to achieve this, one starts with a scalar potential that contains all possible gauge invariant renormalisable operators.
Then one finds constraints on the parameters in the potential such that the potential has a minimum at the vevs in \cref{eq:H1vev,eq:H2vev}.
Then the scalar masses are given by the eigenvalues of the matrix of second derivatives at the minimum.
The most general renormalisable $\trini$-invariant potential constructed from $H_1$, $H_2$ is given by\footnote{The possible terms in the scalar potential can be obtained systematically by following a procedure analogous to the one described in \cref{a:scalarinvariants}. Note that one could also write down terms of the form $\epsilon^{ijk}\epsilon^{ilm}(H_1^*)^j_p(H_1^*)^k_q(H_1)^l_p(H_1)^m_q$ and $\epsilon^{ijk}\epsilon^{ilm}\epsilon_{pqr}\epsilon_{pst}(H_1^*)^j_q(H_1^*)^k_r(H_1)^l_s(H_1)^m_t$. However, using the identity $\epsilon^{ijk}\epsilon^{ilm} = \delta^{jl}\delta^{km} - \delta^{jm}\delta^{kl}$ they can be rewritten as linear combinations of the invariants already listed in \cref{eq:trinificationscalarpotential}. Also note that $\epsilon^{ijk}\epsilon_{lmn}(H_1)^i_l(H_1)^j_m(H_1)^k_n = 3!\cdot\det{H_1}$.}
\begin{align}
V =& \lambda_1\left(\tr{H_1^\dagger H_1}\right)^2 + \lambda_2\tr{H_1^\dagger H_1H_1^\dagger H_1} + \lambda_3\left(\tr{H_2^\dagger H_2}\right)^2 \notag\\
&+ \lambda_4\tr{H_2^\dagger H_2H_2^\dagger H_2} + \lambda_5\tr{H_1^\dagger H_1}\tr{H_2^\dagger H_2} + \lambda_6\tr{H_1^\dagger H_1H_2^\dagger H_2} \notag\\
&+ \lambda_7\tr{H_1^\dagger H_2}\tr{H_2^\dagger H_1} + \lambda_8\tr{H_1^\dagger H_2H_2^\dagger H_1} + \bigg[ \lambda_9\left(\tr{H_1^\dagger H_2}\right)^2 \notag\\
&+ \lambda_{10}\tr{H_1^\dagger H_2H_1^\dagger H_2} + \lambda_{11}\tr{H_1^\dagger H_1}\tr{H_1^\dagger H_2} + \lambda_{12}\tr{H_1^\dagger H_1H_1^\dagger H_2} \notag\\
&+ \lambda_{13}\tr{H_1^\dagger H_2}\tr{H_2^\dagger H_2} + \lambda_{14}\tr{H_1^\dagger H_2H_2^\dagger H_2} \plushc\bigg] \notag\\
&+ \mu_{d1}\left(\det H_1 + \det H_1^\dagger\right) + \mu_{d2}\left(\det H_2 + \det H_2^\dagger\right) \notag\\
&+ \epsilon^{ijk}\epsilon_{lmn} \bigg[ \mu_{112}(H_1)^i_l(H_1)^j_m(H_2)^k_n + \mu_{122}(H_1)^i_l(H_2)^j_m(H_2)^k_n \plushc \bigg] \notag\\
&+ \mu_1^2\tr{H_1^\dagger H_1} + \mu_2^2\tr{H_2^\dagger H_2} + \mu_{12}^2\left( \tr{H_1^\dagger H_2} \plushc \right). \label{eq:trinificationscalarpotential}
\end{align}
Here $\epsilon^{ijk}$ is the completely antisymmetric symbol that satisfied $\epsilon^{123}=+1$.
Alternatively, one can add terms to the potential with a logarithmic dependence on the terms appearing in the above potential, as in ref.~\cite{Stech:2014tla}.
Putting the gradient of the potential at zero at the vev in \cref{eq:H1vev,eq:H2vev}, one can express the dimensionful parameters $\mu^2_1$, $\mu^2_2$, $\mu^2_{12}$, $\mu_{d1}$, $\mu_{d2}$, $\mu_{112}$, $\mu_{122}$, and one of the dimensionless parameters $\lambda_i$ in terms of the thirteen remaining dimensionless parameters and the eight vev parameters.
We do not list these expressions here, as some of them are quite large, but they are easily found using mathematics software such as \emph{Mathematica} \cite{Mathematica10}.
However, a complete analysis of the mass matrix is challenging due to the large number of field components and free parameters.
One can simplify the analysis by considering only benchmark points in which the vev parameters in \cref{eq:H1vev,eq:H2vev} are closely related, e.g.\ $v_1=v_2$, $b_3\sim b_1$, $M_1=M$, $M_2=b_2=0$.
But even in these cases, a complete analysis of the scalar masses and mixings is challenging, and one has to resort to finding numerical benchmark points with an interesting phenomenology.

\subsection{Fermion sector}
In $E_6$-based models, all known fermions are grouped into the fundamental representation $\rep{27}$ of $E_6$.
They are two-component left-handed Weyl spinors with respect to the Lorentz group.
After the gauge symmetry is broken from $E_6$ to $G_{333}$, the fermion field decomposes as in \cref{eq:27}.
The fermions are grouped into a left-handed quark field~$Q_L$, a right-handed quark field~$Q_R$, and a lepton field~$L$.
These are assigned to the representations of $\trini$ in the decomposition of $\rep{27}$ as follows:
\begin{equation}
L \sim (\rep{1},\brep{3},\rep{3}),\qquad
Q_L \sim (\brep{3},\rep{3},\rep{1}),\qquad
Q_R \sim (\rep{3},\rep{1},\brep{3}). \label{eq:trinifermionreps}
\end{equation}
We can write their components into matrix notation, as we did for the Higgs fields.
In this notation, $Q_L$ is a column vector, $Q_R$ is a row vector, and $L$ is a $3\times3$ matrix:
\begin{equation}
Q_L^b = \left(\begin{array}{c} u^b \\ d^b \\ D^b \end{array}\right),\quad
Q_R^b = \left(\begin{array}{ccc} \hat{u}^b & \hat{d}^b & \hat{D}^b \end{array}\right),\quad
L = \left(\begin{array}{ccc} L^1_1 & E^- & e^- \\
E^+ & L^2_2 & \nu \\
e^+ & \hat\nu & L^3_3 \end{array}\right). \label{eq:trinifermions}
\end{equation}
Here $b=1,2,3$ is a colour index.
The components $u$ and $d$ are the left-handed up and down quarks that we know from the Standard Model.
The trinification model introduces a new left-handed quark $D$ with electromagnetic charge $-\frac13$.
The components $\hat{u}$, $\hat{d}$, $\hat{D}$ are the right-handed counterparts of $u$, $d$, $D$ respectively.
The lepton field contains the charged leptons $e^\pm$ and the left-handed neutrino $\nu$.
It contains several new states: a right-handed neutrino $\hat\nu$; three neutral states $L^1_1$, $L^2_2$, $L^3_3$; and a pair of charged leptons $E^\pm$.
There are three copies of each fermion in \cref{eq:trinifermions}; the generation indices $\alpha=1,2,3$ have been suppressed.

Note that the assignments of the $SU(3)_C$ representations of the quarks in \cref{eq:trinifermionreps} are interchanged with respect to the Standard Model, in which the quarks (antiquarks) transform as a $\rep{3}$ ($\brep{3}$) under $SU(3)_C$.
However, this assignment has no physical consequences, and therefore it is an arbitrary choice.
The gauge group $E_6$ has a cyclical symmetry that changes quarks into leptons, leptons into antiquarks, and antiquarks into quarks.
The representations in \cref{eq:trinifermionreps} have been assigned such that this symmetry becomes manifest, while ensuring that the left-handed quarks obtain the correct transformation behaviour under $SU(2)_L$ at low energies.

The fermion covariant derivatives following from the assignments in \cref{eq:trinifermionreps} are given by
\begin{align}
D_\mu Q_L^b =& \partial_\mu Q_L^b - ig_LW_{L\mu}^aT_L^aQ_L^b - ig_CG_\mu^a(\overline{T}_C^aQ_L)^b, \notag\\
D_\mu Q_R^b =& \partial_\mu Q_R^b - ig_RW_{R\mu}^aQ_R^b\overline{T}_R^{aT} - ig_CG_\mu^a(Q_RT_C^{aT})^b, \notag\\
D_\mu L =& \partial_\mu L - ig_LW_{L\mu}^a\overline{T}_L^aL - ig_RW_{R\mu}^aLT_R^{aT}.
\end{align}

Besides describing the gauge boson and scalar sectors in accordance with experiment, any new-physics model has the task of describing the Standard-Model-fermion masses and mixings correctly (see table~\ref{t:fermionmasses}).
The gauge-boson sector is fixed by the choice of gauge group, and the scalar sector is straightforward to work out after the fields and their gauge-group representations have been chosen.
However, the Yukawa sector allows for more freedom: one can choose which scalar fields couple to which fermions, as well as choose the form of these couplings.

\begin{table}[t]
\begin{center}
\begin{tabular}{|c|c|c|}
\hline
$m_u = (1.24 \pm 0.06)$ MeV	&	$m_c = (624 \pm 14)$ MeV	&	$m_t = (171.55 \pm 0.90)$ GeV	\\\hline
$m_d = (2.69 \pm 0.09)$ MeV	&	$m_s = (53.8 \pm 1.4)$ MeV	&	$m_b = (2.85 \pm 0.023)$ GeV	\\\hline
$m_e = 0.510$ MeV		&	$m_\mu = 105.4$ MeV		&	$m_\tau = 1.7725$ GeV		\\\hline
\end{tabular}
\end{center}
\begin{equation*}
|V_\text{CKM}| = \begin{pmatrix} 0.97427 \pm 0.00015	&	0.22534 \pm 0.00065	&	0.00351^{+0.00015}_{-0.00014}	\\
0.22520 \pm 0.00065	&	0.97344 \pm 0.00016	&	0.0412^{+0.0011}_{-0.0005}	\\
0.00867^{+0.00029}_{-0.00031}	& 0.0404^{+0.0011}_{-0.0005}	&	0.999146^{+0.000021}_{-0.000046} \end{pmatrix}
\end{equation*}
\begin{equation*}
\alpha = (89.0^{+4.4}_{-4.2})^\circ,\quad
\beta = (21.4\pm0.8)^\circ,\quad
\gamma = (68^{+10}_{-11})^\circ,\quad
\alpha+\beta+\gamma = (178^{+11}_{-12})^\circ.
\end{equation*}
\caption{Fermion masses in the $\overline{MS}$ scheme at $\mu = M_Z$, magnitudes of the CKM-matrix elements and angles of the unitarity triangle. The fermion masses have been taken from ref.~\cite{Stech:2014tla}, whereas the CKM-matrix elements and angles of the unitarity triangle have been taken from ref.~\cite{Agashe:2014kda}.}\label{t:fermionmasses}
\end{table}

If both $H_1$ and $H_2$ were coupled to fermions, then each of them would have its own Yukawa coupling matrix.
Without making additional assumptions, it is not possible to diagonalise these matrices simultaneously.
This in turn leads to flavour-changing neutral current (FCNC) processes, which are severely restricted by experiment.
In order to suppress FCNC interactions, the existence of a $Z_2$-symmetry is assumed under which $H_1$ ($H_2$) is even (odd).
This implies that $H_2$ does not couple to fermions, and therefore tree-level FCNC diagrams are avoided.
It also means that $H_2$ only contributes to the gauge boson masses, whereas $H_1$ gives mass to both the fermions and the gauge bosons.

The Higgs couplings to the fermions are of the form
\begin{align}
\lag_Y =& -g_tG_{\alpha\beta} \left( Q_R^\alpha H_1^TQ_L^\beta + \frac12\epsilon^{ijk}\epsilon_{lmn}L^i_lL^j_m(H_1)^k_n \right) \notag\\
&- A_{\alpha\beta}\left( Q_R^\alpha H_{Aq}^T Q_L^\beta + \epsilon^{ijk}L^i_lL^j_m(H_{Al})^k_{\{lm\}} \right) \plushc \label{eq:trinificationYukawa}
\end{align}
The first line is a Yukawa interaction built from the fields we have already introduced: the parameter $g_t$ is a dimensionless coupling, $G_{\alpha\beta}$ is a symmetric $3\times3$ generation matrix, and $\epsilon$ is a totally antisymmetric symbol with $\epsilon^{123} = \epsilon_{123} = +1$.
This interaction is sufficient to reproduce the up-quark masses: the matrix $G_{\alpha\beta}$ can be diagonalised by choosing an appropriate generation basis, and its diagonal components can be fit to the masses of the up quarks.
However, without the second line in \cref{eq:trinificationYukawa} it would lead to the same mass hierarchies for the up quarks, down quarks, charged leptons, and neutrinos.
Hence an additional coupling to new scalar fields $H_{Aq} \sim (\rep{1},\brep{3},\rep{3})$ (for the quarks) and $H_{Al} \sim (\rep{1},\brep{3},\brep{6})$ (for the leptons) is introduced, with a Hermitian antisymmetric (and thus imaginary) generation matrix $A_{\alpha\beta}$.
The fields $H_{Aq}$ and $H_{Al}$ are viewed as components of a scalar $H_A$ in the antisymmetric $\rep{351_A}$ representation of $E_6$ (see \cref{a:Yukawatrini}).
It is assumed that they have negligible mixing with $H_1$ and $H_2$ in order to simplify the analysis of the scalar spectrum.

The matrices $G_{\alpha\beta}$ and $A_{\alpha\beta}$ can be viewed respectively as the real and imaginary component of the vev of a scalar field called the `flavon' \cite{Stech:2008wd}.
In this picture, the Yukawa interactions are effective interactions arising from dimension-five operators, which in turn arise from interactions with gauge-singlet fermions.
However, the components of these matrices are simply considered as free parameters of the trinification model.

The second line in \cref{eq:trinificationYukawa} leads to mixing among the $d$-quarks and the $D$-quarks via the seesaw mechanism.
This leads to small masses for the $d$-quarks and large masses for the $D$-quarks.
Using only four parameters (in addition to those of the first Yukawa term), all the quark masses, the CKM-matrix elements, and the angles of the unitarity triangle are reproduced within error limits~\cite{Stech:2014tla}.
Thus $G_{\alpha\beta}$ is responsible for the quark mass hierarchy, and $A_{\alpha\beta}$ is responsible for the quark mixings and $CP$-violation.
Similarly, the third term in \cref{eq:trinificationYukawa} leads to mixing among the charged Standard-Model-leptons and their heavy partners.
A good fit for the charged-lepton masses is obtained with three extra parameters \cite{Stech:2014tla}.

At this stage, neutrinos are still Dirac particles with masses comparable to the other fermion masses.
In order to obtain neutrino masses in accordance with experiment, a dimension-five Yukawa interaction is added to the Lagrangian:
\begin{equation}
\lag_Y^\text{eff} = -\frac{1}{M_N} (G^2)_{\alpha\beta} \tr{L^\alpha H_1^\dagger} \tr{H_2^\dagger L^\beta} \plushc
\end{equation}
This interaction is considered as an effective one that could originate from the exchange of a new heavy Dirac fermion that is a trinification singlet \cite{Stech:2008wd}.
This fermion consists of two Weyl fields, one of which is odd under the $Z_2$-symmetry.
A Dirac mass $M_N$ appears after the aforementioned flavon field obtains a vev, breaking the $Z_2$-symmetry.
Hence this effective interaction violates the $Z_2$-symmetry as well.
This term mixes the neutrinos $\nu$, $\hat\nu$ with the other neutral leptons $L^1_1$, $L^2_2$, $L^3_3$, giving rise to a generalised seesaw mechanism.
The light-neutrino mass matrix introduces two additional parameters, which can be fixed by the experimentally observed atmospheric mass-squared difference and the lightest neutrino mass.

\subsection{From trinification to electromagnetism}\label{s:trinitoem}
In order to show how the trinification model reduces to the Standard Model at low energy scales, we will work out the symmetry breaking chain from the trinification group down to the electromagnetic group.
This can be done by working out the action of the generators in \cref{eq:G333generators} on the vevs in \cref{eq:H1vev,eq:H2vev} to find the unbroken generators, keeping in mind the hierarchy $M_1,M_2 > M \gg v_1,v_2,b_1,b_2,b_3$.
The generators of $SU(3)_C$ are not broken in any step, therefore we omit them from the following discussion.

The first symmetry-breaking step is caused by $M_1$; the vev parameter $M_2$ is of the same order of magnitude and breaks the same symmetry generators as $M_1$.
In this step, the generators $T_L^a$, $T_R^a$ for $a=4,5,6,7,8$ are broken.
We can construct one unbroken generator from $T_L^8$ and $T_R^8$.
The following unbroken generators generate $SU(2)_L\times SU(2)_R\times U(1)_{B-L}$:
\begin{equation}
T_L^1,\; T_L^2,\; T_L^3,\qquad T_R^1,\; T_R^2,\; T_R^3,\qquad Q_{B-L}\equiv \frac{2}{\sqrt3}(T_L^8+T_R^8). \label{eq:su2su2u1generators}
\end{equation}
The normalisation of the $U(1)_{B-L}$ generator $Q_{B-L}$ is arbitrary.
However, it is possible to choose it such that the charges of the Standard Model particles coincide with the known quantum number $B-L$ (baryon number minus lepton number), and here we choose to do so.
The Lagrangian in \cref{eq:3x3gbmass} gives rise to mass terms for some of the gauge bosons.
The gauge bosons corresponding to the broken generators obtain masses of order $M_1$, $M_2$, and can be integrated out.

The second symmetry-breaking step is caused by $M$, which breaks the generators $T_R^1$, $T_R^2$, $T_R^3$, and $Q_{B-L}$.
The latter two can be combined into an unbroken generator.
Thus the following four generators are left unbroken, and generate the electroweak group $SU(2)_L\times U(1)_Y$:
\begin{equation}
T_L^1,\; T_L^2,\; T_L^3,\qquad Y\equiv T_R^3 + \frac12Q_{B-L} = T_R^3 + \frac{1}{\sqrt3}(T_L^8+T_R^8).
\end{equation}
After this step, the Lagrangian in \cref{eq:3x3gbmass} yields additional gauge-boson mass terms.
The gauge bosons corresponding to the broken generators obtain masses of order $M$, and can be integrated out.

The third symmetry-breaking step is caused by $v_1$; the other vev parameters $v_2$, $b_1$, $b_2$, $b_3$ break the electroweak symmetry as well, but they break the same generators as $v_1$.
In this step, the generators $T_L^1$, $T_L^2$, $T_L^3$, and $Y$ are all broken.
A single combination of the broken generators remains unbroken, and generates the electromagnetic gauge group $U(1)_\text{em}$:
\begin{equation}
Q_\text{em} \equiv T_L^3 + Y = T_L^3 + T_R^3 + \frac{1}{\sqrt3}(T_L^8 + T_R^8). \label{eq:emchargeoperator}
\end{equation}
We can summarise these steps in the following symmetry-breaking chain.
The symbol over each arrow denotes the vev that is responsible for this breaking step:
\begin{align}
\trini &\stackrel{M_1}{\longrightarrow} \LRgroup \notag\\
&\stackrel{M}{\longrightarrow} \SM \notag\\
&\stackrel{v_1}{\longrightarrow} SU(3)_C\times U(1)_\text{em}. \label{eq:breakingchain}
\end{align}

\clearpage
\section{The low-energy trinification model}\label{s:LETmodel}
An important aspect of studying a new-physics model is to look for phenomenological aspects at experimentally available energy scales: this allows the model to be tested.
This aspect involves not only determining the masses of new particles, but also studying the effects of the new particles on the couplings of Standard-Model particles.
As was discussed in section~\ref{s:triniscalarsector}, a complete analysis of the phenomenology of the trinification model is challenging: the scalar masses and mixings are determined by a $36\times36$ mass matrix that depends on several free parameters.
However, one can simplify the problem greatly by considering a low-energy effective field theory (EFT) based on the trinification model.
In an EFT, fields with a mass much larger than the energy scale under consideration are integrated out from the theory.
This results in a theory that is more convenient to analyse, but has the same low-energy behaviour as the complete high-energy theory.

According to \cref{eq:breakingchain}, the trinification group $G_{333}$ is broken down to the Standard-Model gauge group via the intermediate gauge group $\LRgroup$.
In this process, some of the fields from the trinification model will obtain masses of order $M_1$, and will therefore be too heavy to have measurable effects at the low energy scales we are interested in.
Hence the low-energy phenomenology of the trinification model can be described conveniently by an EFT based on the gauge group $\LRgroup$, from which the fields with masses of order $M_1$ have been integrated out.
In this section we will derive the field content of this effective model, which we will refer to as the low-energy trinification (LET) model.
We will subsequently discuss the gauge boson sector, scalar sector, and fermion sector.
In order to aid our study of the LET model, we introduce a toy model with a simplified scalar sector, which we will refer to as the simplified LET model.

\subsection{Gauge boson sector}
As we described in section~\ref{s:trinitoem}, the gauge symmetry is broken from $G_{333}$ to $\LRgroup$ by the vev parameter $M_1$.
In this process, the gauge bosons $W_L^a$, $W_R^a$ for $a=4,5,6,7$ and the linear combination $W^8\equiv \frac{1}{\sqrt2}(W_L^8-W_R^8)$ obtain masses of order $M_1$, $M_2$ via the Lagrangian in \cref{eq:3x3gbmass}.
These heavy gauge bosons can now be integrated out from the Lagrangian.
The seven gauge bosons corresponding to the generators in \cref{eq:su2su2u1generators} remain massless.
These fields constitute the gauge boson field content of our model:
\begin{align}
W^1_L, W^2_L, W^3_L &\text{ for } SU(2)_L \text{ with gauge coupling } g_L, \notag\\
W^1_R, W^2_R, W^3_R &\text{ for } SU(2)_R \text{ with gauge coupling } g_R, \notag\\
B\equiv\frac{1}{\sqrt2}(W_L^8+W_R^8) &\text{ for } U(1)_{B-L} \text{ with gauge coupling } g'.
\end{align}
We denote the corresponding gauge-group generators by $T_L^i$, $T_R^i$ ($i=1,2,3$), and $Q_{B-L}$.
Note that we use the same names for the $SU(3)_{L,R}$ generators as for the corresponding $SU(2)_{L,R}$ generators; it should always be clear from the context which version the name refers to.
The $SU(2)$ generators for the doublet representation $\rep{2}$ are given by the Pauli matrices: $T_L^i = T_R^i = \frac{\sigma^i}{2}$, where
\begin{equation}
\sigma^1 = \begin{pmatrix} 0 & 1 \\ 1 & 0 \end{pmatrix},\qquad
\sigma^2 = \begin{pmatrix} 0 & -i \\ i & 0 \end{pmatrix},\qquad
\sigma^3 = \begin{pmatrix} 1 & 0 \\ 0 & -1 \end{pmatrix}.
\end{equation}
Note that these generators correspond to the upper left $2\times2$ blocks of the corresponding $SU(3)$ generators in eq.~\eqref{eq:GellMann}.
The generators of the antidoublet representation $\brep{2}$ are given by $\overline{T}_{L,R}^i = -(T_{L,R})^T$.

The charges of the gauge bosons follow from the action of the electromagnetic charge operator in \cref{eq:emchargeoperator}.
The fields $W_L^3$, $W_R^3$, and $B$ are neutral, whereas the remaining fields mix to form charge eigenstates, analogously to the $W$ bosons of the Standard Model:
\begin{equation}
W_{L,R}^\pm \equiv \frac{1}{\sqrt2}(W_{L,R}^1 \mp iW_{L,R}^2). \label{eq:WLWR}
\end{equation}
After spontaneous symmetry breaking, the gauge fields mix to form six massive gauge bosons and a massless photon field.
The charged states in \cref{eq:WLWR} are rotated by an angle $\zeta$ into two pairs of charged mass eigenstates $W^\pm$ and $W^{\prime\pm}$:
\begin{equation}
\begin{pmatrix} W^\pm \\ W^{\prime\pm} \end{pmatrix}
= \begin{pmatrix} \cos\zeta & \sin\zeta \\ -\sin\zeta & \cos\zeta \end{pmatrix}
\begin{pmatrix} W_L^\pm \\ W_R^\pm \end{pmatrix}.
\end{equation}
Here the $W^\pm$ correspond to the charged vector bosons of the Standard Model.
The $W^{\prime\pm}$ bosons are new massive vector bosons.
The $W-W'$ mixing angle $\zeta$ and the $W$, $W'$ masses are derived in \cref{a:gaugebosons}.
The mixing angle is very small: one can expand it in terms of the small parameter $\frac{v}{M}$, which yields
\begin{equation}
\zeta = \frac{g_Lv^2\sin2\beta}{g_RM^2} + \ord{\frac{v^4}{M^4}}. \label{eq:zeta}
\end{equation}
Here $v^2 \equiv v_1^2 + b_1^2$ and $\tan\beta \equiv b_1/v_1$.
The masses of the charged mass eigenstates are given by
\begin{align}
m_W =& \frac{g_Lv}{2}\left( 1 - \frac12\sin^22\beta\frac{v^2}{M^2} + \ord{\frac{v^4}{M^4}} \right), \notag\\
m_{W'} =& \frac{g_RM}{2}\left( 1 + \frac{v^2}{2M^2} + \ord{\frac{v^4}{M^4}} \right). \label{eq:Wmasses}
\end{align}

The three neutral gauge fields mix to form mass eigenstates $A$, $Z$, and $Z'$.
Here $A$ is the massless photon field, $Z$ is the neutral massive vector boson we know from the Standard Model, and $Z'$ is a new massive state.
They can be expressed in terms of the gauge eigenstates $W_L^3$, $W_R^3$, $B$ by a rotation over three mixing angles $\theta_W$, $\theta_W^\prime$, $\eta$:
\begin{equation}
\begin{pmatrix} A \\ Z \\ Z' \end{pmatrix} =
\begin{pmatrix} s_{\theta_W} & c_{\theta_W}s_{\theta_W^\prime} & c_{\theta_W}c_{\theta_W^\prime} \\
c_{\theta_W}c_\eta & c_{\theta_W^\prime}s_\eta - s_{\theta_W}s_{\theta_W^\prime}c_\eta & -(s_{\theta_W}c_{\theta_W^\prime}c_\eta + s_{\theta_W^\prime}s_\eta) \\
-c_{\theta_W}s_\eta & c_{\theta_W^\prime}c_\eta + s_{\theta_W}s_{\theta_W^\prime}s_\eta & s_{\theta_W}c_{\theta_W^\prime}s_\eta - s_{\theta_W^\prime}c_\eta \end{pmatrix}
\begin{pmatrix} W_L^3 \\ W_R^3 \\ B \end{pmatrix}.
\end{equation}
Here we have defined $s_x \equiv \sin{x}$, $c_x \equiv \cos{x}$ for the sake of brevity.
The angle $\theta_W$ is the Weinberg angle we know from the Standard Model; $\theta_W^\prime$ is an analogon of the Weinberg angle for the breaking of the left-right symmetry; and $\eta$ is the $Z-Z'$ mixing angle.
These angles are given in terms of the gauge couplings by
\begin{align}
\sin\theta_W =& \frac{2g'g_R}{\sqrt{4g^{\prime2}(g_L^2+g_R^2)+g_L^2g_R^2}}, \notag\\
\sin\theta_W^\prime =& \frac{2g'}{\sqrt{g_R^2+4g^{\prime2}}}, \notag\\
\tan\eta =& \frac{g_R^2\sqrt{4g^{\prime2}(g_L^2+g_R^2)+g_L^2g_R^2}}{(g_R^2+4g^{\prime2})^2} \frac{v^2}{M^2} + \ord{\frac{v^4}{M^4}}. \label{eq:neutralAngles}
\end{align}
The masses of the neutral states are given by
\begin{align}
m_A =& 0, \notag\\
m_Z =& \frac{g_Lv}{2\cos\theta_W} \left( 1 - \frac{\cos^4\theta_W^\prime}{2}\frac{v^2}{M^2} + \ord{\frac{v^4}{M^4}} \right), \notag\\
m_{Z'} =& \frac{g_RM}{2\cos\theta_W^\prime} \left( 1 + \frac{\cos^4\theta_W^\prime}{2}\frac{v^2}{M^2} + \ord{\frac{v^4}{M^4}} \right). \label{eq:Zmasses}
\end{align}
For the exact expressions of all masses as well as a derivation of these masses and the mixing angles, see \cref{a:gaugebosons}.

\subsection{Scalar sector}
The scalar fields in the LET model are parts of the trinification scalar fields $H_1$, $H_2$.
First consider the $2\times2$ blocks in the upper left corners of these fields.
It is straightforward to work out how these parts transform under the $\LRgroup$-generators in \cref{eq:su2su2u1generators}.
It turns out that they transform as bidoublets $\Phi_i \sim (\rep{1},\brep{2},\rep{2},0)$:
\begin{equation}
\Phi_i = \begin{pmatrix} \Phi_{i,11}^0 & \Phi_{i,21}^- \\
\Phi_{i,12}^+ & \Phi_{i,22}^0 \end{pmatrix}
\leftrightarrow \begin{pmatrix} (H_i)^1_1 & (H_i)^1_2 & 0 \\
(H_i)^2_1 & (H_i)^2_2 & 0 \\
0 & 0 & 0 \end{pmatrix},\qquad
\vev{\Phi_i} = \frac{1}{\sqrt2}\begin{pmatrix} v_i & 0 \\
0 & b_i \end{pmatrix}. \label{eq:Phiivev}
\end{equation}
Next consider the $(3,1)$ and $(3,2)$ components of $H_2$.
It turns out that these transform as a right-handed doublet $\Phi_R \sim (\rep{1},\rep{1},\rep{2},1)$:
\begin{equation}
\Phi_R = \begin{pmatrix} \Phi_R^+ & \Phi_R^0 \end{pmatrix}
\leftrightarrow \begin{pmatrix} 0 & 0 & 0 \\
0 & 0 & 0 \\
(H_2)^3_1 & (H_2)^3_2 & 0 \end{pmatrix},\qquad
\vev{\Phi_R} = \frac{1}{\sqrt2}\begin{pmatrix} 0 & M \end{pmatrix}. \label{eq:PhiRvev}
\end{equation}
The fields $\Phi_1$, $\Phi_2$, $\Phi_R$, and their vevs are sufficient to describe the symmetry breaking from $\LRgroup$ to electromagnetism via the Standard Model.
Note that a right-handed doublet like $\Phi_R$ resides in $H_1$ as well.
Similarly, the $(1,3)$ and $(2,3)$ components of $H_1$ and $H_2$ contain left-handed antidoublets, and the $(3,3)$ components are $SU(2)_L\times SU(2)_R$ singlets.
We assume these field components to obtain large masses of order $M_1$, so that they can be integrated out from the trinification model.
Note that the $(3,3)$ components of $H_1$ and $H_2$ are total gauge singlets, and therefore are dark matter candidates.
Hence the scalar sector of the LET model contains $2\times8+4=20$ real scalars, which is a great simplification with respect to the 36 real scalars of the trinification model.

Remember that the field $H_2$ does not couple to fermions.
Since $\Phi_2$ and $\Phi_R$ originate from $H_2$, these do not couple to fermions either.
Hence only $\Phi_1$ couples to the fermions in the LET model.

Similarly to the trinification model, the vev components of the bidoublet scalars are constrained by the relation $v_1^2+b_1^2+v_2^2+b_2^2 = v^2 = (246\text{ GeV})^2$.
The parameter $M$ is a large mass scale well above the electroweak scale.
However, in contrast to the trinification model, we do not only consider very high scales $M\sim10^{10}$ GeV.
We will also consider the possibility that $M$ lies in the TeV range.
As we will see in \cref{s:lightHiggs}, the latter scenario means that effects of new physics may be measurable at the LHC.

The scalar-vector interactions of the LET model are determined by the gauge-invariant kinetic terms of the Lagrangian.
Using a similar matrix notation as in the trinification model, with $SU(2)_L$ and $SU(2)_R$ indices running vertically and horizontally respectively, these kinetic terms are given by
\begin{equation}
\lag_s = \tr{(D^\mu\Phi_1)^\dagger(D_\mu\Phi_1)} + \tr{(D^\mu\Phi_2)^\dagger(D_\mu\Phi_2)} + (D^\mu\Phi_R)(D_\mu\Phi_R)^\dagger, \label{eq:LagS}
\end{equation}
where the covariant derivatives are given by
\begin{align}
D_\mu\Phi_{1,2} =& \partial_\mu\Phi_{1,2} - ig_LW_{L\mu}^i\overline{T}_L^i\Phi_{1,2} - ig_RW_{R\mu}^i\Phi_{1,2}T_R^{iT}, \notag\\
D_\mu\Phi_R =& \partial_\mu\Phi_R - ig_RW_{R\mu}^i\Phi_RT_R^{iT} - ig'B_\mu\Phi_R. \label{eq:covderivscalar}
\end{align}
As in the trinification model, we build the scalar potential of the LET model from all possible gauge-invariant renormalisable operators consisting of $\Phi_1$, $\Phi_2$, $\Phi_R$.
A systematic derivation of all possible invariants can be found in \cref{a:scalarinvariants}.
However, since we consider our model to be an EFT originating from the trinification model, some of these invariants can be left out.
Firstly, we do not include any invariants involving charge conjugates of the scalar fields.
The reason is the fact that the $\rep{3}$ and $\brep{3}$ representations of $SU(3)$ are inequivalent, so there is no such thing as the charge conjugate of a triplet field.
This implies that there is no operator in the trinification model from which terms involving charge conjugates could originate.
Secondly, some renormalisable operators in the LET model correspond to nonrenormalisable operators in the trinification model.
Since the trinification model is considered to be a renormalisable theory, we do not take these operators into account.

As we will see in \cref{s:completeLET}, the scalar potential of the LET model still contains 15 free parameters after these simplifications.
This is a lot of freedom, which complicates the analysis of the scalar states, and hence makes it challenging to constrain the model.
Note that without $\Phi_2$, we can still describe the breakdown of the $\LRgroup$-symmetry to the Standard Model.
Hence as a further simplification, we consider a toy model in which we omit the fermiophobic bidoublet $\Phi_2$.
We will refer to this setup as the `simplified LET model', although we will see it is no longer an EFT of the trinification model.
The most general scalar potential for the simplified LET model is given by
\begin{align}
V(\Phi_1,\Phi_R) =& \frac{\lambda_1}{2}\tr{\Phi_1^\dagger\Phi_1}^2 + \frac{\lambda_2}{2}\tr{\Phi_1^\dagger\Phi_1\Phi_1^\dagger\Phi_1} + \frac{\lambda_3}{2}\big(\Phi_R\Phi_R^\dagger\big)^2 \notag\\
&+ \lambda_4\tr{\Phi_1^\dagger\Phi_1}(\Phi_R\Phi_R^\dagger) + \lambda_{5}\Phi_R\Phi_1^\dagger\Phi_1\Phi_R^\dagger \notag\\
&+ \mu^2_{11}\tr{\Phi_1^\dagger\Phi_1} + \mu^2_R\Phi_R\Phi_R^\dagger + \left(\mu^2_1\det\Phi_1 \plushc \right). \label{eq:simplifiedLETpotential}
\end{align}
Here $\mu^2_{11}$, $\mu^2_R$, $\mu^2_1$, and the $\lambda_i$ are real parameters.
The dimensionful parameters $\mu^2_{11}$, $\mu^2_R$, $\mu^2_1$ are fixed in terms of the $\lambda_i$, and the vev parameters $v_1$, $b_1$, $M$ by the condition that the potential has an extremum at the vevs in \cref{eq:Phiivev,eq:PhiRvev}.
The expressions for these parameters can be found in \cref{a:scalarspectrum}.
The simplified LET model has only 12 real scalar fields and five free parameters in the scalar potential, which makes this version much easier to handle than the full LET model.
This will make it easier to calculate the scalar masses and mixings as well as their effects on the couplings of Standard-Model particles.
Of course, this model does not accurately describe the low-energy properties of the trinification model anymore: the bidoublet $\Phi_2$ obtains a vev of the order of the electroweak scale, hence its effects will be important at low energies.
As such, the simplified LET model is not a proper EFT of the trinification model, but rather a toy model that facilitates our study of the LET model.
We will study the simplified LET model in \cref{s:comparison,s:gaugebosonconstraints,s:lightHiggs,s:newHiggses}, and examine how the calculations and results would change if we added $\Phi_2$ in \cref{s:completeLET}.

In the simplified LET model, the vev parameters $v_2$, $b_2$ are left out of consideration.
Hence the light vev parameters are now constrained by the condition $v_1^2+b_1^2 = v^2 = (246 \text{ GeV})^2$.
It will be convenient to reparametrise the vevs as
\begin{equation}
v_1 = v\cos\beta,\qquad b_1 = v\sin\beta.
\end{equation}
At the scale where the left-right symmetry is broken, the vevs are related to the top and bottom quark masses by $b_1/v_1 = m_b/m_t$ \cite{Stech:2014tla}.
Hence we have the hierarchy $b_1 \ll v_1 \ll M$, and $\beta = 0.0166$ is not a free parameter of the model.

The scalar fields $\Phi_1$, $\Phi_R$ contain twelve real scalar components in total.
After spontaneous symmetry breaking, six of them become massless Goldstone bosons that give mass to the six massive gauge bosons.
The remaining components mix to form six massive scalars: three $CP$-even scalars $h^0$, $H_1^0$, and $H_2^0$; one $CP$-odd scalar $A^0$; and a pair of charged scalars $H^\pm$.
Here we define $h^0$ as the scalar that is the most $h^0_{1,11}$-like and $H^0_2$ as the scalar that is the most $h^0_R$-like, where $h^0_{1,11} \equiv \Re\left((\Phi_1)_{11}\right)$ and $h^0_R \equiv \Re\left((\Phi_R)_2\right)$.
The scalar masses are given in terms of the model parameters by
\begin{align}
m_{h^0}^2 =& \left( \lambda_1 + \lambda_2\cos^2\beta - \frac{(\lambda_4+\lambda_5\sin^2\beta)^2}{\lambda_3} + \ord{\frac{v^2}{M^2}} \right)v^2, \notag\\
m_{H^0_1}^2 =& \frac12\lambda_5M^2\sec2\beta - \frac{v^2}{2} \left( \lambda_2\cos^22\beta - \frac{\lambda_5^2\sin^22\beta\cos2\beta}{(\lambda_5-2\lambda_3\cos2\beta)} + \ord{\frac{v^2}{M^2}} \right), \notag\\
m_{H^0_2}^2 =& \lambda_3M^2 + v^2\left( \frac{(\lambda_4+\lambda_5\sin^2\beta)^2}{\lambda_3} - \frac{\lambda_5^2\sin^22\beta\cos2\beta}{\lambda_5-2\lambda_3\cos2\beta} + \ord{\frac{v^2}{M^2}} \right), \notag\\
m_{A^0}^2 =& \frac12\lambda_5M^2\sec2\beta - \frac12\lambda_2v^2,\notag\\
m_{H^\pm}^2 =& \frac{\lambda_5}{2}\left( M^2\sec2\beta + v^2\cos2\beta \right). \label{eq:scalarmasses}
\end{align}
We identify $h^0$ with the Standard-Model-like Higgs particle that has been observed at the LHC \cite{Aad:2012tfa,Chatrchyan:2012ufa}, since it is the only scalar that naturally has a mass at the electroweak scale.
A derivation of the scalar masses and the mixing angles can be found in \cref{a:scalarspectrum}.

The scalar parameters $\lambda_i$ are not completely free, as they have to satisfy three theoretical conditions.
These conditions and the constraints they yield are discussed in detail in \cref{a:scalarspectrum}.
The first constraint comes from the condition of vacuum stability, which means that the scalar potential in \cref{eq:simplifiedLETpotential} must be bounded from below for large field values.
This constraint puts lower bounds on the $\lambda_i$, which are given in \cref{a:vacuumstability}.
The second constraint comes from the condition of S-matrix unitarity, which basically means a conservation of probability in scattering processes.
This constraint implies that the $\lambda_i$ cannot be too large.
The corresponding parameter bounds are derived in detail in \cref{a:Smatrixunitarity}.
The third constraint comes from the condition that the potential has a minimum at the vev given in \cref{eq:Phiivev,eq:PhiRvev}.
This implies that the squared masses in \cref{eq:scalarmasses} should be positive.

\subsection{Fermion sector}
In the trinification model, the fermions obtain their masses from $H_1$; additional scalar fields $H_{Aq}$, $H_{Al}$ are introduced in order to obtain the correct mass hierarchies and mixing patterns for the down quarks and leptons.
Consider only the Yukawa coupling to $H_1$ in \cref{eq:trinificationYukawa}:
\begin{equation}
\lag_Y = -G_{\alpha\beta} \left( Q_R^\alpha H_1^TQ_L^\beta + \frac12\epsilon^{ijk}\epsilon_{lmn}L^i_lL^j_m(H_1)^k_n \right) \plushc \label{eq:triniSingleYukawa}
\end{equation}
Here we absorbed the dimensionless parameter $g_t$ into the generation matrix $G_{\alpha\beta}$.
After the breaking of the $G_{333}$ symmetry to $\LRgroup$, this Lagrangian results in mass terms of order $M_1$ for three Dirac fermions:
\begin{equation}
\psi_D = \begin{pmatrix} D \\ \hat{D}^\dagger \end{pmatrix},\quad
\psi_E = \begin{pmatrix} E^- \\ E^{+\dagger} \end{pmatrix},\quad
\psi_L = \begin{pmatrix} L^2_2 \\ L^{1\dagger}_1 \end{pmatrix}.
\end{equation}
These fields are integrated out from the Lagrangian.
As we did for the scalars, we can figure out the $\LRgroup$-representations of the remaining fermion components by working out the action of the gauge group generators in \cref{eq:su2su2u1generators} on them.
This yields the following representation assignments:
\begin{IEEEeqnarray}{rClrClrCl}
\begin{pmatrix} u \\ d \end{pmatrix} &\sim& (\brep{3},\rep{2},\rep{1}, \tfrac13), &
\begin{pmatrix} e^- \\ \nu \end{pmatrix} &\sim& (\rep{1}, \brep{2}, \rep{1}, -1), &
L^3_3 &\sim& (\rep{1}, \rep{1}, \rep{1}, 0), \nonumber\\
\begin{pmatrix} \hat{u} & \hat{d} \end{pmatrix} &\sim& (\rep{3}, \rep{1}, \brep{2}, -\tfrac13),\qquad &
\begin{pmatrix} e^+ & \hat\nu \end{pmatrix} &\sim& (\rep{1}, \rep{1}, \rep{2}, 1).\qquad &&& 
\end{IEEEeqnarray}
The field $L^3_3$ is a total gauge singlet and has no couplings to the Standard Model fields, hence we can omit it from our effective field theory.

Note that in order to combine the leptons and the Higgs field $H_1$ into a gauge singlet, we need to use the antisymmetric tensor (or equivalently $i\sigma_2$).
We can absorb it into a redefinition of the lepton fields, transforming a $\rep{2}$ into a $\brep{2}$ and vice versa.
Absorbing a minus sign into the phase of the $e^\pm$ fields, we define the fermionic field content of our effective field theory as follows:
\begin{IEEEeqnarray}{rCcClrCcCl}
Q_L &\equiv& \begin{pmatrix} u \\ d \end{pmatrix} &\sim& (\brep{3},\rep{2},\rep{1}, \tfrac13), &
L^- &\equiv& \begin{pmatrix} \nu \\ e^- \end{pmatrix} &\sim& (\rep{1},\rep{2}, \rep{1}, -1), \nonumber\\
Q_R &\equiv& \begin{pmatrix} \hat{u} & \hat{d} \end{pmatrix} &\sim& (\rep{3},\rep{1}, \brep{2}, -\tfrac13),\qquad &
L^+ &\equiv& \begin{pmatrix} \hat\nu & e^+ \end{pmatrix} &\sim& (\rep{1},\rep{1}, \brep{2}, 1). \label{eq:fermionreps}
\end{IEEEeqnarray}
Note that we use $Q_L$, $Q_R$ to denote the quark fields in both the trinification model and the LET model.
It should always be clear from the context which version they refer to.
The covariant derivatives of the fermions in eq.~\eqref{eq:fermionreps} are given by
\begin{align}
D_\mu Q_L =& \partial_\mu Q_L - ig_CG_\mu^a(\overline{T}_C^aQ_L)^b - ig_LW_{L\mu}^i T_L^iQ_L - \tfrac13ig'B_\mu Q_L, \notag\\
D_\mu Q_R =& \partial_\mu Q_R - ig_CG_\mu^a(Q_RT_C^{aT})^b - ig_RW_{R\mu}^i Q_R\overline{T}_R^{iT} + \tfrac13ig'B_\mu Q_R, \notag\\
D_\mu L^- =& \partial_\mu L^- - ig_LW_{L\mu}^i T_L^iL^- + ig'B_\mu L^-, \notag\\
D_\mu L^+ =& \partial_\mu L^+ - ig_RW_{R\mu}^i L^+\overline{T}_R^{iT} - ig'B_\mu L^+. \label{eq:LETfermioncovderiv}
\end{align}
After integrating out the heavy fields, the Yukawa Lagrangian in \cref{eq:triniSingleYukawa} becomes
\begin{equation}
\lag_Y = -G_{\alpha\beta}\left( Q_R^\alpha \Phi_1^TQ_L^\beta + L^{+\alpha}\Phi_1^T L^{-\beta} \right) \plushc \label{eq:yukawalag}
\end{equation}
After spontaneous symmetry breaking, this interaction yields Dirac masses for all fermions, where each left-handed fermion is combined with the corresponding right-handed version.
The universal generation matrix $G_{\alpha\beta}$ results in the same mass hierarchies for the up quarks, the down quarks, the neutrinos, and the charged leptons.
Moreover, the up quarks (down quarks) obtain the same masses as the neutrinos (charged leptons), and the CKM matrix is a unit matrix at this point.
Hence the interaction in \cref{eq:yukawalag} is not sufficient to describe the fermion spectrum in accordance with experiment (see \cref{t:fermionmasses}).
We could try to solve this problem by adding low-energy versions of the interactions with the components of $H_{Aq}$, $H_{Al}$.
However, this is a non-trivial task: in the trinification model, mixings with the heavy fermions are important to describe the masses and mixings of the lighter fermions correctly \cite{Stech:2014tla}.
In the LET model, the heavy fermions have been integrated out already, so the mixing needs to be accounted for in some other way.
Also, renormalisation effects could become important at low energy scales.
Accounting for the mixing and renormalisation effects would require us to introduce additional parameters to describe the fermion sector.
Moreover, our analysis would be complicated by the inclusion of the Higgs fields $H_{Aq}$, $H_{Al}$, since we would need to account for mixing with the other Higgs fields, unless this mixing is suppressed for some reason (as is assumed in e.g.\ \cite{Stech:2014tla}).

Rather than adding baggage to the LET model and complicating our analysis with more free parameters, we restrict ourselves to the single Yukawa term in \cref{eq:yukawalag} and fit the free parameters such that the top- and bottom-quark masses are reproduced correctly.
These fermions are the most relevant to our analysis, since experimental searches for new physics often focus on decays involving the heaviest generation (see e.g.\ the searches discussed in \cref{s:gaugebosonconstraints,s:newHiggses}).
Therefore we ignore the lighter generations and the neutrinos, and restrict ourselves to the heaviest generation.
We assume that any additional new physics, necessary to describe the fermion masses and mixings correctly, does not influence the phenomenology of the scalar particles.
Among the Dirac fermions, we consider only the top and bottom:
\begin{equation}
\psi_t = \begin{pmatrix} t \\ \hat{t}^\dagger \end{pmatrix},\qquad
\psi_b = \begin{pmatrix} b \\ \hat{b}^\dagger \end{pmatrix}.
\end{equation}
Fitting the free parameters to the top- and bottom-quark masses, we find $v_1 = 246$ GeV, $b_1 = 4.09$ GeV, or equivalently $v = 246$ GeV, $\beta = 0.0166$ (see \cref{a:fermionmasseig} for the details).

\clearpage
\section{Comparison to similar models beyond the SM}\label{s:comparison}
The literature contains a plethora of models that resemble the simplified LET model.
The most notable ones are the two-Higgs-doublet model (2HDM) \cite{Lee:1973iz} and left-right-symmetric (LR-symmetric) models \cite{Pati:1974yy,Mohapatra:1974hk,Senjanovic:1975rk}, which have both been studied extensively.
However, because of the origin in the trinification model, the simplified LET model has features that distinguish it from the 2HDM and LR-symmetric models.
Most notably, it has fewer parameters than those other models, and therefore it is more predictive.
In this section, we give an overview of the 2HDM and LR-symmetric models, their similarities to the simplified LET model and the features that distinguish the simplified LET model from the other models.

\subsection{The two-Higgs-doublet model}\label{s:2HDM}
The two-Higgs-doublet model (2HDM) \cite{Lee:1973iz} (see \cite{Branco:2011iw} for a recent review) is an extension of the Standard Model in which the scalar sector has been extended by an additional $SU(2)_L$ doublet.
The model allows for additional sources of $CP$-violation to account for the baryon-antibaryon-asymmetry in the universe \cite{Trodden:1998qg}, provides additional neutral scalar particles that are viable dark matter candidates \cite{Gong:2012ri}, and allows for radiative neutrino mass generation \cite{Ma:2006km}.
However, the main motivation for studying this extension is its appearance in many different models of physics beyond the Standard Model.
It provides a low-energy description of various models such as supersymmetry (see e.g.~\cite{Martin:1997ns} for a review), composite Higgs models \cite{Kaplan:1983sm}, and little Higgs models \cite{ArkaniHamed:2001nc}.

The Minimal Supersymmetric Standard Model (MSSM) \cite{Djouadi:2005gj,Carena:2002es} is a well-known example of a 2HDM.
In the Standard Model, a single $SU(2)_L$-doublet scalar field $\phi$ is sufficient to give mass to all particles.
In particular, $\phi$ gives mass to the down-type fermions, and its charge conjugate $\phi^c\equiv i\sigma_2\phi^*$ gives mass to the up-type fermions.
However, the structure of supersymmetric theories does not allow for a charge conjugate to appear.
The non-gauge interactions of such theories are determined by a superpotential, which is an analytical function of the scalar fields.
Hence charge conjugates cannot appear, and a second scalar doublet is necessary to give mass to the up-type fermions \cite{Martin:1997ns}.
Also, the presence of a second scalar doublet is necessary to ensure the cancellation of gauge anomalies.
The conditions for this cancellation include $\tr{T_3^2Y} = 0$ and $\tr{Y^3} = 0$, where $T_3$ is the third component of isospin, $Y$ is the hypercharge, and the trace runs over all fermions.
These conditions are satisfied by the fermion content of the Standard Model.
However, each scalar field introduces a fermionic superpartner with the same isospin and hypercharge.
Therefore, the fermionic superpartner of a single scalar doublet would spoil the cancellation conditions.
However, if one scalar doublet with $Y = \frac12$ and another with $Y = -\frac12$ are present, the cancellation conditions are satisfied \cite{Martin:1997ns}.

Two-Higgs-doublet models can also appear as effective low-energy descriptions of composite-Higgs models \cite{Kaplan:1983sm}.
In composite-Higgs models, the Higgs doublet is not a fundamental field.
Instead, the gauge group is extended to include another strong interaction, and one introduces new heavy fermion fields that are charged under this new interaction.
The electroweak symmetry is broken by condensates of these fermions, the Standard-Model fermions obtain masses through four-fermion interactions with the heavy fermions, and the Higgs boson appears in the model as a bound state of fermions.
For a particular choice of the four-fermion interactions, the effective field theory at low energies is a 2HDM \cite{Luty:1990bg,Burdman:2011fw}.

Little-Higgs models (see e.g.~\cite{Schmaltz:2005ky,Perelstein:2005ka} for reviews) constitute a class of models in which a new strongly interacting sector is introduced as well.
The Higgs boson is presumed to be a pseudo-Goldstone boson corresponding to a spontaneously broken global symmetry of this new sector.
The global symmetry is broken explicitly but only collectively, i.e.\ it is only broken if two or more of the gauge and Yukawa couplings of the Higgs are nonvanishing.
Therefore any diagram that contributes to the Higgs mass must involve at least two of these couplings.
Since there are no quadratically divergent diagrams with two or more couplings, the Higgs mass is stabilised against large radiative corrections.
In this scenario, a light Higgs requires no fine-tuning if the scale of the new strongly interacting physics is of order 10 TeV.
Such light pseudo-Goldstone bosons are known as `little Higgses'.
Little-Higgs models can vary in symmetry and field content, which can be represented graphically as a diagram called a `moose'.
In the Minimal Moose setup \cite{ArkaniHamed:2002qx}, physics below a TeV is described by a 2HDM with an additional complex weak triplet and a complex singlet.

The 2HDM contains two complex Higgs doublets $\phi_1$, $\phi_2$, both in the representation of the Standard-Model gauge group $(\rep{1},\rep{2},\frac12)$ of $\SM$.
Together, the Higgs doublets contain eight fields, which we parametrise as follows:
\begin{equation}
\phi_i = \begin{pmatrix} \phi_i^+ \\ \frac{v_i + h^0_i + ia^0_i}{\sqrt2} \end{pmatrix},\qquad i=1,2. \label{eq:2hdmHiggsParametrisation}
\end{equation}
The most general gauge-invariant $CP$-conserving renormalisable scalar potential is
\begin{align}
V_\text{2HDM}(\phi_1,\phi_2) =& \frac{\Lambda_1}{2}(\phi_1^\dagger\phi_1)^2 + \frac{\Lambda_2}{2}(\phi_2^\dagger\phi_2)^2 + \Lambda_3(\phi_1^\dagger\phi_1)(\phi_2^\dagger\phi_2) + \Lambda_4|\phi_1^\dagger\phi_2|^2 \notag\\
&+ \left( \frac{\Lambda_5}{2}(\phi_1^\dagger\phi_2)^2 + \Lambda_6(\phi_1^\dagger\phi_1)(\phi_1^\dagger\phi_2) + \Lambda_7(\phi_2^\dagger\phi_2)(\phi_1^\dagger\phi_2) \plushc \right) \notag\\
&+ m_{11}^2\phi_1^\dagger\phi_1 + m_{22}^2\phi_2^\dagger\phi_2 - m_{12}^2\left( \phi_1^\dagger\phi_2 \plushc \right), \label{eq:2hdmpotential}
\end{align}
where all coupling parameters are taken to be real.
The parameters $m_{11}^2$, $m_{22}^2$ are fixed by the minimalisation of the scalar potential.
Usually one imposes a $Z_2$-symmetry on the Higgs doublets under which $\phi_1\rightarrow-\phi_1$, $\phi_2\rightarrow\phi_2$.
This symmetry forbids the terms $\Lambda_6$, $\Lambda_7$, and $m_{12}^2$; the latter is usually retained in the potential since it breaks the $Z_2$-symmetry only softly.
Hence, the scalar potential contains six free parameters in this setup.

As in the Standard Model, three of the fields in \cref{eq:2hdmHiggsParametrisation} become the Goldstone modes $G^\pm$, $G^0$ that give mass to the $W^\pm$, $Z^0$ bosons after spontaneous symmetry breaking.
Hence there are five physical scalars: two $CP$-even states $h^0$, $H^0$, one $CP$-odd state $A^0$, and a pair of charged states $H^\pm$.
The state $h^0$ is interpreted as the observed scalar with $m_{h^0} = 126$ GeV.

Both vacuum expectation values $v_{1,2}$ contribute to the gauge boson masses, and they are therefore restricted by the relation $v_1^2 + v_2^2 = v^2 \equiv (246\text{ GeV})^2$.
It is useful to parametrise the vevs as
\begin{equation}
v_1 = v\cos\beta,\qquad v_2 = v\sin\beta.
\end{equation}
The angle $\beta$ is a free parameter of the model.
It is the rotation angle that diagonalises the squared-mass matrices of the $CP$-odd scalars and the charged scalars:
\begin{equation}
\begin{pmatrix} G^\pm \\ H^\pm \end{pmatrix} = R(\beta) \begin{pmatrix} h_1^\pm \\ h_2^\pm \end{pmatrix},\qquad
\begin{pmatrix} G^0 \\ A^0 \end{pmatrix} = R(\beta) \begin{pmatrix} a_1^\pm \\ a_2^\pm \end{pmatrix}. \label{eq:2HDMscalars}
\end{equation}
Here we defined $h_i^- = (h_i^+)^*$ and
\begin{equation}
R(\beta) \equiv \begin{pmatrix} \cos\beta & \sin\beta \\ -\sin\beta & \cos\beta \end{pmatrix}.
\end{equation}
The squared-mass matrix of the $CP$-even states is diagonalised by an angle $\alpha$:
\begin{equation}
\begin{pmatrix} H^0 \\ h^0 \end{pmatrix} = R(\alpha) \begin{pmatrix} h^0_1 \\ h^0_2 \end{pmatrix}. \label{eq:alpha2hdm}
\end{equation}
The angle $\alpha$ is a complicated function of the parameters of the scalar potential and the vevs of the Higgs doublets (see e.g.\ \cite{Kanemura:2004mg} for the complete expression).
However, in phenomenological studies one usually trades the five dimensionless couplings $\Lambda_i$ in the scalar potential for the four scalar masses $m_{h_0}$, $m_{H^0}$, $m_{A^0}$, $m_{H^\pm}$ and the mixing angle $\alpha$.

The presence of an additional Higgs doublet generally leads to additional Yukawa couplings.
These could give rise to tree-level flavour-changing neutral-current (FCNC) processes, which are strongly constrained by experiment.
In order to eliminate tree-level FCNC couplings, one imposes the aforementioned $Z_2$ symmetry on the Higgs doublets.
In addition, one allows each fermion family to couple to only one Higgs doublet.
This results in four different 2HDM setups:
\begin{itemize}
\item \emph{type I}, in which all fermions couple to $\phi_2$ only;
\item \emph{type II}, in which the up-type (down-type) fermions obtain mass from the vev of $\phi_2$ ($\phi_1$);
\item \emph{lepton-specific}, which has a type-I quark sector and a type-II lepton sector;
\item \emph{flipped}, which has a type-II quark sector and a type-I lepton sector.
\end{itemize}

\subsection{Mapping the simplified LET model onto the 2HDM}\label{s:mapping}
As we will see in \cref{s:lightHiggs,s:newHiggses}, the simplified LET model closely resembles the 2HDM, with respect to both particle content and the resulting deviations of several couplings from their Standard-Model values.
In some respects, the model looks like a simplified version of the 2HDM.
It is easy to see why at the Lagrangian level.
To this end, we ignore $\Phi_R$ for the moment: it has a vev $M$ much larger than the vevs $v_1$, $b_1$ of $\Phi_1$, hence mixing of $\Phi_1$-components with $\Phi_2$-components will be of order $\frac{v}{M} \ll 1$.
Then we are left with a $2\times2$ matrix $\Phi_1$ that is an antidoublet under $SU(2)_L$.
Each column of this matrix is in itself an $SU(2)_L$-antidoublet, and can be parametrised in terms of two $SU(2)_L$ doublets $\phi_1$, $\phi_2$ as follows:
\begin{equation}
\Phi_1 \equiv \begin{pmatrix} \Phi^0_{11} & \Phi^-_{12} \\ \Phi^+_{21} & \Phi^0_{22} \end{pmatrix} = (i\sigma_2\phi_1, \phi_2^*),\qquad
\phi_1 \equiv \begin{pmatrix} -\Phi^+_{21} \\ \Phi^0_{11} \end{pmatrix},\quad
\phi_2 \equiv \begin{pmatrix} \Phi^+_{12} \\ (\Phi^0_{22})^* \end{pmatrix}. \label{eq:Phi1doublets}
\end{equation}
Here we defined $(\Phi_{ij}^\pm)^* = \Phi_{ij}^\mp$.
If we set $\Phi_R=0$ in the scalar potential in \cref{eq:scalarpotential}, we can rewrite the entire scalar potential in terms of $\phi_1$, $\phi_2$ only:
\begin{align}
V(\Phi_1,0) =& \frac{\lambda_1+\lambda_2}{2}\left( (\phi_1^\dagger\phi_1)^2 + (\phi_2^\dagger\phi_2)^2 \right) + (\lambda_1+\lambda_2)(\phi_1^\dagger\phi_1)(\phi_2^\dagger\phi_2) - \lambda_2|\phi_1^\dagger\phi_2|^2 \notag\\
&+ \mu^2_{11}(\phi_1^\dagger\phi_1 + \phi_2^\dagger\phi_2) + \mu^2_1(\phi_1^\dagger\phi_2 \plushc). \label{eq:2doubletLRpotential}
\end{align}
By comparing \cref{eq:2hdmpotential,eq:2doubletLRpotential}, we see that the simplified LET model without $\Phi_R$ corresponds to a 2HDM with the following constraints on the scalar parameters:
\begin{IEEEeqnarray}{rClrCl}
\Lambda_1 = \Lambda_2 = \Lambda_3 &=& \lambda_1+\lambda_2,\qquad & \Lambda_4 &=& -\lambda_2,\qquad \Lambda_5 = \Lambda_6 = \Lambda_7 = 0, \notag\\
m_{11}^2 = m_{22}^2 &=& \mu_{11}^2,\qquad & m_{12}^2 &=& -\mu_1^2. \label{eq:2HDMmapping}
\end{IEEEeqnarray}
Note that $\phi_1$ and $\phi_2$ are part of a single field $\Phi_1$, and therefore must have equal mass terms.
For the same reason, the parameter $m_{12}^2$ cannot be eliminated by imposing a $Z_2$-symmetry on the potential, since $\phi_1$ and $\phi_2$ must have the same charge under such a symmetry.

Analogously to $\Phi_1$, we can regard each column in the right-handed doublet $\Phi_R$ as a complex $SU(2)_L$ singlet, one of which has electromagnetic charge $+1$ and the other being neutral:
\begin{equation}
\Phi_R = \begin{pmatrix} S_+ & S_0 \end{pmatrix},\qquad S_+ \sim (\rep{1},\rep{1},1),\quad S_0 \sim (\rep{1},\rep{1},0). \label{eq:PhiRsinglets}
\end{equation}
Rewriting the scalar potential in \cref{eq:scalarpotential} in terms of $\phi_1$, $\phi_2$, $S_+$, $S_0$ we find
\begin{align}
V(\Phi_1,\Phi_R) =& V(\Phi_1,0) + V_2(\Phi_1,\Phi_R), \notag\\
V_2(\Phi_1,\Phi_R) =& \frac{\lambda_3}{2}\big( |S_+|^4 + |S_0|^4 \big) + \lambda_3|S_+|^2|S_0|^2 + \mu_R^2\big( |S_+|^2 + |S_0|^2 \big) \notag\\
&+ (\lambda_4+\lambda_5)\left( (\phi_1^\dagger\phi_1)|S_+|^2 + (\phi_2^\dagger\phi_2)|S_0|^2 \right) \notag\\
&+ \lambda_4\left( (\phi_1^\dagger\phi_1)|S_0|^2 + (\phi_2^\dagger\phi_2)|S_+|^2 \right) \notag\\
&+ \lambda_5\left( (\phi_1^{c\dagger}\phi_2)S_-S_0 \plushc \right).
\end{align}
Here we defined $S_-=(S_+)^*$ and the charge conjugate is defined by $\phi_1^c\equiv i\sigma_2\phi_1^*$.
We see that our model setup can be viewed as a model with two Higgs doublets and two singlets, one of which is charged.
In this picture, the additional structure that originates from the factor $SU(2)_R$ in the gauge group leads to constraints on the scalar potential.

We can rewrite the Yukawa sector of the simplified LET model in terms of $\phi_1$, $\phi_2$ as well.
The Lagrangian in \cref{eq:yukawalag} becomes
\begin{align}
\lag_Y =& -G_{\alpha\beta}\left( (Q_R^\alpha)_j (\Phi_1)^i_j (Q_L^\beta)^i + (L^{+\alpha})_j (\Phi_1)^i_j (L^{-\beta})^i \right) \plushc \notag\\
=& -G_{\alpha\beta}\Big( (Q_R^\alpha)_1(i\sigma_2\phi_1)^i (Q_L^\beta)^i + (Q_R^\alpha)_2 (\phi_2^*)^i (Q_L^\beta)^i \notag\\
&\phantom{-G_{\alpha\beta}\Big(} + (L^{+\alpha})_1(i\sigma_2\phi_1)^i (L^{-\beta})^i + (L^{+\alpha})_2 (\phi_2^*)^i (L^{+\beta})^i \Big) \plushc \notag\\
=& -G_{\alpha\beta}\Big( \hat{u}^\alpha (i\sigma_2\phi_1)^i (Q_L^\beta)^i + \hat{d}^\alpha (\phi_2^*)^i (Q_L^\beta)^i \notag\\
&\phantom{-G_{\alpha\beta}\Big(} + \hat\nu^\alpha (i\sigma_2\phi_1)^i (L^{-\beta})^i + e^{+\alpha} (\phi_2^*)^i (L^{+\beta})^i \Big) \plushc
\end{align}
Note that the up-type fermions couple only to $\phi_1$, whereas the down-type fermions couple only to $\phi_2$.
Hence the simplified LET model resembles a constrained type-II 2HDM setup.

For the sake of completeness, we also map the parameters of the simplified LET model onto the parameters of the 2HDM.
First we consider the angle $\beta$.
From now on, we will label the similar 2HDM parameter as $\beta_\text{2HDM}$ and the 2HDM doublets as $\phi_\text{1,2HDM}$, $\phi_\text{2,2HDM}$ for the sake of unambiguity.
The angle $\beta_\text{2HDM}$ is defined by the relations
\begin{equation}
\vev{\phi_\text{1,2HDM}} = v\cos\beta_\text{2HDM},\qquad \vev{\phi_\text{2,2HDM}} = v\sin\beta_\text{2HDM}.
\end{equation}
Here the scalar doublets are defined such that $\phi_\text{1,2HDM}$ ($\phi_\text{2,2HDM}$) gives mass to the down-type (up-type) fermions.
In the simplified LET model it is the other way around: we define
\begin{equation}
\vev{\phi_1} = v\cos\beta,\qquad \vev{\phi_2} = v\sin\beta,
\end{equation}
but here $\phi_1$ ($\phi_2$) gives mass to up-type (down-type) fermions.
This leads us to identify $\phi_1$ ($\phi_2$) with $\phi_\text{2,2HDM}$ ($\phi_\text{1,2HDM}$) and gives the mapping
\begin{align}
\cos\beta_\text{2HDM} =& \sin\beta,\qquad \sin\beta_\text{2HDM} = \cos\beta, \notag\\
\Rightarrow \beta_\text{2HDM} =& \frac{\pi}{2} - \beta. \label{eq:beta2HDM}
\end{align}

Now consider the mixing angles of the $CP$-even states.
In the 2HDM, the mixing angle $\alpha$ is defined by \cref{eq:alpha2hdm}:
\begin{equation}
\begin{pmatrix} H^0_\text{2HDM} \\ h^0_\text{2HDM} \end{pmatrix} =
\begin{pmatrix} \cos\alpha & \sin\alpha \\ -\sin\alpha & \cos\alpha \end{pmatrix}
\begin{pmatrix} h^0_1 \\ h^0_2 \end{pmatrix}. \label{eq:alpha2HDM0}
\end{equation}
Here we relabelled the $CP$-even mass eigenstates $H^0_\text{2HDM}$, $h^0_\text{2HDM}$ to avoid ambiguities in the following.
We can bring the analogous expression of the simplified LET model into a similar form by using the mixing angles in \cref{eq:CPevenMixingAngles} and neglecting terms of $\ord{\frac{v}{M}}$:
\begin{equation}
\alpha_1 = \beta + \ord{\frac{v^2}{M^2}},\qquad \alpha_{2,3} = \ord{\frac{v}{M}}. \label{eq:alpha2HDM1}
\end{equation}
Using these approximations, the $CP$-even states in \cref{eq:Higgsmasseig} can be written as
\begin{equation}
\begin{pmatrix} h^0 \\ H_1^0 \\ H_2^0 \end{pmatrix} = \begin{pmatrix} \cos\alpha_1 h^0_{1,11} + \sin\alpha_1 h^0_{1,22} \\ -\sin\alpha_1 h^0_{1,11} + \cos\alpha_1 h^0_{1,22} \\ h^0_R \end{pmatrix} + \ord{\frac{v}{M}}. \label{eq:2HDMmassEigMapping}
\end{equation}
In this approximation, the fermiophobic state $H_2^0$ does not mix with the other states.
The states $h^0$ and $H^0_1$ are obtained by rotating the states $h^0_{1,11}$, $h^0_{1,22}$ by an angle $\alpha_1$.
Therefore we can make the following identifications between the $CP$-even fields of the 2HDM and those of the simplified LET model: $h^0_\text{2HDM} \leftrightarrow h^0$, $H^0_\text{2HDM} \leftrightarrow H^0_1$, $h^0_1 \leftrightarrow h^0_{1,22}$, $h^0_2 \leftrightarrow h^0_{1,11}$.
The 2HDM has no equivalent of the field $h^0_R$.
Thus \cref{eq:alpha2HDM0,eq:alpha2HDM1,eq:2HDMmassEigMapping} allows us to map $\alpha_1$ onto the $CP$-even mixing angle $\alpha$ of the 2HDM:
\begin{equation}
\alpha = -\alpha_1 = -\beta + \ord{\frac{v^2}{M^2}}. \label{eq:alpha2HDM2}
\end{equation}

\subsection{2HDM vs.\ simplified LET model}
As we have seen, the Lagrangian of the simplified LET model can be rewritten in such a way that it resembles the Lagrangian of a type-II 2HDM.
The $SU(2)_R$ gauge symmetry leads to constraints that turn our scalar potential into a simplified version of the 2HDM potential.
However, this additional structure leads to important differences with the 2HDM as well.
The additional factor in the gauge group and the right-handed doublet $\Phi_R$ give rise to new features that cannot be encompassed in the 2HDM.
\begin{itemize}
\item In addition to the two $SU(2)_L$-doublets, the simplified LET model includes an $SU(2)_R$-doublet $\Phi_R$. Its charged and $CP$-odd components do not give rise to additional charged and $CP$-odd states; instead they give mass to the new heavy vector bosons $W^{\prime\pm}$, $Z'$ of $SU(2)_R$ (after mixing with the components of $\Phi_1$). However, the $CP$-even component of $\Phi_R$ does give rise to an additional physical particle $H_2^0$. Moreover, since $\Phi_R$ does not couple to fermions, this new state becomes fermiophobic. As can be seen in the scalar mass expressions in \cref{eq:scalarmasses}, its mass can be tuned independently from the masses of the scalars $H_1^0$, $A^0$, $H^\pm$. This results in several possible mass hierarchies that cannot appear in the 2HDM (see \cref{s:benchmarkpoints}).
\item In the 2HDM, there are three distinct Yukawa matrices for the up quarks, down quarks, and charged leptons. However, in the LET model (and hence in the simplified LET model) there is only one Yukawa matrix $G_{\alpha\beta}$. The reason is the fact that the right-handed up-quarks and the right-handed down-quarks belong to different representations of the gauge group in the 2HDM (as in the Standard Model). This allows one to write down separate Yukawa terms for the up and down-quarks. However, in the LET model the right-handed up and down quarks are components of an $SU(2)_R$-antidoublet $Q_R$, just like the left-handed up and down quarks are components of an $SU(2)_L$-doublet $Q_L$. This implies that there is no gauge-invariant way to write down separate Yukawa terms for the components.

For similar reasons, the leptons are coupled to the scalar sector with the same Yukawa matrix as the quarks. We consider the LET model to be a low-energy description of the trinification model, which in turn comes from an $E_6$ GUT. At high energies where the $E_6$ symmetry is unbroken, the quarks and leptons are components of the same representation of the gauge group, and therefore they couple to the scalars with the same Yukawa matrix \cite{Stech:2014tla}. At low energies, the fermions will generally have different Yukawa matrices because of renormalisation-group (RG) running. We assume RG-effects to be negligibly small; an analysis including these effects is beyond the scope of this thesis.
\item In the 2HDM, the ratio $\tan\beta = \frac{v_2}{v_1}$ is a free parameter. However, in the LET model (and hence in the simplified LET model) this ratio is constrained by the quark masses (see \cref{a:fermionmasseig}). The reason is that the LET model has a single Yukawa matrix $G_{\alpha\beta}$. In the flavour basis where this matrix is diagonal, the top-quark mass is given by $m_t = \frac{1}{\sqrt2}G_{33}v_1 = \frac{1}{\sqrt2}G_{33}v\cos\beta$, whereas the bottom-quark mass is given by $m_b = \frac{1}{\sqrt2}G_{33}b_1 = \frac{1}{\sqrt2}G_{33}v\sin\beta$. Hence $\tan\beta$ is equal to the ratio $\frac{m_b}{m_t}$ at tree level.
\item Similarly to the above, the mixing angles of the $CP$-even scalars are not free parameters either. As was mentioned before, the 2HDM allows one to choose the $CP$-even mixing angle $\alpha$ and the scalar masses as the free parameters (rather than the quartic scalar couplings). However, in the simplified LET model the three $CP$-even mixing angles only depend on the quartic scalar couplings at subleading order in $\frac{v}{M}$. At leading order, they are constants (see \cref{a:scalarmasseigenstates}).
\end{itemize}

\subsection{Left-right-symmetric models}
Left-right-symmetric models (LRSMs) \cite{Pati:1974yy,Mohapatra:1974hk,Senjanovic:1975rk} are extensions of the Standard Model in which the gauge group has been enlarged by an additional factor $SU(2)_R$ in order to incorporate a left-right symmetry in the Lagrangian.
This implies a larger particle content as well: the fermion sector is extended by right-handed neutrinos, the Higgs field responsible for electroweak symmetry breaking obtains extra degrees of freedom, and additional scalar fields are introduced to break the left-right symmetry.

The main motivation for studying these models is to address the origin of parity violation in the weak interactions.
In the Standard Model, this experimental fact is accounted for by hand: the left-handed and right-handed fermions are assigned to different representations of the gauge group, such that only the left-handed fermions couple to the $W$-bosons.
In LRSMs however, one starts with a Lagrangian that is LR-symmetric: the left-handed fermions couple to the $W_L$-bosons of $SU(2)_L$, and the right-handed fermions couple to the newly introduced $W_R$-bosons of $SU(2)_R$.
Parity violation occurs through spontaneous symmetry breaking: although the Lagrangian is LR-symmetric, the vacuum is not.
The $W_R$-bosons obtain large masses, and right-handed $V+A$ interactions become suppressed by the $W_R$-mass.
Thus the weak interactions naturally have the observed left-handed $V-A$ structure at low energies.

LRSMs also explain the generation of neutrino masses via the seesaw mechanism \cite{Minkowski:1977sc,Mohapatra:1979ia} (see e.g.\ \cite{2007ConPh..48..195K,Drewes:2013gca} for recent reviews).
The observation of neutrino oscillations (see e.g.\ \cite{Bilenky:2014eza} for a recent review) has established that neutrinos are massive, and their masses are very small (below an eV).
In the Standard Model however, neutrinos are strictly massless: there is no possible Dirac mass term because there are no right-handed neutrinos.
A Majorana mass for the left-handed neutrinos $\nu_L$ cannot be generated by the Higgs mechanism, as the corresponding Yukawa interaction would violate electroweak gauge symmetry as well as lepton number conservation.
Thus the existence of nonzero neutrino masses calls for an extension of the Standard Model.
LRSMs include a right-handed neutrino $\nu_R$, which allows for Dirac mass terms for the neutrinos.
Since the right-handed neutrino is a total singlet with respect to $\SM$, a Majorana mass term is allowed as well.
Hence after electroweak symmetry breaking, the most general neutrino mass term in the Lagrangian looks like \cite{2007ConPh..48..195K}
\begin{equation}
\lag \supset -\frac12 \begin{pmatrix} \overline{\nu_L} & \overline{\nu^c_R} \end{pmatrix}
\begin{pmatrix} 0 & m_{LR} \\ m_{LR}^T & M_{RR} \end{pmatrix}
\begin{pmatrix} \nu_L^c & \nu_R \end{pmatrix} \plushc \label{eq:seesaw}
\end{equation}
Here the superscript `c' denotes $CP$-conjugation, $m_{LR}$ is a Dirac mass matrix, and $M_{RR}$ is a Majorana mass matrix.
The Dirac mass matrix arises from electroweak symmetry breaking and therefore its eigenvalues are of the order of the electroweak scale.
On the other hand, the Majorana mass matrix does not originate from electroweak symmetry breaking, and therefore its eigenvalues can be arbitrarily large.
For large eigenvalues of $M_{RR}$, diagonalisation of the mass matrix in \cref{eq:seesaw} yields effective Majorana masses for the left-handed neutrinos, with a Majorana mass matrix \cite{2007ConPh..48..195K}
\begin{equation}
m_{LL} \approx m_{LR} M_{RR}^{-1} m_{LR}^T.
\end{equation}
Since the eigenvalues of $M_{RR}$ are assumed to be very large, the resulting eigenvalues of $m_{LL}$ are very small.
Hence we have a natural explanation for small neutrino masses.

LRSMs appear in various contexts.
They can appear as low-energy EFTs of non-supersymmetric GUTs that predict gauge coupling unification at a high energy scale \cite{PhysRevD.89.035002}.
The LR-symmetric gauge group can also be used as a basis for supersymmetric inflationary models \cite{PhysRevD.89.065032}.
The left-right-symmetry-breaking scale can be as low as the TeV scale \cite{PhysRevD.89.035002,PhysRevD.88.093010}, which means that new physics may appear at the LHC.
Also, since LRSMs give Majorana masses to the neutrinos, they predict lepton number violation and therefore the possibility of neutrinoless double-$\beta$ decay \cite{Mohapatra:1980yp} (see \cite{Rodejohann:2011mu,Bilenky:2012qi} for recent reviews).
Hence these models are not only theoretically interesting, but can also be put to the test very soon.

The term `minimal LRSM' appears in the literature, but it can refer to different models.
Left-right-symmetric models have in common that they are based on the gauge group $\LRgroup$.
However, the field content varies from model to model.
Usually `minimal LRSM' refers to a setup with one Higgs bidoublet $\phi \sim (\rep{1},\rep{2},\rep{2},0)$, one left-handed triplet $\Delta_L \sim (\rep{1},\rep{3},\rep{1},2)$, and one right-handed triplet $\Delta_R \sim (\rep{1},\rep{1},\rep{3},2)$ \cite{Mohapatra:1979ia,Mohapatra:1980yp,Mohapatra:1977mj}.
These fields obtain the following vevs:
\begin{equation}
\vev{\phi} = \begin{pmatrix} \kappa & 0 \\ 0 & \kappa' \end{pmatrix},\qquad
\vev{\Delta_L} = \begin{pmatrix} 0 \\ 0 \\ 0 \end{pmatrix},\qquad
\vev{\Delta_R} = \begin{pmatrix} 0 & 0 & v \end{pmatrix}.
\end{equation}
Here $\vev{\Delta_R}$ is responsible for breaking the left-right symmetry, after which the Lagrangian is still symmetric under the gauge group $\SM$.
Then $\vev{\phi}$ breaks the electroweak symmetry further down to $SU(3)_C\times U(1)_\text{em}$.
The field $\Delta_L$ does not break any symmetries, but is necessary in order to make the Lagrangian LR-symmetric.

The left-handed quarks are the same as in the Standard Model, but the right-handed quarks are grouped into a right-handed doublet:
\begin{equation}
Q_L = \begin{pmatrix} u_L \\ d_L \end{pmatrix} \sim (\rep{3},\rep{2},\rep{1},\frac13),\qquad
Q_R = \begin{pmatrix} u_R \\ d_R \end{pmatrix} \sim (\rep{3},\rep{1},\rep{2},\frac13).
\end{equation}
Additionally, the model contains right-handed neutrinos $\nu_R$, so that the leptons form left- and right-handed doublets as well:
\begin{equation}
\psi_L = \begin{pmatrix} \nu_L \\ e_L \end{pmatrix},\qquad
\psi_R = \begin{pmatrix} \nu_R \\ e_R \end{pmatrix},
\end{equation}
where $\psi_L \sim (\rep{1},\rep{2},\rep{1},-1)$ and $\psi_R \sim (\rep{1},\rep{1},\rep{2},-1)$.
The vev of $\Delta_R$ creates a Majorana mass term for the right-handed neutrino \cite{Mohapatra:1980yp}, as necessary for the seesaw mechanism.
A discrete left-right-symmetry is imposed on the Lagrangian; the fields transform under this symmetry as
\begin{IEEEeqnarray}{rClrClrCl}
\phi &\leftrightarrow& \phi^\dagger,\qquad& Q_L &\leftrightarrow& Q_R,\qquad& W_L &\leftrightarrow& W_R, \notag\\
\Delta_L &\leftrightarrow& \Delta_R,\qquad& \psi_L &\leftrightarrow& \psi_R.\qquad& &&
\end{IEEEeqnarray}
Here $W_{L,R}$ are the gauge bosons of $SU(2)_{L,R}$.
This symmetry also implies that the $SU(2)_L$ and $SU(2)_R$ gauge couplings are equal: $g_L = g_R$.

Alternatively, one could choose a set of Higgs doublets $\chi_L \sim (\rep{1},\rep{2},\rep{1},1)$, $\chi_R \sim (\rep{1},\rep{1},\rep{2},1)$ instead of the triplets $\Delta_L$, $\Delta_R$ \cite{Senjanovic:1975rk,Akhmedov:1995ip}.
Under the left-right symmetry, they transform as $\chi_L \leftrightarrow \chi_R$.
The most general gauge-invariant, LR-symmetric, renormalisable scalar potential can be found in e.g.\ \cite{Senjanovic:1975rk}.
In part of the parameter space of the model, the minimum of the scalar potential has the form
\begin{equation}
\vev{\chi_L} = \begin{pmatrix} 0 \\ 0 \end{pmatrix},\qquad
\vev{\chi_R} = \begin{pmatrix} 0 & v_R \end{pmatrix},
\end{equation}
where $v_R \gg \kappa, \kappa'$.
Thus $\chi_R$ breaks the left-right symmetry.
Note that these doublets cannot generate a Majorana mass term for the right-handed neutrino, i.e.\ the standard seesaw mechanism does not work here.
Without any additions, the neutrinos would be Dirac particles, with masses of the same order as the electron mass.
However, small neutrino masses can be explained by the inverse seesaw mechanism \cite{Akhmedov:1995ip,PhysRevLett.56.561,Dias:2012xp}: one introduces a gauge-singlet fermion $S_L \sim (\rep{1},\rep{1},\rep{1}, 0)$ with Yukawa couplings to the leptons and $\chi_{L,R}$.
After spontaneous symmetry breaking, they develop the mass terms
\begin{equation}
\lag \supset -\overline{\nu_L} m_D \nu_R - \overline{S_L} M \nu_R - \frac12\overline{S_L} \mu S_L^c \plushc,
\end{equation}
where $m_D$, $M$, $\mu$ are $3\times3$ complex mass matrices.
In the basis $(\nu_L,\nu_R,S_L)$ this gives a $9\times9$ neutrino mass matrix
\begin{equation}
M_\nu = \begin{pmatrix} 0 & m_D^T & 0 \\
m_D & 0 & M^T \\
0 & M & \mu \end{pmatrix}.
\end{equation}
Analogously to the seesaw mechanism, one considers the case where the eigenvalues of the mass matrices satisfy $\mu \ll m_D \ll M$.
Diagonalisation of this mass matrix then gives an effective mass matrix for the left-handed neutrinos:
\begin{equation}
m_\nu = m_D^T (M^T)^{-1} \mu M^{-1} m_D.
\end{equation}
Since $m_\nu$ has a double suppression by $M$, the smallness of the three left-handed-neutrino masses requires a much lower scale of new physics than in the usual seesaw mechanism: neutrinos with sub-eV masses are obtained with $m_D$ at the electroweak scale, $M$ at the TeV scale and $\mu$ at the keV scale.
The other six neutrinos obtain masses at the TeV scale \cite{Dias:2012xp}.

\subsection{Left-right-symmetric models vs.\ simplified LET model}
The LET model is a LR-symmetric model by construction, hence it bears much resemblance to other LR-symmetric models that exist in the literature.
However, since we consider the LET model to be an EFT based on a trinification setup, it has features that distinguish it from other LR-symmetric  models.
These features apply to the simplified LET model as well:
\begin{itemize}
\item The trinification origin of the LET model forbids certain invariants to appear in the scalar potential. The Higgs field in the trinification model does not have a charge conjugate, because the $\rep{3}$ and $\brep{3}$ representations of $SU(3)$ are inequivalent. Since the LET model originates from the trinification model, we do not include any invariants containing charge conjugates\footnote{Note that we do include invariants that can be rewritten in terms of the original Higgs field only, such as $\frac12\tr{\Phi_1^\dagger\Phi_1^c}\plushc = \det\Phi_1\plushc$.} in our scalar potential, as there would be no equivalent in the trinification model from which they could have originated. This excludes two invariants from our analysis of the simplified LET model:
\begin{align}
J_2^c \equiv& \tr{\Phi_1^\dagger\Phi_1\Phi_1^{c\dagger}\Phi_1} \plushc, \notag\\
J_5^c \equiv& \Phi_R(\Phi_1^{c\dagger}\Phi_1 \plushc)\Phi_R^\dagger.
\end{align}
Similarly, several invariants can be omitted from the complete LET model (see \cref{a:fullLETinvariants}).
Hence, the scalar potential of the LET model is simpler than that of a general left-right symmetric model. 
\item We consider the trinification model to be a complete, renormalisable theory. As such, we omit any invariants in the $\LRgroup$-based EFT and its derived toy model that would have to originate from nonrenormalisable operators in the trinification model. This excludes another four invariants from the simplified LET model:
\begin{align}
J_6 =& (\det\Phi_1)^2 \plushc, \notag\\
J_7 =& \det\Phi_1^\dagger\Phi_1, \notag\\
J_8 =& \tr{\Phi_1^\dagger\Phi_1}(\det\Phi_1 \plushc), \notag\\
J_9 =& \Phi_R\Phi_R^\dagger(\det\Phi_1 \plushc).
\end{align}
\item Usually a manifest left-right symmetry of the Lagrangian is assumed in LR-symmetric models. This symmetry implies $g_L = g_R$. However, we do not assume such a symmetry in our setup. We consider the model to be a description of physics at low energies where the left-right symmetry has been broken. Hence we have to allow for the possibility of different values of the left-handed and right-handed gauge couplings. Note that both the simplified and the complete LET model without $\Phi_R$ are LR-symmetric, and only the vev $M$ breaks this symmetry. Thus if $M$ is in the TeV range, renormalisation-group running might not have driven $g_L$ and $g_R$ very far apart. That is, these couplings may still be close to one another, depending on the magnitude of $M$.
\item The scalar field content of both the simplified and the complete LET model differ from the field content usually employed in LRSMs. As we noted in the previous section, one usually introduces a pair of triplets $\Delta_L$, $\Delta_R$ or a pair of doublets $\chi_L$, $\chi_R$ to break the left-right symmetry. However, since we do not impose an explicit left-right symmetry on the Lagrangian, the right-handed doublet $\Phi_R$ is sufficient to describe this symmetry breaking.
\end{itemize}

\clearpage
\section{Constraints from heavy vector boson searches}\label{s:gaugebosonconstraints}
The simplified LET model predicts the existence of a pair of charged vector bosons $W^{\prime\pm}$ and one neutral vector boson $Z'$ in addition to the Standard-Model gauge bosons.
We know how their masses $m_{W'}$, $m_{Z'}$ as well as their mixings $\zeta$, $\theta_W$, $\theta_W^\prime$, $\eta$ depend on the simplified LET model parameters.
However, the $W'$ and $Z'$ bosons have not been observed as of yet.
This fact leads to lower bounds on their masses as well as constraints on their mixings.
This in turn allows us to constrain the underlying parameters of the simplified LET model.
In this section, we review the bounds that are available in the literature.
We give an overview of both direct searches and precision measurements.
Then we use these bounds to obtain constraints on the underlying simplified LET model parameters.

\subsection{Direct searches}\label{s:directsearches}
Many models of physics beyond the Standard Model include new heavy vector bosons $W'$ and $Z'$ (see e.g.\ the reviews in ref.~\cite{Agashe:2014kda} and references therein).
For example, a $W'$ and a $Z'$ boson appear in LR-symmetric models and models with other gauge groups that embed the electroweak symmetry, such as $SU(3)_W\times U(1)$ or $SU(4)_W\times U(1)$.
The $Z'$ boson also appears in models with gauge groups containing an additional $U(1)'$, extra-dimensional theories in which the electroweak gauge bosons can propagate in the extra dimensions, or as a composite state in confining gauge theories.
Because these heavy vector bosons appear in so many different theories, it is important to look for them.
As such, they have been subject to direct searches in various decay channels at LEP, Tevatron, and the LHC.
Lower bounds on the masses of these particles are readily available in the literature (see ref.~\cite{Agashe:2014kda} for an overview).

Searches in the decay channel $W'\rightarrow\ell\nu$ are the most sensitive \cite{Aad:2012ej}, where $\ell$ is an electron or muon.
The ATLAS \cite{Aad:2012dm} and CMS \cite{Chatrchyan:2013lga} collaborations have performed this search, and found the following lower bounds on the $W'$ mass:
\begin{align}
\text{ATLAS:}\qquad m_{W'} >& 2.55\text{ TeV}, \hspace{.9cm}(\text{SSM}) \notag\\
\text{CMS:}\qquad m_{W'} >& 2.90\text{ TeV}. \hspace{1cm}(\text{SSM})
\end{align}
These bounds depend on assumptions on the $W'$ couplings: the $W'WZ$ coupling is set to zero, and the remaining couplings are taken to be those of the Sequential Standard Model (SSM).
In the SSM, the $W'$ couplings are identical to the corresponding $W$ couplings.
However, this assumption is not justified in the simplified LET model: for example the couplings $g_L$, $g_R$ are not equal.
Also, the decay to leptons would be suppressed if the $W'$ mass were smaller than the mass of the right-handed neutrino.
Therefore we will not use these bounds to constrain the simplified LET model.

It is important to search for $W'$ bosons in quark final states as well, since the results of $W'$ searches in leptonic final states may or may not apply depending on the underlying model.
ATLAS has searched for the decay $W'\rightarrow tb$ by looking for $tb$ resonances in the $\ell\nu bb$ final state \cite{Aad:2012ej}.
They found the following lower bound:
\begin{equation}
m_{W'} > 1.13\text{ TeV}. \hspace{1cm}(\text{SSM})
\end{equation}
However, Standard-Model-like couplings of the $W'$ to the fermions were assumed again, which means that the bound does not apply to the simplified LET model.
The CMS collaboration has searched for the decay $W'\rightarrow tb$ as well, with a final state consisting of a single electron or muon, missing transverse energy, and jets, at least one of which is identified as a $b$-jet \cite{Chatrchyan:2012gqa}.
They modeled the $W'$ couplings to Standard-Model fermions by the Lagrangian
\begin{equation}
\lag = \frac{V_{f_if_j}}{2\sqrt2}g_L \overline{f}_i\gamma_\mu \left[ a^R_{f_if_j}(1+\gamma^5) + a^L_{f_if_j}(1-\gamma^5) \right] W^{\prime\mu}f_j \plushc
\end{equation}
Here $a^R_{f_if_j}$, $a^L_{f_if_j}$ are the right-handed and left-handed couplings of the $W'$ to the fermions $f_i$ and $f_j$, and $V_{f_if_j}$ is the CKM matrix for quarks and a unit matrix for leptons.
For a $W'$ boson that couples to right-handed fermions ($a^L_{f_if_j}=0$, $a^R_{f_if_j}$=1), they found the bound
\begin{equation}
m_{W'} > 1.85\text{ TeV}. \hspace{1cm}(\text{SSM})
\end{equation}
However, this bound still rests of the assumption that the left-handed coupling $g_L$ and the right-handed coupling $g_R$ are equal.
In the simplified LET model, these couplings are not necessarily equal, so we should take $a^R_{f_if_j} = g_R/g_L$.
In fig.~5 of ref.~\cite{Chatrchyan:2012gqa}, contour plots for the $m_{W'}$ bound are given in the $(a^L,a^R)$-plane for $0 \leq a^L,a^R \leq 1$.
Thus if we can constrain the right-handed coupling $g_R$ and find $g_R<g_L$, we can read the corresponding $W'$ mass bound from the point $(0,g_R/g_L)$ in this figure.
Note that these constraints would still depend on the assumption that the CKM matrices for the left-handed and right-handed currents are the same.

In other $W'$ searches one looks for the decay $W'\rightarrow WZ$ by looking for narrow resonances in the $WZ$ mass distribution.
Bounds from these searches are complementary to those from $W'\rightarrow\ell\nu$ searches, since in the latter it is usually assumed that the $W'\rightarrow WZ$ decay is strongly suppressed.
Several final states have been considered: ATLAS \cite{Aad:2013wxa} and CMS \cite{Chatrchyan:2012kk} have performed this search in the $\ell\nu jj$ final state ($\ell=e,\mu$), which led to the bounds
\begin{align}
\text{ATLAS:}\qquad m_{W'} >& 950\text{ GeV}, \hspace{1cm}(\text{EGM}) \notag\\
\text{CMS:}\qquad m_{W'} >& 1143\text{ GeV}. \hspace{1cm}(\text{SSM})
\end{align}
The same search in the $\ell\nu\ell\ell$ final state \cite{Aad:2012vs,Chatrchyan:2012rva} yielded
\begin{align}
\text{ATLAS:}\qquad m_{W'} >& 760\text{ GeV}, \hspace{3.9cm}(\text{EGM}) \notag\\
\text{CMS:}\qquad m_{W'} >& 940\text{ GeV} \text{ or } m_{W'} < 700\text{ GeV}. \hspace{1cm}(\text{SSM})
\end{align}
However, these bounds depend on the assumptions made about the $W'WZ$ coupling, which generally depend on the specifics of the underlying model.
The quoted CMS searches assume SSM couplings, in which the $W'$ is assumed to have the same couplings to Standard-Model particles as the $W$.
The quoted ATLAS searches assume Extended Gauge Model (EGM) couplings: the $W'WZ$ coupling is taken to be equal to the Standard-Model $WWZ$ coupling scaled by a factor $c\times(m_W/m_{W'})^2$, with $c$ of order one.
However, the $W'WZ$ coupling of the simplified LET model is given by (see \cref{a:VVVfeynRules})
\begin{equation}
\frac{g_{W'WZ}}{g_{WWZ}^\text{SM}} = \frac12\cos\eta\sin2\zeta = \frac{g_Rm_W^2}{g_Lm_{W'}^2}\sin2\beta + \ord{\frac{v^4}{M^4}}.
\end{equation}
Hence the coefficient $c$ cannot be taken of order one: $\beta = 0.0166$ is a small parameter fixed by the ratio $m_b/m_t$ of the bottom-quark mass and the top-quark mass, and $g_R$ cannot be too large or else it would become nonperturbative.
A plot of $c$ versus the bound on $m_{W'}$ is given in fig.~7 of ref.~\cite{Aad:2013wxa} and fig.~4 of ref.~\cite{Chatrchyan:2012kk}.
The $W'$ mass is basically unconstrained if $c \lesssim 0.2$, so the simplified LET model is not constrained by these searches.

Direct searches for the $Z'$ boson of LR-symmetric models have been performed at LEP and Tevatron.
These searches focus on the dilepton decay modes of the $Z'$.
The CDF collaboration \cite{Abe:1997fd} has performed this search and set lower mass limits for the $Z'$ in various models based on the gauge group $E_6$.
For $Z'$ bosons in LR-symmetric models, they obtained the lower bound
\begin{equation}
m_{Z'} > 630\text{ GeV}. \hspace{1cm}(g_L = g_R)
\end{equation}
The ALEPH \cite{Schael:2006wu}, DELPHI \cite{Abdallah:2005ph}, and OPAL \cite{Abbiendi:2003dh} collaborations have found the following lower bounds:
\begin{align}
\text{ALEPH:}\qquad m_{Z'} >& 600\text{ GeV}, \hspace{1cm}(g_L=g_R) \notag\\
\text{DELPHI:}\qquad m_{Z'} >& 455\text{ GeV}, \hspace{1cm}(g_L=g_R) \notag\\
\text{OPAL:}\qquad m_{Z'} >& 518\text{ GeV}. \hspace{1cm}(g_L=g_R)
\end{align}
Since $g_L = g_R$ was assumed in these analyses, we cannot apply these limits to the simplified LET model.
However, fig.~21 in ref.~\cite{Abbiendi:2003dh} also gives the mass bounds where the right-handed coupling is allowed to vary.
From this figure we read a lower bound for LR-symmetric models in which $g_R$ is a free parameter:
\begin{equation}
m_{Z'} > 440\text{ GeV}.
\end{equation}

More recent results from the LHC are available, but the analyses so far focus on a $Z'$ with Standard-model-like couplings to fermions.
We give an overview of the resulting bounds here, but we do not apply them to the simplified LET model.
ATLAS has searched for $Z'$ bosons in the decay mode $Z'\rightarrow\tau^+\tau^-$ \cite{Aad:2012gm} and found
\begin{equation}
m_{Z'} > 1.40\text{ TeV}. \hspace{1cm}(\text{SSM})
\end{equation}
CMS has looked at the decay $Z'\rightarrow q\bar{q}$ in dijet mass spectra, resulting in the bound \cite{CMS:2012yf}
\begin{equation}
m_{Z'} > 1.47\text{ TeV}. \hspace{1cm}(\text{SSM})
\end{equation}
Both ATLAS \cite{Aad:2012hf} and CMS \cite{Chatrchyan:2012oaa} have searched in the dielectron and dimuon decay channels as well, giving the lower mass bounds
\begin{align}
\text{ATLAS:}\qquad m_{Z'} >& 2.22\text{ TeV}, \hspace{1cm}(\text{SSM}) \notag\\
\text{CMS:}\qquad m_{Z'} >& 2.590\text{ TeV}. \hspace{.9cm}(\text{SSM})
\end{align}
LHC bounds on $Z'$ bosons in LR-symmetric models are not available as of yet to the best of our knowledge.

\subsection{Electroweak precision data}
Another way to constrain the masses and mixings of the heavy vector bosons is via electroweak precision data (EWPD).
Chay et al.\ \cite{Chay:1998hd} obtained constraints on the neutral sector using EWPD from LEP I.
The advantage of studying the neutral sector over the charged sector is that the couplings of neutral vector bosons are independent of the CKM matrices.
Thus the constraints obtained from the neutral sector are less sensitive to the details of the model under consideration.
In an electroweak precision test, one considers the deviations of several quantities from their Standard-Model values, such as the mass ratio $m_W/m_Z$, the leptonic and $b$-quark decay widths $\Gamma_l$, $\Gamma_b$, and the leptonic and $b$-quark forward-backward asymmetries $A^l_{FB}$, $A^b_{FB}$.
These deviations are expressed in terms of the mixing angles $\theta_W^\prime$, $\eta$.\footnote{The $Z'$ mass is not constrained by LEP I data: since the experiment was performed at the $Z$ peak, the mass-dependent effects of the $Z'$ are strongly suppressed for $m_{Z'} \gg m_Z$. Hence the main contributions come from mixing by $\theta_W^\prime$ and $\eta$ \cite{Chay:1998hd}.}
Using a $\chi^2$-fit of the LEP I data, these mixing angles have been constrained.
Then these constraints were combined with low-energy neutral-current data to obtain stronger constraints on the mixing angles as well as the $Z'$ mass.
This encompasses experimental results for various scattering processes in which both $Z$ and $Z'$ can participate, such as $\nu e\rightarrow\nu e$ scattering.
One works out the effective low-energy Lagrangian for these processes, then fits the couplings to the experimental results.
The following bounds were found \cite{Chay:1998hd}:
\begin{equation}
-0.00040 < \eta < 0.0026,\qquad m_{Z'} > 430\text{ GeV}\qquad \atCL{95} \label{eq:Chaybounds}
\end{equation}
The mixing angle $\theta_W^\prime$ was not constrained by the data considered.
However, if the mixing angle $\eta$ were known, one could read off constraints on $\theta_W^\prime$ from fig.~1 in ref.~\cite{Chay:1998hd}.

Another study of electroweak precision data was performed by del Aguila et al.\ \cite{delAguila:2010mx}.
They studied the effects of new vector bosons in various representations of the Standard-Model gauge group.
The effective Lagrangian at low energies was worked out including operators of dimension six, resulting in several four-fermion interactions.
The couplings of these interactions were then fit to EWPD, and constraints on the new vector bosons were obtained.
The EWPD included $Z$-pole observables, the $W$ mass and width, unitarity constraints of the CKM matrix, low-energy effective couplings from neutrino scattering with nucleons and electrons, atomic parity violation, M\o{}ller scattering, and LEP 2 data.
Bounds were given for some of the $W'$ and $Z'$ couplings to scalars and leptons at 95\% C.L.~\cite{delAguila:2010mx}:
\begin{align}
|G^\phi_{W'}| \leq& 0.11 \text{ TeV}^{-1}, \notag\\
|G^\phi_{Z'}| \leq& 0.098 \text{ TeV}^{-1}, \notag\\
|G^\ell_{Z'}| \leq& 0.210 \text{ TeV}^{-1}, \notag\\
|G^e_{Z'}| \leq& 0.300 \text{ TeV}^{-1}. \label{eq:delAguila}
\end{align}
Here the results are given in terms of the effective couplings $G_V^k \equiv g_V^k/M_V$, where $g_V^k$ is the coupling of the vector boson $V$ to particle $k$ and $M_V$ is the vector-boson mass.
The superscript $\phi$ refers to the Standard-Model-like Higgs field, whereas $\ell$ and $e$ refer to a left-handed lepton and right-handed electron respectively.
We obtain the $W'$ and $Z'$ couplings to the Higgs from the Feynman rules in \cref{a:h0VpVpfeynRules}, and the $Z'$ couplings to a fermion $f$ follow from the gauge currents in \cref{a:gaugecurrents}.
Using the $W'$ and $Z'$ masses given in \cref{eq:Wmasses,eq:Zmasses} we find:
{\interdisplaylinepenalty=10000
\begin{align}
G_{W'}^\phi =& \frac{1}{M} \left[ \sin2\beta + \ord{\frac{v^2}{M^2}} \right], \notag\\
G_{Z'}^\phi =& \frac{1}{M} \left[ \cos^2\theta_W^\prime + \ord{\frac{v^2}{M^2}} \right], \notag\\
G_{Z'}^f =& \frac{2\cos^2\theta_W^\prime}{M} \Bigg[ T_R^3(f) - \frac12\tan^2\theta_W^\prime Q_{B-L}(f) + \ord{\frac{v^2}{M^2}} \Bigg].
\end{align}
}
Here $T_R^3(f)$ refers to the eigenvalue of $f$ under the $SU(2)_R$ generator $T_R^3$, and $Q_{B-L}(f)$ is the charge of $f$ under $U(1)_{B-L}$.
Taking the couplings to leading order in $\frac{v}{M}$, the bounds in \cref{eq:delAguila} amount to
\begin{align}
|G^\phi_{W'}| =& \frac{\sin2\beta}{M} \leq 0.11 \text{ TeV}^{-1}, \notag\\
|G^\phi_{Z'}| =& \frac{\cos^2\theta_W^\prime}{M} \leq 0.098 \text{ TeV}^{-1}, \notag\\
|G^\ell_{Z'}| =& \frac{\sin^2\theta_W^\prime}{M} \leq 0.210 \text{ TeV}^{-1}, \notag\\
|G^e_{Z'}| =& \frac{|\cos2\theta_W^\prime|}{M} \leq 0.300 \text{ TeV}^{-1}. \label{eq:Aguila}
\end{align}

\subsection{High-precision measurements}
Another way to constrain heavy vector bosons is via high-precision measurements of parity violation in the weak interaction.
One example of such an experiment is the measurement of the muon decay parameters, which has been performed by the TWIST collaboration \cite{Bueno:2011fq,TWIST:2011aa} and by Barenboim et al.\ using LEP data \cite{Barenboim:1996nd}.
Muons decay to positrons in the process $\mu^+ \rightarrow e^+ \nu_e\bar\nu_\mu$.
At low energies, this decay can be described by a four-fermion interaction.
However, such an interaction is nonrenormalisable, and hence it can only be viewed as an effective interaction.
Four-fermion interactions can arise through the exchange of various particles: they can be mediated by (pseudo)scalar, (axial) vector and tensor couplings, depending on the underlying model.
The differential decay rate of a muon decaying to a positron can be described in terms of the four muon decay parameters \cite{Michel:1949qe,Bouchiat:1957zz,Kinoshita:1957zz,Kinoshita:1957zza}, which are functions of all couplings that could mediate this decay.
In the Standard Model, muon decay occurs through the exchange of a $W$-boson only.
However, new-physics models generally introduce additional contributions to this process.
For example, LR-symmetric models (including the LET) provide an additional contribution from $W'$ boson exchange.
As such, each model leads to different predictions for the muon decay parameters.
Therefore a high-precision measurement of these parameters provides a test for these models.

Such a measurement in the lepton sector has an advantage over measurements in the quark sector: the signature is clean because hadronisation is not involved in muon decay.
Also, the couplings in the leptonic sector are unaffected by the left-handed and right-handed CKM matrices.
This allows one to obtain model-independent constraints.
The TWIST collaboration has measured the polarised muon decay spectrum in order to extract the muon parameters \cite{Bueno:2011fq,TWIST:2011aa} .
They used the results to put the following constraints on right-handed charged currents in generalised LR-symmetric models (stronger constraints were obtained for manifestly LR-symmetric models, in which $g_L=g_R$):
\begin{equation}
\left| \frac{g_R\zeta}{g_L} \right| < 0.020,\qquad \frac{g_Lm_{W'}}{g_R} \geq 578 \text{ GeV}\qquad \atCL{90} \label{eq:twistBounds}
\end{equation}

Barenboim et al.~\cite{Barenboim:1996nd} performed a $\chi^2$ fit of the muon decay parameters using the muon decay data of LEP.
They used these to constrain the right-handed charged currents as well.
Assuming the right-handed neutrino is lighter than the $W'$, they found the following bounds:
\begin{equation}
\frac{g_R}{g_L} = 0.94 \pm 0.09,\qquad m_{W'} \geq 485 \text{ GeV},\qquad |\zeta| \leq 0.0327. \label{eq:BarenboimBounds}
\end{equation}

\begin{figure}[t]
\begin{minipage}{\textwidth}
\begin{minipage}{.49\textwidth}
\begin{equation*}
\begin{fmfgraph*}(100,30)
  \fmfleft{d1,s1}
  \fmfright{s2,d2}
  \fmfv{label=$s$,label.angle=180}{s1}
  \fmfv{label=$\bar{d}$,label.angle=180}{d1}
  \fmfv{label=$d$,label.angle=0}{d2}
  \fmfv{label=$\bar{s}$,label.angle=0}{s2}
  \fmf{fermion}{s1,v11}
  \fmf{fermion,label=$u,,c,,t$,label.side=left,tension=.5}{v11,v12}
  \fmf{fermion}{v12,d2}
  \fmf{fermion}{s2,v22}
  \fmf{fermion,label=$\bar{u},,\bar{c},,\bar{t}$,label.side=left,tension=.5}{v22,v21}
  \fmf{fermion}{v21,d1}
  \fmffreeze
  \fmf{boson,label=$W$,label.side=left}{v21,v11}
  \fmf{boson,label=$W$,label.side=right}{v22,v12}
\end{fmfgraph*}
\end{equation*}
\vskip1ex
\end{minipage}
\begin{minipage}{.49\textwidth}
\begin{equation*}
\begin{fmfgraph*}(100,30)
  \fmfleft{d1,s1}
  \fmfright{s2,d2}
  \fmfv{label=$s$,label.angle=180}{s1}
  \fmfv{label=$\bar{d}$,label.angle=180}{d1}
  \fmfv{label=$d$,label.angle=0}{d2}
  \fmfv{label=$\bar{s}$,label.angle=0}{s2}
  \fmf{fermion}{s1,v11}
  \fmf{fermion,label=$u,,c,,t$,label.side=left,tension=.5}{v11,v12}
  \fmf{fermion}{v12,d2}
  \fmf{fermion}{s2,v22}
  \fmf{fermion,label=$\bar{u},,\bar{c},,\bar{t}$,label.side=left,tension=.5}{v22,v21}
  \fmf{fermion}{v21,d1}
  \fmffreeze
  \fmf{boson,label=$W$,label.side=left}{v21,v11}
  \fmf{boson,label=$W'$,label.side=right}{v22,v12}
\end{fmfgraph*}
\end{equation*}
\vskip1ex
\end{minipage}
\end{minipage}
\caption{Diagrams that form the main contributions to $K^0-\overline{K^0}$ mixing in LR-symmetric models. The contribution from diagrams with the exchange of two $W'$ bosons is suppressed by $m_{W'}$ with respect to these diagrams. Similar diagrams exist with the $W$ and $W'$ interchanged and/or with crossed vector boson propagators. In the Standard-Model, this mixing only obtains a contribution from the left diagram.}\label{f:kaonmixing}
\end{figure}
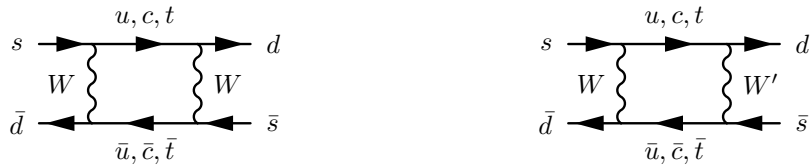

Another constraint on right-handed currents comes from the neutral kaon system.
The mass difference $\Delta m_K \equiv m_{K_L} - m_{K_S}$ is governed by $\Delta S = 2$ strangeness-changing processes induced by box diagrams.
In the Standard Model, the main contribution comes from diagrams involving the exchange of two $W$ bosons.
However, models of new physics may introduce additional contributions to this $K^0-\overline{K^0}$ mixing.
For example, the dominant contributions in LR-symmetric models come from the exchange of either two $W$ bosons or a $W$ and a $W'$ (see \cref{f:kaonmixing}).
The contribution from two $W'$ bosons is suppressed by the large $W'$ mass.
Hence precision measurements of the neutral kaon system allow us to constrain the mass of the $W'$.
Barenboim et al.\ used data on the neutral kaon system to arrive at the constraint \cite{Barenboim:1996nd}
\begin{equation}
m_{W'} \gtrsim 0.7 \text{ TeV}.
\end{equation}

\subsection{Parameter constraints}\label{s:parameterconstraints}
The bounds on the masses of new heavy vector bosons are the strongest in direct searches, putting $W'$ and $Z'$ masses at least in the TeV regime.
However, these bounds apply only to manifestly LR-symmetric models, and as such are not applicable to the simplified LET model.
The strongest bounds for general LR-symmetric models come from analyses of precision experiments.
Moreover, these studies also constrain the mixing angles and gauge couplings.
These allow us to put constraints on the parameters of the simplified LET model.
In order to extract these constraints, we will use the following experimental values from ref.~\cite{Agashe:2014kda} as input: the Weinberg angle $\theta_W$ in the $\overline{MS}$ scheme at $\mu=M_Z$ is given by $\sin^2\theta_W = 0.23126 \pm 0.00005$, the electromagnetic coupling constant is $e=\sqrt{4\pi\alpha(M_Z)} = 0.313402 \pm 0.000017$, where $\alpha(M_Z)$ is the fine-structure constant in the $\overline{MS}$ scheme at $\mu=M_Z$.
This gives a weak coupling constant of $g_L = e/\sin\theta_W = 0.65170 \pm 0.00008$.
We will also use the vev parameter $v = 246$ GeV and the angle $\beta = 0.0166$, which is fixed by the ratio of the bottom- and top-quark masses (see \cref{a:fermionmasseig}).

The ratio of the $SU(2)_R$ coupling $g_R$ to the $SU(2)_L$ coupling $g_L$ is constrained by the LEP muon decay data in \cref{eq:BarenboimBounds}.
Using the experimental value for $g_L$, we obtain a constraint on $g_R$:
\begin{equation}
\frac{g_R}{g_L} = 0.94 \pm 0.09\qquad \Rightarrow\qquad g_R = 0.61 \pm 0.06. \label{eq:gRbound}
\end{equation}
Using the expression for the electromagnetic gauge coupling $e$ in terms of the $\LRgroup$ gauge couplings (see \cref{a:gaugecurrents}), we obtain a constraint on $g'$ as well:
\begin{equation}
g' = \frac{eg_Lg_R}{2\sqrt{g_L^2g_R^2-e^2(g_L^2+g_R^2)}} = 0.22 \pm 0.01. \label{eq:gpbound}
\end{equation}
As we mentioned in \cref{s:directsearches}, we can read off a bound on the $W'$ mass from the point $(a_L,a_R)=(0,g_R/g_L)$ in fig.~5 of ref.~\cite{Chatrchyan:2012gqa} if we know $g_R$.
This gives us the bound
\begin{equation}
m_{W'} \gtrsim 1.4 \text{ TeV}.
\end{equation}

The bounds given in the previous sections yield several constraints on the scale $M$.
For example, the masses of the heavy vector bosons are proportional to $M$, and as such they constrain this parameter.
We use the mass expressions given in \cref{eq:Wmasses,eq:Zmasses} to leading order in $\frac{v}{M}$.
The strongest bounds on the masses of the $W'$ and $Z'$ come from the $W'$ search by CMS, combined with our value for $g_R$, and the LEP I electroweak precision data respectively:
\begin{equation}
\frac{g_RM}{2} = m_{W'} \gtrsim 1.4 \text{ TeV},\qquad \frac{g_RM}{2\cos\theta_W^\prime} = m_{Z'} > 430 \text{ GeV}.
\end{equation}
Note that the simplified LET model predicts $m_{Z'} = m_{W'}/\cos\theta_W^\prime \geq m_{W'}$, so the $m_{Z'}$ bound given here is weaker than the $m_{W'}$ bound, and does not add any information.
Using the upper bound $g_R < 0.67$ obtained above, the $m_{W'}$ mass bound gives the following constraint on $M$:
\begin{equation}
M > 3.6 \text{ TeV}. \label{eq:Mbound}
\end{equation}
Another constraint on $M$ comes from the $W-W'$ mixing angle $\zeta$.
Given the value of $g_R$ in \cref{eq:gRbound}, the strongest constraint on $\zeta$ comes from the TWIST muon-decay parameters (see \cref{eq:twistBounds}).
Using the expression for $\zeta$ in \cref{eq:zeta} to leading order in $\frac{v}{M}$, we obtain the following bound on $M$:
\begin{align}
0.020 >& \left|\frac{g_R\zeta}{g_L}\right| = \frac{v^2\sin2\beta}{M^2}, \notag\\
\Rightarrow M >& v\sqrt{\frac{\sin2\beta}{0.020}} = 317 \text{ GeV}.
\end{align}
This bound is weaker than the one in \cref{eq:Mbound}.
The four bounds on the effective low-energy couplings from Aguila et al.\ in \cref{eq:Aguila} allow us to constrain $M$ as well.
We use $\theta_W^\prime = \arctan(2g'/g_R) = 0.62 \pm 0.05$ to obtain the following constraints:
\begin{align}
M \geq& \frac{\sin2\beta}{0.11\text{ TeV}^{-1}} = 0.302 \text{ TeV}, \notag\\
M \geq& \frac{\cos^2\theta_W^\prime}{0.098\text{ TeV}^{-1}} > 6.2 \text{ TeV}, \notag\\
M \geq& \frac{\sin^2\theta_W^\prime}{0.210\text{ TeV}^{-1}} > 1.4 \text{ TeV}, \notag\\
M \geq& \frac{\cos2\theta_W^\prime}{0.300\text{ TeV}^{-1}} > 0.73 \text{ TeV}.
\end{align}
The second line gives the strongest bound on $M$ we have so far.
Another constraint on $M$ comes from the bound on the mixing angle $\eta$ from Chay et al.
Since this angle is small, it can be rewritten at leading order in $\frac{v}{M}$ as
\begin{equation}
\eta \approx \tan\eta \approx \frac{\sin\theta_W^\prime\cos^3\theta_W^\prime}{\sin\theta_W}\frac{v^2}{M^2} < 0.0026. \label{eq:etaBound1}
\end{equation}
Using the values we have for $\theta_W$, $\theta_W^\prime$, and $v$, we arrive at the bound
\begin{equation}
M > 3.6 \text{ TeV}. \label{eq:etaBound2}
\end{equation}

In summary, the gauge-boson sector of the simplified LET model depends on four new parameters $\beta$, $M$, $g_R$, $g'$ in addition to the Standard-Model parameters.
For the latter, we used the experimental values as input, and the angle $\beta = 0.0166$ is fixed by the mass ratio of the bottom- and top-quark masses.
We have used the available bounds on the masses and mixings of the new heavy vector bosons to constrain the remaining three parameters.
Our strongest bounds come from a combination of low-energy precision measurements; the results are
\begin{equation}
M > 6.2 \text{ TeV},\qquad g_R = 0.61 \pm 0.06,\qquad g' = 0.22 \pm 0.01.
\end{equation}

\clearpage
\section{Couplings of the Standard-Model-like Higgs}\label{s:lightHiggs}
The simplified LET model predicts the existence of new scalar particles.
Their masses depend on several simplified-LET-model parameters.
In parts of parameter space, these masses are low enough such that we could produce them at the LHC.
However, it is also possible that we are not that lucky, because the new scalars are too heavy to detect experimentally in large regions of the simplified-LET-model parameter space.
In that case, the Standard-Model-like scalar with a mass of 126 GeV would be all we could observe in the foreseeable future.

In this scenario, it may still be possible to distinguish the Standard Model from the simplified LET model if we measure the Higgs couplings to all Standard-Model particles.
In the Standard Model, these couplings are fixed by the particle masses and the vev of the Higgs field.
Since the latter are known, an independent measurement of the Higgs couplings provides an important test of the Standard Model.
These couplings are generally modified in the presence of new physics \cite{Lopez-Val:2013yba}: the Standard-Model-like Higgs is a mixture of the scalar gauge eigenstates, and the mixing angles show up in the Standard-Model-like Higgs couplings.
Thus deviations of the Higgs couplings from their Standard-Model values would be an indirect probe of physics beyond the Standard Model.

In this section, we give the Standard-Model Higgs couplings and quantify the Higgs-coupling modifications.
We review how these coupling modifications are measured, and give the values available in the literature.
Then we derive the coupling modifications as predicted by the simplified LET model.
We look at possible patterns in these coupling modifications that may allow us to distinguish the simplified LET model from the Standard Model or some other new-physics model.
Then we see how the measured Higgs-coupling modifications constrain the simplified LET model.
We define a set of benchmark points that represent the possible phenomenological features of the simplified LET model, and compute the coupling modifications for each of them.
We look for scenarios that give a measurable deviation from the Standard-Model Higgs couplings.

\subsection{Standard-Model Higgs couplings}\label{s:SMHiggsCouplings}
The Standard-Model Higgs couplings are not free parameters.
The couplings of the Higgs to vector bosons are completely determined by gauge invariance.
They can be read off the covariant derivative terms for the Higgs field $\phi$:
\begin{align}
\lag_\text{SM} \supset& (D_\mu\phi)^\dagger(D^\mu\phi), \notag\\
D_\mu =& \partial_\mu - igA_\mu^a\tau^a - ig'YB_\mu \notag\\
=& \partial_\mu - \frac{ig}{\sqrt2}(W^+_\mu\tau^+ + W^-_\mu\tau^-) - \frac{ig}{\cos\theta_W}Z_\mu(\tau^3 - \sin^2\theta_WQ) - ieA_\mu Q. \label{eq:SMHiggslag}
\end{align}
Here $g$, $g'$ are the $SU(2)_L$ and $U(1)_Y$ couplings respectively, $A_\mu^a$ are the $SU(2)_L$ gauge fields, $B$ is the hypercharge gauge field, $\tau^a = \frac{\sigma_a}{2}$ are the $SU(2)_L$ generators, and $Y$ is the $U(1)_Y$ generator.
The gauge-boson mass eigenstates are given by
\begin{equation}
W^\pm = \frac{1}{\sqrt2}(A^1\mp iA^2),\qquad
\begin{pmatrix} Z \\ A \end{pmatrix} =
\begin{pmatrix} \cos\theta_W & -\sin\theta_W \\ \sin\theta_W & \cos\theta_W \end{pmatrix}
\begin{pmatrix} A^3 \\ B \end{pmatrix}. \label{eq:SMgaugebosons}
\end{equation}
The corresponding generators are defined by $\tau^\pm = \tau^1\pm i\tau^2$, $Q = \tau^3 + Y$, and their gauge couplings are determined by $\sin\theta_W = g'/\sqrt{g^2+g^{\prime2}}$ and $e=g\sin\theta_W$.
The $W$ bosons obtain a mass $m_W = gv/2$, the $Z$ boson obtains a mass $m_Z = \sqrt{g^2+g^{\prime2}}v/2 = m_W/\cos\theta_W$, whereas the photon $A$ remains massless.
In order to extract the Higgs couplings, we parametrise the Higgs field as follows:
\begin{equation}
\phi = \begin{pmatrix} \phi^+ \\ \frac{v + h + ia}{\sqrt2} \end{pmatrix}.
\end{equation}
Here $h$ is the physical Higgs boson, whereas $\phi^+$, $\phi^- = (\phi^+)^*$, $a$ are the Goldstone bosons that give mass to the $W^\pm$, $Z$ bosons.
Inserting this expression into the Lagrangian in \cref{eq:SMHiggslag}, we obtain the Higgs couplings to the vector bosons:
\begin{align}
g^\text{SM}_{hWW} =& \frac{g^2v}{2} = \frac{2m_W^2}{v}, \notag\\
g^\text{SM}_{hZZ} =& \frac{g^2v}{2\cos^2\theta_W} = \frac{2m_Z^2}{v}. \label{eq:gVSM}
\end{align}
That is, the Higgs coupling to vector bosons is proportional to their squared mass.
The tree-level coupling of the Higgs to photons vanishes.
However, the photon obtains a non-zero effective Higgs coupling through loops involving massive charged particles.
We derive the expression for this coupling for a general theory with any number of scalars, fermions, and vector bosons in \cref{a:Deltagamma}.
For the Standard Model, this expression reduces to
\begin{equation}
g_{h\gamma\gamma}^\text{SM} = \frac{\alpha v}{16\sqrt2\pi} \left( \frac{g_{hWW}}{m_W^2}A_1(\tau_W) + \sum_f \frac{2g_{hff}}{m_f}N_{c,f}Q_f^2A_{1/2}(\tau_f) \right). \label{eq:ggSM}
\end{equation}
Here $\alpha$ is the fine-structure constant, and the sum runs over all fermions $f$ with Higgs coupling $g_{hff}$, mass $m_f$, $N_{c,f}$ colour degrees of freedom, and electromagnetic charge $Q_f$.
The functions $A_{1/2}(x)$, $A_1(x)$ are the spinor and vector loop functions, which are defined in \cref{a:Deltagamma}.
They take the mass ratios $\tau_i \equiv 4m_i^2/m_h^2$ as argument.

Note that a similar effective coupling exists for the gluons as well, which is generated by quark loops.
The main contribution to this coupling comes from the top quark, since it couples the most strongly to the Higgs.
However, the effective gluon coupling is not of relevance to the discussion in this chapter: the LET model does not introduce any new massive coloured states that could modify this coupling.
The only modifications to the Higgs-gluon coupling would be those induced by the coupling modifications of the quarks running in the loop.
Therefore, we omit the gluon coupling from our discussion.

The Higgs couplings to fermions are given by the Yukawa Lagrangian
\begin{equation}
\lag_Y^\text{SM} = - \overline{Q} Y_u (i\sigma_2\phi^*) u_R - \overline{Q} Y_d \phi d_R - \overline{L} Y_\ell \phi \ell_R \plushc
\end{equation}
Here $Q$, $L$ are the quark and lepton doublets, $u_R$, $d_R$, $\ell_R$ are the right-handed fermions, and $Y_u$, $Y_d$, $Y_\ell$ are generation matrices; generation indices are suppressed.
After spontaneous symmetry breaking and diagonalisation of the generation matrices, the quarks and charged leptons obtain masses $m_f = y_fv$, where $y_f$ is the relevant diagonal element of the generation matrix.
The Higgs couplings can be read off as well:
\begin{equation}
g_{hff}^\text{SM} = -y_f = -\frac{m_f}{v}. \label{eq:gfSM}
\end{equation}
Note that all Higgs couplings of the Standard Model are fixed by the corresponding particle masses and the vev $v$.
The latter is fixed by the Fermi coupling constant to be $v = (\sqrt2 G_F)^{-1/2} = 246$ GeV, and all particle masses have been measured.
Hence we know all Standard-Model Higgs couplings indirectly, and a direct measurement of the Higgs couplings provides an important test of the Standard Model.

\subsection{Higgs-coupling modifications}
The Higgs-coupling modifications are defined as the deviations of the Higgs couplings from the Standard-Model values in \cref{eq:gVSM,eq:ggSM,eq:gfSM}.
For any Standard-Model particle $x$, the Higgs couplings $g_x\equiv g_{h^0xx}$ are defined as the coefficient of the operator $h^0xx$ in the Lagrangian.
Then the Higgs-coupling modifications $\Delta_x$ are defined as
\begin{equation}
\frac{g_x}{g_x^\text{SM}} = 1 + \Delta_x. \label{eq:CMdefinition}
\end{equation}
The loop-induced Higgs coupling to photons can be written as follows:
\begin{equation}
\frac{g_\gamma}{g_\gamma^\text{SM}} = 1 + \Delta_\gamma^\text{SM} + \Delta_\gamma.
\end{equation}
Here $\Delta_\gamma^\text{SM}$ is the coupling modification that is induced by coupling modifications of the Standard-Model particles generating the coupling.
The term $\Delta_\gamma$ represents contributions from non-SM particles running in the loops.

\begin{figure}[t]
\begin{center}
\includegraphics[width=.7\textwidth]{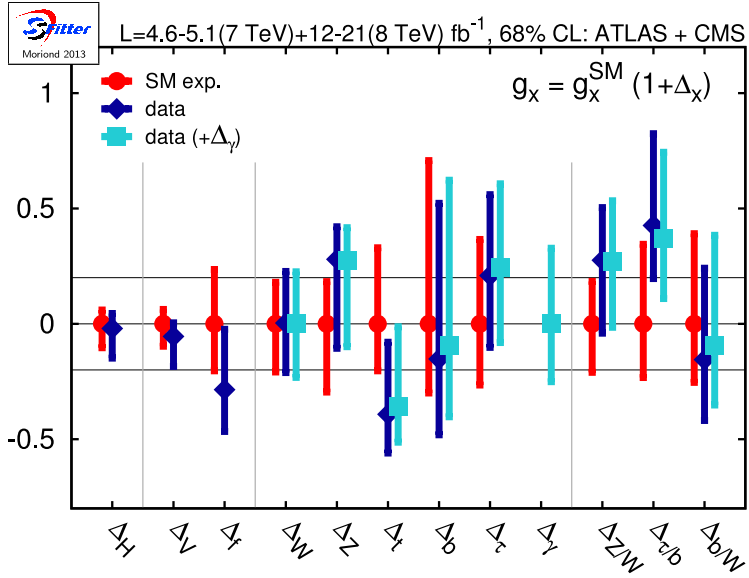}
\end{center}
\caption{Most recent fit of the Higgs-coupling modifications to LHC data. The red points correspond to the expected SM result $\Delta_x = 0$, whereas the dark blue points give the results from the data if the photon coupling is assumed to be determined by the $W$ and $t$ loops only. The light blue points give the results if a free coupling shift in $\Delta_\gamma$ due to new physics is allowed. See the main text for a definition of the various coupling modifications. Figure taken from ref.~\cite{Lopez-Val:2013yba}.}\label{f:Higgscouplingmeasurements}
\end{figure}

The Higgs-coupling modifications have been extracted from LHC data using the tool SFitter \cite{Lafaye:2009vr,Klute:2012pu,Plehn:2012iz,Lopez-Val:2013yba} (see \cref{f:Higgscouplingmeasurements}).
In this analysis, the operators in the Lagrangian are taken to be Standard-Model-like (i.e.\ the Higgs field is a $CP$-even scalar), but with variable Higgs couplings.
The free parameters in the fit of these couplings to data are $\Delta_W$, $\Delta_Z$, $\Delta_t$, $\Delta_b$, $\Delta_\gamma$.
Since the LHC has no sensitivity to the second-generation Yukawa couplings, the latter are linked to the third-generation Yukawa couplings, e.g.\ $g_c = g_tm_c/m_t$.
The top coupling is extracted from the effective gluon coupling, to which the top loop gives the main contribution.
The reason for this indirect determination of $\Delta_t$ is that there is no independent measurement of the top Yukawa coupling yet.
Hence $\Delta_g$ is not included as a free parameter in the fit.\footnote{In the 14 TeV run of the LHC, the top coupling can be probed directly in $t\bar{t}H$ production, so that $\Delta_g$ can be allowed to vary in a future update of the analysis.}
The bottom coupling has not been measured directly yet either: although the Higgs decay to $b\bar{b}$ is the dominant channel, its signal is hard to extract from QCD backgrounds.
Instead, $\Delta_b$ is extracted indirectly from the total Higgs width $\Gamma_\text{tot}$, which is identified with the sum of the observed partial widths (i.e.\ decays into invisible new states are not included).
Hence the unknown $b$ coupling enters the denominator of every $\sigma\times\text{BR}$ prediction, and is extracted from a correlated fit.

The most recent fit of the Higgs-coupling modifications is given in \cref{f:Higgscouplingmeasurements}.
The first three couplings are irrelevant to us: they represent the best fit of a hypothetical universal Higgs-coupling modification $\Delta_x = \Delta_H$ and a two-parameter fit of the scenario $\Delta_{W,Z} = \Delta_V$, $\Delta_{t,b,\tau} = \Delta_f$.
The next six couplings are the best fit when each parameter is taken to be free.
The last three couplings are modifications to ratios of couplings:
\begin{align}
\frac{g_x/g_y}{g_x^\text{SM}/g_y^\text{SM}} =& (1 + \Delta_{x/y}), \notag\\
\Rightarrow \Delta_{x/y} =& \frac{\Delta_x-\Delta_y}{1+\Delta_y}.
\end{align}
These coupling modifications do not introduce any new information, since they are given in terms of the individual coupling modifications in \cref{eq:CMdefinition}.
However, they are included in the analysis because some of the theory and systematic errors are correlated and cancel from the coupling ratio.
The error bars on the $W$, $Z$, $\tau$, and $t$ couplings have similar magnitudes, whereas the error on the $b$ coupling is significantly larger.
The reason is that $\Delta_b$ is determined indirectly from the total Higgs width.

\subsection{Higgs-coupling modifications of the simplified LET model}\label{s:couplingModifications}
The couplings of the SM-like Higgs $h^0$ to the massive vector bosons can be found by substituting the gauge-boson mass eigenstates in \cref{eq:gbmasseigcharged,eq:gbmasseigneutral} and the scalar mass eigenstates in \cref{eq:Higgsmasseig} into the Lagrangian in \cref{eq:LagS}.
The resulting couplings are given by
\begin{align}
g_W =& \frac{v}{2}\Bigg( (g_L^2c_\zeta^2 + g_R^2s_\zeta^2)(c_\beta c_{\alpha_1} + s_\beta s_{\alpha_1}c_{\alpha_2}) - g_Lg_Rs_{2\zeta} (s_\beta c_{\alpha_1} + c_\beta s_{\alpha_1}c_{\alpha_2}) \Bigg) \notag\\
&+ \frac{M}{2}g_R^2s_\zeta^2s_{\alpha_1}s_{\alpha_2} \notag\\
=& \frac{g_L^2v}{2}\left( c_\beta c_{\alpha_1} + s_\beta s_{\alpha_1}c_{\alpha_2} - 2s_{2\beta}(s_\beta c_{\alpha_1} + c_\beta s_{\alpha_1}c_{\alpha_2})\xi^2 + \ord{\xi^3} \right), \notag\\
g_Z =& \frac{v}{2}\Big( (c_\beta c_{\alpha_1} + s_\beta s_{\alpha_1}c_{\alpha_2}) \big(g_Lc_\eta c_{\theta_W} + g_R(c_\eta s_{\theta_W}s_{\theta_W^\prime} - s_\eta c_{\theta_W^\prime}) \big)^2 \Big) \notag\\
&+ \frac{M}{2}s_{\alpha_1}s_{\alpha_2} \big( g_R(c_\eta s_{\theta_W}s_{\theta_W^\prime} - s_\eta c_{\theta_W^\prime}) - 2g'(c_\eta s_{\theta_W}c_{\theta_W^\prime} + s_\eta s_{\theta_W^\prime}) \big)^2 \notag\\
=& \frac{g_L^2v}{2c^2_{\theta_W}}\left( (c_\beta c_{\alpha_1} + s_\beta s_{\alpha_1}c_{\alpha_2})(1 - 2c^4_{\theta_W^\prime} \xi^2) + \ord{\xi^3} \right).
\end{align}
Here we defined $c_x=\cos{x}$, $s_x=\sin{x}$ for the sake of brevity and introduced the dimensionless small parameter $\xi \equiv v/M$.
In the second steps we used the $W-W'$ mixing angle $\zeta$ from \cref{eq:zeta} and the neutral-vector-boson mixing angles $\theta_W$, $\theta_W^\prime$, $\eta$ from \cref{eq:neutralAngles}.
The couplings $g_R$, $g'$ have been eliminated in favour of $g_L$ using the identities $g_R = g_L\tan\theta_W/\sin\theta_W^\prime$, $2g'=g_L\tan\theta_W/\cos\theta_W^\prime$.
Using the SM Higgs couplings given by \cref{eq:gVSM}, with the $W$ and $Z$ masses given in \cref{eq:Wmasses,eq:Zmasses}, the above expressions yield the following Higgs-coupling modifications:
\begin{align}
\Delta_W =& c_\beta c_{\alpha_1} + s_\beta s_{\alpha_1}c_{\alpha_2} - 1 + \left( s^2_{2\beta} - 2s_{2\beta}(s_\beta c_{\alpha_1} + c_\beta s_{\alpha_1}c_{\alpha_2}) \right)\xi^2 + \ord{\xi^3}, \notag\\
\Delta_Z =& c_\beta c_{\alpha_1} + s_\beta s_{\alpha_1}c_{\alpha_2} - 1 - c^4_{\theta_W^\prime}\left( 2(c_\beta c_{\alpha_1} + s_\beta s_{\alpha_1}c_{\alpha_2}) - 1 \right)\xi^2 + \ord{\xi^3}. \label{eq:DeltaV}
\end{align}

The SM Higgs couplings to the fermions are given by \cref{eq:gfSM}.
The corresponding Higgs couplings of the simplified LET model can be read off the Yukawa Lagrangian in mass eigenstates, which is given in \cref{eq:yukawadirac}.
This results in the following coupling modifications:
\begin{align}
\Delta_t =& \frac{\cos\alpha_1}{\cos\beta} - 1, \notag\\
\Delta_b =& \frac{\sin\alpha_1\cos\alpha_2}{\sin\beta} - 1. \label{eq:Deltaf}
\end{align}

The photon coupling is modified indirectly through the modifications of the particles running in the loops that generate this coupling.
However, it also gets a genuine non-SM contribution from loops involving the charged scalar $H^\pm$.
This results in the following photon coupling modification (see \cref{a:Deltagamma} for the full calculation):
\begin{equation}
\Delta_\gamma = \frac{vA_0(\tau_{H^\pm})\lambda_{h^0H^+H^-}}{2m_{H^\pm}^2A_\text{SM}}. \label{eq:Deltagamma}
\end{equation}
Here $A_\text{SM} \equiv A_1(\tau_W) + N_cQ_t^2A_{1/2}(\tau_t) = -6.5$, $\tau_i \equiv 4m_i^2/m_{h^0}^2$, and the $A_s(x)$ are loop functions which are defined in \cref{eq:loopfunctions}.
The coupling $\lambda_{h^0H^+H^-}$ of $h^0$ to the charged scalars is given by
\begin{align}
\lambda_{h^0H^+H^-} =& \lambda_1v(c_\beta c_{\alpha_1} + s_\beta s_{\alpha_1}c_{\alpha_2}) (s_{13}^2 + c_+^2) \notag\\
&+ \lambda_2v\big( (c_\beta c_{\alpha_1} + s_\beta s_{\alpha_1}c_{\alpha_2}) (s_{13}^2 + c_+^2) - (s_\beta c_{\alpha_1} + c_\beta s_{\alpha_1}c_{\alpha_2}) s_{13}c_+ \big) \notag\\
&+ \lambda_3Ms_{\alpha_1}s_{\alpha_2}c_-^2 + \lambda_4(v(c_\beta c_{\alpha_1} + s_\beta s_{\alpha_1}c_{\alpha_2})c_-^2 + Ms_{\alpha_1}s_{\alpha_2}(s_{13}^2 + c_+^2)) \notag\\
&+ \lambda_5\big( v(c_\beta c_{\alpha_1}c_-^2 - s_\beta s_{\alpha_1}s_{\alpha_2}s_{13}c_- + c_\beta s_{\alpha_1}s_{\alpha_2}c_+c_-) \notag\\
&\hspace{1cm}+ M(s_{\alpha_1}s_{\alpha_2}c_+^2 - s_{\alpha_1}c_{\alpha_2}s_{13}c_- + c_{\alpha_1}c_+c_-) \big). \label{eq:lambdah0HpHm}
\end{align}
Here we defined the following combinations of mixing angles of the charged scalar fields, which are given in \cref{eq:chargedMixingAngles}:
\begin{equation}
s_{13} \equiv s_{\gamma_1}s_{\gamma_3},\qquad c_+ \equiv c_{\gamma_1}c_{\gamma_2}s_{\gamma_3} + s_{\gamma_2}c_{\gamma_3},\qquad c_- \equiv c_{\gamma_1}s_{\gamma_2}s_{\gamma_3} - c_{\gamma_2}c_{\gamma_3}. \label{eq:s13cpcmMain}
\end{equation}

The Higgs self-couplings have not been measured so far.
The trilinear coupling could be measured at the 14 TeV run of the LHC from the ratio of cross sections of double-to-single Higgs production \cite{Goertz:2013kp}.
The quartic coupling requires the measurement of triple-Higgs production, which has a low cross-section due to interference between the different diagrams contributing to this process \cite{Plehn:2005nk}.
Hence we have no bounds on the modifications of these couplings yet, and a measurement of the quartic coupling seems to be challenging.
However, we calculate these couplings for the simplified LET model anyway in order to see how precisely we would have to measure these couplings in order to see deviations from the SM values.
In the Standard Model, the scalar potential contains a single quartic and a single quadratic invariant:
\begin{equation}
V_\text{SM} = \lambda(\Phi^\dagger\Phi)^2 - \mu^2\Phi^\dagger\Phi,\qquad \Phi = \begin{pmatrix} \phi^+ \\ \frac{v + h + ia}{\sqrt2} \end{pmatrix},
\end{equation}
where we must set $\mu^2 = \lambda v^2$ in order to have the minimum of the potential at $\vev{\Phi} = (0, v/\sqrt{2})$.
After spontaneous symmetry breaking, this potential gives rise to both trilinear and quartic self-couplings for the Higgs boson $h$:
\begin{align}
\lambda_{3h}^\text{SM} =& 6\lambda v = \frac{3gm_h^2}{2m_W}, \notag\\
\lambda_{4h}^\text{SM} =& 6\lambda = \frac{3g^2m_h^2}{4m_W^2}. \label{eq:SMselfcouplings}
\end{align}
In the simplified LET model, the scalar potential contains several quartic invariants, all of which contribute to the Higgs self-couplings.
Inserting the mass eigenstates defined by \cref{eq:Higgsmasseig,eq:CPevenMixingAngles} into the potential in \cref{eq:simplifiedLETpotential}, we find:
\begin{align}
\frac{\lambda_{3h}}{3v} =& \lambda_1(c_{\alpha_1}^2+s_{\alpha_1}^2c_{\alpha_2}^2)(c_\beta c_{\alpha_1}+s_\beta s_{\alpha_1}c_{\alpha_2}) + \lambda_2(c_\beta c_{\alpha_1}^3+s_\beta s_{\alpha_1}^3c_{\alpha_2}^3) + \lambda_3\frac{M}{v}s_{\alpha_1}^3s_{\alpha_2}^3 \notag\\
&+ \lambda_4\left( \frac{M}{v}s_{\alpha_1}s_{\alpha_2}(c_{\alpha_1}^2+s_{\alpha_1}^2c_{\alpha_2}^2) + s_{\alpha_1}^2s_{\alpha_2}^2(c_\beta c_{\alpha_1}+s_\beta s_{\alpha_1}c_{\alpha_2}) \right) \notag\\
&+ \lambda_5s_{\alpha_1}s_{\alpha_2}\left( \frac{M}{v}s_{\alpha_1}^2c_{\alpha_2}^2 + s_\beta s_{\alpha_1}^2c_{\alpha_2}s_{\alpha_2} \right) \notag\\
\frac{\lambda_{4h}}{3} =& \lambda_1(c_{\alpha_1}^2+s_{\alpha_1}^2c_{\alpha_2}^2)^2 + \lambda_2(c_{\alpha_1}^4+s_{\alpha_1}^4c_{\alpha_2}^4) + \lambda_3s_{\alpha_1}^4s_{\alpha_2}^4 \notag\\
&+ 2\lambda_4s_{\alpha_1}^2s_{\alpha_2}^2 (c_{\alpha_1}^2+s_{\alpha_1}^2c_{\alpha_2}^2) + 2\lambda_5s_{\alpha_1}^4c_{\alpha_2}^2s_{\alpha_2}^2. \label{eq:HiggsSelfCouplings}
\end{align}
Using the Standard-Model couplings in \cref{eq:SMselfcouplings}, the $W$ boson mass in \cref{eq:Wmasses}, and the Higgs mass in \cref{eq:scalarmasses}, we find the following modifications of the Higgs self-couplings:
\begin{align}
\Delta_{\lambda_{3h}} =& \frac{\lambda_{3h}}{\lambda_{3h}^\text{SM}} - 1 = \frac{\lambda_{3h}/(3v)}{\lambda_1 + \lambda_2\cos^2\beta - (\lambda_4+\lambda_5\sin^2\beta)^2/\lambda_3 + \ord{\xi^2}} - 1, \notag\\
\Delta_{\lambda_{4h}} =& \frac{\lambda_{4h}}{\lambda_{4h}^\text{SM}} - 1 = \frac{\lambda_{4h}/3}{\lambda_1 + \lambda_2\cos^2\beta - (\lambda_4+\lambda_5\sin^2\beta)^2/\lambda_3 + \ord{\xi^2}} - 1, \label{eq:selfcouplingmodifications}
\end{align}
where the numerator in these expressions is given in \cref{eq:HiggsSelfCouplings}.

\subsection{The Standard-Model limit}
The Standard Model describes experimental data very successfully so far.
Hence the simplified LET model should reduce to the Standard Model and the new physics should decouple in some limit of parameter space.
Now that we know the effects of the new particles on the Standard-Model particles, we can make this statement more precise.
In the simplified LET model, the scale of new physics is $M$, above which the left-right symmetry is restored.
The effects of new physics at energy scales well below $M$ should therefore be negligible.
Hence we define a small expansion parameter $\xi\equiv\frac{v}{M}$.
We will express the new-physics effects in terms of $\xi$ and consider the limit $\xi\rightarrow0$.

Let us consider the massive gauge bosons first.
Their masses are given in \cref{eq:Wmasses,eq:Zmasses} and can be rewritten in terms of $\xi$ as follows:
\begin{align}
m_W =& \frac{g_Lv}{2} \left( 1 - \frac{\xi^2}{2}\sin^22\beta + \ord{\xi^4} \right), \notag\\
m_{W'} =& \frac{g_Rv}{2} \left( \frac{1}{\xi} + \frac{\xi}{2} + \ord{\xi^3} \right), \notag\\
m_Z =& \frac{g_Lv}{2\cos\theta_W} \left( 1 - \frac{\cos^4\theta_W^\prime}{2}\xi^2 + \ord{\xi^4} \right), \notag\\
m_{Z'} =& \frac{g_Rv}{2\cos\theta_W^\prime} \left( \frac{1}{\xi} + \frac{\cos^4\theta_W^\prime}{2}\xi + \ord{\xi^3} \right).
\end{align}
In the limit of small $\xi$, the new gauge bosons $W'$ and $Z'$ become very heavy, and we can integrate them out.
All higher-dimensional operators that arise from integrating out these fields are suppressed by powers of their mass (that is, they are proportional to powers of $\xi$) and thus vanish in the limit $\xi\rightarrow0$.
Hence the $W'$ and $Z'$ decouple from the Standard-Model fields in the small-$\xi$ limit.

Now let us consider the scalar states.
Their masses are given by \cref{eq:scalarmasses} and can be rewritten in terms of $\xi$ as follows:
\begin{align}
m_{h^0} =& v \left(\sqrt{\lambda_1+\lambda_2\cos^2\beta - \frac{(\lambda_4+\lambda_5\sin^2\beta)^2}{\lambda_3}} + \ord{\xi^2} \right), \notag\\
m_{H_1^0} =& v\sqrt{\frac{\lambda_5\sec2\beta}{2}} \left( \frac{1}{\xi} - \frac{\xi}{2}\left[ \frac{\lambda_2\cos^32\beta}{\lambda_5} - \frac{\lambda_5\sin^22\beta\cos2\beta}{\lambda_5-2\lambda_3\cos2\beta} \right] + \ord{\xi^3} \right), \notag\\
m_{H_2^0} =& v\sqrt{\lambda_3} \left( \frac{1}{\xi} + \frac{\xi}{2}\left[ \left(\frac{\lambda_4+\lambda_5\sin^2\beta}{\lambda_3}\right)^2 - \frac{\lambda_5^2\sin^22\beta\cos2\beta}{\lambda_3(\lambda_5-2\lambda_3\cos2\beta)} \right]  + \ord{\xi^3} \right), \notag\\
m_{A} =& v\sqrt{\frac{\lambda_5\sec2\beta}{2}} \left( \frac{1}{\xi} - \frac{\lambda_2\cos2\beta}{2\lambda_5}\xi + \ord{\xi^3} \right), \notag\\
m_{H^\pm} =& v\sqrt{\frac{\lambda_5\sec2\beta}{2}} \left( \frac{1}{\xi} + \frac{\xi}{2}\cos^22\beta + \ord{\xi^3} \right).
\end{align}
In the limit of small $\xi$, only $h^0$ retains a finite mass.
Like the new vector bosons, all new scalars become very heavy, so we can integrate them out.
The effective operators arising in this process are again suppressed by powers of the corresponding scalar mass (and hence $\xi$) and thus vanish in the limit $\xi\rightarrow0$.
Thus like the new gauge bosons, the new scalars decouple from the Standard-Model fields for $\xi\rightarrow0$.

\begin{table}[t]
{\renewcommand{\arraystretch}{3.2}
\begin{center}
\begin{tabular}{|c|c|}
\hline
$\Delta_W$	&	$ \left( -s^2_{2\beta} + \dfrac{\lambda_2s^2_{4\beta}}{8\lambda_5} + \frac14s_{4\beta}s_{2\beta}\lambda_{453} - \frac12s^2_\beta\lambda_{453}^2 \right)\xi^2 + \ord{\xi^3}$	\\
$\Delta_Z$	&	$ \left( -c^4_{\theta_W^\prime} + \dfrac{\lambda_2s^2_{4\beta}}{8\lambda_5} + \frac14s_{4\beta}s_{2\beta}\lambda_{453} - \frac12s^2_\beta\lambda_{453}^2 \right)\xi^2 + \ord{\xi^3}$	\\\hline
$\Delta_t$	&	$-\left( \dfrac{2\lambda_2s^2_\beta c^2_{2\beta}}{\lambda_5} + 2s^2_\beta c_{2\beta}\lambda_{453} + \frac12\lambda_{453}^2 \right)\xi^2 + \ord{\xi^4}$	\\
$\Delta_b$	&	$-\left( -\dfrac{2\lambda_2c^2_\beta c^2_{2\beta}}{\lambda_5} - 2c^2_\beta c_{2\beta}\lambda_{453} + \frac12\lambda_{453}^2 \right)\xi^2 + \ord{\xi^4}$	\\\hline
$\Delta_\gamma$	&	$\dfrac{\xi^2A_0(\tau_{H^\pm})c_{2\beta}}{A_\text{SM}\lambda_5} \left( \lambda_1 + \lambda_2(1+\frac12s^2_{2\beta}) + \lambda_5c_{2\beta} - \dfrac{\lambda_4(\lambda_4+\lambda_5c^2_\beta)}{\lambda_3} + \ord{\xi^2} \right)$	\\
\hline
$\Delta_{\lambda_{3h}}$	&	$\dfrac{-\lambda_2\lambda_3\sin^2\beta\cos2\beta}{\lambda_3(\lambda_1+\lambda_2\cos^2\beta) - (\lambda_4+\lambda_5\sin^2\beta)^2} + \ord{\xi^2}$ \\
$\Delta_{\lambda_{4h}}$	&	$\dfrac{-\lambda_2\lambda_3\sin^2\beta\cos2\beta + (\lambda_4+\lambda_5\sin^2\beta)^2}{\lambda_3(\lambda_1+\lambda_2\cos^2\beta) - (\lambda_4+\lambda_5\sin^2\beta)^2} + \ord{\xi^2}$	\\\hline
\end{tabular}
\caption{Coupling modifications of the Standard-Model-like Higgs boson $h^0$ to the Standard-Model particles in the limit of small $\xi \equiv v/M$. The last two couplings denote the modifications of the trilinear and quartic Higgs self-couplings. We defined $\lambda_{453} = (\lambda_4+\lambda_5\sin^2\beta)/\lambda_3$, $c_x = \cos{x}$, $s_x = \sin{x}$ for the sake of brevity. The results are given to leading order in $\xi$, and we defined the constant $A_\text{SM} \equiv A_1(\tau_W) + N_cQ_t^2A_{1/2}(\tau_t) = -6.5$.}\label{t:couplingmodifications}
\end{center}}
\end{table}

Now consider the Higgs-coupling modifications in \cref{eq:DeltaV,eq:Deltaf,eq:Deltagamma,eq:selfcouplingmodifications}.
In the small-$\xi$ limit, the mixing angles $\alpha_i$ of the $CP$-even scalars can be approximated by the expressions in \cref{eq:CPevenMixingAngles}.
Thus we can rewrite the Higgs-coupling modifications in terms of $\xi$ as well.
We have listed the resulting expressions in \cref{t:couplingmodifications}.
Note that all coupling modifications to vector bosons and fermions vanish in the limit $\xi\rightarrow0$.
Only the so far unmeasured Higgs self-couplings have a non-decoupling modification.
Hence the simplified LET model could be put to the test by measuring these couplings.

\subsection{Coupling-modification patterns}\label{s:couplingpatterns}
We have worked out the Higgs-coupling modifications for the simplified LET model in terms of the underlying parameters.
We would like to compare these expressions to the measured values of the $\Delta_x$ in order to constrain the parameter space.
However, the coupling modifications cannot be measured with arbitrary precision, and large regions of simplified-LET-model parameter space yield very small coupling modifications.
Hence if the data do not show significant deviations in the Higgs couplings, parts of parameter space would be excluded, but we could not distinguish between the Standard Model and a Standard-Model-like simplified LET model.
In the case that we do find significant deviations from the SM couplings, we can test whether these deviations are compatible with the simplified LET model, or whether another new-physics model is preferred.

Let us consider the possible patterns in the coupling modifications of the simplified LET model.
The question of interest is which regions of parameter space lead to measurable coupling modifications.
Note that the current errors are in the 20\% ballpark (see \cref{f:Higgscouplingmeasurements}) and may improve after the 14 TeV run, so we are looking for coupling modifications in at least the 10\% range.
Let us first consider the couplings to $W$, $Z$ bosons in \cref{eq:DeltaV}.
Note that the coefficients of the $\ord{\xi^2}$ terms cannot become larger than 1.
Given the bound $M > 6.2$ TeV from \cref{s:parameterconstraints}, we have $\xi^2 \lesssim 10^{-3}$.
Thus the $\ord{\xi^2}$ contributions cannot lead to significant deviations of the Higgs couplings.
In order to find regions of parameter space with significant coupling modifications, we can therefore approximate these modifications as
\begin{equation}
\Delta_V \equiv \Delta_{W,Z} = c_\beta c_{\alpha_1} + s_\beta s_{\alpha_1}c_{\alpha_2} - 1 + \ord{\xi^2} \label{eq:approxDeltaV}
\end{equation}
Since the sines and cosines cannot be greater than one, a rough estimate gives
\begin{equation}
\Delta_V \leq c_\beta + s_\beta - 1 \leq s_\beta \sim 10^{-2}.
\end{equation}
Thus our setup does not allow for a positive modification of the $W$, $Z$ couplings larger than about $10^{-2}$.
The measurement of a significant, positive $\Delta_W$ and/or $\Delta_Z$ would therefore falsify the simplified LET model.
Note that the most recent fits (see \cref{f:Higgscouplingmeasurements}) prefer a positive $\Delta_Z$, although the errors are still large enough to allow a small or vanishing coupling modification.
Therefore a more accurate measurement of $\Delta_Z$ would provide a direct test of the simplified LET model.

Another important test comes from the top-coupling modification $\Delta_t$ in \cref{eq:Deltaf}.
Since $\beta \sim 10^{-2}$, we can use $\cos\beta = 1 - \ord{\beta^2}$ to find
\begin{equation}
\Delta_t = \cos\alpha_1 - 1 + \ord{\beta^2} \leq 0.
\end{equation}
That is, the simplified LET model does not allow for a significant positive top coupling modification.
Hence the measurement of $\Delta_t$ provides another direct test of the simplified LET model.
The current data are compatible with the simplified LET model: they prefer a negative $\Delta_t$, where the error bars allow for a vanishing coupling modification but disfavour a positive one.

Now let us consider what coupling patterns are possible if the simplified LET model survives these direct tests.
Comparing the approximate expressions for $\Delta_V$ in \cref{eq:approxDeltaV} to the expressions for $\Delta_t$, $\Delta_b$ in \cref{eq:Deltaf}, we see that they are correlated:
\begin{align}
\Delta_V =& c^2_\beta (1 + \Delta_t) + s^2_\beta (1 + \Delta_b) - 1 + \ord{\xi^2} \notag\\
=& c^2_\beta \Delta_t + s^2_\beta \Delta_b + \ord{\xi^2}.
\end{align}
Although $\Delta_b$ is allowed to be large by the simplified LET model parameter space, a value $|\Delta_b| > 1$ would be in conflict with the measurements of the coupling modifications in \cref{f:Higgscouplingmeasurements}.
Hence for any scenario compatible with the SFitter constraints, we have $|\sin^2\beta \Delta_b| \lesssim 10^{-4}$ and $\cos^2\beta \approx 1$.
This implies that any measurable shift in the Higgs couplings has to satisfy
\begin{equation}
\Delta_V = \Delta_t. \label{eq:measurableDelta}
\end{equation}
Significant differences between the measured values of $\Delta_W$, $\Delta_Z$, and/or $\Delta_t$ would therefore falsify the simplified LET model.
Note that the current data show a tension between $\Delta_Z$ and $\Delta_t$, where the former is preferred to be positive and the latter to be negative.
The error bars have a small overlap around $\Delta_Z = \Delta_t = 0$.
Hence more accurate measurements of the Higgs couplings would provide an important direct test of the simplified LET model.

The last question is whether a significant photon coupling modification is possible in the simplified LET model.
To this end, we consider the leading contributions to the charged-scalar coupling $\lambda_{h^0H^+H^-}$ in \cref{eq:lambdah0HpHm}.
Using the mixing angles of the charged scalars from \cref{eq:chargedMixingAngles}, we find that the quantities in \cref{eq:s13cpcmMain} are given by
\begin{align}
s_{13} =& \sin\beta + \ord{\xi^2} \sim 10^{-2}, \notag\\
c_+ =& -\cos\beta + \ord{\xi^2} \sim -1, \notag\\
c_- =& -(2\cos^2\beta - 1)\xi + \ord{\xi^2} \sim -\xi.
\end{align}
Thus the coupling $\lambda_{h^0H^+H^-}$ can be approximated as:
\begin{align}
\lambda_{h^0H^+H^-} =& (\lambda_1+\lambda_2 )(c_\beta c_{\alpha_1} + s_\beta s_{\alpha_1}c_{\alpha_2})v\left( 1 + \ord{\beta^2}\right) \notag\\
& + \big( \lambda_3\xi^2 + \lambda_4 + \lambda_5 \big)s_{\alpha_1}s_{\alpha_2}M\left( 1 + \ord{\xi^2}\right).
\end{align}
Using the squared mass $m_{H^\pm}^2 = \lambda_5M^2/2(1 + \ord{\xi^2})$ of the charged scalar, this expression yields
\begin{align}
\frac{v\lambda_{h^0H^+H^-}}{m^2_{H^\pm}} =& \frac{2}{\lambda_5} \Big( (\lambda_1 + \lambda_2)(c_\beta c_{\alpha_1} + s_\beta s_{\alpha_1}c_{\alpha_2})\xi^2\big(1+\ord{\beta^2}\big) \notag\\
&\hspace{1cm}+ s_{\alpha_1}s_{\alpha_2} \big( \lambda_3\xi^3 + \lambda_4\xi + \lambda_5\xi \big)\big(1+\ord{\xi^2}\big) \Big).
\end{align}
The numerical factor $A_0(\tau_{H^\pm})/(2A_\text{SM})$ in the expression for $\Delta_\gamma$ in \cref{eq:Deltagamma} is of order $10^{-2}$ for $m_{H^\pm}^2 \gtrsim m_{h^0}^2$.
Hence we need the above expression to be of order 10 to get a measurable effect.
However, this would require $\lambda_5 \sim \xi^2/10$, which would lead to a charged Higgs that is too light.
Hence the simplified LET model does not allow for a substantial $\Delta_\gamma$.
This is compatible with the current SFitter bound, which centers around $\Delta_\gamma = 0$.

\subsection{Benchmark points}\label{s:benchmarkpoints}
The free parameter space of the simplified LET model is spanned by the scale $M$ and the five scalar parameters $\lambda_i$.
A full analysis of this parameter space and the possible signatures is beyond the scope of this work.
In order to see what features could show up in experiment, we define a set of benchmark points that lead to distinct phenomenological features instead.
To get a feel for the possibilities, consider the approximate scalar masses given in \cref{eq:scalarmasses}.
The mass of the Standard-Model-like Higgs $h^0$ can be adjusted independently of the other ones, by changing the values of $\lambda_1$, $\lambda_2$, $\lambda_4$.
For each benchmark point, we tune these parameters such that $m_{h^0} = 126$ GeV.
Note that the leading contributions to the $H^0_1$, $A^0$, and $H^\pm$ masses are all given by $\sqrt{\lambda_5}M$.
Hence we expect them to have similar masses, with $\ord{v}$ mass splittings.
On the other hand, the scalar $H^0_2$ has a mass proportional to $\sqrt{\lambda_3}M$, which can be tuned independently of the other scalar masses.
Thus the simplified LET model allows for different mass hierarchies or compressed spectra, depending on the magnitudes of the parameters $\lambda_3$, $\lambda_5$.

For each benchmark point, we use the parameter values $v=246$ GeV, $\beta=0.0166$ as well as the experimental values $\sin^2\theta_W=0.23126$, $g_L=0.65170$, as in \cref{s:parameterconstraints}.
We also use the best-fit value $\theta_W^\prime = 0.62$ from that section and fix the other gauge couplings by the identities $g_R = g_L\tan\theta_W/\sin\theta_W^\prime$, $2g' = g_L\tan\theta_W/\cos\theta_W^\prime$.
Additionally, we consider two possibilities for the magnitude of the LR-symmetry-breaking scale $M$: a high scale $M=10^{10}$ GeV well outside experimental reach, and a lower scale $M=10^4$ GeV just beyond LHC reach, which is still allowed by the experimental constraints given in \cref{s:gaugebosonconstraints}.
We define six scenarios in terms of the values of the five scalar parameters $\lambda_i$.
For each scenario, we make sure that the constraints from vacuum stability and S-matrix unitarity (see \cref{a:scalarspectrum}) are satisfied.
We describe the distinct phenomenological features of these scenarios below, and summarise the corresponding parameter values in \cref{t:benchmarkpoints}.

Note that the approximate mass expressions for the $CP$-even scalar masses in \cref{eq:scalarmasses} contain some of the scalar parameters in the denominator.
This means that subleading terms in the $\frac{v}{M}$-expansion are potentially large for the benchmarks with small $\lambda_3$ and/or $\lambda_5$.
Therefore, we do not use these approximate expressions to calculate the scalar masses.
Instead, we insert the numerical parameter values directly into the mass matrix in \emph{Mathematica}, calculate its eigensystem, and then extract the masses and mixing angles.
We list the resulting particle masses in \cref{t:benchmarkMasses}.

\begin{table}[t]
\begin{center}
\begin{tabular}{|l|l|c|c|c|c|c|}
\hline
Benchmark point	&	$M$	&	$\lambda_1$	&	$\lambda_2$	&	$\lambda_3$	&	$\lambda_4$	&	$\lambda_5$	\\\hline\hline
SLH-1	&	$10^{10}$	&	0.24	&	0.24	&	0.47	&	0.32	&	0.2	\\
SLH-2	&	$10^4$	&	0.24	&	0.24	&	0.47	&	0.32	&	0.2	\\\hline
2HDM-1	&	$10^{10}$	&	0.41	&	0.4	&	0.44	&	0.49	&	$5\cdot10^{-15}$	\\
2HDM-2	&	$10^4$	&	0.41	&	0.4	&	0.44	&	0.49	&	$5\cdot10^{-3}$	\\\hline
LF-1	&	$10^{10}$	&	0.133	&	0.13	&	$2\cdot10^{-15}$	&	$1\cdot10^{-12}$	&	$3\cdot10^{-7}$	\\
LF-2	&	$10^4$	&	0.14	&	0.14	&	$2\cdot10^{-3}$	&	$5.5\cdot10^{-3}$	&	0.6	\\\hline
Compressed-1	&	$10^{10}$	&	0.133	&	0.13	&	$1.1\cdot10^{-15}$	&	$1\cdot10^{-12}$	&	$5\cdot10^{-15}$	\\
Compressed-2	&	$10^4$	&	0.15	&	0.14	&	$1.1\cdot10^{-3}$	&	$5.2\cdot10^{-3}$	&	$5\cdot10^{-3}$	\\\hline
VLF-1	&	$10^{10}$	&	0.133	&	0.13	&	$1\cdot10^{-20}$	&	$1\cdot10^{-14}$	&	$2\cdot10^{-8}$	\\
VLF-2	&	$10^4$	&	0.13	&	0.13	&	$4.5\cdot10^{-7}$	&	$4\cdot10^{-5}$	&	0.7	\\\hline
Twin-1	&	$10^{10}$	&	0.13	&	0.133	&	$1.58\cdot10^{-16}$	&	$1\cdot10^{-16}$	&	$1\cdot10^{-10}$	\\
Twin-2	&	$10^4$	&	0.131	&	0.131	&	$1.59\cdot10^{-4}$	&	$1\cdot10^{-5}$	&	0.1	\\
\hline
\end{tabular}
\end{center}
\caption{Definitions of the benchmark points in terms of the scalar potential parameters $\lambda_i$ and the LR-symmetry-breaking scale $M$, given in GeV. For each benchmark point, the parameter values $v=246$ GeV, $\beta=0.0166$, $\sin^2\theta_W=0.23126$, $g_L=0.65170$, $\theta_W^\prime=0.62$, $g_R = g_L\tan\theta_W/\sin\theta_W^\prime$, $2g' = g_L\tan\theta_W/\cos\theta_W^\prime$ are kept fixed. We refer to the text for a description of the phenomenological features of each benchmark point.}\label{t:benchmarkpoints}
\end{table}

\begin{table}[t]
\begin{center}
\begin{tabular}{|l|c|c|c|c|}
\hline
Benchmark point	&	$m_{H_1^0}$	&	$m_{A^0}$	&	$m_{H^\pm}$	&	$m_{H_2^0}$	\\\hline\hline
SLH-1	&	$3.2\cdot10^9$	&	$3.2\cdot10^9$	&	$3.2\cdot10^9$	&	$6.9\cdot10^9$	\\
SLH-2	&	$3.2\cdot10^3$	&	$3.2\cdot10^3$	&	$3.2\cdot10^3$	&	$6.9\cdot10^3$	\\\hline
2HDM-1	&	488	&	488	&	500	&	$6.6\cdot10^9$	\\
2HDM-2	&	488	&	488	&	500	&	$6.6\cdot10^3$	\\\hline
LF-1	&	$3.9\cdot10^6$	&	$3.9\cdot10^6$	&	$3.9\cdot10^6$	&	447	\\
LF-2	&	$5.5\cdot10^3$	&	$5.5\cdot10^3$	&	$5.5\cdot10^3$	&	448	\\\hline
Compressed-1	&	496	&	496	&	500	&	332	\\
Compressed-2	&	496	&	496	&	500	&	334	\\\hline
VLF-1	&	$1.0\cdot10^6$	&	$1.0\cdot10^6$	&	$1.0\cdot10^6$	&	1.0	\\
VLF-2	&	$5.9\cdot10^3$	&	$5.9\cdot10^3$	&	$5.9\cdot10^3$	&	0.9	\\\hline
Twin-1	&	$7.1\cdot10^4$	&	$7.1\cdot10^4$	&	$7.1\cdot10^4$	&	126	\\
Twin-2	&	$2.2\cdot10^3$	&	$2.2\cdot10^3$	&	$2.2\cdot10^3$	&	126	\\
\hline
\end{tabular}
\end{center}
\caption{Scalar masses for the benchmark points defined in \cref{t:benchmarkpoints}. All masses are given in GeV. The mass of the Standard-Model-like Higgs $h^0$ has been tuned to 126 GeV in each case.}\label{t:benchmarkMasses}
\end{table}

\paragraph{Single large hierarchy}
If the scalar parameters $\lambda_3$ and $\lambda_5$ are not too small, then all new scalars $H_1^0$, $H_2^0$, $A^0$, $H^\pm$ obtain masses of order $M$, well beyond experimental reach.
An additional hierarchy between $H_1^0$, $A^0$, $H^\pm$ on one side and $H_2^0$ on the other side is possible depending on the relative size of $\lambda_3$ and $\lambda_5$.
However, since all these particles would be well beyond experimental reach, these additional possibilities are of little phenomenological interest.
We denote the benchmark points for a Single Large Hierarchy by SLH-1 ($M=10^{10}$ GeV) and SLH-2 ($M=10^4$ GeV).
The scalar gauge eigenstates barely mix: $h^0$ is almost purely $h^0_{1,11}$-like, $H^0_1$ is almost purely $h^0_{1,22}$-like, and $H^0_2$ is almost purely $h^0_R$-like, with mixings smaller than $10^{-3}$.

\paragraph{2HDM-like hierarchy}
For $\lambda_5 \sim \ord{v^2/M^2}$, the scalars $H_1^0$, $A^0$, $H^\pm$ all have $\ord{v}$ masses.
If $\lambda_3$ remains sizeable, only the fermiophobic $H_2^0$ would have a mass outside of experimental reach.
That is, only the 2HDM-like particles (see \cref{s:2HDM}) could be observed at the LHC.
Hence we denote these benchmark points by 2HDM-1 and 2HDM-2.
We choose the parameters such that the masses of the 2HDM-like particles lie in the $\ord{100\text{ GeV}}$ range.
Note that $H_1^0$, $A^0$ have roughly equal masses, slightly below $m_{H^\pm}$, as we would expect from the scalar mass expressions in \cref{eq:scalarmasses}.
As in the SLH-1 and SLH-2 benchmarks, scalar mixings are smaller than $10^{-3}$.

\paragraph{Light fermiophobic Higgs}
We also consider the reverse of the 2HDM-like hierarchy: if $\lambda_5$ is sizeable but $\lambda_3 \sim \ord{v^2/M^2}$, the fermiophobic state $H_2^0$ can be made light enough to be within experimental reach.
On the other hand, the states $H_1^0$, $A^0$, $H^\pm$ would be out of experimental reach.
Note that other models allow for the existence of a light fermiophobic as well, such as the type-I 2HDM and models with $SU(2)_L$-triplet Higgs fields \cite{Akeroyd:1995hg,Gunion:1989ci,Bamert:1993ah} (we discuss such fermiophobic Higgses in more detail in~\cref{s:fermiophobicHiggs}).
However, in those scenarios the light Higgs $h^0$ is fermiophobic, whereas the benchmark scenario we consider has a fermiophobic light Higgs in addition to the Standard-Model-like Higgs.

We choose the parameters such that $m_{H_2^0}$ lies in the $\ord{100\text{ GeV}}$ range.
Note that the expression for $m_{h^0}^2$ in \cref{eq:scalarmasses} contains the term $-v^2(\lambda_4+\lambda_5\sin^2\beta)^2/\lambda_3$.
Since the small parameter $\lambda_3$ appears in the denominator, we have to choose $\lambda_4$, $\lambda_5$ sufficiently small to ensure that this expression remains positive.
This means that the masses of $H_1^0$, $A^0$, $H^\pm$ lie well below $M$.
We denote the benchmark points with a Light Fermiophobic Higgs by LF-1 and LF-2.
For both benchmarks, mixing is below the percent level.

\paragraph{Compressed spectrum}
A combination of the previous two benchmarks is also possible: if both $\lambda_3$ and $\lambda_5$ are sufficiently small, all new scalars could be within the $\ord{100\text{ GeV}}$ range.
That is, the scalar spectrum does not contain any large hierarchies and is instead compressed.
We denote the corresponding benchmarks by Compressed-1 and Compressed-2.
The mixings for Compressed-1 are smaller than $10^{-3}$, whereas the Compressed-2 benchmark gives a 2\% mixing of $h^0_{1,11}$ and $h^0_R$.

\paragraph{Very light fermiophobic Higgs}
We also consider a special case of the scenario with a light fermiophobic Higgs particle.
Since the mass of $H_2^0$ can be tweaked independently of the other scalar masses by the choice of $\lambda_3$, we could make it even lighter than the Standard-Model-like Higgs $h^0$.
Such a scenario is not necessarily ruled out by experiment, since $H_2^0$ does not decay into fermions.
This means that a very light $H_2^0$ ($m_{H_2^0} \sim \ord{1\text{ GeV}}$) would only decay into pairs of photons.
If the signal strength for its decay is low enough, it could have escaped detection so far.
The experimental constraints on this scenario are given in \cref{s:fermiophobicHiggs}.
We denote the benchmarks for a Very Light Fermiophobic Higgs by VLF-1 and VLF-2.
All scalar mixings of these benchmarks are below $10^{-3}$.

\paragraph{Twin Higgs scenario}
Another interesting special case of the scenario with a light fermiophobic Higgs particle is when $h^0$ and $H^0_2$ are approximately degenerate, i.e.\ if their mass difference is less than their widths.
In that case, both $h^0$ and $H^0_2$ would contribute to the signal strength used for the Higgs discovery, leading to a `twin Higgs'\footnote{Our use of the term `twin Higgs' is not to be confused with `twin Higgs models' in the literature. In those models, each Standard-Model particle has a corresponding particle that transforms under a mirror copy of the SM gauge group (see e.g.~refs.~\cite{Chacko:2005pe,Chacko:2005vw}). The copies are related by a $Z_2$ symmetry called `twin parity', and the twin Higgs is the partner of the Standard-Model Higgs.} \cite{Stech:2013pda,Heikinheimo:2013cua}.
The latter state does not decay to fermions, hence the measured Higgs couplings would deviate from their Standard-Model values.

For these benchmarks, we tweak the parameters such that $m_{h^0} = m_{H_2^0} = 126$ GeV.
We denote these benchmarks by Twin-1 and Twin-2.
Scalar mixings are again small for Twin-1, but mixing among the $h^0_{1,11}$ and $h^0_R$ states is large for Twin-2: the SM-like Higgs $h^0$ is 62\% $h^0_{1,11}$ and 38\% $h^0_R$, whereas the fermiophobic Higgs $H^0_2$ is 38\% $h^0_{1,11}$ and 62\% $h^0_R$.
The mixing with $h^0_{1,22}$ is negligible.

\subsection{Parameter constraints from coupling modifications}
We give the numerical values for the coupling modifications for all benchmark points in \cref{t:benchmarkCMs1,t:benchmarkCMs2}.
As we mentioned in the previous section, we have to be careful with the benchmarks where some of the scalar parameters are very small, because they appear in some of the denominators of the approximate expressions for the masses and mixing angles.
Therefore we have calculated all mixing angles numerically and inserted these into the coupling modifications in \cref{eq:DeltaV,eq:Deltaf,eq:Deltagamma,eq:selfcouplingmodifications}.

\begin{table}[p]
\begin{center}
\begin{tabular}{|l|c|c|c|c|}
\hline
Benchmark point	&	$\Delta_W$	&	$\Delta_Z$	&	$\Delta_t$	&	$\Delta_b$	\\\hline\hline
SLH-1	&	$-1.4\cdot10^{-16}$	&	0.0	&	$-1.4\cdot10^{-16}$	&	0.0	\\
SLH-2	&	$-1.4\cdot10^{-4}$	&	$-4.1\cdot10^{-4}$	&	$-1.4\cdot10^{-4}$	&	$2.1\cdot10^{-3}$	\\\hline
2HDM-1	&	$-1.6\cdot10^{-6}$	&	$-1.6\cdot10^{-6}$	&	$-3.2\cdot10^{-5}$	&	0.11	\\
2HDM-2	&	$-3.8\cdot10^{-4}$	&	$-6.4\cdot10^{-4}$	&	$-4.1\cdot10^{-4}$	&	0.11	\\\hline
LF-1	&	$-6.2\cdot10^{-7}$	&	$-6.2\cdot10^{-7}$	&	$-6.2\cdot10^{-7}$	&	$-6.2\cdot10^{-7}$	\\
LF-2	&	$-2.9\cdot10^{-3}$	&	$-3.1\cdot10^{-3}$	&	$-2.9\cdot10^{-3}$	&	$1.1\cdot10^{-3}$	\\\hline
Compressed-1	&	$-1.6\cdot10^{-7}$	&	$-1.6\cdot10^{-7}$	&	$-9.6\cdot10^{-6}$	&	$3.4\cdot10^{-2}$	\\
Compressed-2	&	$-9.1\cdot10^{-3}$	&	$-9.4\cdot10^{-3}$	&	$-9.1\cdot10^{-3}$	&	$3.5\cdot10^{-2}$	\\\hline
VLF-1	&	$-3.6\cdot10^{-7}$	&	$-3.6\cdot10^{-7}$	&	$-3.6\cdot10^{-7}$	&	$-3.6\cdot10^{-7}$	\\
VLF-2	&	$-6.6\cdot10^{-4}$	&	$-9.3\cdot10^{-4}$	&	$-6.6\cdot10^{-4}$	&	$-2.2\cdot10^{-3}$	\\\hline
Twin-1	&	$-1.9\cdot10^{-7}$	&	$-1.9\cdot10^{-7}$	&	$-1.9\cdot10^{-7}$	&	$1.4\cdot10^{-6}$	\\
Twin-2	&	-0.21	&	-0.21	&	-0.21	&	-0.18	\\
\hline
\end{tabular}
\end{center}
\caption{Numerical results for the Higgs-coupling modifications of tree-level couplings for the benchmark points defined in \cref{t:benchmarkpoints}.}\label{t:benchmarkCMs1}
\end{table}

\begin{table}[p]
\begin{center}
\begin{tabular}{|l|c|c|c|}
\hline
Benchmark point	&	$\Delta_\gamma$	&	$\Delta_{\lambda_{3h}}$	&	$\Delta_{\lambda_{4h}}$	\\\hline\hline
SLH-1	&	$-5.0\cdot10^{-17}$	&	0.0	&	0.83	\\
SLH-2	&	$-5.0\cdot10^{-5}$	&	$-4.2\cdot10^{-4}$	&	0.83	\\\hline
2HDM-1	&	$-1.7\cdot10^{-3}$	&	$-9.7\cdot10^{-5}$	&	2.1	\\
2HDM-2	&	$-1.6\cdot10^{-3}$	&	$-1.2\cdot10^{-3}$	&	2.1	\\\hline
LF-1	&	$-2.6\cdot10^{-11}$	&	$-1.9\cdot10^{-6}$	&	$1.2\cdot10^{-5}$	\\
LF-2	&	$5.1\cdot10^{-5}$	&	$-8.6\cdot10^{-3}$	&	$5.5\cdot10^{-2}$	\\\hline
Compressed-1	&	$-1.6\cdot10^{-3}$	&	$-9.6\cdot10^{-6}$	&	$-1.4\cdot10^{-5}$	\\
Compressed-2	&	$-1.5\cdot10^{-3}$	&	$-2.7\cdot10^{-2}$	&	$7.1\cdot10^{-2}$	\\\hline
VLF-1	&	$-4.1\cdot10^{-10}$	&	$-1.1\cdot10^{-6}$	&	$-2.2\cdot10^{-6}$	\\
VLF-2	&	$-8.8\cdot10^{-5}$	&	$-2.0\cdot10^{-3}$	&	$-4.0\cdot10^{-3}$	\\\hline
Twin-1	&	$-8.1\cdot10^{-8}$	&	$-5.6\cdot10^{-7}$	&	$-7.5\cdot10^{-7}$	\\
Twin-2	&	$6.9\cdot10^{-4}$	&	$-0.52$	&	$-0.61$	\\
\hline
\end{tabular}
\end{center}
\caption{Numerical results for the Higgs-coupling modifications of the loop-induced photon coupling and the Higgs self-couplings for the benchmark points defined in \cref{t:benchmarkpoints}.}\label{t:benchmarkCMs2}
\end{table}

For the benchmark points SLH-1 and SLH-2, all modifications of couplings to vector bosons and fermions are at most of the permille level.
Hence these modifications will not be measurable at the LHC.
Interestingly, the modification of the quartic Higgs self-coupling is as large as 83\%, whereas the trilinear coupling is not modified.
Thus if we could measure both Higgs self-couplings, a simplified LET model with a large hierarchy would distinguish itself from the Standard Model by the strength of only the quartic Higgs self-coupling, even if the LR-symmetry-breaking scale $M$ is very large.

For the benchmark points 2HDM-1 and 2HDM-2, there is an 11\% increase of the Higgs coupling to $b$ quarks.
Such a substantial modification is to be expected from the approximations in \cref{t:couplingmodifications}: the main contribution to $\Delta_b$ in this scenario is proportional to $\xi^2/\lambda_5 \sim 0.1$.
The same is true for $\Delta_t$, but the corresponding term is suppressed by a factor $\sin^2\beta \sim 10^{-4}$, making this coupling modification small.
The $W$, $Z$ coupling modifications also contain a term proportional to $\xi^2/\lambda_5$, but it is suppressed by $\sin^24\beta \sim 10^{-3}$.
Hence the 2HDM-like hierarchy is characterised by an increase in only the $b$ coupling.
Moreover, the quartic Higgs self-coupling is enhanced by a factor 3, whereas the trilinear self-coupling hardly changes.

The LF-1 and LF-2 benchmarks all show very small coupling modifications, at or below the percent level.
Hence these scenarios would not be distinguishable from the Standard Model via the Higgs-coupling modifications.
These scenarios would be characterised by the discovery of a scalar with a mass in the 100 GeV range decaying only into bosons.

The Compressed-1 and Compressed-2 benchmarks both lead to coupling modifications of at most a few percent.
Hence these scenarios would not show themselves via the coupling modifications.
However, in these scenarios there should be several new scalar states in the range of a few hundred GeV, possibly allowing detection at the LHC.

The VLF-1 and VLF-2 benchmarks have small coupling modifications as well, at most at the permille level.
Moreover, a light fermiophobic particle with a mass of 1 GeV would be hard to detect.
Hence these scenarios would be indistinguishable from the Standard Model.

The Twin-1 benchmark has very small deviations from the Standard-Model couplings.
However, the Twin-2 benchmark shows significant modifications: all tree-level couplings are about 20\% weaker than in the Standard Model.
Moreover, both the trilinear and quartic Higgs self-couplings are diminished by 50-60\%.
Currently, the errors on the Higgs couplings are still large enough to allow a 20\% deviation (see \cref{f:Higgscouplingmeasurements}).
If the errors can be reduced after the 14 TeV run, the Twin-2 benchmark can be put to the test.

Note that all benchmark points satisfy $\Delta_W \approx \Delta_Z \approx \Delta_t$ as required by \cref{eq:measurableDelta}, and that $\Delta_\gamma$ is very small as we argued at the end of \cref{s:couplingpatterns}.
The current errors on the Higgs couplings in \cref{f:Higgscouplingmeasurements} are still too large to exclude any of the above scenarios.
However, the twin-Higgs scenario with a low LR-breaking scale could clearly be excluded when the uncertainties shrink.
A diminished bottom coupling with unmodified couplings to other fermions and vector bosons would point in the direction of a 2HDM-like setup, regardless of the magnitude of $M$.
A more interesting scenario is if the Higgs self-couplings can be probed: this would mean that even the single-large-hierarchy scenario can be tested.

\clearpage
\section{Bounds on new scalars}\label{s:newHiggses}
In the last section, we focused on the properties of the Standard-Model-like Higgs boson.
We considered which scenarios could be distinguished from the Standard-Model in experiment and which ones are very Standard-Model-like, looking only at the couplings of $h^0$.
However, the simplified LET model also predicts the existence of additional scalar states.
The observation of a new scalar would be unambiguous proof of physics beyond the Standard Model.
Therefore, it is important to know whether we could observe the new scalar states predicted by the simplified LET model, and how.
This means that we need to study their production cross-sections as well as their branching ratios into Standard-Model particles.

As we have seen in \cref{s:mapping}, the simplified LET model resembles a type-II 2HDM setup.
However, it contains additional physics beyond the Standard Model in the form of an additional gauge group factor $SU(2)_R$ and an $SU(2)_R$-doublet scalar field.
The 2HDM has been studied extensively, so many bounds on the simplified LET model can be drawn from the literature.
Before we can do this, we must know to what extent the non-2HDM physics modifies the 2HDM setup.
In this section, we start with an analysis of the sources that couple this new physics to the Standard-Model fields.
Then we consider the available bounds on the new scalars, and see to what extent they apply to the simplified LET model.
We look at the production and decay channels that are relevant to discovery, and consider a few benchmark examples.
We look for benchmarks that lead to signatures clearly different from the type-II 2HDM, and comment on the prospects for discovery in these scenarios.

\subsection{Couplings to the SM fields}\label{s:couplingsToSM}
As we have shown in \cref{s:mapping}, the simplified LET model looks like a type-II 2HDM with additional constraints on the scalar-potential parameters and the vev ratio $\tan\beta$.
The simplified LET model distinguishes itself from the type-II 2HDM through the presence of three additional gauge bosons and one additional $CP$-even scalar state.
Moreover, the scalar field $\Phi_R$ is fermiophobic, which leads to the possibility of fermiophobic scalar particles.
We are looking for phenomenological scenarios that, if realised in nature, would clearly distinguish the simplified LET model from the Standard Model and the type-II 2HDM.
We have already encountered examples of these scenarios in the previous section: a light fermiophobic Higgs with mass in the $\ord{100\text{ GeV}}$ range, a very light fermiophobic scalar with a mass in the $\ord{1\text{ GeV}}$ range, and a twin Higgs at 126 GeV.
Another interesting possibility would be a scenario with fermiophobic charged scalars.

Let us see what possibilities can be accommodated in the simplified LET model, and in which regions of parameter space these scenarios manifest themselves.
Then we can figure out how these scenarios would manifest themselves in experiment.
Without the right-handed doublet $\Phi_R$, the simplified LET model is a type-II 2HDM with constraints on the parameters and a fixed $\tan\beta$ (see \cref{s:mapping}).
The field $\Phi_R$ does not couple to fermions, hence it can only affect the 2HDM-like fields through its couplings to vector bosons and scalars.
This implies that we can identify three sources through which the non-2HDM physics could couple to the Standard-Model fields:
\begin{itemize}
\item \textbf{$\bm{W-W'}$ mixing.} The field $\Phi_R$ is an $SU(2)_L$ singlet, so it does not couple to the gauge eigenstates $W_L^\pm$. However, it does couple to $W_R^\pm$. The physical $W^\pm$ bosons are mixtures of the gauge eigenstates $W_L^\pm$ and $W_R^\pm$. Hence the components of $\Phi_R$ couple to the $W$ bosons through this mixing. According to \cref{eq:zeta}, the size of the $W-W'$ mixing angle $\zeta$ is set by $\xi^2$, which we found to be of order $10^{-4}$ using the experimental bounds reviewed in \cref{s:gaugebosonconstraints}. Therefore, this mixing will not lead to a significant coupling of $\Phi_R$ to the Standard-Model fields.
\item \textbf{$\bm{Z-Z'}$ mixing.} The neutral gauge bosons are mixtures of $W_L^3$, $W_R^3$, and $B$. At the scale $M$ where the left-right symmetry is broken, the latter two states mix into a massless hypercharge gauge boson $B_Y$ and a massive $\widetilde{Z}'$ (see \cref{a:gaugebosons}). The massless $W_L^3$, $B_Y$ can be rotated into the states $A$, $\widetilde{Z}$ by the Weinberg angle, in analogy to \cref{eq:SMgaugebosons}. After electroweak symmetry breaking, the $\widetilde{Z}$ boson becomes massive. The states $\widetilde{Z}$, $\widetilde{Z}'$ have the same gauge quantum numbers and are therefore mixed into the physical states $Z$, $Z'$ by a rotation over the $Z-Z'$ mixing angle $\eta$. The neutral component of $\Phi_R$ is an $U(1)_Y$ and $SU(2)_L$ singlet, so it couples to neither $B_Y$ nor $\widetilde{Z}$, but it does couple to $\widetilde{Z}'$.\footnote{The charged component of $\Phi_R$ becomes the Goldstone mode that gives mass to the $Z'$ boson, see \cref{eq:Higgsmasseig,eq:chargedMixingAngles}.} This means that $Z-Z'$ mixing is another source through which $\Phi_R$ couples to the Standard Model. However, the $Z-Z'$ mixing angle $\eta$ is of order $\xi^2$ (see \cref{eq:neutralAngles}), which we found to be small using the experimental bounds reviewed in \cref{s:gaugebosonconstraints}. Hence the resulting coupling is not significant either.
\item \textbf{Scalar mixing.} The mixing among the $CP$-even scalars depends on the scalar-potential parameters $\lambda_i$. If all of them have similar magnitudes, then this mixing is small (see \cref{eq:CPevenMixingAngles}) because of the hierarchy $v\ll M$ in the mass matrix. Hence in most of the parameter space, the non-2HDM neutral scalar $H_2^0$ is almost 100\% $h^0_R$-like (see \cref{eq:Higgsmasseig,eq:CPevenMixingAngles}), in which case it has no significant couplings to the Standard-Model fields. However, in some regions of parameter space, this mixing becomes substantial (see e.g.\ the Twin-2 benchmark point). Through this mixing, $H_2^0$ is allowed to couple to the SM particles.
\end{itemize}
Hence the non-2HDM physics only couples to the Standard-Model fields if the mixing among the $CP$-even scalars is substantial.
Therefore scenarios with large scalar mixing are the most interesting when we determine whether the simplified LET model could be tested experimentally.

\subsection{Light fermiophobic Higgs particles}\label{s:fermiophobicHiggs}
Fermiophobic Higgs particles are not unique to the simplified LET model (see e.g.\ ref.~\cite{Akeroyd:1995hg}).
For example, from \cref{eq:alpha2hdm} it follows that in a type-I 2HDM (in which the fermions couple only to $h_2^0$) the Standard-Model-like scalar $h^0$ does not couple to fermions if the $CP$-even mixing angle $\alpha_\text{2HDM}$ equals $\pi/2$ \cite{Akeroyd:1995hg}.
Fermiophobic Higgs particles also appear in models with $SU(2)_L$-triplet Higgs fields \cite{Gunion:1989ci,Bamert:1993ah}.
Note that fermiophobic Higgs particles are incompatible with the type-II 2HDM: the up- and down-type fermions couple to different Higgs doublets, and therefore \cref{eq:alpha2hdm} tells us that $h^0$ couples to at least one type of fermions.
Hence the discovery of a fermiophobic Higgs would have important consequences: apart from indicating the existence of physics beyond the Standard Model, it would rule out the type-II 2HDM, and more specifically the MSSM.
Searches for a fermiophobic Higgs boson have already been performed at particle colliders, and bounds on its mass are readily available in the literature \cite{Agashe:2014kda}.

The signatures of a fermiophobic Higgs are different from those of a Higgs with Standard-Model couplings \cite{Akeroyd:1995hg}.
In the Standard Model, the Higgs coupling to a particle $X$ is determined by the mass $m_X$ (see \cref{s:SMHiggsCouplings}).
That is, the Higgs decays the most often to the heaviest particles that are kinematically accessible.
It also means that the tree-level coupling to photons is zero, and the Higgs can only decay to a pair of photons through loop diagrams involving massive charged particles.
The main contributions to this decay come from a top-quark loop and a $W$-boson loop (see \cref{a:Deltagamma}), which interfere negatively.
Due to this loop-suppression and the negative interference, the branching ratio of the decay $h\rightarrow\gamma\gamma$ is below a percent \cite{Djouadi:1997yw}.
The picture is different for a fermiophobic Higgs.
Such a particle cannot decay into fermions at tree-level.
If the couplings to vector bosons are the same as for the Standard-Model Higgs, the branching ratios into $WW$, $ZZ$, and $\gamma\gamma$ are therefore enhanced with respect to the Standard Model.
Moreover, since the decay into photons is only mediated by $W$ bosons, there is no negative interference from the top-quark loop anymore, so that the photon decay rate is enhanced even further.
For a fermiophobic Higgs with a mass of 120 GeV, the branching ratio into photons is enhanced by an order of magnitude \cite{Aad:2012yq}.

Another consequence of the absence of tree-level couplings to fermions is a reduced production cross-section: gluon fusion, which occurs through a top-quark loop, is turned off.
Hence a fermiophobic Higgs can only be produced through vector-boson fusion or associated production with a vector boson \cite{Akeroyd:1995hg}.
For fermiophobic Higgses with a mass below 120 GeV, the increased branching ratio into photons more than compensates for the diminished production cross-section, leading to an increased signal strength.
The signal strength decreases for larger Higgs masses \cite{Aad:2012yq}.
This makes the diphoton channel an attractive search channel for light fermiophobic Higgses.
Note that these results do not hold any more if the Higgs couplings to $W$, $Z$ bosons are modified with respect to the Standard Model as well.

\subsubsection{Experimental bounds}\label{s:FPexperimentalbounds}
The four LEP experiments \cite{Abreu:2001ib,Heister:2002ub,Achard:2002jh,Abbiendi:2002yc} have searched for fermiophobic Higgs particles in the Higgs-strahlung process $e^+e^-\rightarrow hZ^0$, where the Higgs boson decays into two photons.
Upper limits on $\sigma(e^+e^-\rightarrow hZ) \times BR(H\rightarrow\gamma\gamma)$ were derived as a function of the Higgs mass.
Mass bounds on the fermiophobic Higgs were derived under the assumption that it has the same cross sections as the SM Higgs.
The strongest LEP limit comes from DELPHI \cite{Abreu:2001ib}:
\begin{equation}
m_{H_2^0} > 107\text{ GeV}.\qquad (\text{SM cross sections})
\end{equation}
The other collaborations found lower limits of 105.4 GeV (ALEPH \cite{Heister:2002ub} as well as L3 \cite{Achard:2002jh}) and 105.5 GeV (OPAL \cite{Abbiendi:2002yc}).
However, the cuts on the energy of the photon pair in these analyses make these searches insensitive to fermiophobic Higgs particles with masses below 10 GeV.

More recent searches at the LHC extend the LEP bounds.
Both ATLAS and CMS have searched for fermiophobic Higgs particles produced in either vector-boson fusion or VH associated production and decaying into a pair of photons.
The searches have been performed in the mass range 110-150 GeV.
A fermiophobic Higgs was excluded at 95\% C.L.\ by ATLAS for the mass ranges 110.0-118.0 GeV and 119.5-121.0 GeV \cite{Aad:2012yq}.
CMS excluded a fermiophobic Higgs in the mass range 110-147 GeV \cite{Chatrchyan:2013sfs}.
Combined with the LEP exclusion limits, these exclusion limits result in the mass bound
\begin{equation}
m_{H_2^0} > 147\text{ GeV}.\qquad (\text{SM cross sections})
\end{equation}
Additionally, CMS has explored the mass range 110-300 GeV by combining results for the $\gamma\gamma$, $WW$, and $ZZ$ decay channels.
A fermiophobic Higgs boson was excluded at 95\% C.L.\ in the mass range 110-194 GeV \cite{Chatrchyan:2012vva}.
Thus combined with the previous limits, we have the following lower bound on the mass of the fermiophobic Higgs:
\begin{equation}
m_{H_2^0} > 194\text{ GeV}.\qquad (\text{SM cross sections})
\end{equation}

\subsubsection{Simplified-LET-model limits}\label{s:LETlimitsFP}
In order to see to what extent these mass bounds apply to the fermiophobic Higgs of the simplified LET model, we need to check whether the assumptions made in these searches are valid.
To this end, we need to examine the couplings of $H_2^0$ to the Standard-Model particles.
As we argued in \cref{s:couplingsToSM}, the only possible source of significant couplings of $H_2^0$ to the Standard Model is scalar mixing.
We neglect $\ord{\xi}$ corrections in the following, i.e.\ we set $\zeta = \eta = 0$.

We define the coupling $g_{H_2^0xx}$ of $H_2^0$ to a Standard-Model particle $x$ as the coefficient of the operator $H_2^0xx$ in the Lagrangian.
Using the corresponding Standard-Model coupling $g_{h^0xx}^\text{SM}$ as a normalisation, we define the coupling modifications $\Delta_x^\text{FP}$ of the fermiophobic Higgs in analogy to \cref{eq:CMdefinition}:
\begin{equation}
1 + \Delta_x^\text{FP} \equiv \frac{g_{H_2^0}xx}{g_{h^0xx}^\text{SM}}.
\end{equation}
The Feynman rules for the tree-level couplings of $H_2^0$ to the Standard-Model vector bosons and fermions are given in \cref{a:feynmanrules}.
These give us:
\begin{align}
1 + \Delta_W^\text{FP} =& c_\beta s_{\alpha_1}s_{\alpha_3} - s_\beta(c_{\alpha_1}c_{\alpha_2}s_{\alpha_3}+s_{\alpha_2}c_{\alpha_3}) + \ord{\xi^2}, \notag\\
1 + \Delta_Z^\text{FP} =& c_\beta s_{\alpha_1}s_{\alpha_3} - s_\beta(c_{\alpha_1}c_{\alpha_2}s_{\alpha_3}+s_{\alpha_2}c_{\alpha_3}) + \ord{\xi^2}, \notag\\
1 + \Delta_t^\text{FP} =& \frac{s_{\alpha_1}s_{\alpha_3}}{c_\beta}, \notag\\
1 + \Delta_b^\text{FP} =& \frac{c_{\alpha_1}c_{\alpha_2}s_{\alpha_3}+s_{\alpha_2}c_{\alpha_3}}{s_\beta}. \label{eq:H20couplingmodifications}
\end{align}
We are only interested in cases where these couplings become substantial, which means we can neglect the terms proportional to $s_\beta$ in the $W$ and $Z$ coupling modifications.
Since $c_\beta \approx 1$, the couplings of $H_2^0$ to $W$, $Z$, $t$ are all roughly scaled by $s_{\alpha_1}s_{\alpha_3}$ with respect to the coupling of the Standard-Model Higgs.
Looking at the scalar mass eigenstates in \cref{eq:Higgsmasseig}, we see that this factor only becomes substantial if there is substantial mixing between the gauge eigenstates $h^0_{1,11}$ and $h^0_R$.
Of all our benchmark points in \cref{s:benchmarkpoints}, this is only the case in the Twin-2 benchmark.
On the other hand, the coupling to $b$ is of order $\xi$ unless there is significant mixing between $h^0_{1,22}$ and $h^0_R$.
This is not the case in any of the benchmark points.
Note that the main contributions to the loop-induced coupling of $H_2^0$ to photons (given by \cref{eq:effectivePhotonCoupling}, substituting $H_2^0$ for $h^0$) are given by the $W$ and top loops.\footnote{The $W'$ boson does not contribute significantly to this coupling: its contribution is proportional to $g_{H_2^0W'W'}/m_{W'}^2 \sim 1/M$ and is therefore suppressed by a factor $\xi$ with respect to the $W$ contribution.}
Hence $g_{H_2^0\gamma\gamma}$ scales with $s_{\alpha_1}s_{\alpha_3}$ as well.
The same is true for the effective gluon coupling, since its main contribution comes from the top-quark loop.

In most of the parameter space, the mixing between $h^0_{1,11}$ and $h^0_R$ is negligible.
In those cases, the couplings of $H_2^0$ to the Standard-Model vector bosons are suppressed by $s_{\alpha_1}s_{\alpha_3} \ll 1$.
However, the searches quoted in the previous section all assume that the fermiophobic Higgs has Standard-Model couplings to all bosons.
Hence for most of the simplified-LET-model parameter space, the bounds on the mass of the fermiophobic Higgs are evaded trivially.

The situation is different for the benchmark point Twin-2.
In this scenario, both $h^0$ and $H_2^0$ contribute to the signal strength of the resonance at 126 GeV.
The signal strength $\mu_x(S^0)$ (for $S^0 = h^0, H^0_2$) of a particular decay channel $S^0 \rightarrow xx$ is defined as the observed cross section times branching ratio at a given Higgs mass, normalised by the Standard-Model value:
\begin{equation}
\mu_x(S^0) = \frac{\sigma\times BR(S^0 \rightarrow xx)_\text{obs}}{\sigma\times BR(h^0 \rightarrow xx)_\text{SM}}.
\end{equation}
A deviation from unity could indicate the presence of new physics.
The Higgs signal strength in each channel has been measured at the LHC and can be found in refs.~\cite{Aad:2013wqa} (ATLAS) and \cite{Chatrchyan:2012ufa,Chatrchyan:2014nva} (CMS).
The signal strength in the diphoton channel lies more than one standard deviation from the Standard-Model value.
This may be a statistical fluctuation, but could also be explained by new physics, such as a fermiophobic Higgs boson accompanying the Standard-Model like Higgs, leading to an enhanced diphoton decay rate.
In order to compare the Twin-2 benchmark to these experimental results, we have to consider the change in production cross section as well as the branching ratios for both $h^0$ and $H_2^0$.
Then we need to add the signal strengths coming from both states.
In order to estimate the magnitude of the deviations from their Standard-Model values, we neglect loop corrections in the following discussion.

At the LHC, the Higgs can be produced in vector-boson fusion, VH associated production, gluon fusion, and production in association with $t\bar{t}$ pairs \cite{Dittmaier:2011ti}.
The former two processes are proportional to the Higgs coupling to vector bosons, whereas the latter two scale with the top coupling (the main contribution to gluon fusion) comes from a top-quark loop).
Recall from \cref{s:couplingpatterns} that for $h^0$, we have $\Delta_1 \equiv \Delta_W = \Delta_Z = \Delta_t = c_{\alpha_1} - 1$ (neglecting $\ord{\xi^2}$ corrections and using $c_\beta \approx 1$).
Hence as a tree-level approximation we have
\begin{equation}
\frac{\sigma(pp \rightarrow h^0)}{\sigma(pp \rightarrow h^0)_\text{SM}} = c_{\alpha_1}^2. \label{eq:h0crosssection}
\end{equation}
Similarly, \cref{eq:H20couplingmodifications} tells us that $\Delta_1^\text{FP} \equiv \Delta_W^\text{FP} = \Delta_Z^\text{FP} = \Delta_t^\text{FP} = s_{\alpha_1}s_{\alpha_3} - 1$, plus $\ord{\xi^2}$ and $\ord{\beta^2}$ corrections.
Hence at tree level we have
\begin{equation}
\frac{\sigma(pp \rightarrow H_2^0)}{\sigma(pp \rightarrow h^0)_\text{SM}} = (s_{\alpha_1}s_{\alpha_3})^2. \label{eq:H20crosssection}
\end{equation}

As for the branching ratios, we only take into account the decay channels listed in \cref{t:branchingratios}; all other channels have negligibly small branching ratios.
The given Standard-Model values were calculated with HDecay \cite{Djouadi:1997yw} using $m_h = 126$ GeV.
We estimate the corresponding branching ratios of the simplified LET model using the Higgs-coupling modifications.

Recall that the LET model does not give a correct description of the masses of the lightest two generations of down-quarks nor the leptons (see \cref{a:fermionmasseig}).
The description of their masses and mixings requires an extension of the Yukawa sector.
That means that we currently do not have a good expression for the Yukawa couplings to these fermions.
More specifically, we need to make assumptions about the $\tau$ Yukawa coupling in order to calculate the relevant branching ratios of the simplified LET model.
To this end, we note that the $b$-coupling modifications in \cref{eq:Deltaf,eq:H20couplingmodifications} are determined by the mixings and the vev of $h^0_{1,22}$.
This field component couples in the same way to $b$ and $\tau$ (see \cref{eq:yukawalag}).
Neglecting contributions to the coupling modification from any extensions of the Yukawa sector, we assume that $\Delta_\tau = \Delta_b \equiv \Delta_2$ and $\Delta_\tau^\text{FP} = \Delta_b^\text{FP} \equiv \Delta_2^\text{FP}$.

Now we are ready to calculate the branching ratios of the simplified LET model.
Note that the partial decay widths for the $h^0$ and $H_2^0$ decays into $WW$, $ZZ$, $gg$, $\gamma\gamma$, and $cc$ all scale with respectively $(1+\Delta_1)^2$ and $(1+\Delta_1^\text{FP})^2$.
Similarly, the partial decay widths for the $bb$ and $\tau\tau$ decay channels for $h^0$ and $H_2^0$ scale with respectively $(1+\Delta_2)^2$ and $(1+\Delta_2^\text{FP})^2$.
Thus the branching ratios for the simplified LET model are given by
\begin{align}
BR(h^0 \rightarrow xx) =& \frac{\Gamma(h^0 \rightarrow xx)}{\sum_{y_1}\Gamma(h^0 \rightarrow y_1y_1) + \sum_{y_2}\Gamma(h^0 \rightarrow y_2y_2)} \notag\\
=& \frac{(1+\Delta_x)^2 \Gamma(h^0 \rightarrow xx)_\text{SM}}{(1+\Delta_1)^2 \sum_{y_1}\Gamma(h^0 \rightarrow y_1y_1)_\text{SM} + (1+\Delta_2)^2 \sum_{y_2}\Gamma(h^0 \rightarrow y_2y_2)_\text{SM}} \notag\\
=& \frac{(1+\Delta_x)^2 BR(h^0 \rightarrow xx)_\text{SM}}{(1+\Delta_1)^2 \sum_{y_1}BR(h^0 \rightarrow y_1y_1)_\text{SM} + (1+\Delta_2)^2 \sum_{y_2}BR(h^0 \rightarrow y_2y_2)_\text{SM}}. \label{eq:h0branchingratio}
\end{align}
Here the sums run over $y_1 = W, Z, g, \gamma, c$ and $y_2 = b, \tau$.
The corresponding expression for $BR(H_2^0 \rightarrow xx)$ is obtained by substituting each $\Delta_x$ by $\Delta_x^\text{FP}$.
Recall from \cref{t:benchmarkCMs1} that the Higgs-coupling modifications of the Twin-2 benchmark are given by
\begin{equation}
1 + \Delta_1 = 0.79,\qquad 1 + \Delta_2 = 0.82.
\end{equation}
For the Twin-2 benchmark, the coupling modifications for $H_2^0$ are given by
\begin{equation}
1 + \Delta_1^\text{FP} = 0.62,\qquad 1 + \Delta_2^\text{FP} = -0.58.
\end{equation}
The resulting branching ratios for $h^0$ and $H_2^0$ are listed in \cref{t:branchingratios}.

\begin{table}[t]
\begin{center}
\begin{tabular}{|c|c|c|c|c|c|c|c|}
\hline
	&	$WW$	&	$ZZ$	&	$gg$	&	$\gamma\gamma$	&	$cc$	&	$bb$	&	$\tau\tau$	\\\hline\hline
$BR(h^0 \rightarrow xx)_\text{SM}$	&	0.216	&	0.027	&	0.077	&	0.002	&	0.026	&	0.594	&	0.057	\\\hline\hline
$BR(h^0 \rightarrow xx)_\text{LET}$	&	0.206	&	0.026	&	0.073	&	0.002	&	0.025	&	0.610	&	0.059	\\
$BR(H_2^0\rightarrow xx)_\text{LET}$	&	0.235	&	0.029	&	0.084	&	0.002	&	0.028	&	0.566	&	0.054	\\\hline\hline
$\mu_x(h^0)$	&	0.59	&	0.59	&	0.59	&	0.59	&	0.59	&	0.64	&	0.64	\\
$\mu_x(H_2^0)$	&	0.41	&	0.41	&	0.41	&	0.41	&	0.41	&	0.36	&	0.36	\\\hline
$\mu_x(h^0)+\mu_x(H^0_2)$	&	1.0	&	1.0	&	1.0	&	1.0	&	1.0	&	1.0	&	1.0	\\\hline
\end{tabular}
\end{center}
\caption{Branching ratios and signal strengths of the various $h^0$ and $H_2^0$ decays in the Twin-2 benchmark. The Standard-Model values of the branching ratios were calculated with HDecay \cite{Djouadi:1997yw} for $m_h = 126$ GeV. We neglect the branching ratios for the $\mu\mu$, $ss$, $tt$, $Z\gamma$ decay channels. We refer to the text for details on the approximation of the branching ratios.}\label{t:branchingratios}
\end{table}

The total signal strength is given by the sum of the contributions from $h^0$ and $H_2^0$.
Combining \cref{eq:h0crosssection,eq:H20crosssection,eq:h0branchingratio}, we find
\begin{align}
\mu_{x,\text{tot}} =& \mu_x(h^0) + \mu_x(H_2^0) \notag\\
=& \frac{c_{\alpha_1}^2 (1 + \Delta_x)^2}{(1 + \Delta_1)^2 \sum_{y_1}BR(h^0 \rightarrow y_1y_1)_\text{SM} + (1 + \Delta_2)^2 \sum_{y_2}BR(h^0 \rightarrow y_2y_2)_\text{SM}} \notag\\
&+ \frac{(s_{\alpha_1}s_{\alpha_3})^2 (1 + \Delta_x^\text{FP})^2}{(1 + \Delta_1^\text{FP})^2 \sum_{y_1}BR(h^0 \rightarrow y_1y_1)_\text{SM} + (1 + \Delta_2^\text{FP})^2 \sum_{y_2}BR(h^0 \rightarrow y_2y_2)_\text{SM}}.
\end{align}
The numerical values have been summarised in \cref{t:branchingratios}.
Note that none of the decay channels has a total signal strength that deviates significantly from 1.
In particular, there is no enhancement in the diphoton channel, which would be the case for the fermiophobic scenarios usually considered in the literature.
In order to see whether one could distinguish the twin-Higgs scenario of the simplified LET model from the Standard Model, an analysis of the loop corrections to these signal strengths is necessary.
This is beyond the scope of this work.
However, the lack of any change in the signal strengths is not surprising.
In most of the parameter space, $h^0$ is a Standard-Model-like scalar with negligible deviations from the Standard-Model couplings, whereas $H_2^0$ is a fermiophobic Standard-Model singlet.
The latter can only couple significantly to the Standard Model through mixing with the other scalars.
In case of significant $h^0-H_2^0$ mixing, there is therefore no additional contribution to the signal strength, but rather the Standard-Model contributions to the signal strength are divided among the two scalars.
This scenario is different from fermiophobic scenarios in the literature, in which there is a single, fermiophobic Higgs state that is simply assumed to have couplings different from the Standard-Model Higgs.

Note that the situation may change in the complete LET model, in which there is a second fermiophobic bidoublet $\Phi_2$.
Since its vevs are of order $v$, we expect that there are scenarios in which the components of $\Phi_1$ and $\Phi_2$ mix significantly, and could produce a twin Higgs at 126 GeV.
However, $\Phi_2$ does have its own couplings to the Standard-Model fields, since it is an $SU(2)_L$-antidoublet.
This means that the couplings of the twin state to vector bosons would change.
We explore this possibility in \cref{s:completeLET}.

To summarise, the properties of the fermiophobic Higgs boson as assumed in direct searches do not hold in the simplified LET model.
In most of the parameter space, the couplings of $H_2^0$ to the Standard-Model particles are strongly suppressed.
An exception is the Twin-2 benchmark, in which there is significant mixing between $h^0_{1,11}$ and $h^0_R$.
However, the fermiophobic Higgs in this scenario is very different from the fermiophobic Higgs usually assumed in direct searches: because of the large mixing, the twin state does actually couple to fermions.
Both $h^0$ and $H_2^0$ contribute to the signal strengths used for the Higgs discovery.
However, the total signal strength is not affected significantly.

\subsection{Charged Higgses}\label{s:chargedScalars}
Charged scalar particles are a general feature of 2HDMs, and supersymmetric models in particular (see \cref{s:2HDM} and the references therein).
They also appear in models with $SU(2)_L$-triplet Higgs fields \cite{Gunion:1989ci,Bamert:1993ah}.
Since the Standard Model does not contain any charged scalar particles, the observation of such a particle would be a clear sign of physics beyond the Standard Model.
Therefore charged scalars have been the subject of various collider searches, and limits on their masses are readily available in the literature \cite{Agashe:2014kda}.

In LEP searches, the charged Higgses are assumed to be pair-produced in the process $e^+e^-\rightarrow H^+H^-$ \cite{Abbiendi:2013hk}.
In the currently available LHC searches, the charged Higgs is assumed to be produced in $t\bar{t}$ events, where the top decays as $t\rightarrow H^+b$ and the antitop decays as $\bar{t}\rightarrow\bar{b}W^-$ \cite{Aad:2012tj,Aad:2013hla,Chatrchyan:2012vca}.
In all searches it is assumed that $BR(H^+\rightarrow \tau^+\nu) + BR(H^+\rightarrow c\bar{s}) = 1$, which is valid as long as $\tan\beta_\text{2HDM}$ is larger than a few units \cite{Abbiendi:2013hk}.

In some 2HDM scenarios, the charged scalars can become fermiophobic.
An example is the inert 2HDM \cite{Ma:2006km,Ma:2008uza}, in which a $Z_2$ symmetry is postulated under which the Standard-Model fields and $\phi_1$ are even but $\phi_2$ is odd.
Such a symmetry forbids linear interactions of $\phi_2$ with the SM fields, thus making the second doublet inert.
The $Z_2$ symmetry implies that $\vev{\phi_2}=0$, hence $\beta_\text{2HDM}=0$, so that $H^\pm$ is the unmixed charged component of the fermiophobic $\phi_2$ (see \cref{eq:2HDMscalars}).
Fermiophobic charged scalars can also occur in the type-I 2HDM: all couplings of $H^\pm$ to fermions are then suppressed by $\tan\beta_\text{2HDM}$, which makes the charged scalars fermiophobic for $\tan\beta_\text{2HDM} \gg 1$ \cite{Branco:2011iw}.

Note that if the charged scalars are fermiophobic, the bounds from direct searches are evaded trivially.
In that case, other production and decay channels need to be considered.
This has been done for relatively light charged scalars ($m_W \leq m_{H^\pm} \leq m_W + m_Z$) in ref.~\cite{Ilisie:2014hea}.
It was found that the loop-induced decay $H^+\rightarrow W^+\gamma$ becomes very relevant in this case.
The most important production channels were found to be associated production with either a neutral scalar or a charged $W$.

\subsubsection{Experimental bounds}
The four LEP collaborations have searched for pair-produced charged scalars in the framework of 2HDMs.
The various search channels of all four collaborations have recently been statistically combined, and the results have been interpreted in terms of a type-I or type-II 2HDM \cite{Abbiendi:2013hk}.
It is assumed that $BR(H^+ \rightarrow \tau^+\nu_\tau) + BR(H^+ \rightarrow c\bar{s}) = 1$, i.e.\ the charged Higgs decays as either $H^+\rightarrow\tau^+\nu_\tau$ or $H^+\rightarrow c\bar{s}$.
The combined data result in the following lower bound at 95\% C.L.:
\begin{equation}
m_{H^\pm} > 80\text{ GeV}.\qquad (\text{type-II 2HDM})
\end{equation}
This bound holds for any value of $BR(H^+ \rightarrow \tau^+\nu_\tau)$ between 0 and 1.
Stronger bounds are given for two limiting cases: for $BR(H^+ \rightarrow \tau^+\nu_\tau) = 0$, a charged Higgs is excluded for $m_{H^\pm}$ below 80.5 GeV and in the interval between 83 and 88 GeV.
For $BR(H^+ \rightarrow \tau^+\nu_\tau) = 1$, the lower bound $m_{H^\pm} > 94$ GeV is given.

ATLAS has performed a search in the mass range 90-160 GeV in the $\tau\nu$ channel \cite{Aad:2012tj}, and another one in the mass range 90-150 GeV in the $c\bar{s}$ channel \cite{Aad:2013hla}.
Both searches set limits on $BR(t\rightarrow H^+b)$ in the $1-5\%$ range.
A mass bound on $H^\pm$ was only given in the context of the MSSM.
CMS has performed a search for $m_{H^\pm}$ in the range 80-160 GeV in the $\tau\nu$ channel as well \cite{Chatrchyan:2012vca}.
This resulted in upper limits on the branching ratio $BR(t\rightarrow H^+b)$ in the 2-4\% range.

\subsubsection{Simplified-LET-model limits}\label{s:HpmLimits}
Let us first discuss whether the simplified LET model allows for a fermiophobic charged Higgs.
Note that the mixings of the charged-scalar components are independent of the scalar potential: the mixing angles in \cref{eq:chargedMixingAngles} are completely determined by the vev parameters $v$, $\beta$, $M$.
Inserting the mixing angles into the mass eigenstates in \cref{eq:Higgsmasseig}, we find
\begin{equation}
H^\pm = \sin\beta h^\pm_{1,21} + \cos\beta h^\pm_{1,12} + \xi\cos2\beta h^\pm_R + \ord{\xi^2}.
\end{equation}
That is, the charged Higgs contains only a very small component of the fermiophobic $h^\pm_R$.
Thus the simplified LET model does not allow for a fermiophobic charged scalar, and therefore does not evade the aforementioned bounds from $H^\pm$ searches trivially.
This may change in the presence of a second Higgs bidoublet; we will comment on this possibility in \cref{s:completeLET}.

Since $H^\pm$ contains only a very small $h^\pm_R$-component, we expect the charged scalar of the simplified LET model to be very similar to the charged scalar of the 2HDM.
We can see how large the deviations from the $H^\pm$-couplings of the 2HDM are by considering the Feynman rules for the $H^\pm$ couplings to Standard-Model particles in \cref{a:HpmToSMbosonCouplings,a:HpmWmpSfeynRules} and comparing them to the corresponding Feynman rules of the 2HDM in \cref{a:2HDMfeynmanrules}.
It turns out that these deviations are negligibly small indeed:
\begin{itemize}
\item The tree-level $H^\pm W^\mp \gamma$ and $H^\pm W^\mp Z$ couplings vanish in the 2HDM.
In the simplified LET model, they do not vanish but are suppressed by $\xi^2$ (the quantities $s_{13}$, $c_+$ were defined in \cref{eq:s13cpcmMain}):
\begin{align}
g_{H^\pm W^\mp \gamma} =& -\frac{g_Lev}{2}(s_\beta c_+ + c_\beta s_{13} + \ord{\xi^2}) =  -\frac{g_Lev}{2} \cdot \ord{\xi^2}, \notag\\
g_{H^\pm W^\mp Z} =& \frac{g_L^2vs^2_{\theta_W}}{2c_{\theta_W}} (s_\beta c_+ + c_\beta s_{13}) + \ord{\xi^2}) = \frac{g_L^2vs^2_{\theta_W}}{2c_{\theta_W}} \cdot \ord{\xi^2}.
\end{align}
\item The $H^+H^-\gamma$ and $H^+H^-Z$ couplings are identical (see \cref{a:HpmToSMbosonCouplings,a:2HDMHpmcouplingsToBosons}).
\item Recall from \cref{s:mapping} that we have to set $\beta_\text{2HDM} = \pi/2 - \beta$ when we map the simplified LET model onto the 2HDM. Using the relation $m_b/m_t = \tan\beta$, we see that the $H^+t\bar{b}$ coupling of the simplified LET model in \cref{a:ScalarToFermions} is equal to the 2HDM coupling (see \cref{a:2HDMfeynmanrules}) plus $\ord{\xi^2}$ corrections.
\item Also recall from \cref{s:mapping} that $\alpha_\text{2HDM} = -\alpha_1$.
Hence we see that the $H^\pm W^\mp h^0$ coupling of the 2HDM can be written as
\begin{equation}
g_{H^\pm W^\mp h^0}^\text{2HDM} = \pm\frac{g_L}{2}\cos(\alpha_\text{2HDM} - \beta_\text{2HDM}) = \mp\frac{g_L}{2}\sin(\alpha_1-\beta).
\end{equation}
In most of the parameter space, $\alpha_2 = \ord{\xi^2}$.
Using this approximation, we find that the $H^\pm W^\mp h^0$ coupling of the simplified LET model yields the same expression as the 2HDM with $\ord{\xi^2}$ corrections.
This approximation holds for all of our benchmark points except Twin-2.
A deviation from the 2HDM value of this coupling is to be expected in the Twin-2 scenario because of significant mixing among the $CP$-even scalars.
\item The scalar-potential parameters of the simplified LET model can be mapped onto the 2HDM as in \cref{eq:2HDMmapping}.
Inserting these values into the $h^0H^+H^-$ coupling of the 2HDM (see \cref{a:2HDMfeynmanrules}), we find
\begin{equation}
\lambda_{h^0H^+H^-}^\text{2HDM} = v(\lambda_1 + \lambda_2(1 + \frac12\sin^22\beta)).
\end{equation}
In most of the parameter space, we have $c_{\alpha_1} = c_\beta + \ord{\xi^2}$, $s_{\alpha_1}c_{\alpha_2} = s_\beta + \ord{\xi^2}$.
These approximations do not hold if there is significant mixing between $h^0_{1,11}$ and $h^0_{1,22}$, which is not the case in any of our benchmark scenarios.
Inserting these approximations into the $h^0H^+H^-$ coupling of the simplified LET model (see \cref{a:HpmToSMbosonCouplings} combined with \cref{eq:s13cpcmMain}), we find
\begin{align}
\lambda_{h^0H^+H^-} =& v\big( \lambda_1 + \lambda_2(1 + \frac12\sin^22\beta) + (\lambda_4 + \lambda_5)\frac{s_{\alpha_1}s_{\alpha_2}}{\xi} + \ord{\xi^2} \big).
\end{align}
For all benchmarks except Twin-2, the mixing between $h^0_{1,11}$ and $h^0_R$ is very small: $s_{\alpha_1}s_{\alpha_2} = \ord{\xi^2}$.
This means that the $h^0H^+H^-$ coupling of the simplified LET model is the same as in the 2HDM, up to $\ord{\xi}$ corrections.
Again, a deviation is to be expected for the Twin-2 benchmark because of significant mixing among the $CP$-even scalars.
\end{itemize}
Hence we treat the charged scalars of the simplified LET model as 2HDM-like charged scalars in the following.
Note that we need to make assumptions about their couplings to leptons and the two lightest quark generations, as we have done for the fermiophobic Higgs in \cref{s:LETlimitsFP}.
Since the mixing of the charged-scalar components with the right-handed doublet is negligible, we postulate that the $H^\pm$ couplings to these fermions can be approximated by the corresponding Feynman rules of the 2HDM (see \cref{a:2HDMfermioncouplings}).
That is, we neglect the contributions of additional scalars that would be needed to describe the fermion masses and mixings correctly.

In order to see to what extent the $H^\pm$ searches apply to the simplified LET model, we calculate the $H^\pm$ branching ratios using HDecay \cite{Djouadi:1997yw}.
The program calculates the 2HDM branching ratios from the input parameters $\tan\beta_\text{2HDM}$, $m_{12}^2$, $\alpha_\text{2HDM}$, $m_{h^0}$, $m_{H^0}$, $m_{A^0}$, and $m_{H^\pm}$.
In order to be able to use HDecay for our setup, we map the simplified LET model onto the 2HDM as in \cref{s:mapping}, using the values
\begin{align}
\alpha_\text{2HDM} =& -\beta = -0.0166, \notag\\
\tan\beta_\text{2HDM} =& \cot\beta = 60.2.
\end{align}
We set $m_{h^0} = 126$ GeV and vary $m_{H^\pm}$ from 100 GeV to 500 GeV in steps of 5 GeV.
In order to choose input values for $m_{12}^2$, $m_{H^0}$, and $m_{A^0}$, we note that these parameters are not independent in the simplified LET model.
Using the mapping in \cref{eq:2HDMmapping} as well as the minimum condition in \cref{eq:minimumcondition}, we find
\begin{equation}
m_{12}^2 = \frac14\left( \lambda_5M^2\tan2\beta - \lambda_2v^2\sin2\beta \right) = \frac12m_{A^0}^2\sin2\beta. \label{eq:HDecayApproximation1}
\end{equation}
Furthermore, the mass expressions in \cref{eq:scalarmasses} yield
\begin{align}
m_{H^\pm}^2 =& \frac{\lambda_5}{2}M^2(\sec2\beta + \ord{\xi^2}), \notag\\
m_{A^0}^2 =& m_{H^\pm}^2 - \frac{\lambda_2}{2}v^2 + v^2\cdot\ord{\xi^2}, \notag\\
m_{H_1^0}^2 =& m_{A^0}^2 + v^2\cdot\ord{\beta^2}. \label{eq:HDecayApproximation2}
\end{align}
We neglect the $\ord{\xi^2}$ and $\ord{\beta^2}$ corrections and keep $\lambda_2$ fixed at 0.2, which means $\lambda_2v^2/2 = 6\cdot10^3\text{ GeV}^2$.

\begin{figure}[t]
\begin{center}
\includegraphics[width=.8\textwidth]{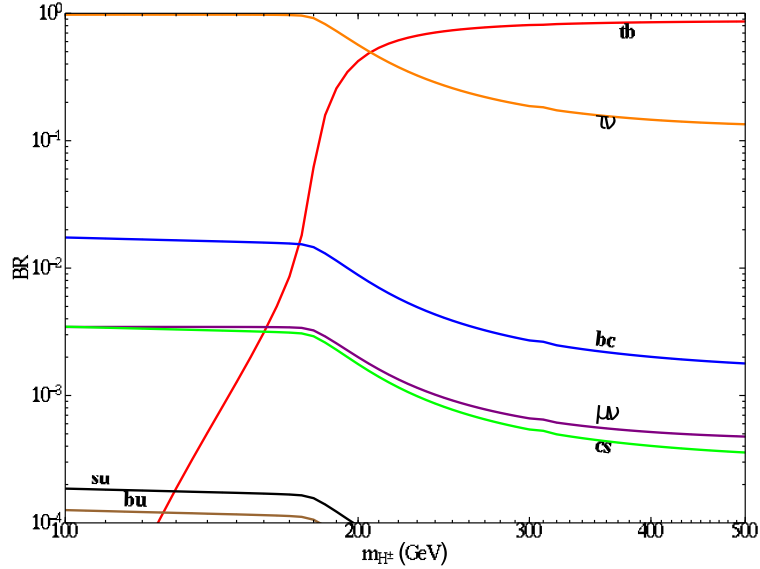}
\end{center}
\caption{$H^\pm$ Branching ratios of the simplified LET model given by HDecay for $m_{H^\pm}$ between 100 and 500 GeV, varied in steps of 5 GeV. We used the values $\tan\beta_\text{2HDM} = 60.2$ and $\alpha_\text{2HDM} = -0.0166$. For an estimate of the $H^0_1$ and $A^0$ masses as well as the 2HDM-parameter $m_{12}^2$, we keep $\lambda_2 = 0.2$ fixed and use the approximations in \cref{eq:HDecayApproximation1,eq:HDecayApproximation2}.}\label{f:HpmBRs}
\end{figure}

The resulting branching ratios are given in \cref{f:HpmBRs}.
Our results agree qualitatively with ref.~\cite{Logan:2010ag,}, in which the $H^\pm$ branching ratios were given for large $\tan\beta_\text{2HDM}$ (see e.g.\ figure 4 in this reference, where the branching ratios are given for $\tan\beta_\text{2HDM} = 50$).
For $m_{H^\pm} > m_t$, the $t\bar{b}$ decay channel is dominant.
For $m_{H^\pm} < m_t$, the $\tau\nu$ channel is the dominant one, whereas the $c\bar{s}$ channel is suppressed.

Note that the simplified LET model evades the mass bounds from the searches in which $BR(H^+ \rightarrow c\bar{s}) = 1$ was assumed.
Instead, we can apply the bound from ref.~\cite{Abbiendi:2013hk} for $BR(H^+ \rightarrow \tau\nu) = 1$:
\begin{equation}
m_{H^\pm} > 94\text{ GeV}.
\end{equation}

\clearpage
\section{The complete LET model}\label{s:completeLET}
Until now we have considered a simplified version of the LET model, in which the scalar sector consists only of the bidoublet $\Phi_1$ and the right-handed doublet $\Phi_R$.
These two scalar fields are sufficient to describe the breakdown of the gauge symmetry from $\LRgroup$ to the Standard-Model gauge group.
More importantly, this simplified setup has allowed us to analyse the phenomenological features of the LET model without being overwhelmed by a plethora of scalar particles and their mixings.
We have discussed the properties of the heavy $W'$, $Z'$ bosons as well as the bounds on their masses that are available in the literature.
Moreover, we have been able to obtain analytical expressions on the scalar masses and mixings, and used these to help us figure out the possible phenomenological features of the model.
We have worked out the modifications of the couplings of the Standard-Model-like Higgs at 126 GeV, and defined a set of benchmark scenarios to get a feel for the magnitude of the deviations from the Standard Model.
We have also discussed experimental bounds on the new scalar particles, and to what extent they apply to the simplified LET model.
Because of the reduced number of scalar fields with respect to the complete LET model, we were able to discern the regions of parameter space that lead to potentially measurable deviations from Standard-Model predictions at the LHC.

Recall that our goal of studying the LET model was to figure out the low-energy phenomenology of the trinification model.
Of course, the simplified LET model is only partly a substitute for the trinification model, since we have neglected the effects of the second scalar bidoublet $\Phi_2$.
This scalar field is not a Standard-Model singlet, so we have to include it in order to get a good grasp on the trinification model.

In this section, we finally include $\Phi_2$ into our analysis, and discuss the consequences it has for the results from the previous sections.
First we discuss the changes in field content as well as the new scalar potential.
Then we discuss the effects of the inclusion of $\Phi_2$ on the gauge-boson masses and mixings, and how the bounds from \cref{s:gaugebosonconstraints} change as a result.
We look at the phenomenological scenarios that are possible in the complete LET model.
Analogously to \cref{s:benchmarkpoints}, we subsequently define a set of benchmark points that lead to these phenomenological scenarios.
Then we calculate to what extent these scenarios lead to measurable deviations from the Standard Model.

\subsection{The second bidoublet}
With the second bidoublet $\Phi_2$ added back, the complete LET model has eight more scalar fields than the simplified setup.
The number of Goldstone bosons remains unchanged, so the complete LET model contains eight additional physical particles: two $CP$-even states, two $CP$-odd states, and two pairs of charged states.
This makes a total of 14 physical scalars:
\begin{align}
CP-\text{even}:\qquad& h^0, H^0_1, H^0_2, H^0_3, H^0_4, \notag\\
CP-\text{odd}:\qquad& A^0_1, A^0_2, A^0_3, \notag\\
\text{charged}:\qquad& H^\pm_1, H^\pm_2, H^\pm_3.
\end{align}
These states are linked to the gauge eigenstates through three $5\times5$ rotation matrices, each of which is described by 10 mixing angles.
These mixings are defined in \cref{a:completeLETmodel}.
As in the simplified LET model, we define $h^0$ as the most $h^0_{1,11}$-like scalar, which naturally has a mass of order $v$.
We define the other $CP$-even scalars $H^0_1$, $H^0_2$, $H^0_3$, $H^0_4$ respectively as the most $h^0_{2,11}$-, $h^0_{1,22}$-, $h^0_{2,22}$-, $h^0_R$-like scalars.
We define the massive $CP$-odd states $A^0_1$, $A^0_2$, $A^0_3$ respectively as the most $a^0_{1,22}$-, $a^0_{2,11}$-, $a^0_R$-like scalars.
The charged states $H^\pm_1$, $H^\pm_2$, $H^\pm_3$ are defined respectively as the most $h^\pm_{2,21}$-, $h^\pm_{1,12}$-, $h^\pm_{2,12}$-like scalars.

Again, the scalar masses and mixings are determined by the scalar potential.
The most general renormalisable scalar potential for the complete LET model is given by
\begin{align}
V_2(\Phi_1,\Phi_2,\Phi_R) =& V(\Phi_1,\Phi_R) + \frac{\widetilde\lambda_1}{2}\tr{\Phi_2^\dagger\Phi_2}^2 + \frac{\widetilde\lambda_2}{2}\tr{\Phi_2^\dagger\Phi_2\Phi_2^\dagger\Phi_2} \notag\\
&+ \widetilde\lambda_3\tr{\Phi_2^\dagger\Phi_2}(\Phi_R\Phi_R^\dagger) + \widetilde\lambda_4\Phi_R\Phi_2^\dagger\Phi_2\Phi_R^\dagger \notag\\
&+ \widetilde\lambda_5\tr{\Phi_1^\dagger\Phi_1}\tr{\Phi_2^\dagger\Phi_2} + \widetilde\lambda_6\left|\tr{\Phi_1^\dagger\Phi_2}\right|^2 \notag\\
&+ \frac{\widetilde\lambda_7}{2}\left( \tr{\Phi_1^\dagger\Phi_2}^2 \plushc\right) + \widetilde\lambda_8\tr{\Phi_1^\dagger\Phi_1\Phi_2^\dagger\Phi_2} \notag\\
&+ \widetilde\lambda_9\tr{\Phi_1^\dagger\Phi_2\Phi_2^\dagger\Phi_1} + \frac{\widetilde\lambda_{10}}{2}\left( \tr{\Phi_1^\dagger\Phi_2\Phi_1^\dagger\Phi_2} \plushc\right) \notag\\
&+ \mu^2_{22}\tr{\Phi_2^\dagger\Phi_2} + \left(\mu^2_2\det\Phi_2 \plushc \right). \label{eq:completeLETpotential}
\end{align}
Here $V(\Phi_1,\Phi_R)$ is the scalar potential of the simplified LET model as given in \cref{eq:simplifiedLETpotential}.
As in the simplified LET model, all parameters appearing in the scalar potential are assumed to be real in order to avoid tree-level $CP$-violation.
This scalar potential should have a minimum at the vev
\begin{equation}
\vev{\Phi_1} = \begin{pmatrix} v_1 & 0 \\ 0 & b_1 \end{pmatrix},\qquad
\vev{\Phi_2} = \begin{pmatrix} v_2 & 0 \\ 0 & b_2 \end{pmatrix},\qquad
\vev{\Phi_R} = \begin{pmatrix} 0 & M \end{pmatrix}.
\end{equation}
If the potential is to have an extremum at this vev, the five dimensionful parameters $\mu_{11}^2$, $\mu_{22}^2$, $\mu_R^2$, $\mu_1^2$, $\mu_2^2$ are fixed in terms of the dimensionless parameters $\lambda_i$, $\widetilde{\lambda}_j$ and the vev parameters $v_1$, $b_1$, $v_2$, $b_2$, $M$.
These five conditions are given in \cref{eq:completeLETminimum}.

The vev parameters are restricted by the condition $v_1^2 + b_1^2 + v_2^2 + b_2^2 = v^2 = (246\text{ GeV})^2$.
In analogy to the simplified LET model, it will be convenient to reparametrise the vev parameters as follows:
\begin{align}
v_1 =& v\cos\alpha\cos\beta_1,\qquad v_2 = v\sin\alpha\cos\beta_2, \notag\\
b_1 =& v\cos\alpha\sin\beta_1,\qquad b_2 = v\sin\alpha\sin\beta_2. \label{eq:completeLETvevRedefinition}
\end{align}
As in the simplified LET model, the vev parameters of $\Phi_1$ are restricted by the ratio of the bottom and top mass: $\tan\beta_1 = b_1/v_1 = m_b/m_t = 0.0166$.
Since $\Phi_2$ does not contribute to the fermion masses, there is no such restriction on $\beta_2$.

In order for the extremum of the scalar potential to be a minimum, the eigenvalues of the mass matrix (except those corresponding to Goldstone bosons) must be positive.
Approximate expressions for the scalar masses to leading order in $\xi^2$ are given in~\cref{eq:completeLETscalarMasses}.
The $CP$-even states $h^0$, $H^0_1$ have masses of order $v$, whereas $H^0_2$, $H^0_3$, $H^0_4$ have masses of order $M$.
This is not surprising: if we decouple the bidoublet $\Phi_2$ from the model, we get one light state $h^0$ and two heavy states.
Since $\Phi_1$ and $\Phi_2$ are copies of the same representation, we expect that $\Phi_2$ adds one light and one heavy scalar to the spectrum as well.
The $CP$-odd state $A^0_1$ is light, whereas $A^0_2$, $A^0_3$ are heavy.
Again, this is not surprising.
In the simplified LET model, the $CP$-odd components of the bidoublet $\Phi_1$ give rise to one Goldstone and one heavy state.
Thus we would expect $\Phi_2$ to contribute one heavy state as well.
Since there are no more would-be Goldstones, the other $CP$-odd component of $\Phi_2$ becomes a massive state with a mass of order $v$.
For the same reason, $H^\pm_1$ is light while $H^\pm_2$, $H^\pm_3$ have masses of order $M$.

Due to the large number of fields and scalar parameters, it is very challenging to analyse all the possible scenarios for the scalar masses and mixings in the complete LET model.
In the rest of this chapter, we will limit ourselves to a study of phenomenologically interesting numerical benchmark points.

\subsection{Gauge-boson mixing}
As we have seen in \cref{s:couplingsToSM}, there are three sources through which the new physics couples to the Standard Model: $W-W'$ mixing, $Z-Z'$ mixing, and scalar mixing.
For the simplified LET model, it turned out that only the latter can lead to significant couplings.
With the inclusion of the second scalar bidoublet $\Phi_2$, the gauge bosons receive additional contributions to the mass matrix.
Hence we need to check the size of $W-W'$ and $Z-Z'$ mixing for the complete LET setup.
The gauge-boson masses and mixings are derived in \cref{a:gbCompleteLET}.
We find that the masses and mixings are not affected significantly by the inclusion of $\Phi_2$: the masses of the charged vector bosons are given by
\begin{align}
m_W =& \frac{g_Lv}{2}\left( 1 - \frac12(\cos^2\alpha\sin2\beta_1 + \sin^2\alpha\sin2\beta_2)^2\frac{v^2}{M^2} + \ord{\frac{v^4}{M^4}} \right), \notag\\
m_{W'} =& \frac{g_RM}{2}\left( 1 + \frac{v^2}{M^2} + \ord{\frac{v^4}{M^4}} \right).
\end{align}
The $W-W'$ mixing angle is given by
\begin{equation}
\zeta = \frac{g_Lv^2(\cos^2\alpha\sin2\beta_1 + \sin^2\alpha\sin2\beta_2)}{g_RM^2} + \ord{\frac{v^4}{M^4}}.
\end{equation}
Note that these expressions reduce to the analogous expressions for the simplified LET model in \cref{eq:zeta,eq:Wmasses} for $\alpha\rightarrow0$.
The $Z$, $Z'$ masses and mixing angles are the same as in the simplified LET model (see \cref{eq:neutralAngles,eq:Zmasses}), with the understanding that $v$ is now composed of all four vev parameters $v_1$, $b_1$, $v_2$, $b_2$.

As for the experimental constraints on heavy vector bosons in \cref{s:gaugebosonconstraints}, we see that most of these bounds are unchanged.
The exceptions are the constraints derived from bounds on the $W-W'$ mixing angles $\zeta$, because the size of $\zeta$ depends on the vev parameters through the combination $v_1b_1 + v_2b_2$.
Apart from the relation $v^2 = v_1^2 + b_1^2 + v_2^2 + b_2^2$, the parameters $v_2$, $b_2$ are unconstrained.
Hence the mixing angle $\zeta$ could be made to vanish, and the experimental bounds on $\zeta$ in \cref{eq:twistBounds,eq:BarenboimBounds} become useless unless we know more about the vev parameters.
Additionally, the constraints from the bounds on $G_{W'}^\phi$, $G_{Z'}^\phi$ in \cref{eq:Aguila} may not apply either, since the mixing angles of the Standard-Model-like scalar are not necessarily the same as in the simplified LET model.
Analytical expressions for the scalar mixing matrices are necessary before we can apply these bounds.

Fortunately, the bound on the $Z-Z'$ mixing angle $\eta$ in \cref{eq:etaBound1} remains unaffected.
The values we obtained on the gauge couplings $g_R$, $g'$ in \cref{eq:gRbound,eq:gpbound}, and hence our value for $\theta_W^\prime$, are unaffected as well.
Hence the bound in \cref{eq:etaBound2} still holds for the complete LET model:
\begin{equation}
M > 3.6\text{ TeV}.
\end{equation}
This means that $W-W'$ and $Z-Z'$ mixing, which are of order $\xi^2$, can still be neglected in the complete LET model.
Only scalar mixing can lead to significant couplings between the Standard-Model particles and the new scalar fields.
This means that as far as the scalar phenomenology is concerned, we can set the $W-W'$ and $Z-Z'$ mixing to zero.
We do so in the rest of this section.

\subsection{Benchmark points}
Like the first bidoublet $\Phi_1$, the second bidoublet consists of two $SU(2)_L$ doublets.
Hence the complete LET model resembles a four-Higgs-doublet model, with two doublets that couple like a type-II 2HDM, two fermiophobic doublets, and two $SU(2)_L$ singlets from $\Phi_R$.
Hence we no longer expect the LET model to look like a type-II 2HDM.
In this section, we qualitatively explore the phenomenological features that may appear in the complete LET model.

In the simplified LET setup, we explored several benchmark scenarios, each with distinct phenomenological features.
However, it turned out that the new physics decouples in most of the parameter space: with all scalar parameters of order 1, the simplified LET model looks like the single-large-hierarchy scenarios SLH-1, SLH-2.
The reason is the fact that there is a large hierarchy $v \ll M$ among the vev parameters of $\Phi_1$, $\Phi_R$.
Significant scalar mixing and non-decoupling were only obtained if one or more of the scalar parameters that appear in denominators were set to $\ord{\xi^2}$ values.
This makes these scenarios very unnatural.

The picture changes drastically with the inclusion of the bidoublet $\Phi_2$.
Its vev introduces two additional mass scales $v_2$, $b_2$, which are bounded by the electroweak scale because they contribute to the $W$ mass.
This means that in the absence of a large hierarchy between the vev components of $\Phi_1$ and $\Phi_2$, we expect that the complete LET model allows for large mixing between the components of both bidoublets if the dimensionless scalar-potential parameters have $\ord{1}$ values.
The result is that the model naturally contains new scalar particles with masses of order $v$ (see \cref{eq:completeLETscalarMasses}).
On the other hand, as in the simplified LET model, we expect that the components of the right-handed doublet $\Phi_R$ can only mix significantly with the other scalar fields if some of the parameters are fine-tuned to small values, since the natural mass scale of $\Phi_R$ is $M$.
These insights have important consequences for the benchmark scenarios we defined in \cref{s:benchmarkpoints}:
\begin{itemize}
\item Since the components of $\Phi_2$ result in new particles with masses of order $v$, we expect that there will always be new scalar particles with a mass in the $\ord{100\text{ GeV}}$ range.
That is, the scenario with a single large hierarchy (see the SLH-1 and SLH-2 benchmark points) is not possible in the complete LET model.
This increases the predictivity of the LET model, since there should be new physics within experimental reach in any case.
\item The mass scales of the 2HDM-like scalars can naturally be of order $v$.
Thus we expect that the 2HDM-like scenario is still possible in the parameter space of the LET model, but without the need for fine-tuning the scalar-potential parameters.
Depending on the mixing of $\Phi_1$ and $\Phi_2$, the 2HDM-like scalars may become fermiophobic, thus distinguishing this scenario from the usual 2HDM.
\item Likewise, we expect the scenario with a light fermiophobic Higgs to be much more natural in the complete LET model than in the simplified setup.
In the simplified model, a fermiophobic Higgs would need to have a large component of $\Phi_R$, which has an associated mass scale $M$.
However, in the complete LET model, a fermiophobic Higgs could also be made to contain significant portions of $\Phi_2$, which has a lower mass scale.
Moreover, a fermiophobic Higgs containing significant portions of $\Phi_2$ would have significant couplings to $W$, $Z$, since $\Phi_2$ is an $SU(2)_L$ doublet.
This means that the bounds on fermiophobic Higgses in \cref{s:FPexperimentalbounds} would not be trivially evaded any longer.
\item We have argued that the complete LET model is expected to have at least some new scalar particles within experimental reach.
With some fine-tuning, all scalars could be made light enough to be within experimental reach, as we did in the simplified LET setup.
Thus a compressed spectrum is still a possibility, albeit an unnatural one compared to the other scenarios.
Moreover, it remains to be seen whether this scenario is of phenomenological interest compared to the other scenarios: the only difference would be whether there is an additional $h^0_R$-like scalar within experimental reach or not.
\item Like the LF scenario, we expect that a very light fermiophobic Higgs with a mass of a few GeV becomes a natural possibility.
If it is $\Phi_2$-like, its mass would be determined mainly by $v_2$, $b_2$, which could be a few GeV like $b_1$.
Moreover, such a very light fermiophobic Higgs would be unlike the one in the simplified LET setup, which has negligible couplings to the Standard Model.
Instead, it would have significant couplings to $W$, $Z$ since $\Phi_2$ is an $SU(2)_L$ doublet.
Hence such a state could have significant production rates at the LHC.
\item The twin-Higgs scenario is a special case of the light-fermiophobic-Higgs scenario, so our comments on the latter apply to the former as well.
That is, it becomes a natural possibility.
Moreover, we would expect deviations from the Standard-Model signal strength in this case.
In the simplified setup, the twin Higgs would arise from mixing between a scalar with Standard-Model-like couplings and a scalar with strongly suppressed couplings to Standard-Model particles.
However, in the complete LET model, the fermiophobic components would still have couplings to $W$, $Z$.
This would give rise to additional contributions to the signal strength of the resonance at 126 GeV, and hence change our analysis in \cref{s:LETlimitsFP}.
\end{itemize}
Moreover, the inclusion of $\Phi_2$ allows for some entirely new phenomenological scenarios.
In \cref{s:chargedScalars} we discussed the possibility of having a fermiophobic charged scalar, which would trivially evade all experimental bounds.
Such a charged scalar is not possible in the simplified LET model: apart from $\ord{\xi}$ mixings, the charged components of $\Phi_R$ become the Goldstone bosons that give mass to the $W'$ bosons.
However, a fermiophobic charged scalar is conceivable in the complete LET model, because $\Phi_2$ contains two pairs of fermiophobic charged scalars.
Depending on the scalar mixing, the lightest charged-scalar mass eigenstate may be either fermiophobic or fermiophilic.
If there is large mixing between the fermiophilic and fermiophobic states, the lightest charged state may have suppressed couplings to fermions.
This means that the experimental bounds on charged scalars would be weakened.

Hence we expect the LET model to allow for more interesting phenomenological scenarios.
More importantly, these scenarios should be testable experimentally, since the new physics does not decouple from the Standard Model in the large-$M$ limit.
The LET model becomes predictive without the need to tune the scalar parameters to values of order $\xi^2$.
In the following, we do not consider such scenarios with unnaturally small scalar parameters.

In the previous two chapters, we have discussed two methods for probing this new physics in the simplified LET setup: measuring deviations from the Standard-Model Higgs couplings and direct searches for new scalars.
We will discuss the prospects for measuring the Higgs-coupling modifications in the context of the complete LET model as well.
To this end, we define a new set of benchmark points, inspired by the considerations given above.
As a starting point for choosing the parameter values, we observe the following about the scalar masses in~\cref{eq:completeLETscalarMasses}:
\begin{itemize}
\item The mass of the Standard-Model-like $h^0$ is mainly determined by $\lambda_1+\lambda_2\cos^2\beta_1$ and to a lesser extent by $\lambda_4$, $\lambda_5$ (see also the expression in~\cref{eq:scalarmasses} for the simplified LET model), but with an overall scaling factor $\cos^2\alpha$ due to the presence of the second bidoublet. We tune the parameters such that $m_{h^0} = 126$ GeV is fixed. This means that the choice of a smaller $\alpha$ is generally accompanied by a smaller value for $\lambda_1+\lambda_2\cos^2\beta_1$.
\item Similarly, the mass of $H^0_1$ is mainly determined by $\widetilde\lambda_1+\widetilde\lambda_2\cos^2\beta_2$ and to a lesser extent by $\widetilde\lambda_3$, $\widetilde\lambda_4$, with an overall factor $\sin^2\alpha$. This means that smaller values of $\alpha$ should be compensated by larger values of $\widetilde\lambda_1+\widetilde\lambda_2\cos^2\beta_2$.
\item The mass difference between $h^0$ and $H^0_1$ and their mixing are governed by $\sin2\alpha$ as well as the scalar parameters $\widetilde\lambda_{5,6,7,8,9,10}$. Hence we can tune these parameters to obtain a parameter set that corresponds to the desired benchmark scenario.
\item The mass of the scalar $H^0_4$ (which is the equivalent of the fermiophobic $H^0_2$ in the simplified LET model) is given by $\sqrt{\lambda_3}M$. Hence we should take a positive $\lambda_3$.
\item The squared masses of the light scalars $A^0_1$, $H^\pm_1$ are determined by the scalar parameters $\widetilde\lambda_{6,7,9,10}$ with an overall minus sign. Hence we take these parameters to be negative to ensure that the mass matrix corresponds to a minimum of the scalar potential.
\item We have $m^2_{H^0_2,A^0_2,H^\pm_2} \sim \lambda_5\sec2\beta_1M^2$. Since we will not tune the parameters to unnaturally small values such that $H^0_2$, $A^0_2$, $H^\pm_2$ become light, it is sufficient to ensure that $\lambda_5 > 0$.
\item Similarly, $m^2_{H^0_3,A^0_3,H^\pm_3} \sim \widetilde\lambda_4\sec2\beta_2M^2$, which means that we should take $\widetilde\lambda_4\sec2\beta_2 > 0$.
\end{itemize}
Additionally, we ensure that the chosen benchmark points satisfy the vacuum stability conditions as described in \cref{a:completeLETmodel}.
The chosen parameter values for the benchmark points are given in \cref{t:completeLETbenchmarkPoints}, and the resulting scalar masses are given in \cref{t:completeLETscalarMasses}.

\begin{table}[t]
{\renewcommand{\arraystretch}{1.3}
\begin{center}
\begin{tabular}{|c|c|c|c|c|c|}
\hline
	&	2HDM-3	&	2HDM-4	&	VLF-3	&	Twin-3	&	Twin-4	\\\hline\hline
$\sin\alpha$	&	0.93	&	0.43	&	0.50	&	0.72	&	0.33	\\
$\sin\beta_2$	&	0.17	&	0.65	&	0.17	&	0.16	&	0.11	\\\hline
$\lambda_1$	&	1.0	&	0.20	&	0.13	&	0.34	&	0.17	\\
$\lambda_2$	&	1.0	&	0.18	&	0.13	&	0.35	&	0.16	\\
$\lambda_3$	&	0.50	&	0.50	&	0.50	&	0.50	&	0.42	\\
$\lambda_4$	&	0.010	&	0.12	&	0.13	&	0.27	&	0.12	\\
$\lambda_5$	&	0.20	&	0.50	&	0.20	&	0.20	&	0.50	\\\hline
$\widetilde\lambda_1$	&	0.40	&	1.3	&	0.27	&	0.34	&	1.3	\\
$\widetilde\lambda_2$	&	0.40	&	1.3	&	0.27	&	0.34	&	1.2	\\
$\widetilde\lambda_3$	&	0.27	&	0.20	&	0.27	&	0.27	&	0.19	\\
$\widetilde\lambda_4$	&	0.20	&	0.10	&	0.20	&	0.20	&	1.0	\\
$\widetilde\lambda_5$	&	0.20	&	0.046	&	0.54	&	0.32	&	0.060	\\
$\widetilde\lambda_6$	&	-0.40	&	-0.50	&	-0.43	&	-0.42	&	-0.30	\\
$\widetilde\lambda_7$	&	-0.24	&	-0.20	&	-0.24	&	-0.24	&	-0.30	\\
$\widetilde\lambda_8$	&	0.84	&	0.95	&	0.83	&	0.84	&	1.0	\\
$\widetilde\lambda_9$	&	-0.050	&	-0.10	&	-0.04	&	-0.053	&	-0.10	\\
$\widetilde\lambda_{10}$	&	-0.30	&	-0.30	&	-0.30	&	-0.30	&	-0.30	\\\hline
\end{tabular}
\end{center}}
\caption{Defining parameters of the benchmark points for the complete LET model. All benchmark points use the values $v = 246$ GeV, $M = 10^{10}$ GeV, $\beta_1 = \arctan(m_b/m_t) = 0.0166$.}\label{t:completeLETbenchmarkPoints}
\end{table}

\begin{table}[t]
{\renewcommand{\arraystretch}{1.3}
\begin{center}
\begin{tabular}{|c|c|c|c|c|c|}
\hline
	&	2HDM-3	&	2HDM-4	&	VLF-3	&	Twin-3	&	Twin-4	\\\hline\hline
$m_{h^0}$	&	126	&	126	&	126	&	126	&	126	\\
$m_{H^0_1}$	&	182	&	148	&	3.9	&	126	&	126	\\
$m_{H^0_2}$	&	$3.2\cdot10^9$	&	$5.0\cdot10^9$	&	$3.2\cdot10^9$	&	$3.2\cdot10^9$	&	$5.0\cdot10^9$	\\
$m_{H^0_3}$	&	$3.3\cdot10^9$	&	$5.8\cdot10^9$	&	$3.3\cdot10^9$	&	$3.2\cdot10^9$	&	$7.2\cdot10^9$	\\
$m_{H^0_4}$	&	$7.1\cdot10^9$	&	$7.1\cdot10^9$	&	$7.1\cdot10^9$	&	$7.1\cdot10^9$	&	$6.5\cdot10^9$	\\\hline
$m_{A^0_1}$	&	179	&	134	&	179	&	179	&	190	\\
$m_{A^0_2}$	&	$3.2\cdot10^9$	&	$5.0\cdot10^9$	&	$3.2\cdot10^9$	&	$3.2\cdot10^9$	&	$5.0\cdot10^9$	\\
$m_{A^0_3}$	&	$3.3\cdot10^9$	&	$5.8\cdot10^9$	&	$3.3\cdot10^9$	&	$3.2\cdot10^9$	&	$7.2\cdot10^9$	\\\hline
$m_{H^\pm_1}$	&	171	&	135	&	173	&	173	&	173	\\
$m_{H^\pm_2}$	&	$3.2\cdot10^9$	&	$5.0\cdot10^9$	&	$3.2\cdot10^9$	&	$3.2\cdot10^9$	&	$5.0\cdot10^9$	\\
$m_{H^\pm_3}$	&	$3.3\cdot10^9$	&	$5.8\cdot10^9$	&	$3.3\cdot10^9$	&	$3.2\cdot10^9$	&	$7.2\cdot10^9$	\\\hline
\end{tabular}
\end{center}}
\caption{Scalar masses for each of the benchmark points defined in \cref{t:completeLETbenchmarkPoints}. All masses are in GeV.}\label{t:completeLETscalarMasses}
\end{table}

\paragraph{2HDM-3}
For this benchmark point, the scalars $H_1^0$, $A_1^0$, and $H^\pm$ all have masses in the $\ord{100\text{ GeV}}$ range.
The Standard-Model-like scalar $h^0$ is almost purely $h^0_{1,11}$, with permille level mixing with the other states.
The light $CP$-even state $H_1^0$ is 97\% $h^0_{2,11}$ and 3\% $h^0_{2,22}$; mixing with the other states is negligible.
The lightest $CP$-odd state $A_1^0$ is 87\% $a^0_{1,11}$ and 13\% $a^0_{2,11}$.
Similarly, the lightest charged state $H_1^\pm$ is 87\% $h^\pm_{1,21}$ and 13\% $h^\pm_{2,21}$.

We call this scenario 2HDM-3, because as in the 2HDM-1 and 2HDM-2 scenarios from \cref{s:benchmarkpoints}, the particle content within experimental reach resembles that of the 2HDM.
However, we expect that this scenario could be distinguished from a type-II 2HDM in experiment: the $CP$-even state $H^0_1$ is completely fermiophobic and would therefore dominantly decay into pairs of vector bosons.
Also, the lightest $CP$-odd and charged states have significant mixing with components of the fermiophobic field $\Phi_2$.
Thus they would have reduced couplings to fermions compared to the 2HDM.

\paragraph{2HDM-4}
Like the previous benchmark point, 2HDM-4 has a 2HDM-like set of scalars with masses in the $\ord{100\text{ GeV}}$ range.
Mixing among the $CP$-even scalars is substantial: the Standard-Model-like scalar $h^0$ is 74\% $h^0_{1,11}$, with a 15\% admixture of $h^0_{2,11}$ and 11\% of $h^0_{2,22}$.
The next-to-lightest $CP$-even state $H^0_1$ is 25\% $h^0_{1,11}$, 43\% $h^0_{2,11}$ and 32\% $h^0_{2,22}$.
The lightest $CP$-odd state $A^0_1$ is a mixture of fermiophilic and fermiophobic scalars as well: it is 18\% $a^0_{1,11}$, 47\% $a^0_{2,11}$, and 35\% $a^0_{2,22}$.
The lightest charged scalar $H^\pm_1$ is a mixture of fermiophilic and fermiophobic components too: it is 18\% $h^\pm_{1,21}$, 47\% $h^\pm_{2,21}$ and 35\% $h^\pm_{2,22}$.

Like the 2HDM-3 scenario, this benchmark has a fermiophobic $H^0_1$, distinguishing it from a type-II 2HDM.
The scalars $A^0_1$ and $H^\pm_1$ are mostly fermiophobic as well.
They have suppressed couplings to fermions, so that current bounds on their masses may be evaded.

\paragraph{VLF-3}
This benchmark point distinguishes itself from the previous ones by having a very light fermiophobic Higgs $H^0_1$, with a mass of a few GeV.
The states $A^0_1$ and $H^\pm_1$ again have masses within experimental reach.
The Standard-Model-like Higgs $h^0$ is a mixture of 65\% $h^0_{1,11}$, 34\% $h^0_{2,11}$, and 1\% $h^0_{2,22}$.
Thus it contains a significant portion of fermiophobic scalars, and we expect it to have reduced couplings to fermions with respect to the Standard Model.
The very light scalar $H^0_1$ is 35\% $h^0_{1,11}$, 63\% $h^0_{2,11}$, and 2\% $h^0_{2,22}$.
It is light and mainly fermiophobic, which means it could have evaded the LEP searches.
The lightest $CP$-odd scalar $A^0_1$ is 25\% $a^0_{1,11}$, 73\% $a^0_{2,11}$, and 2\% $a^0_{2,22}$, so it is mostly fermiophobic.
Similarly, the lightest charged scalar $H^\pm_1$ is 25\% $h^\pm_{1,21}$, 73\% $h^\pm_{2,21}$, and 2\% $h^\pm_{2,22}$, and is therefore mostly fermiophobic as well.

\paragraph{Twin-3}
For this benchmark point, both $h^0$ and $H^0_1$ have a mass of 126 GeV, similarly to the Twin-1 and Twin-2 benchmarks in \cref{s:benchmarkpoints}.
The former is 87\% $h^0_{1,11}$ with a 13\% admixture of $h^0_{2,11}$.
The latter is 13\% $h^0_{1,11}$, 85\% $h^0_{2,11}$ and 2\% $h^0_{2,22}$.
The lightest $CP$-odd and charged states are almost 50-50 mixtures of fermiophilic and fermiophobic states: $A^0_1$ ($H^\pm_1$) is 51\% $a^0_{1,11}$ ($h^\pm_{1,21}$), 47\% $a^0_{2,11}$ ($h^\pm_{2,21}$), and 1\% $a^0_{2,22}$ ($h^\pm_{2,12}$).

\paragraph{Twin-4}
Like Twin-3, this benchmark contains a twin Higgs state at 126 GeV.
Here $h^0$ is almost purely $h^0_{1,11}$, with only permille-level admixtures of other states.
The other state $H^0_1$ is 99\% $h^0_{2,11}$ with a 1\% admixture of $h^0_{2,22}$.
The lightest $CP$-odd and charged states are mostly fermiophobic: $A^0_1$ ($H^\pm_1$) is 11\% $a^0_{1,11}$ ($h^\pm_{1,21}$), 88\% $a^0_{2,11}$ ($h^\pm_{2,21}$), and 1\% $a^0_{2,22}$ ($h^\pm_{2,12}$).

\subsection{Higgs-coupling modifications}
The complete LET model contains eight more scalar fields and ten more free scalar-potential parameters than the simplified setup.
The mixing matrices for the $CP$-even, $CP$-odd, and charged scalar components are now $5\times5$ matrices, which contain 15 mixing angles in total.
This makes it challenging and cumbersome to find analytical expressions for the scalar masses and mixings.
Therefore, we restrict ourselves to a numerical analysis of the Higgs-coupling modifications for the benchmark points defined in the previous section.
Below we describe how we extract the relevant scalar couplings for these benchmark points.

As we did in for the simplified LET model in chapter 6, we calculate the scalar mass matrix numerically in \emph{Mathematica}.
Then we calculate the eigenvalues and eigenvectors.
For the rotation matrices, we use the conventions described in \cref{a:completeLETmodel} for the order of the rows and the phases of the mass eigenstates.
Using the mixing matrices, we express the gauge eigenstates in mass eigenstates with numerical coefficients.
We insert these into the Lagrangian of \cref{eq:LagS} and extract the couplings as numerical coefficients of the operators.

The modifications of the Standard-Model-like Higgs couplings to $W$, $Z$, $t$, $b$ can be obtained easily by comparing the numerical values to the Standard-Model values, which are calculated from \cref{eq:gVSM,eq:gfSM}.
For the photon-coupling modification $\Delta_\gamma$, we take into account the contributions of all three charged scalars $H^\pm_{1,2,3}$, each of which is given by \cref{eq:DeltaGammaContribution}.

All coupling modifications have been summarised in \cref{t:completeLEThiggsCouplingModifications}.
Note that contrary to the simplified LET model, the photon-coupling modification not negligibly small any more: the charged scalar $H^\pm_1$ has a mass of order $v$ and hence yields a sizeable contribution to the effective photon coupling.

\begin{table}[t]
{\renewcommand{\arraystretch}{1.3}
\begin{center}
\begin{tabular}{|c|c|c|c|c|c|}
\hline
	&	2HDM-3	&	2HDM-4	&	VLF-3	&	Twin-3	&	Twin-4	\\\hline\hline
$\Delta_W$	&	-0.69	&	-0.01	&	-0.01	&	-0.61	&	-0.06	\\
$\Delta_Z$	&	-0.69	&	-0.001	&	-0.001	&	-0.60	&	-0.05	\\
$\Delta_t$	&	1.8	&	-0.05	&	-0.07	&	0.34	&	0.06	\\
$\Delta_b$	&	1.8	&	-0.05	&	-0.07	&	0.34	&	0.06	\\
$\Delta_\gamma$	&	-0.05	&	-0.09	&	-0.07	&	-0.03	&	-0.05	\\\hline
\end{tabular}
\end{center}}
\caption{Higgs-coupling modifications for the complete LET model, as defined in~\cref{eq:CMdefinition}.}\label{t:completeLEThiggsCouplingModifications}
\end{table}

The 2HDM-3 benchmark has large coupling modifications: the couplings to $V=W, Z$ are suppressed by a factor 0.3, whereas the couplings to $t$, $b$ are enhanced by almost a factor 3.
This is not surprising: the state $h^0$ is almost purely $h^0_{1,11}$.
In the simplified LET model, its coupling to $W$, $Z$ is equal to $g_L^2v/2$, $g_L^2v/(2\cos\theta_W)$ respectively.
In the complete LET model, we have to make the replacement $v\rightarrow v\cos\alpha$ (see \cref{eq:completeLETvevRedefinition}).
This means that the $W$, $Z$ couplings are suppressed by $\cos\alpha = 0.36$.
On the other hand, the $t$, $b$ couplings are given by $m_f/v$ in the simplified LET model.
This implies that these couplings are enhanced by a factor $1/\cos\alpha = 2.8$ in the complete LET model.
This benchmark point is clearly incompatible with the measured coupling modifications in \cref{f:Higgscouplingmeasurements}.
Note that for a given amount of $\Phi_1-\Phi_2$ mixing, the measured coupling modifications can be used to constrain the parameter $\alpha$.

In contrast, the 2HDM-4 benchmark has smaller but still sizeable coupling modifications.
The $V$ couplings are suppressed by $\cos\alpha = 0.90$, but the total coupling is a few percent higher because there is an additional contribution from the second bidoublet.
On the other hand, the quark couplings are enhanced by a factor $1/\cos\alpha = 1.11$.
Still, $\Delta_t$, $\Delta_b$ are smaller than 0.11 because $h^0$ contains a significant admixture of the fermiophobic second bidoublet.
This tension results in coupling modifications that are actually negative.
All coupling modifications for this benchmark point are consistent with the measured coupling modifications.

The VLF-3 scenario has large mixing between the fermiophilic and fermiophobic scalar gauge eigenstates.
The vector-boson couplings obtain contributions from both.
The fermion couplings are enhanced by a factor $1/\cos\alpha = 1.2$ but at the same time suppressed by the large scalar mixing.
The resulting coupling modifications are at the percent level, all compatible with the measured values.

Like the 2HDM-3 benchmark, the Twin-3 scenario has large coupling modifications: the $W$, $Z$ couplings are reduced by about 60\% whereas the $t$, $b$ couplings are enhanced by 34\%.
The strong suppression of the vector-boson couplings is mostly due to interference between the contributions of the fermiophobic and fermiophilic scalar components.
The fermion couplings are enhanced by a factor $1/\cos\alpha = 1.4$, which is slightly reduced by $h^0_{1,11}-h^0_{2,11}$ mixing.
This benchmark point is incompatible with the data.

The Twin-4 scenario has negligible $h^0_{1,11}$-$h^0_{2,11}$ mixing.
Hence the $V$ couplings are suppressed by $\cos\alpha = 0.94$ whereas the fermion couplings are enhanced by $1/\cos\alpha = 1.06$.
This results in percent-level coupling modifications, compatible with the measured values.

We have illustrated that the complete LET model allows for various phenomenological scenarios, and is predictive at the same time.
The model allows for large coupling modifications as well as moderate ones that can be expected to be measurable.
Both the 2HDM-like and the twin scenario can be excluded in some parts of parameter space, and are compatible with experiment in other regions.
Thus the complete LET model is testable, and a more thorough analysis of the parameter space is required to see which parameter values are preferred by experiment.
Also, it would be interesting to see whether we can expect general patterns among the coupling modifications, as was the case for the simplified LET model in \cref{s:couplingpatterns}.

\clearpage
\section{Conclusions}\label{s:conclusions}
In this work we have studied the low-energy phenomenology of the trinification model as described in refs.~\cite{Achiman:1978vg,Stech:2003sb,Stech:2008wd,Stech:2010gf,Stech:2012zr,Stech:2014tla,}, which is based on the trinification group $\trini$.
In order to simplify our study, we have integrated out the fields that obtain masses of the order of the trinification scale.
This resulted in a left-right-symmetric model with two scalar bidoublets $\Phi_1$, $\Phi_2$ and one right-handed doublet $\Phi_R$, where only $\Phi_1$ couples to fermions, as well as constraints on the form of the scalar potential.
We call this effective model the low-energy trinification (LET) model.
In order to simplify our analysis further, we have introduced a toy model (the simplified LET model) in which the fermiophobic bidoublet $\Phi_2$ has been omitted.
We have studied this toy model in the first part of this work, and used the results to help us understand the complete LET model.

The simplified LET model contains three heavy vector bosons $W^{\prime\pm}$, $Z'$ and five massive scalars $H^0_1$, $H^0_2$, $A^0$, $H^\pm$ on top of the Standard-Model particles.
We have worked out the masses and mixings of these new particles.
Our Ansatz for the Yukawa sector was based on the Yukawa Lagrangian of the trinification model.
Here we left out a discussion of the first and second fermion generations, where a description of the mixing requires the introduction of new Higgs fields.
The free parameters were used to fix the masses of the top and bottom quarks only, since these are the most important for comparison to experimental searches.
In the trinification model, the fermion masses and mixings can be described using only a few parameters due to the presence of new heavy quarks as well as interactions with additional Higgs fields.
An improved version of the LET model would require a better understanding of the Yukawa sector at low energy scales, where renormalisation-group effects and mixing with the heavy quark states become important.

We have reviewed direct searches as well as precision measurements that lead to lower bounds on heavy-vector-boson masses.
Then we discussed to what extent these bounds apply to the $W'$, $Z'$ bosons of the LET model, and used them to obtain a lower bound on the left-right-symmetry-breaking scale $M$ in the TeV range.
Values for the $SU(2)_R$ gauge coupling $g_R$ and the $U(1)_{B-L}$ coupling $g'$ were obtained as well.

We have given approximate expressions for the Higgs-coupling modifications in the simplified LET model.
In order to showcase the possible phenomenological scenarios in the simplified LET model, we have defined a set of benchmark points.
The new scalars turned out to decouple in most of the parameter space.
By tuning some of the scalar parameters to $\ord{v^2/M^2}$ values, other scenarios were obtained: a scenario similar to the two-Higgs-doublet model (2HDM) where only $H^0_1$, $A^0$, $H^\pm$ are within experimental reach; a scenario with a light fermiophobic $H^0_2$ in which all other new scalars are heavy; a scenario where $H^0_2$ has a mass of a few GeV, thereby escaping direct searches; a scenario where all new scalars have masses within experimental reach; and a scenario with a degenerate (`twin') Higgs state at 126 GeV.
We calculated the Higgs-coupling modifications for these benchmarks to show to what extent these scenarios can be distinguished from the Standard Model in experiment.
In most cases, the coupling modifications are negligibly small.
In those parameter-space regions where the coupling modifications become measurably large, the $W$, $Z$, and $t$ coupling modifications are fully correlated.
Moreover, a measurement of the quartic Higgs self-coupling would allow us to distinguish some of these benchmarks from the Standard Model, since modifications of this coupling can be as large as $\ord{1}$.

We have reviewed the available bounds on charged scalars and fermiophobic neutral scalars, and discussed to what extent these bounds apply to the simplified LET model.
The fermiophobic $H^0_2$ evades these bounds trivially, since it has negligible couplings to Standard-Model particles.
The charged scalars have 2HDM-like couplings to Standard-Model particles.
Using additional assumptions on the Yukawa couplings to the lightest two fermion generations, we found that the charged scalars decay dominantly to $\tau\nu$ below the $t\bar{b}$ threshold.
This part of our analysis would benefit from an improved understanding of the Yukawa sector as well.

The twin-Higgs scenario was found to have large scalar mixing.
However, we expect no significant deviations of the Higgs signal strength from Standard-Model predictions, since the twin state $h^0$, $H^0_2$ arises from a basis rotation of a scalar with Standard-Model-like couplings and a scalar that does not couple to the Standard-Model vector bosons and fermions.

In the last section, we have added back the fermiophobic bidoublet $\Phi_2$ to the model, and discussed how the phenomenology of the complete LET model changes with respect to the simplified setup.
We found that $M$ is still at least in the TeV range, and that the effects of $W-W'$ and $Z-Z'$ mixing are negligibly small.
The complete LET model introduces four new $CP$-even states $H^0_{1,2,3,4}$, three $CP$-odd states $A^0_{1,2,3}$, and three pairs of charged states $H^\pm_{1,2,3}$.
We have found approximate expressions for their masses in terms of the underlying model parameters.
There are always at least one other $CP$-even scalar, one $CP$-odd scalar, and a pair of charged scalars with masses in the $\ord{100\text{ GeV}}$ range, without tuning the scalar parameters to very small values.
Hence interesting scenarios like a 2HDM-like spectrum at low energies or a twin Higgs are natural possibilities.
Additionally, the lightest charged scalar can become fermiophobic.

In order to illustrate these scenarios, we defined a set of numerical benchmark points and calculated the corresponding Higgs-coupling modifications.
We found that the various phenomenological scenarios lead to sizeable effects on the Higgs couplings.
Parts of parameter space can already be excluded using the available values in the literature.

The LET model is an interesting extension of the Standard Model with various phenomenological possibilities.
The model is predictive and can be tested using LHC data.
We have only calculated a few benchmark points to illustrate the possible phenomenological scenarios of the LET model.
A more thorough analysis of the scalar mixing is necessary to constrain the parameter space systematically.
Furthermore, approximate expressions for the scalar mixing angles would help us find patterns in the Higgs-coupling modifications.
These are also necessary to figure out what regions of parameter space lead to significant deviations of the Higgs signal strength from the Standard-Model predictions.
Most interestingly, we could test whether the observed scalar state at 126 GeV is a twin state or not.
We leave these improvements to future work.

\clearpage
\appendix

\section{Gauge-boson mass eigenstates}\label{a:gaugebosons}
The low-energy trinification model contains seven gauge bosons: $W_{L,R}^i$, $i=1,2,3$ for $SU(2)_{L,R}$ with gauge coupling $g_{L,R}$ and $B$ for $U(1)_{B-L}$ with gauge coupling $g'$.
The fields $W_L^3$, $W_R^3$, and $B$ are neutral, whereas the remaining fields mix to form the charge eigenstates $W_{L,R}^\pm=\frac{1}{\sqrt2}(W_{L,R}^1\mp iW_{L,R}^2)$.
After spontaneous symmetry breaking, one linear combination of the neutral fields forms the massless photon, whereas the remaining six gauge bosons obtain mass.
In this appendix, we work out their masses and mixings.

The gauge-boson mass terms are determined by the scalar-scalar-vector-vector interactions, which follow from the gauge-invariant kinetic terms for the scalars (see \cref{eq:LagS}).
For the simplified LET model, these terms are given by
\begin{align}
\lag_s =& \tr{(D^\mu\Phi_1)^\dagger(D_\mu\Phi_1)} + (D^\mu\Phi_R)(D_\mu\Phi_R)^\dagger \notag\\
\supset& g_L^2W_L^{i\mu}W_{L\mu}^j \tr{\Phi_1^\dagger\overline{T}_L^i\overline{T}_L^j\Phi_1} + g_R^2W_R^{i\mu}W_{R\mu}^j \tr{\Phi_1^\dagger \Phi_1T_R^{iT}T_R^{jT}} \notag\\
&+ g_R^2W_R^{i\mu}W_{R\mu}^j \Phi_RT_R^{iT}T_R^{jT}\Phi_R^\dagger + g'^2B^\mu B_\mu\Phi_R\Phi_R^\dagger \notag\\
&+ 2g_Lg_RW_L^{i\mu}W_{R\mu}^j \tr{\Phi_1^\dagger\overline{T}_L^i\Phi_1T_R^{jT}} + 2g_Rg'W_R^{i\mu}B_\mu \Phi_RT_R^{iT}\Phi_R^\dagger,  \label{eq:scalargauge}
\end{align}
where the covariant derivatives are given by
\begin{align}
D_\mu\Phi_1 =& \partial_\mu\Phi_1 - ig_LW_{L\mu}^i\overline{T}_L^i\Phi_1 - ig_RW_{R\mu}^i\Phi_1T_R^{iT}, \notag\\
D_\mu\Phi_R =& \partial_\mu\Phi_R - ig_RW_{R\mu}^i\Phi_RT_R^{iT} - ig'B_\mu\Phi_R.
\end{align}
The Higgs fields obtain the vevs
\begin{equation}
\vev{\Phi_1} = \begin{pmatrix} v_1 & 0 \\ 0 & b_1 \end{pmatrix},\qquad
\vev{\Phi_R} = \begin{pmatrix} 0 & M \end{pmatrix}. \label{eq:vevs}
\end{equation}
The vev parameters of $\Phi_1$ are constrained by the condition $v_1^2+b_1^2 = v^2$, hence it will often be convenient to reparametrise them as $v_1 = v\cos\beta$, $b_1 = v\sin\beta$.
After spontaneous symmetry breaking, the interaction terms in \cref{eq:scalargauge} give rise to mass terms for the gauge bosons.
These are easily obtained using mathematics software such as \emph{Mathematica} \cite{Mathematica10}.
In the subsequent sections, we will study how the charged and neutral gauge bosons mix to form mass eigenstates.
We will also derive the Weinberg angle from the expressions of the neutral mass eigenstates.

\subsection{Charged gauge bosons}
There are two pairs $W_L^\pm$, $W_R^\pm$ of charged gauge bosons.
After spontaneous symmetry breaking, they will mix into two pairs of mass eigenstates $W^\pm$, $W^{\prime\pm}$.
We identify the former with the charged vector bosons from the Standard Model; the $W^{\prime\pm}$ are new massive states.
After the scalar fields obtain the vevs in \cref{eq:vevs}, the interactions in \cref{eq:scalargauge} give rise to the following mass matrix for the charged gauge bosons in the basis $\{W_L^\pm,W_R^\pm\}$:
\begin{equation}
M^2_\pm = \begin{pmatrix} \frac14g_L^2v^2 & -\frac14g_Lg_Rv^2\sin2\beta \\ -\frac14g_Lg_Rv^2\sin2\beta & \frac14g_R^2(M^2+v^2) \end{pmatrix}. \label{eq:WWpmassmatrix}
\end{equation}
The squared masses of the $W$ and $W'$ bosons are given by the eigenvalues of $M^2_\pm$; we find them to be
\begin{align}
m_W^2 =& \frac18\left( g_R^2M^2 + (g_L^2+g_R^2)v^2 - g_R^2M^2\sqrt{ \left(1 + \frac{(g_R^2-g_L^2)v^2}{g_R^2M^2}\right)^2 + \frac{4g_L^2v^4\sin^22\beta}{g_R^2M^4} }\right) \notag\\
=& \frac14g_L^2v^2\left( 1 - \sin^22\beta\frac{v^2}{M^2} + \ord{\frac{v^4}{M^4}}\right), \notag\\
m_{W'}^2 =& \frac18\left( g_R^2M^2 + (g_L^2+g_R^2)v^2 + g_R^2M^2\sqrt{ \left(1 + \frac{(g_R^2-g_L^2)v^2}{g_R^2M^2}\right)^2 + \frac{4g_L^2v^4\sin^22\beta}{g_R^2M^4} }\right) \notag\\
=& \frac14g_R^2M^2\left( 1 + \frac{v^2}{M^2} + \ord{\frac{v^4}{M^4}}\right).
\end{align}
The mass eigenstates $W$ and $W'$ are given by the eigenvectors of $M^2_\pm$; these turn out to be
\begin{equation}
\begin{pmatrix} W^\pm \\\\ W^{\prime\pm} \end{pmatrix} =
\begin{pmatrix} \dfrac{1}{\sqrt{N_W}}\left( g_R^2\left(1 + \dfrac{v^2}{M^2}\right) - 4\dfrac{m_W^2}{M^2} \right) & \dfrac{1}{\sqrt{N_W}}\dfrac{g_Lg_Rv^2\sin2\beta}{M^2} \\
\dfrac{1}{\sqrt{N_{W'}}}\left( 4\dfrac{m_W^2}{M^2} - g_L^2\dfrac{v^2}{M^2}\right) & \dfrac{1}{\sqrt{N_{W'}}}\dfrac{g_Lg_Rv^2\sin2\beta}{M^2} \end{pmatrix}
\begin{pmatrix} W_L^\pm \\\\ W_R^\pm \end{pmatrix}. \label{eq:gbmasseigcharged}
\end{equation}
Here we defined the normalisation factors
\begin{align}
N_W \equiv& \left(g_R^2\left(1+\frac{v^2}{M^2}\right) - 4\frac{m_W^2}{M^2}\right)^2 + \left(\frac{g_Lg_Rv^2\sin2\beta}{M^2}\right)^2, \notag\\
N_{W'} \equiv& \left(4\frac{m_W^2}{M^2} - g_L^2\frac{v^2}{M^2}\right)^2 + \left(\frac{g_Lg_Rv^2\sin2\beta}{M^2}\right)^2,
\end{align}
Since this is a two-dimensional basis rotation, \cref{eq:gbmasseigcharged} is usually expressed in terms of a mixing angle $\zeta$:
\begin{equation}
\begin{pmatrix} W^\pm \\ W^{\prime\pm} \end{pmatrix}
= \begin{pmatrix} \cos\zeta & \sin\zeta \\ -\sin\zeta & \cos\zeta \end{pmatrix}
\begin{pmatrix} W_L^\pm \\ W_R^\pm \end{pmatrix}. \label{eq:Wmasseigenstates}
\end{equation}
Expanding \cref{eq:gbmasseigcharged} in the small parameter $\frac{v}{M}$ yields the following expression for $\zeta$:
\begin{equation}
\zeta = \frac{g_Lv^2\sin2\beta}{g_RM^2} + \ord{\frac{v^4}{M^4}}. \label{eq:Wmixingangle}
\end{equation}
Thus the mixing among the charged gauge bosons is very small.
The Standard-Model $W$ bosons are almost completely left-handed gauge bosons, and the new $W'$ bosons are almost completely right-handed gauge bosons:
\begin{equation}
W^\pm = W_L^\pm + \ord{\frac{v^2}{M^2}},\qquad W^{\prime\pm} = W_R^\pm + \ord{\frac{v^2}{M^2}}.
\end{equation}

\subsection{Neutral gauge bosons}
There are three neutral gauge bosons $W_L^3$, $W_R^3$, $B$.
After spontaneous symmetry breaking, they will mix into the mass eigenstates $A$, $Z$, $Z'$.
Here $A$ is the photon, $Z$ is the neutral massive vector boson we know from the Standard Model, and $Z'$ is a new massive state.
The gauge-boson mass terms give rise to the following mass matrix for the neutral gauge bosons in the basis $\{W_L^3,W_R^3,B\}$:
\begin{equation}
M_0^2 = \begin{pmatrix} \frac14g_L^2v^2 & -\frac14g_Lg_Rv^2 & 0 \\
-\frac14g_Lg_Rv^2 & \frac14g_R^2(M^2+v^2) & -\frac12g'g_RM^2 \\
0 & -\frac12g'g_RM^2 & g^{\prime2}M^2 \end{pmatrix}. \label{eq:AZZpmassmatrix}
\end{equation}
The squared masses of $A$, $Z$, $Z'$ are given by the eigenvalues of $M_0^2$; we find them to be
\begin{align}
m_A^2 =& 0, \notag\\
m_Z^2 =& \frac18\Bigg( (g_R^2+4{g'}^2)M^2 + (g_L^2+g_R^2)v^2 \notag\\
&- (g_R^2+4{g'}^2)M^2\sqrt{1 + \frac{\left(4g_R^4-2(g_L^2+g_R^2)(g_R^2+4{g'}^2)\right)v^2}{(g_R^2+4{g'}^2)^2M^2} + \frac{(g_L^2+g_R^2)^2v^4}{(g_R^2+4{g'}^2)^2M^4} }\Bigg) \notag\\
=& \frac{v^2}{4} \frac{4{g'}^2(g_L^2+g_R^2)+g_L^2g_R^2}{g_R^2+4{g'}^2} \left( 1 - \frac{g_R^4}{(g_R^2+4g^{\prime2})^2}\frac{v^2}{M^2} + \ord{\frac{v^4}{M^4}} \right), \notag\\
m_{Z'}^2 =& \frac18\Bigg( (g_R^2+4{g'}^2)M^2 + (g_L^2+g_R^2)v^2 \notag\\
&+ (g_R^2+4{g'}^2)M^2\sqrt{1 + \frac{\left(4g_R^4-2(g_L^2+g_R^2)(g_R^2+4{g'}^2)\right)v^2}{(g_R^2+4{g'}^2)^2M^2} + \frac{(g_L^2+g_R^2)^2v^4}{(g_R^2+4{g'}^2)^2M^4} }\Bigg) \notag\\
=& \tfrac14(g_R^2+4{g'}^2)M^2\left( 1 + \frac{g_R^4}{(g_R^2+4g^{\prime2})^2}\frac{v^2}{M^2} + \ord{\frac{v^4}{M^4}}\right).
\end{align}
The mass eigenstates $A$, $Z$, $Z'$ are given by the eigenvectors of $M_0^2$; these turn out to be
\begin{equation}
\begin{pmatrix} A \\\\ Z \\\\ Z' \end{pmatrix} =
\begin{pmatrix} \dfrac{2g'g_R}{\sqrt{N_A}} & \dfrac{2g'g_L}{\sqrt{N_A}} & \dfrac{g_Lg_R}{\sqrt{N_A}} \\
\dfrac{g_L}{\sqrt{N_Z}}\left(g_R^2+4{g'}^2-4\dfrac{m_Z^2}{M^2}\right) & -\dfrac{4g_R}{\sqrt{N_Z}}\left({g'}^2-\dfrac{m_Z^2}{M^2}\right) & -\dfrac{2g'g_R^2}{\sqrt{N_Z}} \\
\dfrac{-g_L}{\sqrt{N_{Z'}}} \dfrac{(g_L^2+g_R^2)v^2-4m_Z^2}{M^2} & \dfrac{g_R}{\sqrt{N_{Z'}}}\left(g_R^2 + \dfrac{(g_L^2+g_R^2)v^2-4m_Z^2}{M^2}\right) & -\dfrac{2g'g_R^2}{\sqrt{N_{Z'}}} \end{pmatrix}
\begin{pmatrix} W_L^3 \\\\ W_R^3 \\\\ B \end{pmatrix}. \label{eq:gbmasseigneutral}
\end{equation}
Here we defined the normalisation factors
\begin{align}
N_A \equiv& 4{g'}^2(g_L^2+g_R^2)+g_L^2g_R^2, \notag\\
N_Z \equiv& N_A(g_R^2+4{g'}^2) - 8N_A\tfrac{m_Z^2}{M^2} + 16(g_L^2+g_R^2)\tfrac{m_Z^4}{M^4}, \notag\\
N_{Z'} \equiv& g_R^4(g_R^2+4{g'}^2) + 2g_R^4\big([g_L^2+g_R^2]\tfrac{v^2}{M^2}-4\tfrac{m_Z^2}{M^2}\big) + (g_L^2+g_R^2)\big((g_L^2+g_R^2)\tfrac{v^2}{M^2}-4\tfrac{m_Z^2}{M^2}\big)^2,
\end{align}
We express this basis rotation in terms of three mixing angles $\theta_W$, $\theta_W^\prime$, $\eta$, following the conventions in \cite{Chay:1998hd}:
\begin{equation}
\begin{pmatrix} A \\ Z \\ Z' \end{pmatrix} =
\begin{pmatrix} s_{\theta_W} & c_{\theta_W}s_{\theta_W^\prime} & c_{\theta_W}c_{\theta_W^\prime} \\
c_{\theta_W}c_\eta & c_{\theta_W^\prime}s_\eta - s_{\theta_W}s_{\theta_W^\prime}c_\eta & -(s_{\theta_W}c_{\theta_W^\prime}c_\eta + s_{\theta_W^\prime}s_\eta) \\
-c_{\theta_W}s_\eta & c_{\theta_W^\prime}c_\eta + s_{\theta_W}s_{\theta_W^\prime}s_\eta & s_{\theta_W}c_{\theta_W^\prime}s_\eta - s_{\theta_W^\prime}c_\eta \end{pmatrix}
\begin{pmatrix} W_L^3 \\ W_R^3 \\ B \end{pmatrix}. \label{eq:AZZpmasseigenstates}
\end{equation}
Here we have written $s_x\equiv\sin{x}$, $c_x\equiv\cos{x}$ for the sake of brevity.
The mixing angles are given by
\begin{align}
\sin\theta_W =& \frac{2g'g_R}{\sqrt{4g^{\prime2}(g_L^2+g_R^2)+g_L^2g_R^2}}, \notag\\
\sin\theta_W^\prime =& \frac{2g'}{\sqrt{g_R^2+4g^{\prime2}}}, \notag\\
\tan\eta =& \frac{(g_L^2+g_R^2)v^2-4m_Z^2}{(g_R^2+4g^{\prime2})M^2-4m_Z^2} \sqrt{\frac{N_Z}{N_{Z'}}} \notag\\
=& \frac{g_R^2\sqrt{4g^{\prime2}(g_L^2+g_R^2)+g_L^2g_R^2}}{(g_R^2+4g^{\prime2})^2} \frac{v^2}{M^2} + \ord{\frac{v^4}{M^4}} \notag\\
=& \frac{\sin\theta_W^\prime\cos^3\theta_W^\prime}{\sin\theta_W} \frac{v^2}{M^2} + \ord{\frac{v^4}{M^4}}. \label{eq:Zmixingangles}
\end{align}
The above mixing angles can be interpreted as follows.
After the left-right symmetry is broken by the vev parameter $M$, the mass matrix in \cref{eq:AZZpmassmatrix} yields mass terms of order $M$ for $W_R^3$, $B$.
Their mixing results in a massless boson $B_Y$, corresponding to the unbroken $U(1)_Y$ symmetry, and a massive boson $\widetilde{Z}'$ with mass $m_{\widetilde{Z}'} = g_RM/(2\cos\theta_W^\prime)$:
\begin{equation}
\begin{pmatrix} B_Y \\ \widetilde{Z}' \end{pmatrix} =
\begin{pmatrix} \sin\theta_W^\prime & \cos\theta_W^\prime \\ -\cos\theta_W^\prime & \sin\theta_W^\prime \end{pmatrix}
\begin{pmatrix} W_R^3 \\ B \end{pmatrix}. \label{eq:neutralRotation1}
\end{equation}
At this stage, both $W_L^3$ and $B_Y$ are massless.
We can perform a basis rotation in analogy to the Standard Model:
\begin{equation}
\begin{pmatrix} A \\ \widetilde{Z} \end{pmatrix} \equiv
\begin{pmatrix} \sin\theta_W & \cos\theta_W \\ -\cos\theta_W & \sin\theta_W \end{pmatrix}
\begin{pmatrix} W_L^3 \\ B_Y \end{pmatrix}. \label{eq:neutralRotation2}
\end{equation}
After the electroweak symmetry is broken by the vev parameters $v_1$, $b_1$, mass terms appear for $\widetilde{Z}$.
The field $A$ remains massless and therefore corresponds to the photon.
A comparison of the above basis rotation to the analogous Standard-Model expression shows that $\theta_W$ is the familiar Weinberg angle.
The field $\widetilde{Z}$ does not correspond to the physical $Z$ boson, since the mass matrix in \cref{eq:AZZpmassmatrix} yields mixing terms between $\widetilde{Z}$ and $\widetilde{Z}'$ as well.
The mass eigenstates $Z$, $Z'$ are given by a small rotation over $\eta$:
\begin{equation}
\begin{pmatrix} Z \\ Z' \end{pmatrix} =
\begin{pmatrix} -\cos\eta & -\sin\eta \\ \sin\eta & -\cos\eta \end{pmatrix}
\begin{pmatrix} \widetilde{Z} \\ \widetilde{Z}' \end{pmatrix}. \label{eq:neutralRotation3}
\end{equation}
A successive combination of the three rotations in \cref{eq:neutralRotation1,eq:neutralRotation2,eq:neutralRotation3} leads to the rotation matrix in \cref{eq:AZZpmasseigenstates}.
That is, we identify $\theta_W$ with the Weinberg angle; $\theta_W^\prime$ is an analogon of the Weinberg angle for the breaking of the left-right symmetry; and $\eta$ is the $Z-Z'$ mixing angle \cite{Chay:1998hd}.

\subsection{Gauge bosons in the complete LET model}\label{a:gbCompleteLET}
In the complete LET model, there is an additional scalar field $\Phi_2$ in the same representation as $\Phi_1$.
It obtains a vev
\begin{equation}
\vev{\Phi_2} = \begin{pmatrix} v_2 & 0 \\ 0 & b_2 \end{pmatrix}.
\end{equation}
This vev contributes to the gauge-boson mass matrix the same way that $\Phi_1$ does.
Hence we find the $W$ mass matrix for the complete LET model to be
\begin{equation}
M^2_\pm = \begin{pmatrix} \frac14g_L^2v^2 & -\frac12g_Lg_R(v_1b_1+v_2b_2) \\ -\frac12g_Lg_R(v_1b_1+v_2b_2) & \frac14g_R^2(M^2+v^2) \end{pmatrix}.
\end{equation}
Here it is understood that $v^2 = v_1^2+b_1^2+v_2^2+b_2^2$.
Note that this mass matrix is the same as the one for the simplified LET model in \cref{eq:WWpmassmatrix}, with the substitution
\begin{equation}
\sin2\beta \rightarrow \frac{2(v_1b_1+v_2b_2)}{v^2}.
\end{equation}
This means that we can copy the results from the simplified LET model, provided we apply this substitution.
Thus the $W$, $W'$ masses are given by
\begin{align}
m_W^2 =& \frac14g_L^2v^2 \left( 1 - \frac{4(v_1b_1+v_2b_2)^2}{v^2M^2} + \ord{\frac{v^4}{M^4}} \right), \notag\\
m_{W'}^2 =& \frac14g_R^2M^2 \left( 1 + \frac{v^2}{M^2} + \ord{\frac{v^4}{M^4}} \right).
\end{align}
The $W-W'$ mixing angle is given by
\begin{equation}
\zeta = \frac{2g_L(v_1b_1+v_2b_2)}{g_RM^2} + \ord{\frac{v^4}{M^4}}.
\end{equation}

As for the neutral gauge bosons, we find their mass matrix to be the same as the one for the simplified LET model in \cref{eq:AZZpmassmatrix}, with the understanding that $v^2 = v_1^2 + b_1^2 + v_2^2 + b_2^2$.
Hence the $Z$, $Z'$ masses and mixing angles are identical to those of the simplified LET model.

\clearpage
\section{Scalar spectrum}\label{a:scalarspectrum}
In order to study the scalar sector of the LET model and its  effects on Standard-Model couplings, we need to know the masses and mixings of the scalars.
We work them out in this appendix.
We start with the most general scalar potential for the simplified LET model and ensure it has a minimum at the envisioned vev.
Then we work out the theoretical constraints on the scalar parameters that come from vacuum stability and S-matrix unitarity.
After that, we work out the scalar masses and extract additional parameter constraints from the condition that the squared masses should be positive.
We also extract the mixing angles from the mass eigenstates.
We conclude with a brief discussion of the scalar spectrum for the complete LET model.

\subsection{Scalar potential}
We consider the scalar sector of the simplified LET model first, which contains only the fields $\Phi_1\sim(\rep{1},\brep{2},\rep{2},0)$ and $\Phi_R\sim(\rep{1},\rep{1},\rep{2},1)$.
We parametrise their components as follows:
\begin{align}
\Phi_1 =& \begin{pmatrix} \dfrac{v_1+h^0_{1,11}+ia^0_{1,11}}{\sqrt2} & h^-_{1,12} \\ h^+_{1,21} & \dfrac{b_1+h^0_{1,22}+ia^0_{1,22}}{\sqrt2} \end{pmatrix}, \notag\\
\Phi_R =& \begin{pmatrix} h_R^+ & \dfrac{M+h^0_R+ia^0_R}{\sqrt2} \end{pmatrix}. \label{eq:phidef}
\end{align}
Here the components $h^0_x$ are $CP$-even gauge eigenstates, and the components $a^0_x$ are $CP$-odd gauge eigenstates.
For the charged components, we define their conjugates as $h^\mp_x \equiv (h^\pm_x)^\dagger$.
As before, $SU(2)_L$ indices run vertically and $SU(2)_R$ indices run horizontally, hence $\Phi_1$ is a $2\times2$ matrix and $\Phi_R$ is a two-dimensional row vector.
The vev parameters $v_1$, $b_1$ are constrained by the relation $v_1^2+b_1^2=v^2=(246\text{ GeV})^2$.
Hence we reparametrise them as $v_1=v\cos\beta$, $b_1=v\sin\beta$.
The most general scalar potential for the simplified LET model is
\begin{align}
V(\Phi_1,\Phi_R) =& \frac{\lambda_1}{2}\tr{\Phi_1^\dagger\Phi_1}^2 + \frac{\lambda_2}{2}\tr{\Phi_1^\dagger\Phi_1\Phi_1^\dagger\Phi_1} + \frac{\lambda_3}{2}\big(\Phi_R\Phi_R^\dagger\big)^2 \notag\\
&+ \lambda_4\tr{\Phi_1^\dagger\Phi_1}(\Phi_R\Phi_R^\dagger) + \lambda_{5}\Phi_R\Phi_1^\dagger\Phi_1\Phi_R^\dagger \notag\\
&+ \mu^2_{11}\tr{\Phi_1^\dagger\Phi_1} + \mu^2_R\Phi_R\Phi_R^\dagger + \left(\mu^2_1\det\Phi_1 \plushc \right). \label{eq:scalarpotential}
\end{align}
Here the $\mu^2_i$ and $\lambda_i$ are real parameters.
In general LR-symmetric models, additional invariants could appear in the potential; see \cref{a:scalarinvariants} for a systematic derivation of all possible scalar invariants, as well as a discussion of which ones cannot arise from the trinification model.

The potential in \cref{eq:scalarpotential} should have a minimum at the vev
\begin{equation}
\vev{\Phi_1} = \begin{pmatrix} v_1 & 0 \\ 0 & b_1 \end{pmatrix},\qquad
\vev{\Phi_R} = \begin{pmatrix} 0 & M \end{pmatrix}.
\end{equation}
We obtain contraints on the scalar parameters from the condition that the first derivatives of the potential with respect to the component fields in \cref{eq:phidef} vanish at the vev.
This yields three conditions, which allow us to express the three dimensionful scalar parameters in terms of the dimensionless scalar parameters and the vevs:
\begin{align}
\mu_{11}^2 =&  -\frac12(\lambda_1+\lambda_2)v^2 - \frac{\lambda_4}{2}M^2 + \frac{\lambda_5}{2}M^2\frac{\tan^2\beta}{1-\tan^2\beta}, \notag\\
\mu_R^2 =& -\frac{\lambda_3}{2}M^2 - \frac{\lambda_4}{2}v^2 - \frac{\lambda_5}{2}v^2\sin^2\beta, \notag\\
\mu_1^2 =& \frac{\lambda_2}{4}v^2\sin 2\beta - \frac{\lambda_5}{4}M^2\tan 2\beta. \label{eq:minimumcondition}
\end{align}

\subsection{Vacuum stability}\label{a:vacuumstability}
The minimised scalar potential still has five free parameters, namely the dimensionless scalar couplings $\lambda_i$.
Their values are restricted by the condition of vacuum stability.
This condition requires the potential in \cref{eq:scalarpotential} to be positive for large field values $|\Phi_1|, |\Phi_R| \rightarrow\infty$.
In this section we work out some necessary constraints on the $\lambda_i$ arising from this condition.
To this end, we follow the same procedure as described in \cite{ElKaffas:2006nt}.

First, we rewrite the scalar potential in terms of two $SU(2)_L$ doublets $\phi_1$, $\phi_2$ and two singlets $S_+$, $S_0$ as in \cref{eq:Phi1doublets,eq:PhiRsinglets}.
Then we reparametrise these fields as follows:
\begin{equation}
\phi_1 = \|\phi_1\| \hat\phi_1,\quad \phi_2 = \|\phi_2\| \hat\phi_2,\quad S_+ = r_+e^{i\theta_+},\quad S_0 = r_0e^{i\theta_0},
\end{equation}
where $\hat\phi_1$, $\hat\phi_2$ are spinors of unit norm, $r_+,r_0\geq0$, and $\theta_+,\theta_0\in[0,2\pi)$.
Furthermore, we parametrise the norms of these fields as
\begin{IEEEeqnarray}{rClrCl}
\|\phi_1\| &=& R\cos\gamma_1,\quad & \|\phi_2\| &=& R\sin\gamma_1\cos\gamma_2,\quad \notag\\
r_+ &=& R\sin\gamma_1\sin\gamma_2\cos\gamma_3,\quad & r_0 &=& R\sin\gamma_1\sin\gamma_2\sin\gamma_3.
\end{IEEEeqnarray}
where $R\geq0$ and $\gamma_1,\gamma_2,\gamma_3\in[0,\frac{\pi}{2}]$.
We parametrise the unit spinor products as
\begin{equation}
\frac{\hat\phi_1^\dagger\hat\phi_2}{\|\phi_1\| \|\phi_2\|} = \rho_1e^{i\theta_1},\quad \frac{\hat\phi_1^{c\dagger}\hat\phi_2}{\|\phi_1\| \|\phi_2\|} = \rho_2e^{i\theta_2},
\end{equation}
where $\rho_1,\rho_2\in[0,1]$ and $\theta_1,\theta_2\in[0,2\pi)$.
In terms of this reparametrisation, the scalar potential becomes
\begin{align}
V(\Phi_1,\Phi_R) =& V_4R^4 + V_2R^2, \notag\\
V_4 \equiv& \frac{\lambda_1+\lambda_2}{2}(1 - \sin^2\gamma_1\sin^2\gamma_2)^2 - \lambda_2\rho_1^2\cos^2\gamma_1\sin^2\gamma_1\cos^2\gamma_2 \notag\\
&+ \frac{\lambda_3}{2}\sin^4\gamma_1\sin^4\gamma_2 + \lambda_4\sin^2\gamma_1\sin^2\gamma_2(1 - \sin^2\gamma_1\sin^2\gamma_2) \notag\\
&+ \lambda_5\sin^2\gamma_1\sin^2\gamma_2\big( \cos^2\gamma_1\cos^2\gamma_3 + \sin^2\gamma_1\cos^2\gamma_2\sin^2\gamma_3 \notag\\
&\quad+ \rho_2\cos\gamma_1\sin\gamma_1\cos\gamma_2\cos\gamma_3\sin\gamma_3\cos(\theta_0-\theta_++\theta_2)\big), \notag\\
V_2 \equiv& \mu_{11}^2(\cos^2\gamma_1 + \sin^2\gamma_1\cos^2\gamma_2) + \mu_1^2\rho_1\cos\gamma_1\sin\gamma_1\cos\gamma_2\cos\theta_1 \notag\\
&+ \mu_R^2\sin^2\gamma_1\sin^2\gamma_2. \label{eq:vacStabReparametrisation}
\end{align}
At large $R$, the $R^4$ term becomes dominant over the $R^2$ term.
Hence in order to ensure vacuum stability, it is a necessary and sufficient condition to ensure $V_4>0$ for all possible combinations of $\gamma_1,\gamma_2,\gamma_3\in[0,\frac{\pi}{2}]$; $\theta_+,\theta_0,\theta_1,\theta_2\in[0,2\pi)$; $\rho_1,\rho_2\in[0,1]$.
By considering specific points in this parameter space, we obtain very simple necessary (but not sufficient) stability conditions:
\begin{itemize}
\item {\boldmath $V_4(\gamma_1=0) = \frac{\lambda_1+\lambda_2}{2}$}, which gives the condition $$\lambda_1+\lambda_2 > 0.$$
\item {\boldmath $V_4(\gamma_1=\frac{\pi}{4}, \gamma_2=0, \rho_1=1) = \frac{2\lambda_1+\lambda_2}{4}$}, which gives the condition $$\lambda_2 > -2\lambda_1.$$
\item {\boldmath $V_4(\gamma_1=\gamma_2=\frac{\pi}{2}) = \frac{\lambda_3}{2}$}, which gives the condition $$\lambda_3 > 0.$$
\item {\boldmath $V_4(\gamma_1=\frac{\pi}{4}, \gamma_2=\gamma_3=\frac{\pi}{2}) = \frac{\lambda_1 + \lambda_2 + \lambda_3 + 2\lambda_4}{8}$}, which gives the condition $$\lambda_4 > -\frac{\lambda_1+\lambda_2+\lambda_3}{2}.$$
\item {\boldmath $V_4(\gamma_1=\gamma_3=\frac{\pi}{2}, \gamma_2=\frac{\pi}{4}, \rho_1=\rho_2=0) = \frac18(\lambda_1+\lambda_2+\lambda_3+2\lambda_4+2\lambda_5)$}, which gives the condition $$\lambda_5 > -\frac{\lambda_1+\lambda_2+\lambda_3+2\lambda_4}{2}.$$
\end{itemize}

\subsection{Unitarity of the S-matrix}\label{a:Smatrixunitarity}
Other constraints on the scalar parameters come from the requirement of S-matrix unitarity.
This basically means a conservation of probability in scattering processes.
Tree-level unitarity of the S-matrix is a necessary condition for perturbative renormalisability \cite{Cornwall:1974km}.
It implies that the scalar couplings cannot be too large.
We will make this statement more precise in this section, following the technique described in refs.~\cite{Kanemura:1993hm,Akeroyd:2000wc,Arhrib:2000is}.

One can impose tree-level unitarity for scalar-scalar, vector-vector, and scalar-vector scattering processes.
However, the search for these constraints can be simplified by using the Goldstone boson equivalence theorem \cite{Cornwall:1974km,Vayonakis:1976vz}.
This theorem states that in the massless limit (i.e.\ at very high energies), scattering amplitudes for processes involving longitudinal vector bosons can be approximated by the scattering amplitudes in which the vector bosons have been replaced by their corresponding Goldstone bosons.
Hence unitarity constraints can be obtained by considering only scalar scattering processes.

Consider such a process $\phi_1\phi_2\rightarrow\phi_3\phi_4$.
The amplitude $\mathcal{M}$ for this process can be written as a partial wave decomposition:
\begin{equation}
\mathcal{M}(s,t,u) = 16\pi\sum_{l=0}^\infty (2l+l)P_l(\cos\theta)a_l(s),
\end{equation}
where $s,t,u$ are the Mandelstam variables, $P_l(x)$ are the Legendre polynomials, and $a_l(s)$ are partial wave amplitudes.
Using the orthogonality of the Legendre polynomials, the partial waves can be expressed as
\begin{equation}
a_l(s) = \frac{1}{32\pi}\int_{-1}^1 \d(\cos\theta) P_l(\cos\theta) \mathcal{M}(s,t,u).
\end{equation}
Unitarity of the S-matrix implies \cite{Marciano:1989ns}
\begin{equation}
\left| \Re(a_l(s)) \right| \leq \frac12\qquad \forall l. \label{eq:SmatrixUnitarity}
\end{equation}
In the limit of high-energy scattering, $\mathcal{M}$ depends only on the quartic scalar coupling $\lambda_{\phi_1\phi_2\phi_3\phi_4}$: by dimensional analysis, the diagrams involving trilinear couplings are suppressed by the energy of the scattering process.
Note that $\mathcal{M}$ (and hence $a_l$) is real at tree-level.
Considering only the $l=0$ partial waves, we find the condition
\begin{align}
\frac12 \geq& \left|a_0\right|  = \frac{1}{16\pi} \left| \mathcal{M} \right|, \notag\\
\Rightarrow 8\pi \geq& \left| \mathcal{M} \right|. \label{eq:Smatrixunitarity}
\end{align}
If we consider all possible scalar-scattering processes, $\mathcal{M}$ becomes a matrix of scattering amplitudes.
In that case, the condition in \cref{eq:SmatrixUnitarity} applies to all eigenvalues of this matrix \cite{Kanemura:1993hm}.

The quartic couplings of the Lagrangian in terms of the physical fields of our model are quite complicated expressions.
However, our analysis is simplified by the fact that the S-matrix in terms of the physical fields is related to the S-matrix in terms of the gauge-eigenstate fields by a unitary transformation.
Since we are only interested in the eigenvalues, we can obtain the constraints in any basis we want.
The quartic couplings in terms of gauge eigenstates are easily obtained from the scalar potential in \cref{eq:scalarpotential}.

For the simplified LET model, the matrix $\mathcal{M}$ can be written as a block diagonal matrix composed of 10 submatrices.
The first block $\mathcal{M}_1$ corresponds to scattering where the initial and final states are elements of the basis $\{ \frac{h^0_{1,11}h^0_{1,11}}{\sqrt2},$ $\frac{h^0_{1,22}h^0_{1,22}}{\sqrt2},$ $\frac{h^0_Rh^0_R}{\sqrt2},$ $\frac{a^0_{1,11}a^0_{1,11}}{\sqrt2},$ $\frac{a^0_{1,22}a^0_{1,22}}{\sqrt2},$ $\frac{a^0_Ra^0_R}{\sqrt2},$ $h^+_{1,12}h^-_{1,12},$ $h^+_{1,21}h^-_{1,21},$ $h^+_Rh^-_R \}$.
The factors $\frac{1}{\sqrt2}$ are there to account for identical-particle statistics.
\emph{Mathematica} gives us
\begin{equation}
\mathcal{M}_1 =\\ \begin{pmatrix}
 6 \lambda_{12} & 2 \lambda_1 & 2 \lambda_4 & 2 \lambda_{12} & 2 \lambda_1 & 2 \lambda_4 & \sqrt{2} \lambda_{12} & \sqrt{2} \lambda_{12} & \sqrt{2} \lambda_{45} \\
 2 \lambda_1 & 6 \lambda_{12} & 2 \lambda_{45} & 2 \lambda_1 & 2 \lambda_{12} & 2 \lambda_{45} & \sqrt{2} \lambda_{12} & \sqrt{2} \lambda_{12} & \sqrt{2} \lambda_4 \\
 2 \lambda_4 & 2 \lambda_{45} & 6 \lambda_3 & 2 \lambda_4 & 2 \lambda_{45} & 2 \lambda_3 & \sqrt{2} \lambda_4 & \sqrt{2} \lambda_{45} & \sqrt{2} \lambda_3 \\
 2 \lambda_{12} & 2 \lambda_1 & 2 \lambda_4 & 6 \lambda_{12} & 2 \lambda_1 & 2 \lambda_4 & \sqrt{2} \lambda_{12} & \sqrt{2} \lambda_{12} & \sqrt{2} \lambda_{45} \\
 2 \lambda_1 & 2 \lambda_{12} & 2 \lambda_{45} & 2 \lambda_1 & 6 \lambda_{12} & 2 \lambda_{45} & \sqrt{2} \lambda_{12} & \sqrt{2} \lambda_{12} & \sqrt{2} \lambda_4 \\
 2 \lambda_4 & 2 \lambda_{45} & 2 \lambda_3 & 2 \lambda_4 & 2 \lambda_{45} & 6 \lambda_3 & \sqrt{2} \lambda_4 & \sqrt{2} \lambda_{45} & \sqrt{2} \lambda_3 \\
 \sqrt{2} \lambda_{12} & \sqrt{2} \lambda_{12} & \sqrt{2} \lambda_4 & \sqrt{2} \lambda_{12} & \sqrt{2} \lambda_{12} & \sqrt{2} \lambda_4 & 2 \lambda_{12} & \lambda_1 & \lambda_{45} \\
 \sqrt{2} \lambda_{12} & \sqrt{2} \lambda_{12} & \sqrt{2} \lambda_{45} & \sqrt{2} \lambda_{12} & \sqrt{2} \lambda_{12} & \sqrt{2} \lambda_{45} & \lambda_1 & 2 \lambda_{12} & \lambda_4 \\
 \sqrt{2} \lambda_{45} & \sqrt{2} \lambda_4 & \sqrt{2} \lambda_3 & \sqrt{2} \lambda_{45} & \sqrt{2} \lambda_4 & \sqrt{2} \lambda_3 & \lambda_{45} & \lambda_4 & 2 \lambda_3 
\end{pmatrix},
\end{equation}
where we have written $\lambda_{ij}\equiv \lambda_i+\lambda_j$ for the sake of brevity.
Three of the eigenvalues of $\mathcal{M}_1$ are simple expressions:
\begin{equation}
4(\lambda_1+\lambda_2),\; (2\times)\qquad 4\lambda_3.
\end{equation}
The remaining six eigenvalues are roots of a sixth-degree polynomial, which we could not solve analytically.

The second block $\mathcal{M}_2$ of the S-matrix corresponds to scattering with initial and final states contained in $\{ h^0_{1,11}h^0_{1,22},$ $h^0_{1,11}a^0_{1,22},$ $h^0_{1,22}a^0_{1,11},$ $a^0_{1,11}a^0_{1,22},$ $h^+_{1,21}h^-_{1,12},$ $h^+_{1,12}h^-_{1,21} \}$.
It is given by
\begin{equation}
\mathcal{M}_2 = \begin{pmatrix}
 \lambda_1 & 0 & 0 & 0 & \frac{\lambda_2}{2} & \frac{\lambda_2}{2} \\
 0 & \lambda_1 & 0 & 0 & \frac{i \lambda_2}{2} & -\frac{i\lambda_2}{2} \\
 0 & 0 & \lambda_1 & 0 & \frac{i \lambda_2}{2} & -\frac{i\lambda_2}{2} \\
 0 & 0 & 0 & \lambda_1 & -\frac{\lambda_2}{2} & -\frac{\lambda_2}{2} \\
 \frac{\lambda_2}{2} & -\frac{i\lambda_2}{2} & -\frac{i\lambda_2}{2} & -\frac{\lambda_2}{2} & \lambda_1 & 0 \\
 \frac{\lambda_2}{2} & \frac{i \lambda_2}{2} & \frac{i \lambda_2}{2} & -\frac{\lambda_2}{2} & 0 & \lambda_1 
\end{pmatrix}.
\end{equation}
The eigenvalues of $\mathcal{M}_2$ are
\begin{equation}
\lambda_1,\; (2\times)\quad \lambda_1\pm\lambda_2\; (2\times).
\end{equation}

The third block $\mathcal{M}_3$ of the S-matrix corresponds to scattering with initial and final states contained in $\{ h^0_{1,11}h^0_R,$ $h^0_{1,11}a^0_R,$ $h^0_Ra^0_{1,11},$ $a^0_{1,11}a^0_R,$ $h^+_{1,12}h^-_R,$ $h^+_Rh^-_{1,12} \}$.
We find
\begin{equation}
\mathcal{M}_3 = \begin{pmatrix}
 \lambda_4 & 0 & 0 & 0 & \frac{\lambda_5}{2} & \frac{\lambda_5}{2} \\
 0 & \lambda_4 & 0 & 0 & -\frac{i\lambda_5}{2} & \frac{i \lambda_5}{2} \\
 0 & 0 & \lambda_4 & 0 & -\frac{i\lambda_5}{2} & \frac{i \lambda_5}{2} \\
 0 & 0 & 0 & \lambda_4 & -\frac{\lambda_5}{2} & -\frac{\lambda_5}{2} \\
 \frac{\lambda_5}{2} & \frac{i \lambda_5}{2} & \frac{i \lambda_5}{2} & -\frac{\lambda_5}{2} & \lambda_4 & 0 \\
 \frac{\lambda_5}{2} & -\frac{i\lambda_5}{2} & -\frac{i\lambda_5}{2} & -\frac{\lambda_5}{2} & 0 & \lambda_4 
\end{pmatrix}.
\end{equation}
The eigenvalues of $\mathcal{M}_3$ are
\begin{equation}
\lambda_4,\; (2\times)\qquad \lambda_4 \pm \lambda_5\; (2\times).
\end{equation}

The fourth block $\mathcal{M}_4$ of the S-matrix corresponds to scattering with initial and final states contained in $\{ h^0_{1,22}h^0_R,$ $h^0_{1,22}a^0_R,$ $h^0_Ra^0_{1,22},$ $a^0_{1,22}a^0_R,$ $h^+_{1,21}h^-_R , h^+_Rh^-_{1,21} \}$.
It is given by
\begin{equation}
\mathcal{M}_4 = \begin{pmatrix}
 \lambda_{45} & 0 & 0 & 0 & \frac{\lambda_5}{2} & \frac{\lambda_5}{2} \\
 0 & \lambda_{45} & 0 & 0 & -\frac{i\lambda_5}{2} & \frac{i \lambda_5}{2} \\
 0 & 0 & \lambda_{45} & 0 & \frac{i \lambda_5}{2} & -\frac{i\lambda_5}{2} \\
 0 & 0 & 0 & \lambda_{45} & \frac{\lambda_5}{2} & \frac{\lambda_5}{2} \\
 \frac{\lambda_5}{2} & \frac{i \lambda_5}{2} & -\frac{i\lambda_5}{2} & \frac{\lambda_5}{2} & \lambda_{45} & 0 \\
 \frac{\lambda_5}{2} & -\frac{i\lambda_5}{2} & \frac{i \lambda_5}{2} & \frac{\lambda_5}{2} & 0 & \lambda_{45}
\end{pmatrix}.
\end{equation}
The eigenvalues of $\mathcal{M}_4$ are
\begin{equation}
\lambda_4\; (2\times),\qquad \lambda_4+\lambda_5\; (2\times),\qquad \lambda_4 + 2\lambda_5\; (2\times).
\end{equation}

The block $\mathcal{M}_5$ in the basis $\{ h^0_{1,11}a^0_{1,11},$ $h^0_{1,22}a^0_{1,22},$ $h^0_Ra^0_R \}$ is diagonal, hence its eigenvalues are simply the diagonal elements:
\begin{equation}
\mathcal{M}_5 = diag(\lambda_{12} , \lambda_{12} , \lambda_3).
\end{equation}

The block $\mathcal{M}_6$ corresponds to scattering with initial and final states contained in $\{ h^+_{1,21}h^0_{1,11},$ $h^+_{1,12}h^0_{1,22},$ $h^+_{1,21}a^0_{1,11},$ $h^+_{1,12}a^0_{1,22} \}$.
We find
\begin{equation}
\mathcal{M}_6 = \begin{pmatrix}
 \lambda_{12} & \frac{\lambda_2}{2} & 0 & \frac{i \lambda_2}{2} \\
 \frac{\lambda_2}{2} & \lambda_{12} & -\frac{i\lambda_2}{2} & 0 \\
 0 & \frac{i \lambda_2}{2} & \lambda_{12} & -\frac{\lambda_2}{2} \\
 -\frac{i\lambda_2}{2} & 0 & -\frac{\lambda_2}{2} & \lambda_{12}
\end{pmatrix}.
\end{equation}
Its eigenvalues are
\begin{equation}
\lambda_1,\qquad \lambda_1+\lambda_2\; (2\times),\qquad \lambda_1+2\lambda_2.
\end{equation}

The block $\mathcal{M}_7$ corresponds to scattering with initial and final states contained in $\{ h^+_{1,21}h^0_{1,22},$ $h^+_{1,12}h^0_{1,11},$ $h^+_Rh^0_R,$ $h^+_{1,21}a^0_{1,22},$ $h^+_{1,12}a^0_{1,11},$ $h^+_Ra^0_R \}$, and is given by
\begin{equation}
\mathcal{M}_7 = \begin{pmatrix}
 \lambda_{12} & \frac{\lambda_2}{2} & \frac{\lambda_5}{2} & 0 & \frac{i \lambda_2}{2} & -\frac{i\lambda_5}{2} \\
 \frac{\lambda_2}{2} & \lambda_{12} & \frac{\lambda_5}{2} & -\frac{i\lambda_2}{2} & 0 & -\frac{i\lambda_5}{2} \\
 \frac{\lambda_5}{2} & \frac{\lambda_5}{2} & \lambda_3 & -\frac{i\lambda_5}{2} & \frac{i \lambda_5}{2} & 0 \\
 0 & \frac{i \lambda_2}{2} & \frac{i \lambda_5}{2} & \lambda_{12} & -\frac{\lambda_2}{2} & \frac{\lambda_5}{2} \\
 -\frac{i\lambda_2}{2} & 0 & -\frac{i\lambda_5}{2} & -\frac{\lambda_2}{2} & \lambda_{12} & -\frac{\lambda_5}{2} \\
 \frac{i \lambda_5}{2} & \frac{i \lambda_5}{2} & 0 & \frac{\lambda_5}{2} & -\frac{\lambda_5}{2} & \lambda_3 
\end{pmatrix}.
\end{equation}
Its eigenvalues are
\begin{multline}
\lambda_1,\qquad \lambda_1+\lambda_2\; (2\times),\qquad \lambda_3,\\ \frac12\left(\lambda_1+2\lambda_2+\lambda_3 \pm \sqrt{(\lambda_1+2\lambda_2)^2 - \lambda_3(2\lambda_1+4\lambda_2-\lambda_3) + 8\lambda_5^2}\right).
\end{multline}

The block $\mathcal{M}_8$ corresponds to scattering with initial and final states contained in $\{ h^+_{1,21}h^0_R,$ $h^+_Rh^0_{1,22},$ $h^+_{1,21}a^0_R,$ $h^+_R a^0_{1,22} \}$, and is given by
\begin{equation}
\mathcal{M}_8 = \begin{pmatrix}
 \lambda_4 & \frac{\lambda_5}{2} & 0 & -\frac{i\lambda_5}{2} \\
 \frac{\lambda_5}{2} & \lambda_4 & -\frac{i\lambda_5}{2} & 0 \\
 0 & \frac{i \lambda_5}{2} & \lambda_4 & \frac{\lambda_5}{2} \\
 \frac{i \lambda_5}{2} & 0 & \frac{\lambda_5}{2} & \lambda_4
\end{pmatrix}.
\end{equation}
Its eigenvalues are
\begin{equation}
\lambda_4\; (2\times),\qquad \lambda_4 \pm \lambda_5.
\end{equation}

The block $\mathcal{M}_9$ corresponds to scattering with initial and final states contained in $\{ h^+_{1,12}h^0_R,$ $h^+_Rh^0_{1,21},$ $h^+_{1,12}a^0_R,$ $h^+_Ra^0_{1,21} \}$. It is given by
\begin{equation}
\mathcal{M}_9 = \begin{pmatrix}
 \lambda_{45} & \frac{\lambda_5}{2} & 0 & -\frac{i\lambda_5}{2} \\
 \frac{\lambda_5}{2} & \lambda_{45} & \frac{i \lambda_5}{2} & 0 \\
 0 & -\frac{i\lambda_5}{2} & \lambda_{45} & -\frac{\lambda_5}{2} \\
 \frac{i \lambda_5}{2} & 0 & -\frac{\lambda_5}{2} & \lambda_{45} \\
\end{pmatrix}.
\end{equation}
Its eigenvalues are
\begin{equation}
\lambda_4,\qquad \lambda_4+\lambda_5\; (2\times),\qquad \lambda_4+2\lambda_5.
\end{equation}

The block $\mathcal{M}_{10}$ in the basis $\{ \frac{h^+_{1,21}h^+_{1,21}}{\sqrt2},$ $\frac{h^+_{1,12}h^+_{1,12}}{\sqrt2},$ $\frac{h^+_Rh^+_R}{\sqrt2},$ $h^+_{1,21}h^+_{1,12},$ $h^+_{1,21}h^+_R,$ $h^+_{1,12}h^+_R \}$ is diagonal, hence its eigenvalues are simply the diagonal elements:
\begin{equation}
\mathcal{M}_{10} = diag(4\lambda_{12} , 4\lambda_{12} , 4\lambda_3 , \lambda_1 , \lambda_{45} , \lambda_4).
\end{equation}

All eigenvalues need to satisfy the S-matrix-unitarity condition in \cref{eq:Smatrixunitarity}.
This yields the following constraints on the scalar parameters:
\begin{align}
2\pi \geq& |\lambda_3|, |\lambda_1+\lambda_2|, \notag\\
8\pi \geq& |\lambda_1|, |\lambda_4|, |\lambda_1-\lambda_2|, |\lambda_4+\lambda_5|, |\lambda_4-\lambda_5|, |\lambda_1+2\lambda_2|, |\lambda_4+2\lambda_5|, \notag\\
16\pi \geq& \left|\lambda_1 + 2\lambda_2 + \lambda_3 \pm \sqrt{(\lambda_1+2\lambda_2)^2 - \lambda_3(2\lambda_1+4\lambda_2-\lambda_3) + 8\lambda_5^2} \right|.
\end{align}

\subsection{Scalar masses and mass eigenstates}\label{a:scalarmasseigenstates}
In our parametrisation of the scalar field components in \cref{eq:phidef}, the gauge eigenstates contain three pairs of charged states $h^\pm_{1,21}$, $h^\pm_{1,12}$, $h^\pm_R$, three $CP$-odd states $a^0_{1,11}$, $a^0_{1,22}$, $a^0_R$, and three $CP$-even states $h^0_{1,11}$, $h^0_{1,22}$, $h^0_R$.
After spontaneous symmetry breaking, the potential in \cref{eq:scalarpotential} gives rise to scalar mass terms.
The scalar mass matrix is the matrix of second derivatives of the potential with respect to the scalar field components at the vev.
The eigenvectors of this matrix become the physical states we observe, and their eigenvalues are their squared masses.
The mass eigenstates can be parametrised by a rotation of the gauge eigenstates.
These rotations are described by three $3\times3$ matrices, each of which we parametrise by three Euler angles:
\begin{equation}
R(\alpha_1,\alpha_2,\alpha_3) \equiv \begin{pmatrix}
c_1 & s_1c_2 & s_1s_2 \\
-s_1c_3 & c_1c_2c_3 - s_2s_3 & c_1s_2c_3 + c_2s_3 \\
s_1s_3 & -s_2c_3 - c_1c_2s_3 & c_2c_3 - c_1s_2s_3 \label{eq:rotationmatrix}
\end{pmatrix}.
\end{equation}
Here we defined $c_i \equiv \cos\alpha_i$ and $s_i \equiv \sin\alpha_i$ for the sake of brevity.
The ranges for the angles are $\alpha_1\in[0,\frac{\pi}{2}]$, $\alpha_2\in(-\pi,\pi]$, and $\alpha_3\in[-\frac{\pi}{2},\frac{\pi}{2}]$.\footnote{\label{fn:Eulerambiguity} For general three-dimensional rotations, the ranges for the Euler angles are $\alpha_1\in[0,\pi]$ and $\alpha_2,\alpha_3\in(-\pi,\pi]$. However, there is an ambiguity in the definition of the mixing angles because the phase of the mass eigenstates is arbitrary: if for example $h^0\equiv c_1h^0_{1,11} + s_1c_2h^0_{1,22} + s_1s_2h^0_R$ is a mass eigenstate with mass $m_{h^0}$, then so is $-h^0$. Hence we can redefine $\alpha_{1,2,3}$ by multiplying any pair of rows in $R(\alpha_1,\alpha_2,\alpha_3)$ by a minus sign, keeping the determinant at $+1$. Multiplying the first two rows by $-1$ corresponds to the redefinition $\alpha_1\rightarrow\pi-\alpha_1$, $\alpha_2\rightarrow\pi+\alpha_2$, $\alpha_3\rightarrow\pi-\alpha_3$, whereas multiplying the last two rows by $-1$ corresponds to $\alpha_1\rightarrow\alpha_1$, $\alpha_2\rightarrow\alpha_2$, and $\alpha_3\rightarrow\pi+\alpha_3$. Thus we have the freedom to choose our mass eigenstates such that $\alpha_1\in[0,\frac{\pi}{2}]$, $\alpha_3\in[-\frac{\pi}{2},\frac{\pi}{2}]$. In practice, if we want to extract these rotation angles from a rotation matrix $R$, we first multiply the first two rows by $-1$ if $R_{(1,1)}<0$, and then multiply the last two rows by $-1$ if $R_{(2,1)}>0$. Only then do we extract $\alpha_1$, $\alpha_2$, $\alpha_3$ from \cref{eq:rotationmatrix}}
In terms of this parametrisation, we define the scalar mass eigenstates as follows:
\begin{align}
\begin{pmatrix} h^0 \\ H_1^0 \\ H_2^0 \end{pmatrix}
=& R(\alpha_1,\alpha_2,\alpha_3) \begin{pmatrix} h_{1,11}^0 \\ h_{1,22}^0 \\ h_R^0 \end{pmatrix},\qquad \begin{pmatrix} G^0 \\ G^{\prime0} \\ A^0 \end{pmatrix}
= R(\beta_1,\beta_2,\beta_3) \begin{pmatrix} a_{1,11}^0 \\ a_{1,22}^0 \\ a_R^0 \end{pmatrix}, \notag\\
\begin{pmatrix} G^\pm \\ G^{\prime\pm} \\ H^\pm \end{pmatrix}
=& R(\gamma_1,\gamma_2,\gamma_3) \begin{pmatrix} h_{1,21}^\pm \\ h_{1,12}^\pm \\ h_R^\pm \end{pmatrix}. \label{eq:Higgsmasseig}
\end{align}
Here the $CP$-odd states $G^0$ and $G^{\prime0}$ are defined as the Goldstones that give mass to the $Z$ and $Z'$ bosons respectively.
Similarly, the charged states $G^\pm$ and $G^{\prime\pm}$ are the Goldstones that give mass to the $W^\pm$ and $W^{\prime\pm}$ bosons respectively.
Note that the simplified LET model predicts the existence of a massive $CP$-odd scalar $A^0$ and a pair of massive charged scalars $H^\pm$.

The simplified LET model contains three massive $CP$-even scalars $h^0$, $H^0_1$, $H^0_2$, of which only one has been observed.
Of course, the definition of the mixing angles $\alpha_1$, $\alpha_2$, $\alpha_3$ in \cref{eq:Higgsmasseig} is ambiguous unless we specify how to distinguish the three scalars from each other.
We define $h^0$ as the state that is the most $h^0_{1,11}$-like, and $H^0_2$ as the state that is the most $h^0_R$-like.
It will turn out that $h^0$ has a mass proportional to $v$ whereas $H_1^0$, $H_2^0$ have masses proportional to $M$.
That is, $h^0$ is the only $CP$-even scalar that naturally has a mass of the order of the electroweak scale.
Therefore we identify $h^0$ with the scalar particle that has been observed at the LHC.
This means that the Standard-Model-like Higgs in our model is given by
\begin{equation}
h^0 = \cos\alpha_1 h^0_{1,11} + \sin\alpha_1\cos\alpha_2 h^0_{1,22} + \sin\alpha_1\sin\alpha_2 h^0_R.
\end{equation}
Note that if $h^0$ contains a nonzero component of the fermiophobic $h^0_R$, the decay rate to fermions will be less than the rate predicted by the Standard Model.
If the coefficient $\sin\alpha_1\sin\alpha_2$ is not too small, this reduction may be observable experimentally.

\paragraph{Charged mass eigenstates}
After spontaneous symmetry breaking, the charged states in the basis $\{h^\pm_{1,21},h^\pm_{1,12},h^\pm_R\}$ have the following mass matrix:
\begin{equation}
\mathcal{M}^2_\pm = \frac12\lambda_5 \begin{pmatrix} M^2\sin^2\beta\sec2\beta & \frac12M^2\tan2\beta & Mv\sin\beta \\
\frac12M^2\tan2\beta & M^2\cos^2\beta\sec2\beta & Mv\cos\beta \\
Mv\sin\beta & Mv\cos\beta & v^2\cos2\beta \end{pmatrix}.
\end{equation}
The eigenvalues can easily be solved by \emph{Mathematica}.
Two of them are zero: they correspond to the charged Goldstones $G^\pm$, $G^{\prime\pm}$.
The third eigenvalue gives the mass of $H^\pm$:
\begin{equation}
m_{H^\pm}^2 = \frac{\lambda_5}{2}\left( M^2\sec2\beta + v^2\cos2\beta \right). \label{eq:HpmMass}
\end{equation}
In order for the scalar potential to be at a minimum, the eigenvalues of the mass matrix that do not correspond to Goldstones must be positive.
This yields a new constraint on the scalar parameter $\lambda_5$:
\begin{equation}
\lambda_5 > 0.
\end{equation}
The mixing angles $\gamma_1$, $\gamma_2$, $\gamma_3$ can be solved analytically as well.
We find:
\begin{align}
\gamma_1 =& \arctan\left( \tan\beta\sqrt{1 + \frac{2M^2v^2\sin^22\beta}{(M^2+v^2\cos2\beta)^2}} \right) \notag\\
=& \beta + \frac12\sin^32\beta\frac{v^2}{M^2} + \ord{\frac{v^4}{M^4}}, \notag\\
\gamma_2 =& \arctan\left( \frac{2Mv\cos\beta}{M^2-v^2\cos2\beta} \right) - \pi \notag\\
=& -\pi + 2\cos\beta\frac{v}{M} + \ord{\frac{v^3}{M^3}}, \notag\\
\gamma_3 =& \arccos\left( v\sqrt{\frac{M^2+v^2\cos^22\beta}{M^4+2M^2v^2+v^4\cos^22\beta}} \right) \notag\\
=& \frac{\pi}{2} - \frac{v}{M} + \ord{\frac{v^3}{M^3}}. \label{eq:chargedMixingAngles}
\end{align}

\paragraph{$CP$-odd mass eigenstates}
The $CP$-odd states in the basis $\{a^0_{1,11},a^0_{1,22},a^0_R\}$ have the following mass matrix:
\begin{equation}
\mathcal{M}^2_\text{odd} = \frac12(\lambda_5M^2-\lambda_2v^2\cos2\beta) \begin{pmatrix} \sin^2\beta\sec2\beta & \frac12\tan2\beta & 0 \\
\frac12\tan2\beta & \cos^2\beta\sec2\beta & 0 \\
0 & 0 & 0 \end{pmatrix}.
\end{equation}
Two of its eigenvalues are zero: they correspond to the $CP$-odd Goldstones $G^0$, $G^{\prime0}$.
The third eigenvalue gives the mass of $A^0$:
\begin{equation}
m_A^2 = \frac12\left( \lambda_5M^2\sec2\beta - \lambda_2v^2 \right). \label{eq:A0Mass}
\end{equation}
This eigenvalue should be positive, hence we obtain another constraint on the scalar parameters:
\begin{equation}
\lambda_5 > \lambda_2\frac{v^2}{M^2}\cos2\beta.
\end{equation}
The mixing angles $\beta_1$, $\beta_2$, $\beta_3$ can be solved analytically as well.
They have the following simple form:
\begin{equation}
\beta_1 = \beta,\qquad \beta_2 = \pi,\qquad\beta_3 = \frac{\pi}{2}. \label{eq:CPoddMixingAngles}
\end{equation}

\paragraph{$CP$-even mass eigenstates}\label{p:solveEigenvalues}
The $CP$-even states in the basis $\{h^0_{1,11},h^0_{1,22},h^0_R\}$ have the following mass matrix:
\begin{align}
\mathcal{M}^2_\text{even} =& \begin{pmatrix}
C_1+C_2\cos2\beta+2C_3\sin^2\beta & (C_1-C_3)\sin2\beta & \lambda_4Mv\cos\beta \\
(C_1-C_3)\sin2\beta & C_1-C_2\cos2\beta+2C_3\cos^2\beta & (\lambda_4+\lambda_5)Mv\sin\beta \\
\lambda_4Mv\cos\beta & (\lambda_4+\lambda_5)Mv\sin\beta & \lambda_3M^2
\end{pmatrix}, \notag\\
C_1 \equiv& \frac14(2\lambda_1+\lambda_2)v^2,\quad C_2 \equiv \frac14(2\lambda_1+3\lambda_2)v^2,\quad C_3 \equiv \frac14\lambda_5M^2\sec2\beta.
\end{align}
The three eigenvalues $m_i^2$ are the roots of the characteristic polynomial $p(x) = \det(xI_3 - \mathcal{M}^2_\text{even})$.
We have no analytical solution for these eigenvalues, since $p(x)$ is a third-degree polynomial with large expressions as its coefficients.
However, it is straightforward to calculate the $m_i^2$ as an expansion in $\frac{v}{M}$:
\begin{equation}
m_i^2(M,v) = M^2\left( x_{i0} + x_{i1}\frac{v}{M} + \frac12x_{i2}\frac{v^2}{M^2} + \ord{\frac{v^3}{M^3}} \right), \notag\\
\end{equation}
The expansion coefficients $x_{in}$ only depend on $\beta$ and the $\lambda_i$.
We obtain these coefficients by solving the eigenvalue equation $p(m_i^2)=0$ order by order in $\frac{v}{M}$.
This yields
\begin{align}
m_{h^0}^2 =& \left( \lambda_1 + \lambda_2\cos^2\beta - \frac{(\lambda_4+\lambda_5\sin^2\beta)^2}{\lambda_3} + \ord{\frac{v^2}{M^2}} \right)v^2, \notag\\
m_{H^0_1}^2 =& \frac12\lambda_5M^2\sec2\beta - \frac{v^2}{2} \left( \lambda_2\cos^22\beta - \frac{\lambda_5^2\sin^22\beta\cos2\beta}{\lambda_5-2\lambda_3\cos2\beta} + \ord{\frac{v^2}{M^2}} \right), \notag\\
m_{H^0_2}^2 =& \lambda_3M^2 + v^2\left( \frac{(\lambda_4+\lambda_5\sin^2\beta)^2}{\lambda_3} - \frac{\lambda_5^2\sin^22\beta\cos2\beta}{\lambda_5-2\lambda_3\cos2\beta} + \ord{\frac{v^2}{M^2}} \right). \label{eq:CPevenMass}
\end{align}
Note that since $\beta$ is small, $H_1^0$ is approximately degenerate in mass with $A^0$.
For $M\gg v$, the vacuum stability condition $\lambda_3>0$ and the minimum condition $\lambda_5>0$ guarantee that $m^2_{H^0_1}$, $m^2_{H^0_2}$ are positive.
The positivity of $m_{h^0}^2$ gives an additional constraint (using $\lambda_3>0$):
\begin{equation}
(\lambda_1+\lambda_2\cos^2\beta)\lambda_3 > (\lambda_4 + \lambda_5\sin^2\beta)^2.
\end{equation}
The mixing angles $\alpha_1$, $\alpha_2$, $\alpha_3$ can be solved order by order in $\frac{v}{M}$ as well.
Let $v_i$ be the eigenvector of $\mathcal{M}^2_\text{even}$ with eigenvalue $m_i^2$.
We then write
\begin{equation}
v_i = (v_i)_0 + (v_i)_1\frac{v}{M} + \frac12(v_i)_2\frac{v^2}{M^2} + \ord{\frac{v^3}{M^3}}.
\end{equation}
Again, the expansion coefficients $(v_i)_n$ depend only on $\beta$ and the $\lambda_i$.
We obtain them by solving the eigenvector equation $\mathcal{M}^2_\text{even}v_i = m_i^2v_i$ order by order in $\frac{v}{M}$, using the masses in \cref{eq:CPevenMass}.
The rotation matrix $R(\alpha_1,\alpha_2,\alpha_3)$ has the $v_i$ as its rows.
Using \cref{eq:Higgsmasseig}, we extract the following mixing angles from this matrix:
\begin{align}
\alpha_1 =& \beta + \Bigg( \frac12\cot\beta \left(\frac{\lambda_4+\lambda_5\sin^2\beta}{\lambda_3}\right)^2 + \frac12\sin4\beta \left(\frac{\lambda_4+\lambda_5\sin^2\beta}{\lambda_3}\right) \notag\\
&\hspace{4cm} + \frac{\lambda_2\sin2\beta\cos^22\beta}{\lambda_5} \Bigg)\frac{v^2}{M^2} + \ord{\frac{v^3}{M^3}}, \notag\\
\alpha_2 =& -\csc\beta \left(\frac{\lambda_4 + \lambda_5\sin^2\beta}{\lambda_3}\right) \frac{v}{M} + \ord{\frac{v^2}{M^2}}, \notag\\
\alpha_3 =& \left( \frac{\lambda_4\cot\beta}{\lambda_3} + \frac{\lambda_5^2\tan2\beta}{2\lambda_3(\lambda_5\sec2\beta-2\lambda_3)} \right)\frac{v}{M} + \ord{\frac{v^2}{M^2}}. \label{eq:CPevenMixingAngles}
\end{align}

\subsection{The complete LET model}\label{a:completeLETmodel}
Now let us consider the scalar sector of the complete LET model.
That is, we add a second bidoublet $\Phi_2 \sim (\rep{1},\brep{2},\rep{2},0)$ to the scalar sector.
We parametrise it the same way as $\Phi_1$:
\begin{equation}
\Phi_2 = \begin{pmatrix} \dfrac{v_2+h^0_{2,11}+ia^0_{2,11}}{\sqrt2} & h^-_{2,12} \\ h^+_{2,21} & \dfrac{b_2+h^0_{2,22}+ia^0_{2,22}}{\sqrt2} \end{pmatrix}. \label{eq:phi2def}
\end{equation}
The vev parameters of $\Phi_1$ and $\Phi_2$ are now constrained by $v_1^2+b_1^2+v_2^2+b_2^2 = v^2 = (246\text{ GeV})^2$.
In analogy to the simplified LET model, it will be convenient to reparametrise them as
\begin{align}
v_1 =& v\cos\alpha\cos\beta_1,\qquad v_2 = v\sin\alpha\cos\beta_2, \notag\\
b_1 =& v\cos\alpha\sin\beta_1,\qquad b_2 = v\sin\alpha\sin\beta_2.
\end{align}
In order to avoid FCNC processes in the fermion sector, we introduced a $Z_2$-symmetry under which $\Phi_1$ ($\Phi_2$) is even (odd).
The most general scalar potential for the complete LET model is
\begin{align}
V_2(\Phi_1,\Phi_2,\Phi_R) =& V(\Phi_1,\Phi_R) + \frac{\widetilde\lambda_1}{2}\tr{\Phi_2^\dagger\Phi_2}^2 + \frac{\widetilde\lambda_2}{2}\tr{\Phi_2^\dagger\Phi_2\Phi_2^\dagger\Phi_2} \notag\\
&+ \widetilde\lambda_3\tr{\Phi_2^\dagger\Phi_2}(\Phi_R\Phi_R^\dagger) + \widetilde\lambda_4\Phi_R\Phi_2^\dagger\Phi_2\Phi_R^\dagger \notag\\
&+ \widetilde\lambda_5\tr{\Phi_1^\dagger\Phi_1}\tr{\Phi_2^\dagger\Phi_2} + \widetilde\lambda_6\left|\tr{\Phi_1^\dagger\Phi_2}\right|^2 \notag\\
&+ \frac{\widetilde\lambda_7}{2}\left( \tr{\Phi_1^\dagger\Phi_2}^2 \plushc\right) + \widetilde\lambda_8\tr{\Phi_1^\dagger\Phi_1\Phi_2^\dagger\Phi_2} \notag\\
&+ \widetilde\lambda_9\tr{\Phi_1^\dagger\Phi_2\Phi_2^\dagger\Phi_1} + \frac{\widetilde\lambda_{10}}{2}\left( \tr{\Phi_1^\dagger\Phi_2\Phi_1^\dagger\Phi_2} \plushc\right) \notag\\
&+ \mu^2_{22}\tr{\Phi_2^\dagger\Phi_2} + \left(\mu^2_2\det\Phi_2 \plushc \right).  \label{eq:scalarpotential2}
\end{align}
Here $V(\Phi_1,\Phi_R)$ is the scalar potential for the simplified LET model, which was given in \cref{eq:scalarpotential}.
Again, all coupling parameters in the potential are assumed to be real.
For a derivation of all possible scalar invariants, including the ones we omitted because they violate the $Z_2$-symmetry, see \cref{a:scalarinvariants}.

The potential in \cref{eq:scalarpotential2} should have a minimum at the vev
\begin{equation}
\vev{\Phi_1} = \begin{pmatrix} v_1 & 0 \\ 0 & b_1 \end{pmatrix},\qquad
\vev{\Phi_2} = \begin{pmatrix} v_2 & 0 \\ 0 & b_2 \end{pmatrix},\qquad
\vev{\Phi_R} = \begin{pmatrix} 0 & M \end{pmatrix}.
\end{equation}
If we demand that the first derivatives of the potential with respect to the component fields vanish at the vev, we obtain five conditions.
These allow us to express the five dimensionful scalar parameters in terms of the dimensionless scalar parameters and the vevs:
\begin{align}
\mu^2_{11} =& -\frac12(\lambda_1+\lambda_2)(v_1^2+b_1^2) - \frac{\lambda_4}{2}M^2 + \frac{\lambda_5}{2}\frac{M^2b_1^2}{v_1^2-b_1^2} \notag\\
&- \frac{\widetilde\lambda_5}{2}(v_2^2+b_2^2) - \frac12(\widetilde\lambda_6 + \widetilde\lambda_7 + \widetilde\lambda_8 + \widetilde\lambda_9 + \widetilde\lambda_{10})\frac{v_1^2v_2^2-b_1^2b_2^2}{v_1^2-b_1^2}, \notag\\
\mu^2_{22} =& -\frac12(\widetilde\lambda_1+\widetilde\lambda_2)(v_2^2+b_2^2) - \frac{\widetilde\lambda_3}{2}M^2 + \frac{\widetilde\lambda_4}{2}\frac{M^2b_2^2}{v_2^2-b_2^2} \notag\\
&- \frac{\widetilde\lambda_5}{2}(v_1^2+b_1^2) - \frac12(\widetilde\lambda_6 + \widetilde\lambda_7 + \widetilde\lambda_8 + \widetilde\lambda_9 + \widetilde\lambda_{10})\frac{v_1^2v_2^2-b_1^2b_2^2}{v_2^2-b_2^2}, \notag\\
\mu^2_R =& -\frac12\left( \lambda_3M^2 + \lambda_4(v_1^2+b_1^2) + \lambda_5b_1^2 + \widetilde\lambda_3(v_2^2+b_2^2) + \widetilde\lambda_4b_2^2 \right), \notag\\
\mu^2_1 =& \frac{\lambda_2}{2}v_1b_1 - \frac{\lambda_5}{2}\frac{M^2v_1b_1}{v_1^2-b_1^2} + \frac12(\widetilde\lambda_6 + \widetilde\lambda_7)\left( v_1b_1\frac{v_2^2-b_2^2}{v_1^2-b_1^2} - v_2b_2\right) \notag\\
&+ \frac12(\widetilde\lambda_8 + \widetilde\lambda_9 + \widetilde\lambda_{10})v_1b_1\frac{v_2^2-b_2^2}{v_1^2-b_1^2}, \notag\\
\mu^2_2 =& \frac{\widetilde\lambda_2}{2}v_2b_2 - \frac{\widetilde\lambda_4}{2}\frac{M^2v_2b_2}{v_2^2-b_2^2} + \frac12(\widetilde\lambda_6 + \widetilde\lambda_7)\left( v_2b_2\frac{v_1^2-b_1^2}{v_2^2-b_2^2} - v_1b_1\right) \notag\\
&+ \frac12(\widetilde\lambda_8 + \widetilde\lambda_9 + \widetilde\lambda_{10})v_2b_2\frac{v_1^2-b_1^2}{v_2^2-b_2^2}. \label{eq:completeLETminimum}
\end{align}

The scalar mass eigenstates are eigenstates of the mass matrix, which is the matrix of second derivatives of $V_2(\Phi_1,\Phi_2,\Phi_R)$.
These eigenstates are rotations of the gauge eigenstates in \cref{eq:phidef,eq:phi2def}, which amount to five $CP$-even scalars, five $CP$-odd scalars and five pairs of charged scalars.
Their mixings are given by three $5\times5$ matrices.
We parametrise these matrices using the 5-dimensional analogue of the Euler angles.
In a general $N$-dimensional vector space, a rotation over angle $\alpha$ in the plane spanned by the $m$th and $n$th basis vectors is given by the $N\times N$ matrix
\begin{equation}
R_{mn}(\alpha) \equiv \begin{array}{cc}
{\begin{array}{ccccccccccc} & & & m\text{th column} & & & & n\text{th column} & & & \end{array}} & \\
{\left(\begin{array}{cccrccccccc}
 1 & & & & & & & & & & \\
 & \ddots & & & & & & & & & \\
 & & 1 & & & & & & & & \\
 & & & \cos\alpha & & & & \sin\alpha & & & \\
 & & & & 1 & & & & & & \\
 & & & & & \ddots & & & & & \\
 & & & & & & 1 & & & & \\
 & & & -\sin\alpha & & & & \cos\alpha & & & \\
 & & & & & & & & 1 & & \\
 & & & & & & & & & \ddots & \\
 & & & & & & & & & & 1 
\end{array}\right)}
& {\begin{array}{c} \phantom{1} \\ \phantom{\ddots} \\ \phantom{1} \\ m\text{th row} \\ \phantom{1} \\ \phantom{\ddots} \\ \phantom{1} \\ n\text{th row} \\ \phantom{1} \\ \phantom{\ddots} \\ \phantom{1} \end{array}}
\end{array}
\end{equation}
where the empty entries are zeroes.
In five dimensions there are $\binom{5}{2} = 10$ such planes, hence we need 10 rotation angles to describe a general five-dimensional rotation.
Using the analog of the Euler angles, the corresponding rotation matrix can be parametrised as
\begin{multline}
R(\alpha_1,\ldots,\alpha_{10}) \equiv  R_{45}(\alpha_{10}) R_{34}(\alpha_8) R_{45}(\alpha_9) R_{23}(\alpha_5) R_{34}(\alpha_6) \cdot \\ R_{45}(\alpha_7) R_{12}(\alpha_1) R_{23}(\alpha_2) R_{34}(\alpha_3) R_{45}(\alpha_4). \label{eq:5DEuler}
\end{multline}
We omit the full expression of the rotation matrix, as it is quite large.
It can easily be generated in \emph{Mathematica}.
The ranges for the angles are $\alpha_i\in[0,\frac{\pi}{2}]$ for $i=1,5,8$; $\alpha_i\in[0,\pi]$ for $i=2,3,6$; $\alpha_i\in(-\pi,\pi]$ for $i=4,7,9$; and $\alpha_{10}\in[-\frac{\pi}{2},\frac{\pi}{2})$.\footnote{The usual ranges for the Euler angles in five dimensions are $\alpha_i\in[0,\pi]$ for $i=1,2,3,5,6,8$ and $\alpha_i\in(-\pi,\pi]$ for $i=4,7,9,10$. However, as in the three-dimensional case there is an ambiguity due to the phase freedom of the mass eigenstates, which allows for a redefinition of the angles by multiplying any pair of rows of the rotation matrix by $-1$ (see footnote~\ref{fn:Eulerambiguity} on page~\pageref{fn:Eulerambiguity}). Multiplying the first two rows by $-1$ corresponds to the redefinition $\alpha_i\rightarrow\pi-\alpha_i$ for $i=1,2,3,5,6,7$ and $\alpha_4\rightarrow\pi+\alpha_4$; the other angles are left unchanged. Multiplying the second and third rows by $-1$ corresponds to $\alpha_i\rightarrow\pi-\alpha_i$ for $i=5,6,8,9$ and $\alpha_7\rightarrow\pi+\alpha_7$, with the other angles unchanged. The third and fourth rows correspond to $\alpha_i\rightarrow\pi-\alpha_i$ for $i=8,10$ and $\alpha_9\rightarrow\pi+\alpha_9$, with all other angles unchanged. The last two rows only change $\alpha_{10}\rightarrow\pi+\alpha_{10}$. Thus, given a rotation matrix $R$, we have the freedom to choose $\alpha_1\in[0,\frac{\pi}{2}]$ by multiplying the first two rows by $-1$ if $R_{(1,1)}<0$. Then we can choose $\alpha_5\in[0,\frac{\pi}{2}]$ by multiplying the second and third rows by $-1$ if $R_{(2,1)}>0$. If $R_{(3,1)}<0$, we multiply the third and fourth rows by $-1$ to let $\alpha_8\in[0,\frac{\pi}{2}]$. Then if $R_{(4,1)}>0$, we multiply the last two rows by $-1$ to let $\alpha_{10}\in[-\frac{\pi}{2},\frac{\pi}{2}]$. Only after these operations, we extract the mixing angles from \cref{eq:5DEuler}.}
In terms of this parametrisation, we write for the Higgs mass eigenstates:
\begin{align}
\begin{pmatrix} h^0 \\ H_1^0 \\ H_2^0 \\ H_3^0 \\ H_4^0 \end{pmatrix}
=& R(\alpha_1,\ldots,\alpha_{10}) \begin{pmatrix} h_{1,11}^0 \\ h_{1,22}^0 \\ h_{2,11}^0 \\ h_{2,22}^0 \\ h_R^0 \end{pmatrix},\qquad \begin{pmatrix} G^0 \\ G^{\prime0} \\ A_1^0 \\ A_2^0 \\ A_3^0 \end{pmatrix}
= R(\beta_1,\ldots,\beta_{10}) \begin{pmatrix} a_{1,11}^0 \\ a_{1,22}^0 \\ a_{2,11}^0 \\ a_{2,22}^0 \\ a_R^0 \end{pmatrix}, \notag\\
\begin{pmatrix} G^\pm \\ G^{\prime\pm} \\ H_1^\pm \\ H_2^\pm \\ H_3^\pm \end{pmatrix}
=& R(\gamma_1,\ldots,\gamma_{10}) \begin{pmatrix} h_{1,21}^\pm \\ h_{1,12}^\pm \\ h_{2,21}^\pm \\ h_{2,12}^\pm \\ h_R^\pm \end{pmatrix}.
\end{align}
As in the simplified LET model, we have to define the mass eigenstates before these expressions makes sense.
Recall that in the simplified setup, the $CP$-even part of the scalar spectrum contains a $h^0_{1,11}$-like state $h^0$ with a mass of order $v$, a $h^0_{1,22}$-like state $H^0_1$ with a mass of order $M$, and a $h^0_R$-like state with a mass of order $M$.
Note that the scalar field $\Phi_2$ is in the same representation as $\Phi_1$.
Hence if we decoupled $\Phi_1$ from the complete LET model, we would expect a $h^0_{2,11}$-like state with a mass of order $v$, as well as a $h^0_{2,22}$-like state and a $h^0_R$-like state with masses of order $M$.
Likewise, we would expect a massive $CP$-odd state that is $a^0_{2,22}$-like, whereas the Goldstones $G^0$, $G^{\prime0}$ would be $a^0_{2,11}$- and $a^0_R$-like respectively.
The massive charged state would be $h^\pm_{2,21}$-like, and the Goldstones $G^\pm$, $G^{\prime\pm}$ would be $h^\pm_{2,12}$- and $h^\pm_R$-like respectively.

In the complete LET model, we get a mixture of these two scenarios, and we use the above considerations to define the mass eigenstates.
Again, we define $h^0$ as the most $h^0_{1,11}$-like state, and we define $H^0_1$ as the most $h^0_{2,11}$-like.
Both will turn out to have masses of order $v$, so they are naturally light.
We identify $h^0$ with the observed scalar at 126 GeV, since $h^0_{1,11}$ couples to fermions whereas $h^0_{2,11}$ does not.
In terms of mixing angles, the expression for the Standard-Model-like Higgs boson is
\begin{equation}
h^0 = c_1h_{1,11}^0 + s_1c_2h_{1,22}^0 + s_1s_2c_3h_{2,21}^0 + s_1s_2s_3c_4h_{2,22}^0 + s_1s_2s_3s_4h_R^0,
\end{equation}
where $c_i\equiv\cos\alpha_i$ and $s_i\equiv\sin\alpha_i$.
We define $H^0_2$, $H^0_3$, $H^0_4$ respectively as the most $h^0_{1,22}$-, $h^0_{2,22}$-, $h^0_R$-like states.
They will turn out to have masses of order $M$.

As for the $CP$-odd states, we define $A^0_1$, $A^0_2$, $A^0_3$ respectively as the most $a^0_{2,11}$-, $a^0_{1,22}$-, $a^0_{2,22}$-like states.
The Goldstones $G^0$, $G^{\prime0}$ turn out to be $a^0_{1,11}$- and $a^0_R$-like respectively.
Similarly, we define the charged states $H^\pm_1$, $H^\pm_2$, $H^\pm_3$ respectively as the most $h^\pm_{2,21}$-, $h^\pm_{1,12}$-, $h^\pm_{2,12}$-like states.
The Goldstones $G^\pm$, $G^{\prime\pm}$ turn out to be $h^\pm_{1,21}$- and $h^\pm_R$-like respectively.

The scalar masses can be found by solving for the eigenvalues of the scalar mass matrix as described in the previous section.
This amounts to finding the eigenvalues of one $5\times5$ matrix and two $3\times3$ matrices.
This can be done in an expansion in $\frac{v}{M}$, as described on \pageref{p:solveEigenvalues}.
We find the following expressions:
\begin{align}
m_{h^0,H^0_1}^2 =& \frac{v^2}{2}\Bigg( \Lambda_1\cos^2\alpha + \Lambda_2\sin^2\alpha \notag\\
&\hspace{3cm} \pm \sqrt{\left( \Lambda_1\cos^2\alpha - \Lambda_2\sin^2\alpha \right)^2 + \Lambda_3^2\sin^22\alpha} + \ord{\frac{v^2}{M^2}} \Bigg) \notag\\
m_{H^0_2}^2 =& M^2\left( \frac{\lambda_5}{2}\sec2\beta_1 + \ord{\frac{v^2}{M^2}} \right), \notag\\
m_{H^0_3}^2 =& M^2\left( \frac{\widetilde\lambda_4}{2}\sec2\beta_2 + \ord{\frac{v^2}{M^2}} \right), \notag\\
m_{H^0_4}^2 =& M^2\left( \lambda_3 + \ord{\frac{v^2}{M^2}} \right), \notag\\
m_{A^0_1}^2 =& -v^2\Bigg( (\widetilde\lambda_7+\widetilde\lambda_{10})(\cos^2\beta_1\cos^2\beta_2 + \sin^2\beta_1\sin^2\beta_2) \notag\\
&\hspace{6cm} + \frac{\widetilde\lambda_6}{2}\sin2\beta_1\sin2\beta_2 + \ord{\frac{v^2}{M^2}} \Bigg), \notag\\
m_{A^0_2}^2 =& M^2\left( \frac{\lambda_5}{2}\sec2\beta_1 + \ord{\frac{v^2}{M^2}} \right), \notag\\
m_{A^0_3}^2 =& M^2\left( \frac{\widetilde\lambda_4}{2}\sec2\beta_2 + \ord{\frac{v^2}{M^2}} \right), \notag\\
m_{H^\pm_1}^2 =& -\frac{v^2}{2}\left( (\widetilde\lambda_6+\widetilde\lambda_7+\widetilde\lambda_{10})\cos^2(\beta_1-\beta_2) + \widetilde\lambda_9\cos2\beta_1\cos2\beta_2 + \ord{\frac{v^2}{M^2}} \right), \notag\\
m_{H^\pm_2}^2 =& M^2\left( \frac{\lambda_5}{2}\sec2\beta_1 + \ord{\frac{v^2}{M^2}} \right), \notag\\
m_{H^\pm_3}^2 =& M^2\left( \frac{\widetilde\lambda_4}{2}\sec2\beta_2 + \ord{\frac{v^2}{M^2}} \right). \label{eq:completeLETscalarMasses}
\end{align}
Here we have defined the quantities
\begin{align}
\Lambda_1 \equiv& \lambda_1 + \lambda_2\cos^2\beta_1 - \frac{(\lambda_4+\lambda_5\sin^2\beta_1)^2}{\lambda_3}, \notag\\
\Lambda_2 \equiv& \widetilde\lambda_1 + \widetilde\lambda_2\cos^2\beta_2 - \frac{(\widetilde\lambda_3+\widetilde\lambda_4\sin^2\beta_2)^2}{\lambda_3}, \notag\\
\Lambda_3 \equiv& -\frac{(\lambda_4+\lambda_5\sin^2\beta_1)(\widetilde\lambda_3+\widetilde\lambda_4\sin^2\beta_2)}{\lambda_3} + \widetilde\lambda_5 + (\widetilde\lambda_6+\widetilde\lambda_7)\cos^2(\beta_1-\beta_2) \notag\\
&+ (\widetilde\lambda_8+\widetilde\lambda_9+\widetilde\lambda_{10})(\cos^2\beta_1\cos^2\beta_2 + \sin^2\beta_1\sin^2\beta_2).
\end{align}

We can find vacuum-stability conditions for the complete LET model completely analogously to~\cref{a:vacuumstability}.
Because of the large number of fields and parameters, we generate these conditions in \emph{Mathematica}.
To this end, we write the scalar fields in terms of $SU(2)_L$ doublets $\phi_1$, $\phi_2$, $\chi_1$, $\chi_2$ and $SU(2)_L$ singlets $S_+$, $S_0$:
\begin{equation}
\Phi_1 = \begin{pmatrix} i\sigma_2\phi_1, & \phi_2^* \end{pmatrix},\qquad
\Phi_2 = \begin{pmatrix} i\sigma_2\chi_1, & \chi_2^* \end{pmatrix},\qquad
\Phi_R = \begin{pmatrix} S_+, & S_0 \end{pmatrix}.
\end{equation}
We reparametrise these fields in terms of unit spinors $\hat{\phi}_{1,2}$, $\hat{\chi}_{1,2}$ and the norms of the fields:
\begin{equation}
\phi_{1,2} = \|\phi_{1,2}\| \hat{\phi}_{1,2},\qquad 
\chi_{1,2} = \|\chi_{1,2}\| \hat{\chi}_{1,2},\qquad 
S_{+,0} = r_{+,0}e^{i\theta_{+,0}}.
\end{equation}
Here all norms are positive and $\theta_{+,0}\in[0,2\pi)$.
We reparametrise the norms of the $SU(2)_L$ doublets and singlets as follows:
\begin{IEEEeqnarray}{rClrCl}
\|\phi_1\| &=& R\cos\gamma_1,\qquad	&	\|\chi_2\| &=& R\sin\gamma_1\sin\gamma_2\sin\gamma_3\cos\gamma_4,	\nonumber\\
\|\phi_2\| &=& R\sin\gamma_1\cos\gamma_2,\qquad	&	r_+ &=& R\sin\gamma_1\sin\gamma_2\sin\gamma_3\sin\gamma_4\cos\gamma_5,	\nonumber\\
\|\chi_1\| &=& R\sin\gamma_1\sin\gamma_2\cos\gamma_3,\qquad	&	r_0 &=& R\sin\gamma_1\sin\gamma_2\sin\gamma_3\sin\gamma_4\sin\gamma_5.
\end{IEEEeqnarray}
Here $R \geq 0$ and $\gamma_{1,2,3,4,5} \in[0,\pi/2]$.
The products of unit spinors can be parametrised as follows:
\begin{equation}
\frac{(\Phi_1^\dagger\Phi_2)_{ij}}{\|\phi_i\| \|\chi_j\|} = \phi_{ij}e^{i\theta_{ij}},\qquad
\frac{\phi_1^{cT}\phi_2^*}{\|\phi_i\| \|\phi_j\|} = \rho_\phi e^{i\theta_\phi},\qquad
\frac{\chi_1^{cT}\chi_2^*}{\|\chi_i\| \|\chi_j\|} = \rho_\chi e^{i\theta_\chi},
\end{equation}
where $\rho_{ij},\rho_\phi,\rho_\chi \in [0,1]$ and $\theta_{ij},\theta_\phi,\theta_\chi \in [0,2\pi)$.
Inserting these reparametrisations into the scalar potential in \cref{eq:completeLETpotential}, we write the scalar potential as 
\begin{equation}
V(\Phi_1,\Phi_2,\Phi_R) = V_4R^4 + V_2R^2,
\end{equation}
in analogy to~\cref{eq:vacStabReparametrisation}.
We omit the full expression for $V_4$, as it is quite large.
The condition of vacuum stability requires that $V_4 > 0$ for all possible values of $\gamma_{1,2,3,4,5}$, $\rho_{ij}$, $\rho_{\phi,\chi}$, $\theta_{ij}$, $\theta_{+,0,\phi,\chi}$.
We generate a list necessary (but not sufficient) conditions in \emph{Mathematica} by requiring that the condition $V_4 > 0$ is satisfied for all combinations of $\theta_{ij} = \theta_{+,0,\phi,\chi} = 0$, $\gamma_{1,2,3,4,5} \in \{0,\pi/4,\pi/2\}$, and $\rho_{ij},\rho_{\phi,\chi} \in \{0,1\}$.
This results in 367 unique vacuum-stability conditions.
For all benchmark points in \cref{s:completeLET}, we check that the parameters satisfy all 367 vacuum-stability conditions.

\clearpage
\section{Systematic derivation of scalar invariants}\label{a:scalarinvariants}
The scalar masses, mixings, and interactions of a theory are determined by the scalar potential.
This potential contains all possible gauge-invariant combinations of the scalar fields up to mass dimension four.
The most general scalar potential for the simplified LET model was given in \cref{eq:scalarpotential}, and the potential for the full LET model was given in \cref{eq:scalarpotential2}.
In this appendix we derive all possible invariants for the LET model in a systematical way, in order to show that the given scalar potentials are indeed the most general for our setup.
First we will find the invariants for the simplified LET model, constructed from $\Phi_1 \sim (\rep{1},\brep{2},\rep{2},0)$ and $\Phi_R \sim (\rep{1},\rep{1},\rep{2},1)$ only.
Then we will find the additional invariants that are allowed when the field $\Phi_2 \sim (\rep{1},\brep{2},\rep{2},0)$ is added for the complete LET model.

Field components will be labelled with an upper index for the $SU(2)_L$ component and a lower index for the $SU(2)_R$ component.
In the rest of this work, the invariants have been cast into an index-free matrix notation, in which $SU(2)_L$ indices run vertically and $SU(2)_R$ indices run horizontally.
In this appendix, we will show how the expressions in this matrix notation arise from the expressions with all indices restored.
Since this index-free notation obscures the difference between $SU(2)_L$ and $SU(2)_R$ indices, care has to be taken whenever transposes and hermitian conjugates are used.
To avoid confusion, we will indicate the use of this matrix notation by putting the relevant expressions in boldface.

In order to cast all invariants in matrix notation, we employ the following identities.
The determinant of an $n\times n$ matrix $\bm{A}$ can be written as
\begin{equation}
\det\bm{A} = \epsilon_{i_1\ldots i_n} A^{i_1}_1\ldots A^{i_n}_n = \frac{1}{n!}\epsilon_{i_1\ldots i_n}\epsilon_{j_1\ldots j_n} A^{i_1}_{j_1}\ldots A^{i_n}_{j_n}. \label{eq:deta}
\end{equation}
Here $\epsilon$ is the completely antisymmetric symbol that satisfies $\epsilon_{1\ldots n} = +1$.
For $n=2$, the $\epsilon$-symbol can be written in matrix notation as
\begin{equation}
\bm\epsilon = \begin{pmatrix} 0 & 1 \\ -1 & 0 \end{pmatrix}.
\end{equation}
Then \cref{eq:deta} becomes
\begin{equation}
\det\bm{A} = \tfrac12\tr{\bm\epsilon^T\bm{A\epsilon A}^T}
\end{equation}
We also note that $\bm\epsilon=i\bm\sigma_2$, so that the charge conjugate of $\Phi_1$ can be written in terms of $\epsilon$:
\begin{align}
(\Phi_1^c)^i_k \equiv& (i\sigma_2)^{ij}(i\sigma_2)_{kl}(\Phi_1^*)^j_l = \epsilon^{ij}(\Phi_1^*)^j_l(\epsilon^T)_{lk}, \notag\\
\Rightarrow \mathbf{\Phi_1^c} =& \bf{\epsilon\Phi_1^*\epsilon}^T.
\end{align}

\subsection{The simplified LET model}
The scalar fields of the LET model are only in singlet and (anti)doublet representations of $SU(2)_{L,R}$.
We need to combine them in such a way that the whole becomes a singlet with respect to the gauge group.
A doublet and an antidoublet can be combined symmetrically into a singlet.
Also, two doublets or two antidoublets can be combined antisymmetrically into a singlet, using the completely antisymmetric symbols $\epsilon^{ij}$ (for $SU(2)_L$) and $\epsilon_{ij}$ (for $SU(2)_R$).
This implies that each invariant needs to contain an even number of factors $\Phi_1$ and/or $\Phi_1^\dagger$ in order to be an $SU(2)_L$ singlet.
Furthermore, each invariant needs to contain an equal number of factors $\Phi_R$ and $\Phi_R^\dagger$ in order to be a $U(1)_{B-L}$ singlet.

Now let us derive the scalar invariants with couplings of positive mass dimension first.
Using the above observations and the fact that all invariants should be real, we find the following three invariants:
\begin{align}
I_1 \equiv& (\Phi_1^*)^i_j(\Phi_1)^i_j = \tr{\mathbf{\Phi_1^\dagger\Phi_1}}, \notag\\
I_2 \equiv& \tfrac12\epsilon^{ij}\epsilon_{kl}\left((\Phi_1)^i_k(\Phi_1)^j_l + (\Phi_1^*)^i_k(\Phi_1^*)^j_l \right) = \det\mathbf{\Phi_1} + \det\mathbf{\Phi_1^\dagger}, \notag\\
I_3 \equiv& (\Phi_R^*)_i(\Phi_R)_i = \mathbf{\Phi_R\cdot\Phi_R^\dagger}.
\end{align}
Note that for $I_3$, we put $\mathbf{\Phi_R^\dagger}$ on the right since $\mathbf{\Phi_R}$ is a row vector in our notation.
Now we turn to the quartic terms.
We divide them into invariants that do not mix $\Phi_1$ and $\Phi_R$ and those that do.

\paragraph{Unmixed invariants}
First let us write down the terms involving only $\Phi_1$.
We need to combine $i\in\{0,\ldots,4\}$ copies of $\Phi_1$ and $4-i$ copies of $\Phi_1^*$, and we need the antisymmetric symbol whenever we combine the indices of two factors $\Phi_1$ or two factors $\Phi_1^*$.
This gives the following invariants (note that we are not taking real combinations yet).
\begin{itemize}
\item For $i=0$ we have two distinct options of combining the indices, although they lead to the same invariant:
\begin{align}
\tfrac14\epsilon^{ij}\epsilon^{kl}\epsilon_{pq}\epsilon_{rs} (\Phi_1)^i_p(\Phi_1)^j_q(\Phi_1)^k_r(\Phi_1)^l_s =& \tfrac14\tr{\bm{\epsilon}^T\mathbf{\Phi_1}\bm{\epsilon}\mathbf{\Phi_1^T}}^2 \notag\\
=& (\det\mathbf{\Phi_1})^2, \notag\\
\tfrac12\epsilon^{ij}\epsilon^{kl}\epsilon_{pq}\epsilon_{rs} (\Phi_1)^i_p(\Phi_1)^j_r(\Phi_1)^k_q(\Phi_1)^l_s =& \tfrac12\tr{\bm{\epsilon}^T\mathbf{\Phi_1}\bm{\epsilon}\mathbf{\Phi_1^T}\bm\epsilon\mathbf{\Phi_1}\bm\epsilon^T\mathbf{\Phi_1^T}} \notag\\
=& \det(\mathbf{\Phi_1}\bm\epsilon\mathbf{\Phi_1^T}) \notag\\
=& (\det\mathbf{\Phi_1})^2.
\end{align}
\item For $i=1$ we also have two options, leading to two different invariants:
\begin{align}
\tfrac12\epsilon^{jk}\epsilon_{mn} (\Phi_1^*)^i_l(\Phi_1)^i_l(\Phi_1)^j_m(\Phi_1)^k_n =& \tr{\mathbf{\Phi_1^\dagger \Phi_1}}\cdot\tfrac12\tr{\bm\epsilon^T\mathbf{\Phi_1}\bm\epsilon\mathbf{\Phi_1^T}} \notag\\
=& \tr{\mathbf{\Phi_1^\dagger \Phi_1}}\det\mathbf{\Phi_1}, \notag\\
\tfrac12\epsilon^{jk}\epsilon_{mn} (\Phi_1^*)^i_l(\Phi_1)^i_m(\Phi_1)^j_l(\Phi_1)^k_n =& \tfrac12\tr{\bm\epsilon^T\mathbf{\Phi_1\Phi_1^\dagger \Phi_1}\bm\epsilon\mathbf{\Phi_1^T}} \notag\\
=& \tr{\mathbf{\Phi_1^\dagger \Phi_1(\Phi_1^c)^\dagger \Phi_1}}.
\end{align}
\item For $i=2$ we get several options by combining the left-handed and right-handed indices either symmetrically or antisymmetrically:
\begin{align}
(\Phi_1^*)^i_k(\Phi_1^*)^j_l(\Phi_1)^i_k(\Phi_1)^j_l =& \tr{\mathbf{\Phi_1^\dagger \Phi_1}}^2, \notag\\
(\Phi_1^*)^i_k(\Phi_1^*)^j_l(\Phi_1)^i_l(\Phi_1)^j_k =& \tr{\mathbf{\Phi_1^\dagger \Phi_1\Phi_1^\dagger \Phi_1}}, \notag\\
\tfrac12\epsilon^{ij}\epsilon^{kl} (\Phi_1^*)^i_m(\Phi_1^*)^j_n(\Phi_1)^k_m(\Phi_1)^l_n =& \tfrac12\tr{\bm\epsilon^T\mathbf{\Phi_1\Phi_1^\dagger}\bm\epsilon\mathbf{\Phi_1^*\Phi_1^T}} \notag\\
=& \det(\mathbf{\Phi_1^\dagger \Phi_1}), \notag\\
\tfrac12\epsilon_{pq}\epsilon_{rs} (\Phi_1^*)^i_p(\Phi_1^*)^j_q(\Phi_1)^i_r(\Phi_1)^j_s =& \tfrac12\tr{\bm\epsilon^T\mathbf{\Phi_1^\dagger \Phi_1}\bm\epsilon\mathbf{\Phi_1^T\Phi_1^*}} \notag\\
=& \det(\mathbf{\Phi_1^\dagger \Phi_1}), \notag\\
\tfrac14\epsilon_{pq}\epsilon_{rs}\epsilon^{ij}\epsilon^{kl} (\Phi_1^*)^i_p(\Phi_1^*)^j_q(\Phi_1)^k_r(\Phi_1)^l_s =& \tfrac14\tr{\bm\epsilon^T\mathbf{\Phi_1^\dagger}\bm\epsilon\mathbf{\Phi_1^*}}\tr{\bm\epsilon^T\mathbf{\Phi_1}\bm\epsilon\mathbf{\Phi_1^T}} \notag\\
=& \det(\mathbf{\Phi_1^\dagger \Phi_1}).
\end{align}
\item For $i=3$ we have two possibilities, which are the Hermitian conjugates of the combinations for $i=1$:
\begin{align}
\tfrac12\epsilon^{jk}\epsilon_{mn} (\Phi_1^*)^j_m(\Phi_1^*)^k_n(\Phi_1^*)^i_l(\Phi_1)^i_l =& \tr{\mathbf{\Phi_1^\dagger \Phi_1}}\cdot\tfrac12\tr{\bm\epsilon^T\mathbf{\Phi_1^*}\bm\epsilon\mathbf{\Phi_1^\dagger}} \notag\\
=& \tr{\mathbf{\Phi_1^\dagger \Phi_1}}\det\mathbf{\Phi_1^\dagger}, \notag\\
\tfrac12\epsilon^{jk}\epsilon_{mn} (\Phi_1^*)^j_l(\Phi_1^*)^k_n(\Phi_1^*)^i_l(\Phi_1)^i_m =& \tfrac12\tr{\bm\epsilon^T\mathbf{\Phi_1^*\Phi_1^T\Phi_1^*}\bm\epsilon\mathbf{\Phi_1^\dagger}} \notag\\
=& \tr{\mathbf{\Phi_1^\dagger \Phi_1\Phi_1^\dagger \Phi_1^c}},
\end{align}
\item For $i=4$ we get the Hermitian conjugates of the invariants for $i=0$:
\begin{align}
\tfrac14\epsilon^{ij}\epsilon^{kl}\epsilon_{pq}\epsilon_{rs} (\Phi_1^*)^i_p(\Phi_1^*)^j_q(\Phi_1^*)^k_r(\Phi_1^*)^l_s =& \tfrac14\tr{\bm{\epsilon}^T\mathbf{\Phi_1^*}\bm{\epsilon}\mathbf{\Phi_1^\dagger}}^2 \notag\\
=& (\det\mathbf{\Phi_1^\dagger})^2, \notag\\
\tfrac12\epsilon^{ij}\epsilon^{kl}\epsilon_{pq}\epsilon_{rs} (\Phi_1^*)^i_p(\Phi_1^*)^j_r(\Phi_1^*)^k_q(\Phi_1^*)^l_s =& \tfrac12\tr{\bm{\epsilon}^T\mathbf{\Phi_1^*}\bm{\epsilon}\mathbf{\Phi_1^\dagger}\bm\epsilon\mathbf{\Phi_1^*}\bm\epsilon^T\mathbf{\Phi_1^\dagger}} \notag\\
=& \det(\mathbf{\Phi_1^*}\bm\epsilon\mathbf{\Phi_1^\dagger}) \notag\\
=& (\det\mathbf{\Phi_1^\dagger})^2.
\end{align}
\end{itemize}
The above list exhausts the possible options for gauge-invariant quartic terms built from $\Phi_1$ only.
However, the invariants appearing in the scalar potential should be real as well, which yields only six combinations:
\begin{align}
J_1 \equiv& \tr{\mathbf{\Phi_1^\dagger \Phi_1}}^2 = I_1^2, \notag\\
J_2 \equiv& \tr{\mathbf{\Phi_1^\dagger \Phi_1\Phi_1^\dagger \Phi_1}}, \notag\\
J_2^c \equiv& \tr{\mathbf{\Phi_1^\dagger \Phi_1(\Phi_1^c)^\dagger \Phi_1}} \plushc, \notag\\
J_6 \equiv& (\det\mathbf{\Phi_1})^2 + (\det\mathbf{\Phi_1^\dagger})^2, \notag\\
J_7 \equiv& \det(\mathbf{\Phi_1^\dagger \Phi_1}), \notag\\
J_8 \equiv& \tr{\mathbf{\Phi_1^\dagger \Phi_1}}\left( \det\mathbf{\Phi_1}+\det\mathbf{\Phi_1^\dagger} \right) = I_1I_2.
\end{align}
For $\Phi_R$, there is only one unmixed quartic invariant:
\begin{equation}
J_3 \equiv (\Phi_R^*)_i(\Phi_R)_i(\Phi_R^*)_j(\Phi_R)_j = (\mathbf{\Phi_R\cdot\Phi_R^\dagger})^2 = I_3^2.
\end{equation}

\paragraph{Mixed invariants}
Now we find all quartic invariants that contain both $\Phi_1$ and $\Phi_R$.
Recall that each of them must contain one copy of $\Phi_R$ and one copy of $\Phi_R^*$.
We can combine them with either two factors of $\Phi_1$ or $\Phi_1^*$ (with antisymmetrically combined $SU(2)_L$ indices) or with one factor of $\Phi_1$ and $\Phi_1^*$ each (with symmetrically combined $SU(2)_L$ indices).
The possible gauge-invariant (not necessarily real) combinations are given by
\begin{align}
\frac12\epsilon^{ij}\epsilon_{kl}(\Phi_R^*)_m(\Phi_R)_m(\Phi_1)^i_k(\Phi_1)^j_l =& \mathbf{\Phi_R\cdot \Phi_R^\dagger}\det\mathbf{\Phi_1}, \notag\\
\epsilon^{ij}\epsilon_{kl}(\Phi_R^*)_m(\Phi_R)_k(\Phi_1)^i_m(\Phi_1)^j_l =& \mathbf{\Phi_R}\bm\epsilon\mathbf{\Phi_1^T}\bm\epsilon^T\mathbf{\Phi_1\Phi_R^\dagger} \notag\\
=& \mathbf{\Phi_R\Phi_1^{c\dagger}\Phi_1\Phi_R^\dagger}, \notag\\
(\Phi_R^*)_j(\Phi_R)_j(\Phi_1^*)^i_k(\Phi_1)^i_k =& \mathbf{\Phi_R\cdot \Phi_R^\dagger}\tr{\mathbf{\Phi_1^\dagger \Phi_1}}, \notag\\
(\Phi_R^*)_j(\Phi_R)_k(\Phi_1^*)^i_k(\Phi_1)^i_j =& \mathbf{\Phi_R\Phi_1^\dagger \Phi_1\Phi_R^\dagger}, \notag\\
\frac12\epsilon^{ij}\epsilon_{kl}(\Phi_R^*)_m(\Phi_R)_m(\Phi_1^*)^i_k(\Phi_1^*)^j_l =& \mathbf{\Phi_R\cdot \Phi_R^\dagger}\det\mathbf{\Phi_1^\dagger}, \notag\\
\epsilon^{ij}\epsilon_{kl}(\Phi_R^*)_m(\Phi_R)_k(\Phi_1^*)^i_m(\Phi_1^*)^j_l =& \mathbf{\Phi_R}\bm\epsilon\mathbf{\Phi_R^\dagger}\bm\epsilon^T\mathbf{\Phi_1^*\Phi_R^\dagger} \notag\\
=& \mathbf{\Phi_R\Phi_1^\dagger \Phi_1^c\Phi_R^\dagger}.
\end{align}
Taking only real combinations, we end up with four invariants:
\begin{align}
J_4 \equiv& \mathbf{\Phi_R\cdot \Phi_R^\dagger}\tr{\mathbf{\Phi_1^\dagger \Phi_1}} = I_1I_3, \notag\\
J_5 \equiv& \mathbf{\Phi_R\Phi_1^\dagger \Phi_1\Phi_R^\dagger}, \notag\\
J_5^c \equiv& \mathbf{\Phi_R\Phi_1^\dagger \Phi_1^c\Phi_R^\dagger} \plushc, \notag\\
J_9 \equiv& \mathbf{\Phi_R\cdot \Phi_R^\dagger}\left( \det\mathbf{\Phi_1} + \det\mathbf{\Phi_1^\dagger} \right) = I_2I_3.
\end{align}

This concludes the list of all possible invariants built from $\Phi_1$ and $\Phi_R$.
We have three quadratic and eleven quartic invariants.
However, so far we have not taken the trinification origin of the LET model into account.
The fact that our setup is derived from the trinification model puts two restrictions on the possible invariants.

Firstly, charge conjugates are not allowed to appear, which eliminates the invariants $J_2^c$ and $J_5^c$.\footnote{We could write down other invariants containing charge conjugates, but they would be the same as one of the invariants already listed. For example, $\tr{\mathbf{\Phi_1^\dagger \Phi_1^c}} \plushc$ is just another way to write $\det\mathbf{\Phi_1} \plushc$.}
The reason is that the (anti)doublets of the LET model come from (anti)triplets in the trinification model.
Since the $\rep{3}$ and $\brep{3}$ representations of $SU(3)$ are inequivalent, charge conjugates of scalars in the LET model would have no equivalent in the trinification model.
Hence there are no invariants in the trinification model from which the invariants $J_2^c$ and $J_5^c$ could originate.

Secondly, the trinification model is considered to be a renormalisable theory.
This condition eliminates all invariants from the LET model that would have to originate from nonrenormalisable operators.
Note that the $2\times2$ matrix $\mathbf{\Phi_1}$ originates from a $3\times3$ matrix $\mathbf{H_1}$ in the trinification model, so any invariant that contains $\det\mathbf{\Phi_1}$ comes from an invariant of a higher mass dimension.
This implies that the invariants $J_6$, $J_7$, $J_8$, and $J_9$ would have to come from operators of dimension six, six, five, and five respectively in the trinification model.
Therefore we eliminate these invariants from the scalar potential, and our setup contains only five quartic and three quadratic terms.
We summarise them here:
\begin{IEEEeqnarray}{rClrCl}
I_1 &=& \tr{\mathbf{\Phi_1^\dagger\Phi_1}}, & J_1 &=& I_1^2, \nonumber\\
I_2 &=& \det\mathbf{\Phi_1} + \det\mathbf{\Phi_1^\dagger},\qquad & J_2 &=& \tr{\mathbf{\Phi_1^\dagger \Phi_1\Phi_1^\dagger \Phi_1}}, \nonumber\\
I_3 &=& \mathbf{\Phi_R\cdot\Phi_R^\dagger}, & J_3 &=& I_3^2, \nonumber\\
&&& J_4 &=& I_1I_3, \nonumber\\
&&& J_5 &=& \mathbf{\Phi_R\Phi_1^\dagger \Phi_1\Phi_R^\dagger}. \label{eq:scalarinvariants1}
\end{IEEEeqnarray}
Thus the scalar potential in \cref{eq:scalarpotential} is indeed the most general potential for the simplified LET model.

\subsection{The full LET model}\label{a:fullLETinvariants}
Now let us derive the additional invariants that can appear in the scalar potential when we add $\Phi_2$ to the scalar sector.
Note that $\Phi_1$ and $\Phi_2$ are in the same representation of the gauge group, so all we have to do is replace one or more factors of $\Phi_1$ by $\Phi_2$ in the invariants in eq.~\eqref{eq:scalarinvariants1} in every possible way, while checking that the resulting invariants are real.
This procedure results in two quadratic and ten quartic invariants that are even under the $Z_2$-symmetry of the scalar sector:
\begin{IEEEeqnarray}{rClrCl}
\widetilde{I}_1 &=& \tr{\mathbf{\Phi_2^\dagger\Phi_2}},\qquad & \widetilde{J}_1 &=& \tr{\mathbf{\Phi_2^\dagger\Phi_2}}^2, \nonumber\\ 
\widetilde{I}_2 &=& \det\mathbf{\Phi_2} + \det\mathbf{\Phi_2^\dagger},\qquad & \widetilde{J}_2 &=& \tr{\mathbf{\Phi_2^\dagger \Phi_2\Phi_2^\dagger \Phi_2}}, \nonumber\\
&&& \widetilde{J}_3 &=& \tr{\mathbf{\Phi_2^\dagger\Phi_2}}\mathbf{\Phi_R\cdot\Phi_R^\dagger}, \nonumber\\
&&& \widetilde{J}_4 &=& \mathbf{\Phi_R\Phi_2^\dagger\Phi_2\Phi_R^\dagger}, \nonumber\\
&&& \widetilde{J}_5 &=& \tr{\mathbf{\Phi_1^\dagger\Phi_1}}\tr{\mathbf{\Phi_2^\dagger\Phi_2}}, \nonumber\\
&&& \widetilde{J}_6 &=& \left|\tr{\mathbf{\Phi_1^\dagger\Phi_2}}\right|^2, \nonumber\\
&&& \widetilde{J}_7 &=& \tr{\mathbf{\Phi_1^\dagger\Phi_2}}\tr{\mathbf{\Phi_1^\dagger\Phi_2}} \plushc, \nonumber\\
&&& \widetilde{J}_8 &=& \tr{\mathbf{\Phi_1^\dagger \Phi_1\Phi_2^\dagger \Phi_2}}, \nonumber\\
&&& \widetilde{J}_9 &=& \tr{\mathbf{\Phi_1^\dagger \Phi_2\Phi_2^\dagger \Phi_1}}, \nonumber\\
&&& \widetilde{J}_{10} &=& \tr{\mathbf{\Phi_1^\dagger \Phi_2\Phi_1^\dagger \Phi_2}} \plushc
\end{IEEEeqnarray}
In addition, there are one quadratic and six quartic invariants that are odd under the $Z_2$-symmetry.
We do not include them in the scalar potential, but we mention them for the sake of completeness:
{\interdisplaylinepenalty=10000
\begin{IEEEeqnarray}{rClrCl}
\widetilde{I}_{12} &=& \tr{\mathbf{\Phi_1^\dagger\Phi_2}} \plushc,\qquad & \widetilde{J}_{11} &=& \tr{\mathbf{\Phi_1^\dagger\Phi_1}} (\tr{\mathbf{\Phi_1^\dagger\Phi_2}} \plushc), \nonumber\\
&&& \widetilde{J}_{12} &=& \tr{\mathbf{\Phi_2^\dagger\Phi_2}} (\tr{\mathbf{\Phi_1^\dagger\Phi_2}} \plushc), \nonumber\\
&&& \widetilde{J}_{13} &=& \tr{\mathbf{\Phi_1^\dagger \Phi_1\Phi_1^\dagger \Phi_2}} \plushc, \nonumber\\
&&& \widetilde{J}_{14} &=& \tr{\mathbf{\Phi_1^\dagger \Phi_2\Phi_2^\dagger \Phi_2}} \plushc, \nonumber\\
&&& \widetilde{J}_{15} &=& (\tr{\mathbf{\Phi_1^\dagger\Phi_2}} \plushc) \mathbf{\Phi_R\cdot\Phi_R^\dagger}, \nonumber\\
&&& \widetilde{J}_{16} &=& \mathbf{\Phi_R}(\mathbf{\Phi_1^\dagger\Phi_2} \plushc)\mathbf{\Phi_R^\dagger}.
\end{IEEEeqnarray}
}

\clearpage
\section{Yukawa sector}
Any new-physics model should reproduce the properties of all particles that have been observed.
This includes a correct description of the Standard-Model fermion masses and mixings, which we listed in \cref{t:fermionmasses}.
The gauge-boson sector is fixed by the choice of gauge group.
The scalar sector is straightforward to work out after the fields, their gauge-group representations, and any additional symmetries have been chosen.
However, the fermion sector allows for additional freedom.
This includes choosing which scalar fields couple to which fermions, as well as the particular form of these couplings.
In this appendix we propose a form for the Yukawa interactions of the LET model based on the trinification model.
First we describe the Yukawa sector of the trinification model, based on refs.~\cite{Stech:2003sb,Stech:2014tla}.
Then we work out the LET equivalent of the Yukawa Lagrangian and the fermion mass eigenstates that result from our choice, and give the Yukawa interactions in the mass eigenstate basis for the simplified LET model.
We also comment on the description of fermion mixing in charged-current interactions.

\subsection{Yukawa interactions in the trinification model}\label{a:Yukawatrini}
In the trinification model, the scalar sector contains two bitriplets $H_1,H_2 \sim (\rep{1},\brep{3},\rep{3})$.
In order to avoid tree-level FCNC interactions, a $Z_2$ symmetry is postulated under which $H_1$ ($H_2)$ is even (odd).
This symmetry permits only Yukawa couplings to $H_1$.
It couples to the fermions with a symmetric generation matrix $G_{\alpha\beta}$.
This matrix can be described as the real component of the vev of a flavon field, but in the trinification model its entries are considered as free parameters.
The Lagrangian term is given by
\begin{align}
\lag_Y \supset& -g_tG_{\alpha\beta}\left( \psi^{\alpha T}H_1\psi^\beta \right) \plushc \notag\\
=& -g_tG_{\alpha\beta}\left( Q_R^\alpha H_1^T Q_L^\beta + \frac12\epsilon^{ijk}\epsilon_{lmn}L^i_lL^j_m(H_1)^k_n \right) \plushc \label{eq:YukawaGtrini}
\end{align}
Here $g_t$ is a dimensionless coupling, and $\psi$ is a fermion field in the $\rep{27}$ representation of $E_6$, containing all quarks, antiquarks, and leptons.
In the second line, we show how this Yukawa interaction decomposes in terms of the separate fermion types of the trinification model.

The matrix $G_{\alpha\beta}$ can be diagonalised by a biunitary transformation: $G=UYW^\dagger$ where $U$, $W$ are unitary matrices and $Y$ is a diagonal matrix.
Then $U$, $W$ can be absorbed into a redefinition of the fermion fields.
Thus we can take $G$ to be diagonal without loss of generality.
Its diagonal elements can be fit to the up-quark masses.
However, this implies that without additional terms, \cref{eq:YukawaGtrini} yields the same mass hierarchies for the down quarks, charged leptons, and neutrinos as for the up quarks \cite{Stech:2014tla}.
Also, it would mean that the neutrinos (charged leptons) have the same masses as the up quarks (down quarks), and that the CKM matrix is a unit matrix.
All of this is in clear contradiction with the experimental values listed in \cref{t:fermionmasses}.

In order to reproduce these fermion properties correctly, additional scalar fields $H_{Aq} \sim (\rep{1},\brep{3},\rep{3})$, $H_{Al} \sim (\rep{1},\brep{3},\brep{6})$ are introduced.
They originate from a field $H_A$ in the antisymmetric $\rep{351_A}$ representation of $E_6$.
They couple to the fermions with an antisymmetric Hermitian matrix $A_{\alpha\beta}$, which is considered to be the imaginary component of the vev of the aforementioned flavon field.
Including these fields, the Yukawa Lagrangian becomes
\begin{align}
\lag_Y \supset& -g_tG_{\alpha\beta}\left( \psi^{\alpha T}H_1\psi^\beta \right) + A_{\alpha\beta}\left( \psi^{\alpha T} H_A \psi^\beta \right) \plushc  \notag\\
=& -g_tG_{\alpha\beta}\left( Q_R^\alpha H_1^T Q_L^\beta + \frac12\epsilon^{ijk}\epsilon_{lmn}L^i_lL^j_m(H_1)^k_n \right) \notag\\
&- A_{\alpha\beta}\left( Q_R^\alpha H_{Aq}^T Q_L^\beta + \epsilon^{ijk}L^i_lL^j_m(H_{Al})^k_{\{lm\}} \right) \plushc 
\end{align}
Again, the first line contains the Yukawa Lagrangian from the $E_6$ model whereas the second line is written in terms of the separate fermion fields of the trinification model.
This Lagrangian is sufficient to reproduce the down-quark masses, the CKM-matrix, the angles of the unitarity triangle, and the charged lepton masses and mixings.
However, the neutrino masses are still Dirac neutrinos with masses comparable to the quark masses.
Therefore an effective Yukawa interaction is added to the Lagrangian \cite{Stech:2014tla}:
\begin{align}
\lag_Y =& -g_tG_{\alpha\beta}\left( \psi^{\alpha T}H_1\psi^\beta \right) - A_{\alpha\beta}\left( \psi^{\alpha T} H_A \psi^\beta \right) \notag\\
&- \frac{1}{M_N}(G^2)_{\alpha\beta}\left( (\psi^{\alpha T}H_1^\dagger)_1(H_2^\dagger\psi^\beta)_1 \right) \plushc  \notag\\
=& -g_tG_{\alpha\beta}\left( Q_R^\alpha H_1^T Q_L^\beta + \frac12\epsilon^{ijk}\epsilon_{lmn}L^{\alpha i}_lL^{\beta j}_m(H_1)^k_n \right) \notag\\
&- A_{\alpha\beta}\left( Q_R^\alpha H_{Aq}^T Q_L^\beta + \epsilon^{ijk}L^{\alpha i}_lL^{\beta j}_m(H_{Al})^k_{\{lm\}} \right) \notag\\
&- \frac{1}{M_N}(G^2)_{\alpha\beta}\tr{L^\alpha H_1^\dagger}\tr{H_2^\dagger L^\beta} \plushc \label{eq:triniYukawaLag}
\end{align}
Here the subscript `1' in the denotes that the fields $\psi$ and $H_{1,2}^\dagger$ are combined into an $E_6$-singlet.
Since $H_1$ and $H_2$ are bitriplets of $SU(3)_L\times SU(3)_R$, only the leptons are involved in this interaction.
It could arise as an effective interaction via the exchange of a gauge-singlet Dirac fermion with a large mass $M_N \sim M_1$ \cite{Stech:2008wd}.
This fermion has a mass term that violates the $Z_2$ symmetry, hence the effective Yukawa term violates this symmetry as well.
After spontaneous symmetry breaking, this interaction yields additional mass terms for the neutrinos.
These make it possible to give low masses for the Standard-Model neutrinos and large masses to the other neutral leptons via the seesaw mechanism.

\subsection{Yukawa interactions in the LET model}
As we discussed at the beginning of this appendix, we need to determine which scalars in the LET model couple to which fermions, as well as the form of these couplings.
The scalar sector contains three fields $\Phi_1,\Phi_2 \sim (\rep{1},\brep{2},\rep{2},0)$ and $\Phi_R \sim (\rep{1},\rep{1},\rep{2},1)$.
The first one comes from the trinification field $H_1$ whereas the last two come from the fermiophobic $H_2$.
Hence we let only $\Phi_1$ couple to fermions.
The trinification model introduced additional scalar fields $H_{Aq}$, $H_{Al}$, which couple to the fermions as well.
However, we choose not to include their components in the LET model.
One reason is the fact that the description of the fermion masses and mixings via $H_{Aq}$, $H_{Al}$ involves mixing of light fermions with heavy fermions, which have been integrated out in the LET model.
Another reason is the fact that these components would mix with the components of the other scalar fields to form mass eigenstates.
This complicates our analysis of the scalar sector, unless we assume that the components of $\Phi_1$, $\Phi_2$, $\Phi_R$ do not mix with the components of $H_{Aq}$, $H_{Al}$ (this is assumed in the trinification model as well \cite{Stech:2014tla}).
Hence we do not include these additional scalar fields in the LET model.
Instead, we restrict ourselves to the first term of the Yukawa Lagrangian in \cref{eq:triniYukawaLag}.
We assume that any additional new physics, necessary to describe the fermion masses and mixings correctly, does not influence the phenomenology of the scalar particles.
When restricted to the fermion fields of the LET model, the first term in the Lagrangian in \cref{eq:triniYukawaLag} becomes
\begin{equation}
\lag_Y = -G_{\alpha\beta}\left( Q_R^\alpha \Phi_1^TQ_L^\beta + L^{+\alpha}\Phi_1^T L^{-\beta} \right) \plushc \label{eq:LETYukawa}
\end{equation}
Here the dimensionless parameter $g_t$ has been absorbed into $G_{\alpha\beta}$.

\subsection{Mass eigenstate basis}\label{a:fermionmasseig}
The Yukawa Lagrangian in \cref{eq:LETYukawa} determines all couplings of the physical scalars to the fermions.
In order to compare these predictions to experiment, it is necessary to rewrite the interactions in terms of mass eigenstates.
The scalar mass eigenstates have been worked out in \cref{a:scalarmasseigenstates}, so we only need to determine the fermion mass eigenstates.

As we discussed in \cref{a:Yukawatrini}, the generation matrix $G_{\alpha\beta}$ can be diagonalised by a biunitary transformation $G=UYW^\dagger$.
Thus we can write the Yukawa Lagrangian as
\begin{align}
\lag_Y =& -G_{\alpha\beta}\left( Q_R^\alpha \Phi_1^TQ_L^\beta + L^{+\alpha}\Phi_1^T L^{-\beta} \right) \plushc \notag\\
=& -Y_{\alpha\alpha}\left( Q_R^{\prime\alpha}\Phi_1^TQ_L^{\prime\alpha} + L^{+\prime\alpha}\Phi_1^TL^{-\prime\alpha} \right) \plushc \label{eq:yukawalagprime}
\end{align}
Here we have introduced the fermions in the mass eigenstate basis:
\begin{equation}
Q_R^\prime \equiv Q_RU,\quad Q_L^\prime \equiv W^\dagger Q_L,\quad L^{+\prime} \equiv L^+U,\quad L^{-\prime} \equiv W^\dagger L^-.
\end{equation}
After spontaneous symmetry breaking, the primed fermion fields obtain Dirac mass terms.
These combine left-handed and right-handed fermion fields, described by two-component Weyl spinors, into four-component Dirac spinors:
\begin{equation}
\psi_u^\alpha = \begin{pmatrix} u^{\prime\alpha} \\ \hat{u}^{\prime\alpha\dagger} \end{pmatrix},\quad
\psi_d^\alpha = \begin{pmatrix} d^{\prime\alpha} \\ \hat{d}^{\prime\alpha\dagger} \end{pmatrix},\quad
\psi_\nu^\alpha = \begin{pmatrix} \nu^{\prime\alpha} \\ \hat\nu^{\prime\alpha\dagger} \end{pmatrix},\quad
\psi_e^\alpha = \begin{pmatrix} e^{-\prime\alpha} \\ e^{+\prime\alpha\dagger} \end{pmatrix},\quad \label{eq:diracfermions}
\end{equation}
In terms of these Dirac spinors, the Lagrangian in \cref{eq:yukawalagprime} can be rewritten as
\begin{align}
\lag_Y =& -\frac{1}{\sqrt2}Y_{\alpha\alpha}(v_1+h_{1,11}^0) (\overline{\psi_u^\alpha}\psi_u^\alpha + \overline{\psi_\nu^\alpha}\psi_\nu^\alpha) - \frac{1}{\sqrt2}Y_{\alpha\alpha}(b_1+h_{1,22}^0) (\overline{\psi_d^\alpha}\psi_d^\alpha + \overline{\psi_e^\alpha}\psi_e^\alpha) \notag\\
&+ \frac{i}{\sqrt2}Y_{\alpha\alpha}a_{1,11}^0 (\overline{\psi_u^\alpha}\gamma^5\psi_u^\alpha + \overline{\psi_\nu^\alpha}\gamma^5\psi_\nu^\alpha) + \frac{i}{\sqrt2}Y_{\alpha\alpha}a_{1,22}^0 (\overline{\psi_d^\alpha}\gamma^5\psi_d^\alpha + \overline{\psi_e^\alpha}\gamma^5\psi_e^\alpha) \notag\\
&- Y_{\alpha\alpha} h^+_{1,21} \left( \overline{\psi_u^\alpha}\tfrac{1-\gamma^5}{2}\psi_d^\alpha + \overline{\psi_\nu^\alpha}\tfrac{1-\gamma^5}{2}\psi_e^\alpha \right) - Y_{\alpha\alpha} h^-_{1,21} \left( \overline{\psi_d^\alpha}\tfrac{1+\gamma^5}{2}\psi_u^\alpha + \overline{\psi_e^\alpha}\tfrac{1+\gamma^5}{2}\psi_\nu^\alpha \right) \notag\\
&- Y_{\alpha\alpha} h^+_{1,12} \left( \overline{\psi_u^\alpha}\tfrac{1+\gamma^5}{2}\psi_d^\alpha + \overline{\psi_\nu^\alpha}\tfrac{1+\gamma^5}{2}\psi_e^\alpha \right) - Y_{\alpha\alpha} h^-_{1,12} \left( \overline{\psi_d^\alpha}\tfrac{1-\gamma^5}{2}\psi_u^\alpha + \overline{\psi_e^\alpha}\tfrac{1-\gamma^5}{2}\psi_\nu^\alpha \right).
\end{align}
We fit the diagonal elements of $Y$ to the up-quark masses:
\begin{equation}
Y = \begin{pmatrix} y_u & 0 & 0 \\ 0 & y_c & 0 \\ 0 & 0 & y_t \end{pmatrix} = \frac{\sqrt2}{v_1}\begin{pmatrix} m_u & 0 & 0 \\ 0 & m_c & 0 \\ 0 & 0 & m_t \end{pmatrix}.
\end{equation}
We define $m_\alpha \equiv (m_u,m_c,m_t)$, so that we can abbreviate this as $Y_{\alpha\alpha}=\frac{\sqrt2}{v_1}m_\alpha$.
Now we can find the Yukawa interactions between the mass eigenstates of the simplified LET model by inserting the scalar mass eigenstates from \cref{eq:Higgsmasseig} into \cref{eq:yukawalagprime}.
We write $c_x\equiv\cos{x}$, $s_x\equiv\sin{x}$ for the sake of brevity.
Since we are only interested in the couplings of the physical particles, we omit all couplings to Goldstone bosons.
This yields the following physical Yukawa Lagrangian for the simplified LET model:
\begin{align}
\lag_Y^\text{phys} =& -m_\alpha\left( 1 + c_{\alpha_1}\frac{h^0}{v_1} \right) \left( \overline{\psi_u^\alpha}\psi_u^\alpha + \overline{\psi_\nu^\alpha}\psi_\nu^\alpha \right) - m_\alpha\frac{b_1}{v_1} \left( 1 + s_{\alpha_1}c_{\alpha_2}\frac{h^0}{b_1} \right) \left( \overline{\psi_d^\alpha}\psi_d^\alpha + \overline{\psi_e^\alpha}\psi_e^\alpha \right) \notag\\
&- \frac{m_\alpha}{v_1}H_1^0 \left( -s_{\alpha_1}c_{\alpha_3} \left[ \overline{\psi_u^\alpha}\psi_u^\alpha + \overline{\psi_\nu^\alpha}\psi_\nu^\alpha \right] + (c_{\alpha_1}c_{\alpha_2}c_{\alpha_3}-s_{\alpha_2}s_{\alpha_3}) \left[ \overline{\psi_d^\alpha}\psi_d^\alpha + \overline{\psi_e^\alpha}\psi_e^\alpha \right] \right) \notag\\
&- \frac{m_\alpha}{v_1}H_2^0 \left( s_{\alpha_1}s_{\alpha_3} \left[ \overline{\psi_u^\alpha}\psi_u^\alpha + \overline{\psi_\nu^\alpha}\psi_\nu^\alpha \right] - (s_{\alpha_2}c_{\alpha_3} + c_{\alpha_1}c_{\alpha_2}s_{\alpha_3}) \left[ \overline{\psi_d^\alpha}\psi_d^\alpha + \overline{\psi_e^\alpha}\psi_e^\alpha \right] \right) \notag\\
&- \frac{im_\alpha}{v_1}A^0 \left( s_{\beta_1}s_{\beta_3} \left[ \overline{\psi_u^\alpha}\psi_u^\alpha + \overline{\psi_\nu^\alpha}\psi_\nu^\alpha \right] - (s_{\beta_2}c_{\beta_3} + c_{\beta_1}c_{\beta_2}s_{\beta_3}) \left[ \overline{\psi_d^\alpha}\psi_d^\alpha + \overline{\psi_e^\alpha}\psi_e^\alpha \right] \right) \notag\\
&- \frac{\sqrt2m_\alpha}{v_1} H^+ \bigg( s_{\gamma_1}s_{\gamma_3} \left[ \overline{\psi_u^\alpha}\tfrac{1-\gamma^5}{2}\psi_d^\alpha + \overline{\psi_\nu^\alpha}\tfrac{1-\gamma^5}{2}\psi_e^\alpha \right] \notag\\
&\phantom{-\frac{\sqrt2m_\alpha}{v_1}H^+\bigg(} - (c_{\gamma_1}c_{\gamma_2}s_{\gamma_3} + s_{\gamma_2}c_{\gamma_3}) \left[ \overline{\psi_u^\alpha}\tfrac{1+\gamma^5}{2}\psi_d^\alpha + \overline{\psi_\nu^\alpha}\tfrac{1+\gamma^5}{2}\psi_e^\alpha \right] \bigg) \notag\\
&- \frac{\sqrt2m_\alpha}{v_1} H^- \bigg( s_{\gamma_1}s_{\gamma_3} \left[ \overline{\psi_d^\alpha}\tfrac{1+\gamma^5}{2}\psi_u^\alpha + \overline{\psi_e^\alpha}\tfrac{1+\gamma^5}{2}\psi_\nu^\alpha \right] \notag\\
&\phantom{-\frac{\sqrt2m_\alpha}{v_1}H^-\bigg(} - (c_{\gamma_1}c_{\gamma_2}s_{\gamma_3} + s_{\gamma_2}c_{\gamma_3}) \left[ \overline{\psi_d^\alpha}\tfrac{1-\gamma^5}{2}\psi_u^\alpha + \overline{\psi_e^\alpha}\tfrac{1-\gamma^5}{2}\psi_\nu^\alpha \right] \bigg). \label{eq:yukawadirac}
\end{align}
We are interested in the case $M\gg v$, so we also insert the scalar mixing angles from \cref{eq:chargedMixingAngles,eq:CPoddMixingAngles,eq:CPevenMixingAngles} into \cref{eq:yukawadirac}.
This gives us
\begin{align}
\lag_Y^\text{phys} =& -m_\alpha\left( 1 + \cos\beta\frac{h^0}{v_1} \right) \left( \overline{\psi_u^\alpha}\psi_u^\alpha + \overline{\psi_\nu^\alpha}\psi_\nu^\alpha \right) \notag\\
&- m_\alpha\frac{b_1}{v_1} \left( 1 + \sin\beta\frac{h^0}{b_1} \right) \left( \overline{\psi_d^\alpha}\psi_d^\alpha + \overline{\psi_e^\alpha}\psi_e^\alpha \right) \notag\\
&- \frac{m_\alpha}{v_1}H_1^0 \left( -\sin\beta \left[ \overline{\psi_u^\alpha}\psi_u^\alpha + \overline{\psi_\nu^\alpha}\psi_\nu^\alpha \right] + \cos\beta \left[ \overline{\psi_d^\alpha}\psi_d^\alpha + \overline{\psi_e^\alpha}\psi_e^\alpha \right] \right) \notag\\
&- \frac{im_\alpha}{v_1}A^0 \left( \sin\beta \left[ \overline{\psi_u^\alpha}\psi_u^\alpha + \overline{\psi_\nu^\alpha}\psi_\nu^\alpha \right] + \cos\beta \left[ \overline{\psi_d^\alpha}\psi_d^\alpha + \overline{\psi_e^\alpha}\psi_e^\alpha \right] \right) \notag\\
&- \frac{\sqrt2m_\alpha}{v_1} H^+ \bigg( \sin\beta \left[ \overline{\psi_u^\alpha}\tfrac{1-\gamma^5}{2}\psi_d^\alpha + \overline{\psi_\nu^\alpha}\tfrac{1-\gamma^5}{2}\psi_e^\alpha \right] \notag\\
&\phantom{-\frac{\sqrt2m_\alpha}{v_1}H^+\bigg(} + \cos\beta \left[ \overline{\psi_u^\alpha}\tfrac{1+\gamma^5}{2}\psi_d^\alpha + \overline{\psi_\nu^\alpha}\tfrac{1+\gamma^5}{2}\psi_e^\alpha \right] \bigg) \notag\\
&- \frac{\sqrt2m_\alpha}{v_1} H^- \bigg( \sin\beta \left[ \overline{\psi_d^\alpha}\tfrac{1+\gamma^5}{2}\psi_u^\alpha + \overline{\psi_e^\alpha}\tfrac{1+\gamma^5}{2}\psi_\nu^\alpha \right] \notag\\
&\phantom{-\frac{\sqrt2m_\alpha}{v_1}H^-\bigg(} + \cos\beta \left[ \overline{\psi_d^\alpha}\tfrac{1-\gamma^5}{2}\psi_u^\alpha + \overline{\psi_e^\alpha}\tfrac{1-\gamma^5}{2}\psi_\nu^\alpha \right] \bigg) + \ord{\frac{v}{M}}. \label{eq:YukawaLaglimit}
\end{align}
The fermion mass hierarchies are not represented correctly in our model.
Therefore we have to choose which masses we use to fix the free parameters of the model.
We fixed the three Yukawa couplings by the masses of the up-quarks.
Then we can fix $b_1$ and $v_1$ using the running $\overline{MS}$-mass of the $b$-quark at $M_Z$ and the condition $v_1^2+b_1^2 = v^2 = (246\text{ GeV})^2$:
\begin{align}
m_b = m_t\frac{b_1}{v_1} =& m_t\tan\beta,\qquad \Rightarrow \tan\beta = 0.0166, \notag\\
\Rightarrow v_1 =& 246\text{ GeV},\qquad b_1 = 4.09\text{ GeV}.
\end{align}

\subsection{Fermion mixing in charged current interactions}\label{a:CKMmatrix}
After switching to the fermion mass-eigenstate basis in \cref{eq:diracfermions}, the matrices $U$, $W$ have disappeared from the Yukawa Lagrangian.
They cancel from the neutral currents in \cref{eq:gcmasseig} as well, as they should.
In order to reproduce the experimental fermion mixings, they should not cancel from the charged currents.
However, since all left-handed (right-handed) fermions are brought into their mass eigenstates by the same transformation matrix $W^\dagger$ ($U$), these matrices do cancel from the charged currents in our setup.
Unless we include a mechanism for quark mixing, the CKM-matrix is simply a unit matrix, in contradiction with the experimental values listed in table~\ref{t:fermionmasses}.
A full understanding of fermion mixing in the LET model requires us to include mixing with the heavy $D$ quarks from the trinification model as well as renormalisation-group effects.
This is beyond the scope of this work.

\clearpage
\section{Gauge currents}\label{a:gaugecurrents}
The LET model contains fifteen gauge bosons: eight gluons for $SU(3)_C$, four electroweak bosons as in the Standard Model, a new pair of charged states $W^{\prime\pm}$, and a new neutral state $Z'$.
The model should reproduce the correct couplings of the fermions to the Standard-Model vector bosons.
In this appendix we derive the expressions for the gauge currents in terms of four-component Dirac spinors.
The couplings to the new heavy vector bosons will allow us to put constraints on their masses and mixings.
Note that the colour sector of the LET model is unchanged with respect to the Standard Model, so we omit the gluons from the following discussion.

The fermion-gauge-boson couplings are fixed by the covariant derivatives of the fermion fields.
At this stage, we describe the fermions by two-component left-handed Weyl spinors.
Omitting the gluon couplings, the covariant derivatives in \cref{eq:LETfermioncovderiv} become
\begin{align}
D_\mu Q_L =& \partial_\mu Q_L - ig_LW_{L\mu}^i T_L^iQ_L - \tfrac13ig'B_\mu Q_L, \notag\\
D_\mu Q_R =& \partial_\mu Q_R - ig_RW_{R\mu}^i Q_R\overline{T}_R^{iT} + \tfrac13ig'B_\mu Q_R, \notag\\
D_\mu L^- =& \partial_\mu L^- - ig_LW_{L\mu}^i T_L^iL^- + ig'B_\mu L^-, \notag\\
D_\mu L^+ =& \partial_\mu L^+ - ig_RW_{R\mu}^i L^+\overline{T}_R^{iT} - ig'B_\mu L^+, \label{eq:fermioncovderiv}
\end{align}
where $i=1,2,3$.
These covariant derivatives appear in the gauge-invariant kinetic part of the Lagrangian, which is given by
\begin{align}
\lag_f \equiv& iQ_L^\dagger\bar\sigma^\mu(D_\mu Q_L) - i(D_\mu Q_R)\sigma^\mu Q_R^\dagger \notag\\
&+ i(L^-)^\dagger\bar\sigma^\mu(D_\mu L^-) - i(D_\mu L^+)\sigma^\mu(L^+)^\dagger.
\end{align}
Here we have employed the spinor identity $\xi^\dagger\bar\sigma^\mu\eta = -\eta\sigma^\mu\xi^\dagger$ to order the spinors in such a way that they conform to both index-free spinor notation (in which descending undotted indices ${}^\alpha_{\phantom\alpha\alpha}$ and ascending dotted indices ${}_{\dot\alpha}^{\phantom\alpha\dot\alpha}$ can be suppressed) and our matrix notation for left- and right-handed components.
The daggers on the fermion fields imply both spinor conjugation and Hermitian conjugation w.r.t.\ matrix notation.
We can read off the fermion-gauge-boson interactions directly from the Lagrangian after we insert the covariant derivatives from \cref{eq:fermioncovderiv} into the above expression:
\begin{align}
\lag_f =& \sum_f if^\dagger\bar\sigma^\mu\partial_\mu f + g_LW_{L\mu}^i\Big( Q_L^\dagger T_L^iQ_L + (L^-)^\dagger T_L^i\bar\sigma^\mu L^- \Big) \notag\\
&+ g_RW_{R\mu}^i\Big( -Q_R\overline{T}_R^{iT}\sigma^\mu Q_R^\dagger - L^+\overline{T}_R^{iT}\sigma^\mu(L^+)^\dagger \Big) \notag\\
&+ g'B_\mu\sum_f f^\dagger\bar\sigma^\mu Q_{B-L}f.
\end{align}
Here $f$ runs over all two-component spinors $u$, $d$, $\nu$, $e^-$, $\hat{u}$, $\hat{d}$, $\hat\nu$, $e^+$.
As usual, we introduce the gauge-boson charge eigenstates $W_{L,R}^\pm \equiv \frac{1}{\sqrt2}(W_{L,R}^1\mp iW_{L,R}^2)$.
If we define $T_{L,R}^\pm\equiv \frac{1}{\sqrt2}(T_{L,R}^1\pm iT_{L,R}^2)$ as well, we can use the identity
\begin{equation}
\sum_{i=1,2} W_{L,R\mu}^iT_{L,R}^i = W_{L,R}^+T_{L,R}^+ + W_{L,R}^-T_{L,R}^-
\end{equation}
to rewrite the kinetic part of the Lagrangian as
\begin{align}
\lag_f =& \sum_f if^\dagger\bar\sigma^\mu\partial_\mu f + \frac{g_L}{\sqrt2}\Big( W_{L\mu}^+u^\dagger\bar\sigma^\mu d + W_{L\mu}^+ \nu^\dagger\bar\sigma^\mu e^- \plushc \Big) \notag\\
&+ \frac{g_R}{\sqrt2}\Big( W_{R\mu}^+\hat{u}\sigma^\mu\hat{d}^\dagger + W_{R\mu}^+\hat\nu\sigma^\mu(e^+)^\dagger \plushc \Big) \notag\\
&+ g_LW_{L\mu}^3\Big( Q_L^\dagger T_L^3\bar\sigma^\mu Q_L + (L^-)^\dagger T_L^3\bar\sigma^\mu L^- \Big) \notag\\
&+ g_RW_{R\mu}^3\Big( -Q_R\overline{T}_R^{3T}\sigma^\mu Q_R^\dagger - L^+\overline{T}_R^{3T}(L^+)^\dagger \Big) \notag\\
&+ g'B_\mu\sum_f f^\dagger\bar\sigma^\mu Q_{B-L}f.
\end{align}
As we discussed in \cref{a:fermionmasseig}, the two-component Weyl spinors are combined into four-component Dirac spinors after spontaneous symmetry breaking.
These are defined as
\begin{equation}
\psi_u \equiv \begin{pmatrix} u \\ \hat{u}^\dagger \end{pmatrix},\qquad
\psi_d \equiv \begin{pmatrix} d \\ \hat{d}^\dagger \end{pmatrix},\qquad
\psi_\nu \equiv \begin{pmatrix} \nu \\ \hat\nu^\dagger \end{pmatrix},\qquad
\psi_e \equiv \begin{pmatrix} e^- \\ (e^+)^\dagger \end{pmatrix}.
\end{equation}
We can write expressions involving two-component Weyl spinors and Pauli matrices in terms of four-component Dirac spinors and gamma matrices using the identities
\begin{align}
\overline\psi_{f_1}\gamma^\mu\psi_{f_1} =& \hat{f}_1\sigma^\mu\hat{f}_1^\dagger + f_1^\dagger\bar\sigma^\mu f_1, \notag\\
\overline\psi_{f_1}\gamma^\mu P_-\psi_{f_2} =& f_1^\dagger\bar\sigma^\mu f_2, \notag\\
\overline\psi_{f_1}\gamma^\mu P_+\psi_{f_2} =& \hat{f}_1\sigma^\mu\hat{f}^\dagger_2.
\end{align}
Here we define the projection operators $P_\pm \equiv \frac12(1\pm\gamma^5)$ and denote pairs of fermions as $(f_i,\hat{f}_i) = (u,\hat{u}), (d,\hat{d}), (\nu,\hat\nu), (e^-,e^+)$.
In terms of these Dirac spinors, the Lagrangian becomes
\begin{align}
\lag_f =& \sum_{f=u,d,\nu,e}\overline\psi_fi\gamma^\mu\partial_\mu\psi_f \notag\\
&+ \frac{g_L}{\sqrt2}W_{L\mu}^+\Big( \overline\psi_u\gamma^\mu P_-\psi_d + \overline\psi_\nu\gamma^\mu P_-\psi_e \Big) + \frac{g_L}{\sqrt2}W_{L\mu}^-\Big( \overline\psi_d\gamma^\mu P_-\psi_u + \overline\psi_e\gamma^\mu P_-\psi_\nu \Big) \notag\\
&+ \frac{g_R}{\sqrt2}W_{R\mu}^+\Big( \overline\psi_u\gamma^\mu P_+\psi_d + \overline\psi_\nu\gamma^\mu P_+\psi_e \Big) + \frac{g_R}{\sqrt2}W_{R\mu}^-\Big( \overline\psi_d\gamma^\mu P_+\psi_u + \overline\psi_e\gamma^\mu P_+\psi_\nu \Big) \notag\\
&+ \sum_{f=u,d,\nu,e}\overline\psi_f\gamma^\mu\Big( g_LW_{L\mu}^3T_L^3(f) P_- + g_RW_{R\mu}^3T_R^3(f) P_+ + g'B_\mu Q_{B-L}(f) \Big)\psi_f. \label{eq:Diraclag}
\end{align}
Here we defined $T_L^3(f) = T_R^3(f) = +\frac12$ for $f=u,\nu$; $T_L^3(f) = T_R^3(f) = -\frac12$ for $f=d,e$; and $Q_{B-L}(f)$ denotes the $B-L$ quantum number of the fermion $f$.
We will rewrite this expression in terms of gauge-boson mass eigenstates to leading order in $\frac{v^2}{M^2}$.
These mass eigenstates were given in \cref{eq:Wmasseigenstates,eq:AZZpmasseigenstates}:
\begin{align}
\begin{pmatrix} W_L^3 \\ W_R^3 \\ B \end{pmatrix} =&
\begin{pmatrix} s_{\theta_W} & c_{\theta_W}c_\eta & -c_{\theta_W}s_\eta \\
c_{\theta_W}s_{\theta_W^\prime} & c_{\theta_W^\prime}s_\eta - s_{\theta_W}s_{\theta_W^\prime}c_\eta & c_{\theta_W^\prime}c_\eta + s_{\theta_W}s_{\theta_W^\prime}s_\eta \\
c_{\theta_W}c_{\theta_W^\prime} & -(s_{\theta_W}c_{\theta_W^\prime}c_\eta + s_{\theta_W^\prime}s_\eta) & s_{\theta_W}c_{\theta_W^\prime}s_\eta - s_{\theta_W^\prime}c_\eta \end{pmatrix}
\begin{pmatrix} A \\ Z \\ Z' \end{pmatrix}, \notag\\
\begin{pmatrix} W_L^\pm \\ W_R^\pm \end{pmatrix} =&
\begin{pmatrix} \cos\zeta & -\sin\zeta \\ \sin\zeta & \cos\zeta \end{pmatrix}
\begin{pmatrix} W^\pm \\ W^{\prime\pm} \end{pmatrix}.
\end{align}
Inserting these expressions into \cref{eq:Diraclag} and collecting terms proportional to $A_\mu$, we can extract the electromagnetic gauge coupling $e$ in terms of the gauge couplings:
\begin{equation}
e \equiv \frac{2g'g_Lg_R}{\sqrt{4g^{\prime2}(g_L^2+g_R^2)+g_L^2g_R^2}} = g_L\sin\theta_W.
\end{equation}
We rewrite the Lagrangian in \cref{eq:Diraclag} using this expression as well as the mixing angles in \cref{eq:Wmixingangle,eq:Zmixingangles}.
Here we expand the result to first order in $\zeta$, $\eta$, since both angles are of order $\frac{v^2}{M^2}$:
\begin{align}
\lag_f =& \sum_{f=u,d,\nu,e}\overline\psi_fi\gamma^\mu\partial_\mu\psi_f \notag\\
&+ \frac{g_L}{\sqrt2}W_\mu^+\Big( \overline\psi_u\gamma^\mu (P_-+\frac{g_R\zeta}{g_L}P_+)\psi_d + \overline\psi_\nu\gamma^\mu (P_-+\frac{g_R\zeta}{g_L}P_+)\psi_e + \ord{\zeta^2} \Big) \notag\\
&+ \frac{g_L}{\sqrt2}W_\mu^-\Big( \overline\psi_d\gamma^\mu (P_-+\frac{g_R\zeta}{g_L}P_+)\psi_u + \overline\psi_e\gamma^\mu (P_-+\frac{g_R\zeta}{g_L}P_+)\psi_\nu + \ord{\zeta^2} \Big) \notag\\
&+ \frac{g_R}{\sqrt2}W_\mu^{\prime+}\Big( \overline\psi_u\gamma^\mu (P_+ - \frac{g_L\zeta}{g_R}P_-)\psi_d + \overline\psi_\nu\gamma^\mu (P_+ - \frac{g_L\zeta}{g_R}P_-)\psi_e + \ord{\zeta^2} \Big) \notag\\
&+ \frac{g_R}{\sqrt2}W_\mu^{\prime-}\Big( \overline\psi_d\gamma^\mu (P_+ - \frac{g_L\zeta}{g_R}P_-)\psi_u + \overline\psi_e\gamma^\mu (P_+ - \frac{g_L\zeta}{g_R}P_-)\psi_\nu + \ord{\zeta^2} \Big) \notag\\
&+ eA_\mu \sum_{f=u,d,\nu,e}\overline\psi_f\gamma^\mu Q_{em}(f) \psi_f \notag\\
&+ \frac{g_L}{c_{\theta_W}}Z_\mu \sum_{f=u,d,\nu,e}\overline\psi_f\gamma^\mu \Bigg[ \left( 1 + \frac{g_Rc_{\theta_W}s_{\theta_W^\prime}^2\eta}{g_Lc_{\theta_W^\prime}} \right)T_L^3(f) P_- + \frac{s_{\theta_W}\eta}{s_{\theta_W^\prime}c_{\theta_W^\prime}} T_R^3(f)P_+ \notag\\
&\hspace{4cm}- s_{\theta_W}^2\left( 1 + \frac{g_Rc_{\theta_W}s_{\theta_W^\prime}^2\eta}{s_{\theta_W}^2c_{\theta_W^\prime}} \right) Q_{em}(f) + \ord{\eta^2} \Bigg] \psi_f \notag\\
&+ g_Rc_{\theta_W^\prime} Z_\mu^\prime \sum_{f=u,d,\nu,e}\overline\psi_f\gamma^\mu \Bigg[ -\frac{g_Lc_{\theta_W}\eta}{g_Rc_{\theta_W^\prime}} T_L^3(f)P_- + \left( 1 + \frac{g_Rc_{\theta_W}s_{\theta_W^\prime}^2\eta}{g_Lc_{\theta_W^\prime}} \right) T_R^3(f) P_+ \notag\\
&\hspace{2cm}- \frac12t_{\theta_W^\prime}^2 \left( 1 - \frac{g_Rc_{\theta_W}c_{\theta_W^\prime}\eta}{g_L} \right) Q_{B-L}(f) + \ord{\eta^2} \Bigg] \psi_f. \label{eq:gcmasseig}
\end{align}
Here $Q_{em}(f)$ denotes the electromagnetic charge of the fermion $f$, and $t_x \equiv \tan{x}$.
Note that at leading order in $\zeta$, $\eta$ (or equivalently $\frac{v^2}{M^2}$), the LET model reproduces the Standard-Model gauge currents correctly.
The couplings of the fermions to $W'$, $Z'$ can be used to constrain the model, see \cref{s:gaugebosonconstraints}.

\clearpage
\section{Photon coupling modification}\label{a:Deltagamma}
In this section we calculate the photon coupling modification $\Delta_\gamma \equiv (g_\gamma-g_\gamma^\text{SM})/g_\gamma^\text{SM}$ for the simplified LET model.
First we derive a general expression for the photon coupling.
Then we discuss the relevant contributions in the simplified LET model, and derive an expression for $\Delta_\gamma$ in terms of simplified-LET-model parameters.

The Higgs boson has no direct couplings to photons.
However, a Higgs boson can decay into a pair of photons via a loop involving a massive charged particle.
For a general theory with a set of vector bosons $V$, fermions $f$, and scalars $S$, the resulting decay width $\Gamma(h^0\rightarrow\gamma\gamma)$ is given by \cite{Carena:2012xa}
\begin{align}
\Gamma(h^0\rightarrow\gamma\gamma) =& \frac{\alpha^2m_{h^0}^3}{1024\pi^3} \left| \sum_V \frac{g_{h^0VV}}{m_V^2}Q_V^2A_1(\tau_V) + \sum_f \frac{2g_{h^0ff}}{m_f}N_{c,f}Q_f^2A_{1/2}(\tau_f) \right. \notag\\
&\hspace{3cm} \left. + \sum_S N_{c,S}Q_S^2\frac{g_{h^0SS}}{m_S^2}A_0(\tau_S) \right|^2. \label{eq:photondecaywidth}
\end{align}
Here $\alpha$ is the fine-structure constant, $g_{h^0xx}$ denotes the Higgs coupling to the particle $x$, $Q_x$ is the electromagnetic charge of $x$, $N_{c,x}$ is the number of colour degrees of freedom for $x$, and $\tau_x$ is defined as the ratio $\tau_x \equiv 4m_x^2/m_{h^0}^2$.
Furthermore, the functions $A_s(x)$ are the scalar, spinor, and vector loop functions, which are defined as
\begin{align}
A_0(x) \equiv& -x^2(x^{-1}-f(x)), \notag\\
A_{1/2}(x) \equiv& 2x^2(x^{-1}+(x^{-1}-1)f(x)), \notag\\
A_1(x) \equiv& -x^2(2x^{-2}+3x^{-1}+3(2x^{-1}-1)f(x)), \notag\\
f(x) \equiv& \begin{cases} \arcsin^2\sqrt{x^{-1}} & \tau>1, \\
-\frac14\left( \log\frac{1+\sqrt{1-\tau}}{1-\sqrt{1-\tau}} - i\pi \right)^2 & \tau<1. \end{cases} \label{eq:loopfunctions}
\end{align}
We can compute the decay width in \cref{eq:photondecaywidth} in terms of an effective Higgs-photon coupling as well.
This will allow us to express the effective coupling in terms of the underlying couplings and masses.
The Higgs-photon coupling is defined by the effective Lagrangian
\begin{equation}
\lag_{h^0\gamma\gamma} = \frac{g_\gamma}{v}h^0A_{\mu\nu}A^{\mu\nu}.
\end{equation}
The corresponding Feynman rule is given by
\begin{equation}
\parbox{20mm}{\begin{fmfgraph*}(50,30)
  \fmfleft{h}
  \fmfright{nu,mu}
  \fmfv{label=$h^0$,label.angle=180}{h}
  \fmfv{label=$\gamma_\mu$,label.angle=0}{mu}
  \fmfv{label=$\gamma_\nu$,label.angle=0}{nu}
  \fmf{dashes}{h,v}
  \fmf{boson,label=$p_1\swarrow$,label.side=right,label.dist=1}{mu,v}
  \fmf{boson,label=$p_2\nwarrow$,label.side=left,label.dist=1}{nu,v}
\end{fmfgraph*}}
= -\frac{4ig_\gamma}{v}((p_1\cdot p_2)g^{\mu\nu} - p_2^\mu p_1^\nu).
\end{equation}
The resulting matrix element for the decay $h^0\rightarrow\gamma\gamma$ is
\begin{equation}
i\mathcal{M}(m_h\rightarrow\{p_1,p_2\}) = -\frac{4ig_\gamma}{v}((p_1\cdot p_2)g^{\mu\nu} - p_1^\nu p_2^\mu)\epsilon^*_{1\mu}(p_1)\epsilon^*_{2\nu}(p_2).
\end{equation}
Summing over all photon polarisations, we find
\begin{align}
\sum_{\epsilon_1,\epsilon_2}|\mathcal{M}|^2 =& \frac{16g_\gamma^2}{v^2}((p_1\cdot p_2)g^{\mu\nu}-p_1^\nu p_2^\mu)g_{\mu\rho}g_{\nu\sigma}((p_1\cdot p_2)g^{\rho\sigma}-p_1^\sigma p_2^\rho) \notag\\
=& \frac{16g_\gamma^2}{v^2}(2(p_1\cdot p_2)^2+p_1^2p_2^2) \notag\\
=& \frac{32g_\gamma^2}{v^2}(p_1\cdot p_2)^2.
\end{align}
In the last line we used the fact that $p_1^2=p_2^2=0$ for the photons.
The decay width is therefore given by
\begin{align}
\d\Gamma =& \frac{1}{2m_{h^0}} \frac{\d\vec{p}_1}{(2\pi)^3} \frac{\d\vec{p}_2}{(2\pi)^3} \frac{1}{4E_1E_2} |\mathcal{M}(m_h\rightarrow\{p_1,p_2\})|^2(2\pi)^4\delta^{(4)}(p_h-p_1-p_2) \notag\\
=& \frac{1}{2m_{h^0}} \frac{\d\vec{p}_1}{(2\pi)^3} \frac{\d\vec{p}_2}{(2\pi)^3} \frac{1}{4E_1E_2} \frac{32g_\gamma^2}{v^2}(p_1\cdot p_2)^2 (2\pi)^4\delta^{(4)}(p_h-p_1-p_2) \notag\\
\Gamma(h^0\rightarrow\gamma\gamma) =& \frac{4g_\gamma^2}{m_{h^0}v^2} \int \frac{\d\vec{p}_1}{(2\pi)^3} \frac{\d\vec{p}_2}{(2\pi)^3} \frac{(p_1\cdot p_2)^2}{E_1E_2} (2\pi)^4\delta^{(4)}(p_h-p_1-p_2) \notag\\
\stackrel{\text{\tiny (CM)}}{=}& \frac{4g_\gamma^2}{m_{h^0}v^2} \int \frac{\d\vec{p}_1}{(2\pi)^3} \frac{\d\vec{p}_2}{(2\pi)^3} \frac{(p_1\cdot p_2)^2}{E_1E_2} (2\pi)^4\delta^{(3)}(\vec{p}_1+\vec{p}_2)\delta(m_{h^0}-E_1-E_2) \notag\\
=& \frac{4g_\gamma^2}{m_{h^0}v^2} \int \frac{\d\vec{p}_1}{(2\pi)^3} \frac{(2|\vec{p}_1|^2)^2}{|\vec{p}_1|^2} (2\pi)\delta(m_{h^0}-E_1-E_2) \notag\\
=& \frac{4g_\gamma^2}{m_{h^0}v^2} \int \frac{\d\Omega_{CM}\d|\vec{p}_1| |\vec{p}_1|^2}{(2\pi)^3} 4|\vec{p}_1|^4 (2\pi)\delta(m_{h^0}-E_1-E_2) \notag\\
=& \frac{4g_\gamma^2}{m_{h^0}v^2} \int \frac{\d\Omega_{CM}}{2\pi^2}\d E |\vec{p}_1|^4 \delta(m_{h^0}-E) \notag\\
=& \frac{8g_\gamma^2}{\pi m_{h^0}v^2} \left.|\vec{p}_1|^4\right|_{E=m_{h^0}} \notag\\
=& \frac{g_\gamma^2m_{h^0}^3}{2\pi v^2}. \label{eq:effphotoncoupling}
\end{align}
Here we defined $E=E_1+E_2$ and used the fact that $\d E = (\d E_1/\d|\vec{p}_1| + \d E_2/\d|\vec{p}_1|)\d|\vec{p}_1| = 2\d|\vec{p}_1|$ on the third last line.
Comparing \cref{eq:photondecaywidth,eq:effphotoncoupling}, we find the photon coupling
\begin{align}
g_\gamma =& \frac{\alpha v}{16\sqrt2\pi} \left( \sum_V \frac{g_{h^0VV}}{m_V^2}Q_V^2A_1(\tau_V) + \sum_f \frac{2g_{h^0ff}}{m_f}N_{c,f}Q_f^2A_{1/2}(\tau_f) \right. \notag\\
&\hspace{3cm} \left. + \sum_S N_{c,S}Q_S^2\frac{g_{hSS}}{m_S^2}A_0(\tau_S) \right). \label{eq:effectivePhotonCoupling}
\end{align}

Now we use this result to determine the relevant contributions to the photon coupling modification.
In the Standard Model, the leading contributions come from the $W$ loop and the top loop.
The simplified LET model introduces new contributions from a $W'$ loop and a $H^\pm$ loop.
However, the $W'$ contribution turns out to be negligible: from the Feynman rules in \cref{a:SVVfeynrules} we find $g_{h^0W'W'}\sim v$, hence $g_{h^0W'W'}/m_{W'}^2 \sim v/M^2$.
This implies that the $W'$ contribution is suppressed by a factor $\frac{v^2}{M^2}$ with respect to the $W$ contribution.
Therefore we take only the charged scalar contribution into account.
The resulting photon coupling modification reads
\begin{equation}
\Delta_\gamma = \left(\frac{\lambda_{h^0H^+H^-}}{m_{H^\pm}^2}A_0(\tau_{H^\pm})\right) \left(\frac{g_{h^0WW}}{m_W^2}A_1(\tau_W) + \frac{2g_{htt}}{m_t}N_cQ_t^2A_{1/2}(\tau_t)\right)^{-1}.
\end{equation}
The denominator can be simplified by inserting $g_{h^0WW}=2m_W^2/v$ and $g_{h^0tt}=m_t/v$:
\begin{align}
\Delta_\gamma =& \frac{v\lambda_{h^0H^+H^-}A_0(\tau_{H^\pm})}{2m_{H^\pm}^2A_\text{SM}}, \notag\\
A_\text{SM} \equiv& \frac{v}{2} \left(\frac{g_{h^0WW}}{m_W^2}A_1(\tau_W) + \frac{2g_{htt}}{m_t}N_cQ_t^2A_{1/2}(\tau_t)\right) \notag\\
=& A_1(\tau_W) + N_cQ_t^2A_{1/2}(\tau_t) = -6.5. \label{eq:DeltaGammaContribution}
\end{align}
The charged scalar mass is given by \cref{eq:HpmMass}:
\begin{equation}
m_{H^\pm}^2 = \frac{\lambda_5}{2}\left( M^2\sec2\beta + v^2\cos2\beta \right).
\end{equation}
Now the only thing missing is an expression for $\lambda_{h^0H^+H^-}$.
It follows from the scalar potential in \cref{eq:scalarpotential} combined with the minimum condition in \cref{eq:minimumcondition}, rewritten in terms of the mass eigenstates in \cref{eq:Higgsmasseig}.
\emph{Mathematica} gives us
\begin{align}
\lambda_{h^0H^+H^-} =& \lambda_1v(c_\beta c_{\alpha_1} + s_\beta s_{\alpha_1}c_{\alpha_2}) (s_{13}^2 + c_+^2) \notag\\
&+ \lambda_2v\big( (c_\beta c_{\alpha_1} + s_\beta s_{\alpha_1}c_{\alpha_2}) (s_{13}^2 + c_+^2) - (s_\beta c_{\alpha_1} + c_\beta s_{\alpha_1}c_{\alpha_2}) s_{13}c_+ \big) \notag\\
&+ \lambda_3Ms_{\alpha_1}s_{\alpha_2}c_-^2 + \lambda_4(v(c_\beta c_{\alpha_1} + s_\beta s_{\alpha_1}c_{\alpha_2})c_-^2 + Ms_{\alpha_1}s_{\alpha_2}(s_{13}^2 + c_+^2)) \notag\\
&+ \lambda_5\big( v(c_\beta c_{\alpha_1}c_-^2 - s_\beta s_{\alpha_1}s_{\alpha_2}s_{13}c_- + c_\beta s_{\alpha_1}s_{\alpha_2}c_+c_-) \notag\\
&\hspace{1cm}+ M(s_{\alpha_1}s_{\alpha_2}c_+^2 - s_{\alpha_1}c_{\alpha_2}s_{13}c_- + c_{\alpha_1}c_+c_-) \big).
\end{align}
Here we defined the following combinations of mixing angles of the charged scalar fields, which are given in \cref{eq:chargedMixingAngles}:
\begin{equation}
s_{13} \equiv s_{\gamma_1}s_{\gamma_3},\qquad c_+ \equiv c_{\gamma_1}c_{\gamma_2}s_{\gamma_3} + s_{\gamma_2}c_{\gamma_3},\qquad c_- \equiv c_{\gamma_1}s_{\gamma_2}s_{\gamma_3} - c_{\gamma_2}c_{\gamma_3}. \label{eq:s13cpcm}
\end{equation}
In the limit of small $\xi \equiv \frac{v}{M}$, we can use the approximations for the $CP$-even mixing angles in \cref{eq:CPevenMixingAngles}.
This results in the following coupling:
\begin{equation}
\frac{\lambda_{h^0H^+H^-}}{v} = \lambda_1 + \lambda_2(1+\frac12\sin^22\beta) + \lambda_5\cos2\beta - \frac{\lambda_4(\lambda_4+\lambda_5\cos^2\beta)}{\lambda_3} + \ord{\xi^2}.
\end{equation}
Thus in the small-$\xi$ limit, the photon coupling modification is given by
\begin{align}
\Delta_\gamma(\xi\rightarrow0) =& \frac{\xi^2A_0(\tau_{H^\pm})\cos2\beta}{A_\text{SM}\lambda_5} \Bigg( \lambda_1 + \lambda_2(1+\frac12\sin^22\beta) + \lambda_5\cos2\beta \notag\\
&\hspace{3cm} - \frac{\lambda_4(\lambda_4+\lambda_5\cos^2\beta)}{\lambda_3} + \ord{\xi^2} \Bigg).
\end{align}

\clearpage
\section{Feynman rules}\label{a:feynmanrules}
In this appendix, we give the Feynman rules of the simplified LET model that are relevant to our analysis.
In the following, we denote the charged gauge bosons by $V^\pm \equiv \{W^\pm,W^{\prime\pm}\}$ and the neutral gauge bosons by $V^0 \in \{\gamma,Z,Z'\}$.
Also, we write $\Phi^0 \equiv \{h^0, H^0_1, H^0_2, A^0\}$ for the neutral scalars and $S^0 \equiv \{h^0, H^0_1, H^0_2\}$ for the $CP$-even scalars.
For the sake of brevity, we write $s_x \equiv \sin{x}$, $c_x \equiv \cos{x}$, $t_x \equiv \tan{x}$.
The combinations $s_{13}$, $c_\pm$ of mixing angles of the charged states are defined in \cref{eq:s13cpcm}.
Other combinations of parameters are defined when they appear.

\subsection{Triple-vector-boson couplings}\label{a:VVVfeynRules}
\begin{align}
\parbox{20mm}{\begin{fmfgraph*}(40,30)
  \fmfleft{V2,V1}
  \fmfright{V3}
  \fmfv{label=$V_{1\mu}^\pm$,label.angle=180}{V1}
  \fmfv{label=$V_{2\nu}^\pm$,label.angle=180}{V2}
  \fmfv{label=$V_{3\rho}^0$,label.angle=0}{V3}
  \fmf{boson,label=$p\searrow$,label.side=right,label.dist=.1}{V1,v}
  \fmf{boson,label=$\nearrow q$,label.side=right,label.dist=.1}{V2,v}
  \fmf{boson,label=$\leftarrow r$,label.side=right,label.dist=1}{V3,v}
\end{fmfgraph*}}
=& -ig_{V_1^\pm}g_{V_2^\pm}g_{V_3^0}\left( g_{\mu\nu}(p-q)_\rho + g_{\nu\rho}(q-r)_\mu + g_{\rho\mu}(r-p)_\nu \right), \notag\\
g_{V^\pm} =& \{ \cos\zeta , \pm\sin\zeta \}, \notag\\
g_{V^0} =& \{ e , g_L\cos\theta_W\cos\eta , -g_L\cos\theta_W\sin\eta \}.
\end{align}

\subsection{Couplings of SM-like Higgs components to a heavy vector boson}\label{a:h0VpVpfeynRules}
\begin{align}
\parbox{30mm}{\begin{fmfgraph*}(60,40)
  \fmfleft{h0,G}
  \fmfright{Wp}
  \fmfv{label=$h^0$,label.angle=180}{h0}
  \fmfv{label=$G^\mp$,label.angle=180}{G}
  \fmfv{label=$W^{\prime\pm}_\mu$,label.angle=0}{Wp}
  \fmf{dashes,label=$\nearrow p$,label.side=right,label.dist=.1}{h0,v}
  \fmf{dashes,label=$q\searrow$,label.side=right,label.dist=.1}{G,v}
  \fmf{boson}{Wp,v}
\end{fmfgraph*}}
=& \pm\frac{ig_R\sin2\beta}{2}(p-q)_\mu, \notag\\[5ex]
\parbox{30mm}{\begin{fmfgraph*}(60,40)
  \fmfleft{G0,G}
  \fmfright{Wp}
  \fmfv{label=$G^0$,label.angle=180}{G0}
  \fmfv{label=$G^\mp$,label.angle=180}{G}
  \fmfv{label=$W^{\prime\pm}_\mu$,label.angle=0}{Wp}
  \fmf{dashes,label=$\nearrow p$,label.side=right,label.dist=.1}{G0,v}
  \fmf{dashes,label=$q\searrow$,label.side=right,label.dist=.1}{G,v}
  \fmf{boson}{Wp,v}
\end{fmfgraph*}}
=& \frac{g_R\sin2\beta}{2}(p-q)_\mu, \notag\\[5ex]
\parbox{30mm}{\begin{fmfgraph*}(60,40)
  \fmfleft{h0,G}
  \fmfright{Zp}
  \fmfv{label=$h^0$,label.angle=180}{h0}
  \fmfv{label=$G^0$,label.angle=180}{G}
  \fmfv{label=$Z^\prime_\mu$,label.angle=0}{Zp}
  \fmf{dashes,label=$\nearrow p$,label.side=right,label.dist=.1}{h0,v}
  \fmf{dashes,label=$q\searrow$,label.side=right,label.dist=.1}{G,v}
  \fmf{boson}{Zp,v}
\end{fmfgraph*}}
=& \frac{g_R\cos\theta_W^\prime}{2}(p-q)_\mu, \notag\\[5ex]
\parbox{30mm}{\begin{fmfgraph*}(60,40)
  \fmfleft{Gp,Gm}
  \fmfright{Zp}
  \fmfv{label=$G^+$,label.angle=180}{Gp}
  \fmfv{label=$G^-$,label.angle=180}{Gm}
  \fmfv{label=$Z^\prime_\mu$,label.angle=0}{Zp}
  \fmf{dashes,label=$\nearrow p$,label.side=right,label.dist=.1}{Gp,v}
  \fmf{dashes,label=$q\searrow$,label.side=right,label.dist=.1}{Gm,v}
  \fmf{boson}{Zp,v}
\end{fmfgraph*}}
=& \frac{ig_R\cos\theta_W^\prime}{2}(p-q)_\mu.
\end{align}

\subsection{$H^\pm$ Couplings to SM bosons}\label{a:HpmToSMbosonCouplings}
\begin{align}
\parbox{20mm}{\begin{fmfgraph*}(40,30)
  \fmfleft{Hpm}
  \fmfright{gam,W}
  \fmfv{label=$H^\pm$,label.angle=180}{Hpm}
  \fmfv{label=$W^\pm_\mu$,label.angle=0}{W}
  \fmfv{label=$\gamma_\nu$,label.angle=0}{gam}
  \fmf{dashes}{Hpm,v}
  \fmf{boson}{W,v,gam}
\end{fmfgraph*}}
=& -\frac{ig_{\mu\nu}}{2}g_Le \Bigg( v\left( c_\zeta[c_\beta s_{13} + s_\beta c_+] - \frac{s_\zeta t_{\theta_W}}{s_{\theta_W^\prime}}[c_\beta c_+ + s_\beta s_{13}] \right) \notag\\
&\hspace{6cm}+ M\frac{s_\zeta t_{\theta_W}c_-}{s_{\theta_W^\prime}} \Bigg), \notag\\[3ex]
\parbox{20mm}{\begin{fmfgraph*}(40,30)
  \fmfleft{Hpm}
  \fmfright{Z,W}
  \fmfv{label=$H^\pm$,label.angle=180}{Hpm}
  \fmfv{label=$W^\pm_\mu$,label.angle=0}{W}
  \fmfv{label=$Z_\nu$,label.angle=0}{Z}
  \fmf{dashes}{Hpm,v}
  \fmf{boson}{W,v,Z}
\end{fmfgraph*}}
=& \frac{ig_{\mu\nu}}{2}g_L^2t_{\theta_W} \Bigg( v\Bigg( c_\zeta[c_\beta s_{13} + s_\beta c_+][c_\eta s_{\theta_W} - \frac{s_\eta}{t_{\theta_W^\prime}}] \notag\\
&\hspace{1cm}+ \frac{s_\zeta c_\eta c_{\theta_W}}{s_{\theta_W^\prime}}[c_\beta c_+ + s_\beta s_{13}] \Bigg) + M\frac{s_\zeta c_-t_{\theta_W}}{s_{\theta_W^\prime}}(c_\eta s_{\theta_W} + s_\eta t_{\theta_W^\prime}) \Bigg), \notag\\[2ex]
\parbox{20mm}{\begin{fmfgraph*}(40,30)
  \fmfleft{Hm,Hp}
  \fmfright{V0}
  \fmfv{label=$H^+$,label.angle=180}{Hp}
  \fmfv{label=$H^-$,label.angle=180}{Hm}
  \fmfv{label=$V^0_\mu$,label.angle=0}{V0}
  \fmf{boson}{V0,v}
  \fmf{dashes,label=$p\searrow$,label.side=right,label.dist=.1}{Hp,v}
  \fmf{dashes,label=$\nearrow q$,label.side=right,label.dist=.1}{Hm,v}
\end{fmfgraph*}}
=& ig_{V^0}(p-q)_\mu,\hspace{2cm} g_\gamma = e,\quad g_Z = \frac{g_L\cos2\theta_W}{2\cos\theta_W} \notag\\[5ex]
\parbox{20mm}{\begin{fmfgraph*}(40,30)
  \fmfleft{Hm,Hp}
  \fmfright{h0}
  \fmfv{label=$H^+$,label.angle=180}{Hp}
  \fmfv{label=$H^-$,label.angle=180}{Hm}
  \fmfv{label=$h^0$,label.angle=0}{h0}
  \fmf{dashes}{h0,v}
  \fmf{dashes}{Hp,v,Hm}
\end{fmfgraph*}}
=& -i\lambda_{h^0H^+H^-}, \notag\\
\lambda_{h^0H^+H^-} =& \lambda_1v(c_\beta c_{\alpha_1} + s_\beta s_{\alpha_1}c_{\alpha_2}) (s_{13}^2 + c_+^2) \notag\\
&+ \lambda_2v\big( (c_\beta c_{\alpha_1} + s_\beta s_{\alpha_1}c_{\alpha_2}) (s_{13}^2 + c_+^2) - (s_\beta c_{\alpha_1} + c_\beta s_{\alpha_1}c_{\alpha_2}) s_{13}c_+ \big) \notag\\
&+ \lambda_3Ms_{\alpha_1}s_{\alpha_2}c_-^2 + \lambda_4(v(c_\beta c_{\alpha_1} + s_\beta s_{\alpha_1}c_{\alpha_2})c_-^2 + Ms_{\alpha_1}s_{\alpha_2}(s_{13}^2 + c_+^2)) \notag\\
&+ \lambda_5\big( v(c_\beta c_{\alpha_1}c_-^2 - s_\beta s_{\alpha_1}s_{\alpha_2}s_{13}c_- + c_\beta s_{\alpha_1}s_{\alpha_2}c_+c_-) \notag\\
&\hspace{1cm}+ M(s_{\alpha_1}s_{\alpha_2}c_+^2 - s_{\alpha_1}c_{\alpha_2}s_{13}c_- + c_{\alpha_1}c_+c_-) \big).
\end{align}

\vskip1cm
\subsection{$H^\pm$ Couplings to a $W$ and a neutral scalar}\label{a:HpmWmpSfeynRules}
\begin{align}
\parbox{20mm}{\begin{fmfgraph*}(40,30)
  \fmfleft{Hpm}
  \fmfright{S0,Wpm}
  \fmfv{label=$H^\pm$,label.angle=180}{Hpm}
  \fmfv{label=$W^\pm_\mu$,label.angle=0}{Wpm}
  \fmfv{label=$\Phi^0$,label.angle=0}{S0}
  \fmf{dashes,label=$\overrightarrow{p}$,label.side=right}{Hpm,v}
  \fmf{boson}{Wpm,v}
  \fmf{dashes,label=$\nwarrow q$,label.side=right,label.dist=1}{S0,v}
\end{fmfgraph*}}
=& \pm\frac{i}{2}(p-q)_\mu g_{\Phi^0}, \notag\\
g_{h^0} =& -g_Lc_\zeta( c_{\alpha_1}s_{13} + s_{\alpha_1}c_{\alpha_2}c_+) + g_Rs_\zeta( c_{\alpha_1}c_+ + s_{\alpha_1}c_{\alpha_2}s_{13} ), \notag\\
g_{H^0_1} =& g_Lc_\zeta\big( s_{\alpha_1}c_{\alpha_3}s_{13} - (c_{\alpha_1}c_{\alpha_2}c_{\alpha_3} - s_{\alpha_2}s_{\alpha_3})c_+ \big) \notag\\
&+ g_Rs_\zeta\big( (c_{\alpha_1}c_{\alpha_2}c_{\alpha_3}-s_{\alpha_2}s_{\alpha_3})s_{13} - s_{\alpha_1}c_{\alpha_3}c_+ \big), \notag\\
g_{H^0_2} =& g_Lc_\zeta\big( -s_{\alpha_1}s_{\alpha_3}s_{13} + (c_{\alpha_1}c_{\alpha_2}s_{\alpha_3}+s_{\alpha_2}c_{\alpha_3})c_+ \big) \notag\\
&+ g_Rs_\zeta\big( -(c_{\alpha_1}c_{\alpha_2}s_{\alpha_3}+s_{\alpha_2}c_{\alpha_3})s_{13} + s_{\alpha_1}s_{\alpha_3}c_+ \big), \notag\\
g_{A^0} =& \pm i\big( g_Lc_\zeta( s_\beta s_{13} - c_\beta c_+) + g_Rs_\zeta( -c_\beta s_{13} + s_\beta c_+) \big). \label{eq:HpWmS0}
\end{align}

\subsection{Couplings of neutral scalars to pairs of vector bosons}\label{a:SVVfeynrules}
\begin{align}
\parbox{20mm}{\begin{fmfgraph*}(40,30)
  \fmfleft{S0}
  \fmfright{Wm,Wp}
  \fmfv{label=$S^0$,label.angle=180}{S0}
  \fmfv{label=$V^\pm_{1\mu}$,label.angle=0}{Wp}
  \fmfv{label=$V^\mp_{2\nu}$,label.angle=0}{Wm}
  \fmf{dashes}{S0,v}
  \fmf{boson}{Wp,v,Wm}
\end{fmfgraph*}}
=& \frac{ig_{\mu\nu}}{2} \Big( g_L^2vC^1_{S^0}C^1_{V_1^\pm}C^1_{V_2^\mp} + g_R^2(vC^1_{S^0} + MC^2_{S^0})C^2_{V_1^\pm}C^2_{V_2^\mp}  \notag\\
&- g_Lg_RvC^3_{S^0}(C^1_{V_1^\pm}C^2_{V_2^\mp} + C^2_{V_1^\pm}C^1_{V_2^\mp}) \Big), \notag\\
\parbox{20mm}{\begin{fmfgraph*}(40,30)
  \fmfleft{S0}
  \fmfright{Z2,Z1}
  \fmfv{label=$S^0$,label.angle=180}{S0}
  \fmfv{label=$V^0_{1\mu}$,label.angle=0}{Z1}
  \fmfv{label=$V^0_{2\nu}$,label.angle=0}{Z2}
  \fmf{dashes}{S0,v}
  \fmf{boson}{Z1,v,Z2}
\end{fmfgraph*}}
=& \frac{ig_L^2g_{\mu\nu}}{2\cos^2\theta_W} \Big( vC^1_{S^0}C^1_{V_1^0}C^1_{V_2^0} + \frac{4Ms_{\theta_W}^2}{s^2_{2\theta_W^\prime}}C^2_{S^0}C^2_{V_1^0}C^2_{V_2^0} \Big), \notag\\[5ex]
\parbox{20mm}{\begin{fmfgraph*}(40,30)
  \fmfleft{A0}
  \fmfright{W,Wp}
  \fmfv{label=$A^0$,label.angle=180}{A0}
  \fmfv{label=$W^\pm$,label.angle=0}{W}
  \fmfv{label=$W^{\prime\mp}$,label.angle=0}{Wp}
  \fmf{dashes}{A0,v}
  \fmf{boson}{W,v,Wp}
\end{fmfgraph*}}
=& \mp\frac{g_Lg_Rvc_{2\beta}}{2}g_{\mu\nu}, \notag
\end{align}
\begin{align}
C_{V^\pm}^1 =& \{ c_\zeta , -s_\zeta \}, \notag\\
C_{V^\pm}^2 =& \{ s_\zeta , c_\zeta \}, \notag\\
C_{V^0}^1 =& \{ 0 , -(c_\eta - s_\eta s_{\theta_W}\cot\theta_W^\prime), s_\eta + c_\eta s_{\theta_W}\cot\theta_W^\prime \}, \notag\\
C_{V^0}^2 =& \{ 0 , s_\eta , c_\eta \}, \notag\\
C_{S^0}^1 =& \big\{ c_\beta c_{\alpha_1} + s_\beta s_{\alpha_1}c_{\alpha_2}, -\big( c_\beta s_{\alpha_1}c_{\alpha_3} - s_\beta( c_{\alpha_1}c_{\alpha_2}c_{\alpha_3} - s_{\alpha_2}s_{\alpha_3}) \big), \notag\\
&\hspace{1cm} c_\beta s_{\alpha_1}s_{\alpha_3} - s_\beta(c_{\alpha_1}c_{\alpha_2}s_{\alpha_3} + s_{\alpha_2}c_{\alpha_3}) \big\}, \notag\\
C_{S^0}^2 =& \big\{ s_{\alpha_1}s_{\alpha_2}, c_{\alpha_1}s_{\alpha_2}c_{\alpha_3} + c_{\alpha_2}s_{\alpha_3}, -(c_{\alpha_1}s_{\alpha_2}s_{\alpha_3} - c_{\alpha_2}c_{\alpha_3}) \big\}, \notag\\
C_{S^0}^3 =& \big\{ c_\beta s_{\alpha_1}c_{\alpha_2} + s_\beta c_{\alpha_1}, c_\beta(c_{\alpha_1}c_{\alpha_2}c_{\alpha_3} - s_{\alpha_1}s_{\alpha_3}) - s_\beta s_{\alpha_1}c_{\alpha_3}, \notag\\
&\hspace{1cm} -\big( c_\beta(c_{\alpha_1}c_{\alpha_2}s_{\alpha_3} + s_{\alpha_2}c_{\alpha_3}) - s_\beta s_{\alpha_1}s_{\alpha_3}\big) \big\}.
\end{align}


\subsection{Scalar couplings to fermions}\label{a:ScalarToFermions}
For the sake of brevity, we write $u^\alpha = \{u,c,t\}$, $m_\alpha = \{m_u, m_c, m_t\}$.
We only give couplings of neutral scalars to the up-type quarks and the bottom quark, and couplings of charged scalars to the heaviest quark generation.
The other couplings involve fermions of which the LET model does not reproduce the masses correctly.
However, they can be ignored since the most important contributions in our analysis come from the heaviest quarks.
\begin{align}
\parbox{20mm}{\begin{fmfgraph*}(40,30)
  \fmfleft{S0}
  \fmfright{u2,u1}
  \fmfv{label=$\Phi^0$,label.angle=180}{S0}
  \fmfv{label=$u^\alpha$,label.angle=0}{u1}
  \fmfv{label=$u^\alpha$,label.angle=0}{u2}
  \fmf{dashes}{S0,v}
  \fmf{fermion}{u2,v,u1}
\end{fmfgraph*}}
=& -\frac{im_\alpha}{v} \left\{ \frac{c_{\alpha_1}}{c_\beta} , -\frac{s_{\alpha_1}c_{\alpha_3}}{c_\beta} , \frac{s_{\alpha_1}s_{\alpha_3}}{c_\beta}, i\tan\beta \right\}, \notag\\[5ex]
\parbox{20mm}{\begin{fmfgraph*}(40,30)
  \fmfleft{S0}
  \fmfright{d2,d1}
  \fmfv{label=$\Phi^0$,label.angle=180}{S0}
  \fmfv{label=$b$,label.angle=0}{d1}
  \fmfv{label=$b$,label.angle=0}{d2}
  \fmf{dashes}{S0,v}
  \fmf{fermion}{d2,v,d1}
\end{fmfgraph*}}
=& -\frac{im_b}{v} \bigg\{ \frac{s_{\alpha_1}c_{\alpha_2}}{s_\beta} , \frac{c_{\alpha_1}c_{\alpha_2}c_{\alpha_3}-s_{\alpha_2}s_{\alpha_3}}{s_\beta}, \notag\\
&\hspace{3cm} \frac{c_{\alpha_1}c_{\alpha_2}s_{\alpha_3}+s_{\alpha_2}c_{\alpha_3}}{s_\beta} , i\cot\beta \bigg\}, \notag\\
\parbox{20mm}{\begin{fmfgraph*}(40,30)
  \fmfleft{Hp}
  \fmfright{d,u}
  \fmfv{label=$H^+$,label.angle=180}{Hp}
  \fmfv{label=$t$,label.angle=0}{u}
  \fmfv{label=$b$,label.angle=0}{d}
  \fmf{dashes}{Hp,v}
  \fmf{fermion}{d,v,u}
\end{fmfgraph*}}
=& -\frac{i\sqrt2}{v} \left( m_b\frac{s_{13}}{s_\beta} \frac{1-\gamma^5}{2} - m_t\frac{c_+}{c_\beta} \frac{1+\gamma^5}{2} \right), \notag\\[5ex]
\parbox{20mm}{\begin{fmfgraph*}(40,30)
  \fmfleft{Hm}
  \fmfright{d,u}
  \fmfv{label=$H^-$,label.angle=180}{Hm}
  \fmfv{label=$t$,label.angle=0}{u}
  \fmfv{label=$b$,label.angle=0}{d}
  \fmf{dashes}{Hm,v}
  \fmf{fermion}{u,v,d}
\end{fmfgraph*}}
=& -\frac{i\sqrt2}{v} \left( -m_t\frac{c_+}{c_\beta} \frac{1-\gamma^5}{2} + m_b\frac{s_{13}}{s_\beta} \frac{1+\gamma^5}{2} \right).
\end{align}

\clearpage
\section{Charged-scalar couplings of the 2HDM}\label{a:2HDMfeynmanrules}
The $H^\pm$ branching ratios of the simplified LET model have been calculated using the 2HDM-modus of HDecay (see \cref{s:HpmLimits}).
To make sure that the results apply to the simplified LET model, we need to compare the $H^\pm$ couplings of the simplified LET model to the corresponding 2HDM-couplings.
In this appendix, we list the relevant Feynman rules, assuming a type-II Yukawa sector.
We write $V^0 = \{\gamma,Z\}$ and $\Phi^0 = (h^0, H^0, A^0)$ for the sake of brevity.

\subsection{Couplings to bosons}\label{a:2HDMHpmcouplingsToBosons}
\begin{align}
\parbox{20mm}{\begin{fmfgraph*}(40,30)
  \fmfleft{Hm,Hp}
  \fmfright{V0}
  \fmfv{label=$H^+$,label.angle=180}{Hp}
  \fmfv{label=$H^-$,label.angle=180}{Hm}
  \fmfv{label=$V^0_\mu$,label.angle=0}{V0}
  \fmf{dashes,label=$\searrow p$,label.side=left,label.dist=1}{Hp,v}
  \fmf{dashes,label=$\nearrow q$,label.side=right,label.dist=1}{Hm,v}
  \fmf{boson}{V0,v}
\end{fmfgraph*}}
=& i(p-q)_\mu \left\{ e , \frac{g\cos2\theta_W}{2\cos\theta_W} \right\}, \notag\\[5ex]
\parbox{20mm}{\begin{fmfgraph*}(40,30)
  \fmfleft{Hm,Hp}
  \fmfright{h0}
  \fmfv{label=$H^+$,label.angle=180}{Hp}
  \fmfv{label=$H^-$,label.angle=180}{Hm}
  \fmfv{label=$h^0$,label.angle=0}{h0}
  \fmf{dashes}{Hp,v,Hm}
  \fmf{dashes}{v,h0}
\end{fmfgraph*}}
=& -iv\big( -\Lambda_3\cos^3\beta\sin\alpha + (\Lambda_2-\Lambda_4-\Lambda_5)\cos\alpha\sin\beta\cos^2\beta \notag\\
&\phantom{-iv\big(} - (\Lambda_1-\Lambda_4-\Lambda_5)\sin\alpha\sin^2\beta\cos\beta + \Lambda_3\cos\alpha\sin^3\beta \big), \notag\\[3ex]
\parbox{20mm}{\begin{fmfgraph*}(40,30)
  \fmfleft{Hpm}
  \fmfright{S0,Wpm}
  \fmfv{label=$H^\pm$,label.angle=180}{Hpm}
  \fmfv{label=$\Phi^0$,label.angle=0}{S0}
  \fmfv{label=$W^\pm_\mu$,label.angle=0}{Wpm}
  \fmf{dashes,label=$\underrightarrow{p}$,label.side=left}{Hpm,v}
  \fmf{dashes,label=$q\nwarrow$,label.side=left,label.dist=1}{S0,v}
  \fmf{boson}{Wpm,v}
\end{fmfgraph*}}
=& \frac{ig}{2}(p-q)_\mu \left\{ \pm\cos(\alpha-\beta) , \pm\sin(\alpha-\beta) , -i \right\}.
\end{align}


\vskip1cm
\subsection{Couplings to fermions}\label{a:2HDMfermioncouplings}
\begin{align}
\parbox{20mm}{\begin{fmfgraph*}(40,30)
  \fmfleft{Hp}
  \fmfright{b,t}
  \fmfv{label=$H^+$,label.angle=180}{Hp}
  \fmfv{label=$t^\alpha$,label.angle=0}{t}
  \fmfv{label=$b^\alpha$,label.angle=0}{b}
  \fmf{dashes}{Hp,v}
  \fmf{fermion}{b,v,t}
\end{fmfgraph*}}
=& \frac{i\sqrt2}{v}\left( m_t^\alpha\cot\beta\tfrac{1-\gamma^5}{2} + m_b^\alpha\tan\beta\tfrac{1+\gamma^5}{2} \right), \notag\\[5ex]
\parbox{20mm}{\begin{fmfgraph*}(40,30)
  \fmfleft{Hm}
  \fmfright{t,b}
  \fmfv{label=$H^-$,label.angle=180}{Hm}
  \fmfv{label=$t^\alpha$,label.angle=0}{t}
  \fmfv{label=$b^\alpha$,label.angle=0}{b}
  \fmf{dashes}{Hm,v}
  \fmf{fermion}{t,v,b}
\end{fmfgraph*}}
=& \frac{i\sqrt2}{v}\left( m_b^\alpha\tan\beta\tfrac{1-\gamma^5}{2} + m_t^\alpha\cot\beta\tfrac{1+\gamma^5}{2} \right), \notag\\[5ex]
\parbox{20mm}{\begin{fmfgraph*}(40,30)
  \fmfleft{Hp}
  \fmfright{e,nu}
  \fmfv{label=$H^+$,label.angle=180}{Hp}
  \fmfv{label=$\nu^\alpha$,label.angle=0}{nu}
  \fmfv{label=$e^\alpha$,label.angle=0}{e}
  \fmf{dashes}{Hp,v}
  \fmf{fermion}{e,v,nu}
\end{fmfgraph*}}
=& \frac{i\sqrt2}{v}m_e^\alpha\tan\beta\tfrac{1+\gamma^5}{2}, \notag\\[5ex]
\parbox{20mm}{\begin{fmfgraph*}(40,30)
  \fmfleft{Hm}
  \fmfright{nu,e}
  \fmfv{label=$H^-$,label.angle=180}{Hm}
  \fmfv{label=$\nu^\alpha$,label.angle=0}{nu}
  \fmfv{label=$e^\alpha$,label.angle=0}{e}
  \fmf{dashes}{Hm,v}
  \fmf{fermion}{nu,v,e}
\end{fmfgraph*}}
=& \frac{i\sqrt2}{v}m_e^\alpha\tan\beta\tfrac{1-\gamma^5}{2}.
\end{align}

\printbibliography[heading=bibintoc]

\end{fmffile}
\end{document}